\documentclass[a4paper,12pt]{book}

\usepackage{ifthen}                      
\usepackage{natbib}                      
\usepackage{txfonts}                     
\usepackage{fancyhdr}                    
\usepackage{util}                        

\usepackage{graphicx}                    
\usepackage[draft=true]{hyperref}        

\renewcommand{\chaptermark}[1]{       
  \markboth{\chaptername\ \thechapter.\ #1}{}}

\makeatletter
\def\cleardoublepage{\clearpage\if@twoside \ifodd\c@page\else
    \hbox{}
    \thispagestyle{empty}
    \newpage
    \if@twocolumn\hbox{}\newpage\fi\fi\fi}
\makeatother

\newcommand{\teff}   {T$_{\rm eff}$}
\newcommand{\logg}   {log\,$g$}
\newcommand{\vsini}  {{\it v}\,sin\,{\it i}}
\newcommand{\vrad}   {{\it v\,rad}}
\newcommand{\kms}    {km\,s$^{-1}$}
\newcommand{\ms}     {m\,s$^{-2}$}
\newcommand{\deltav} {$\Delta v$}

\begin{document}

\bibliographystyle{thesis}

\defcitealias{mora2002}{\mbox{Paper I}}
\defcitealias{mora2004}{\mbox{Paper II}}

\begin{titlepage}
\pdfbookmark{Cover page}{cover}

\begin{center}

{\large 
UNIVERSIDAD AUT\'ONOMA DE MADRID

\vspace{0.1cm}

Facultad de Ciencias

\vspace{0.2cm}

Departamento de F\'{\i}sica Te\'orica
}
\vspace{1cm}

\centerline{\includegraphics[clip=true,width=2.7cm]{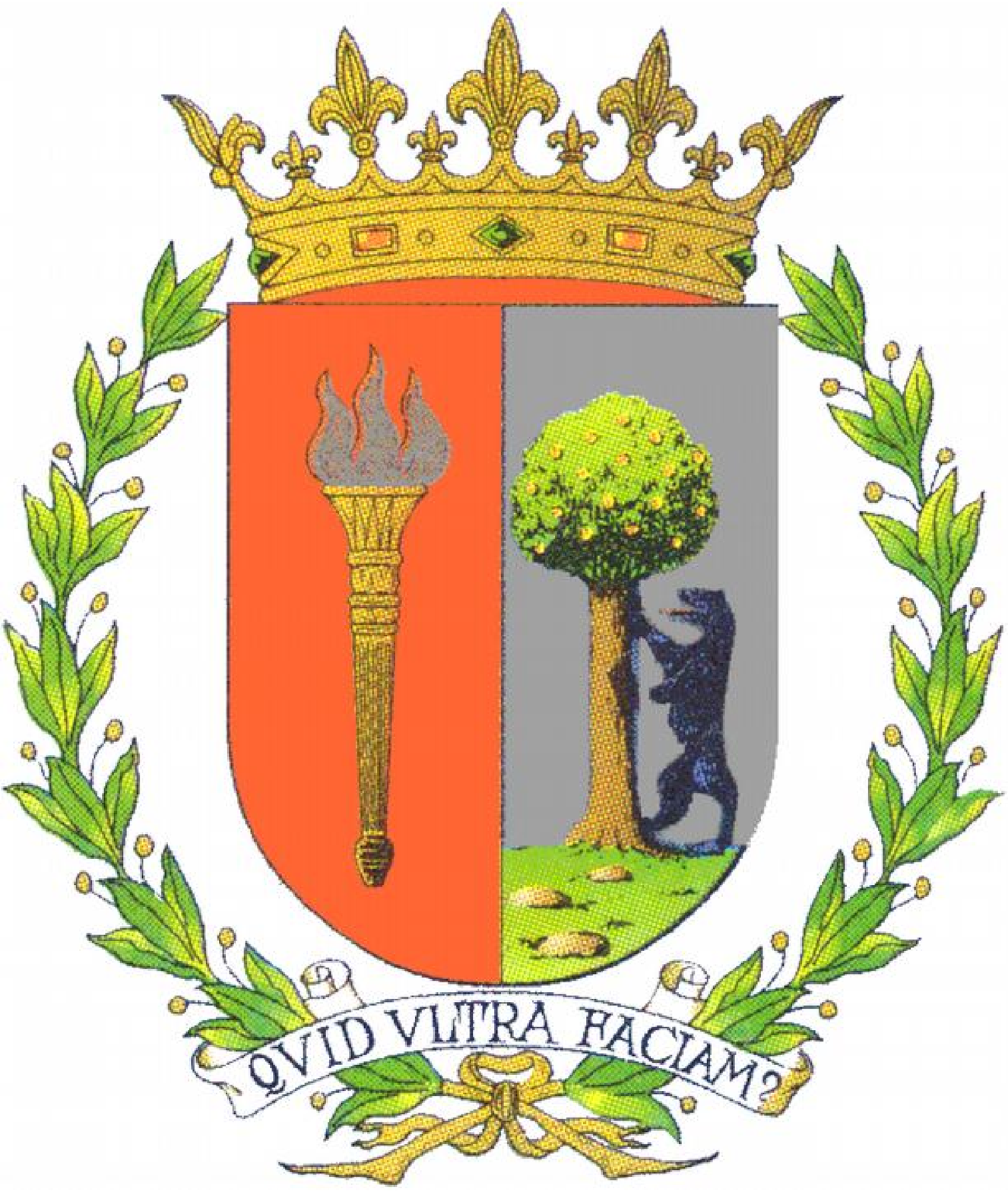}}
\vspace{3cm}

{\Huge \bf Kinematics of the circumstellar gas around UXOR stars}
\vspace{3cm}

{\large
PhD dissertation submitted by
}
\vspace{0.2cm}

{\Large \bf Alcione Mora Fern\'andez}
\vspace{0.2cm}

{\large
for the degree of Doctor in Physics
}
\vspace{1.5cm}

{\large
Supervised by
\vspace{0.2cm}

{\Large \bf Dr. Carlos Eiroa de San Francisco}
\vspace{0.2cm}

University lecturer, Universidad Aut\'onoma de Madrid
\vspace{3.4cm}

Madrid, April 6$^{\rm th}$ 2004
}
\end{center}

\end{titlepage}

\cleardoublepage

\begin{titlepage}
\pdfbookmark{Cover page (Spanish)}{portada}

\begin{center}

{\large 
UNIVERSIDAD AUT\'ONOMA DE MADRID

\vspace{0.1cm}

Facultad de Ciencias

\vspace{0.2cm}

Departamento de F\'{\i}sica Te\'orica
}
\vspace{1cm}

\centerline{\includegraphics[clip=true,width=2.7cm]{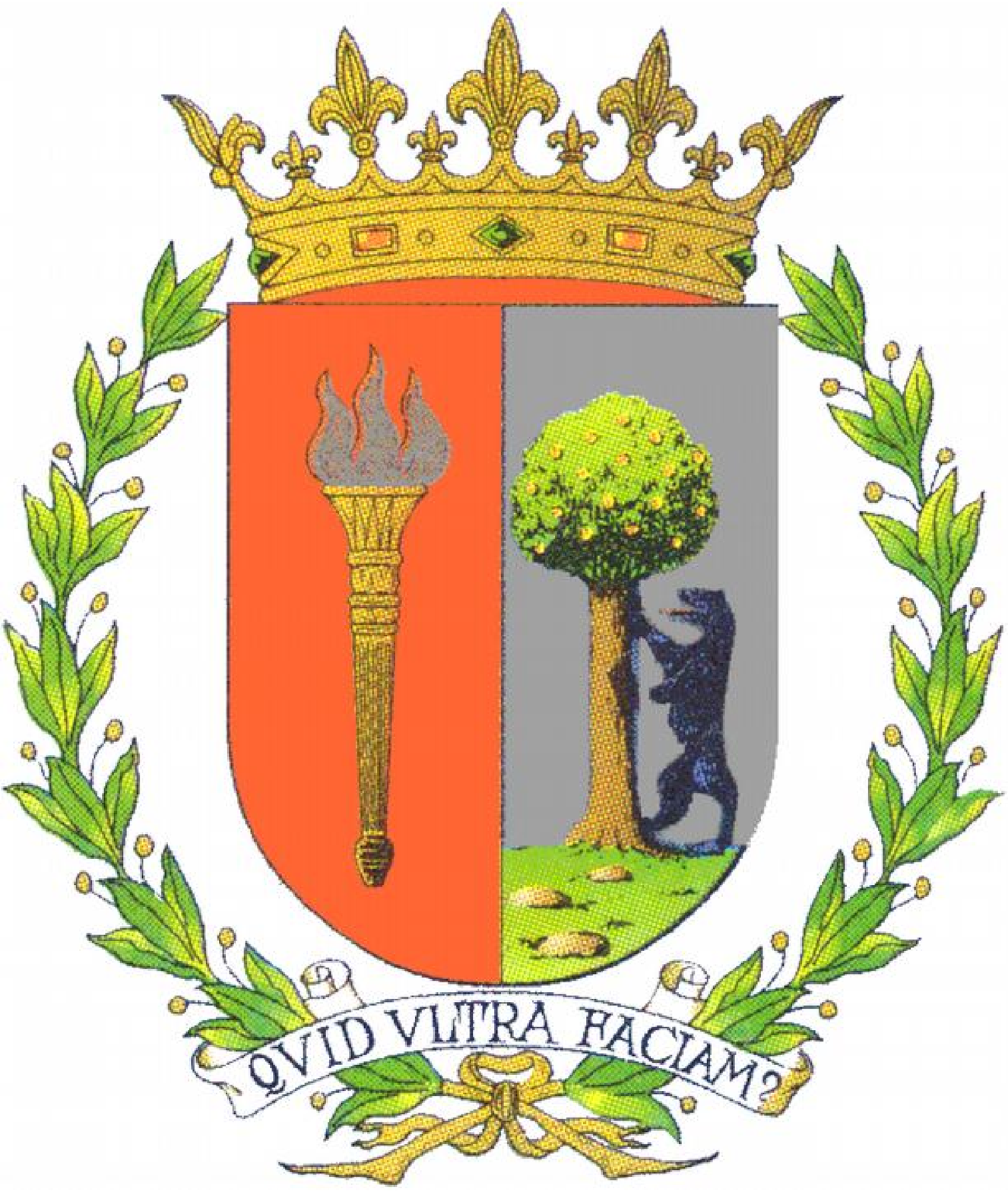}}
\vspace{3cm}

{\Huge \bf Cinem\'atica del gas circunestelar en estrellas UXOR}
\vspace{3cm}

{\large
Memoria de tesis doctoral presentada por
}
\vspace{0.2cm}

{\Large \bf Alcione Mora Fern\'andez}
\vspace{0.2cm}

{\large
para optar al grado de Doctor en Ciencias F\'{\i}sicas
}
\vspace{1.5cm}

{\large
Trabajo dirigido por el
\vspace{0.2cm}

{\Large \bf Dr. Carlos Eiroa de San Francisco}
\vspace{0.2cm}

Profesor Titular de la Universidad Aut\'onoma de Madrid
\vspace{3.4cm}

Madrid, a 6 de abril de 2004
}
\end{center}

\end{titlepage}

\cleardoublepage

\begin{titlepage}

\parbox{\hsize}{
\begin{flushright}

{\it
\vspace{8cm}
A Nuria y a mi familia, \\
por su amor, inspiraci\'on y apoyo
}
\end{flushright}
}

\end{titlepage}

\cleardoublepage

\begin{titlepage}
\pdfbookmark{Quote}{quote}

\parbox[t]{7cm}
{\it
\vspace{6cm}
The Road goes ever on and on       \\
Down from the door where it began. \\
Now far ahead the Road has gone,   \\
And I must follow, if I can,       \\
Pursuing it with eager feet,       \\
Until it joins some larger way     \\
Where many paths and errands meet. \\
And whither then? I cannot say.    \\
                                   \\
\rm J.R.R. Tolkien}
\hspace{1cm}
\parbox[t]{7cm}
{\it
\vspace{6cm}
El Camino sigue y sigue                                  \\
desde la puerta.                                         \\
El Camino ha ido muy lejos,                              \\
y si es posible he de seguirlo                           \\
recorri\'endolo con pie decidido                         \\
hasta llegar a un camino m\'as ancho                     \\
donde se encuentran senderos y cursos.                   \\
?`Y de ah\'{\i} ad\'onde ir\'e? No podr\'{\i}a decirlo.  \\
                                                         \\
\rm J.R.R. Tolkien}

\end{titlepage}

\cleardoublepage

\pagenumbering{roman} \setcounter{page}{1}

\chapter*{Agradecimientos\markboth{Agradecimientos}{Agradecimientos}}
\pdfbookmark{Agradecimientos}{agradecimientos}

He recibido el cari\~no, apoyo, inspiraci\'on y amistad de muchas personas durante la ela\-bo\-ra\-ci\'on de esta tesis doctoral.
Quiero desde aqu\'{\i} darle las gracias a todas ellas.

En primer lugar quiero decirle a Nuria que la quiero, que no habr\'{\i}a sido capaz de acabar sin su ayuda y que me hace muy feliz el saber que puedo contar con ella, tanto para los buenos como para los malos ratos.

Mi familia me ha apoyado en todo momento.
En realidad, si he llegado a iniciar una carrera investigadora ha sido gracias a la curiosidad que mis padres me han despertado desde la infancia.
A mi padre le quiero agradecer esas charlas sobre f\'{\i}sica, qu\'{\i}mica e ingenier\'{\i}a con las que me aleccionaba desde peque\~no.
A mi madre ese amor por la lectura que a\'un perdura.
A mi hermano le quiero agradecer esa confianza que siempre ha depositado en m\'{\i}.
A Paqui ese inter\'es y cari\~no que siempre ha tenido por m\'{\i}.

Tengo muchas cosas que agradecerle a Carlos Eiroa, mi director de tesis, tanto en el \'ambito cient\'{\i}fico como en el personal.
Por una parte el haberme seleccionado como doctorando, haber exigido siempre lo m\'aximo de mi capacidad y proporcionarme todos los medios materiales para realizar una tesis de calidad.
Por otra parte su manera de dirigir, gracias a la cual he gozado de libertad para enfocar mi trabajo de la manera m\'as adecuada, su trato amable y cercano y, finalmente, su paciencia y comprensi\'on que me han permitido simultanear esta tesis con los estudios de ingenier\'{\i}a de materiales, lo cual ha redundado en un retraso considerable en la fecha de finalizaci\'on

El tema de estudio finalmente escogido para esta tesis se debe en buena medida a la ayuda proporcionada por Antonella Natta, quien con su experiencia sugiri\'o los m\'etodos de an\'alisis m\'as adecuados para los datos obtenidos y particip\'o intensamente en la interpretaci\'on f\'{\i}sica de los resultados.
Tambi\'en le quiero agradecer a Antonella el ofre\-ci\-mien\-to y las gestiones que me permitieron tanto realizar dos estancias breves en el Observatorio Astrof\'{\i}sico de Arcetri (Florencia, Italia), como recibir apoyo econ\'omico durante la segunda estancia.
Por este mismo motivo quiero agradecer tambi\'en la hospitalidad brindada por la instituci\'on del Observatorio.

A Benjam\'{\i}n Montesinos le quiero agradecer la amistad y el apoyo cient\'{\i}fico que me ha otorgado desde el mismo comienzo de esta tesis.
Especialmente le quiero agradecer el haberme ayudado con la traducci\'on al ingl\'es de toda la tesis.

Quiero expresar mi agradecimiento a todos aquellos investigadores con los que he colaborado.
En especial a Bruno Mer\'{\i}n con quien he podido compartir amistad, in\-quie\-tu\-des y confidencias,
a Javier Palacios quien me dio una c\'alida acogida inicial y me introdujo en la administraci\'on de m\'aquinas UNIX,
a Enrique Solano con quien di mis primeros pasos en la reducci\'on de espectros \'echelle, s\'{\i}ntesis espectral y medida de velocidades de rotaci\'on,
a Dolf de Winter bajo cuya direcci\'on efectu\'e mis observaciones,
a Rene D. Oudmaijer que ha efectuado numerosas sugerencias en los art\'{\i}culos y, por \'ultimo,
a John K. Davies y Alan W. Harris quienes me han ayudado con el ingl\'es de los art\'{\i}culos.

De entre el resto de astr\'onomos que tambi\'en me han ayudado quiero destacar a Vla\-di\-mir P. Grinin por las fruct\'{\i}feras discusiones cient\'{\i}ficas mantenidas en Arcetri,
a John R. Barnes por la ayuda prestada en la reducci\'on de espectros \'echelle en la Universidad de St. Andrews (Reino Unido) y
a Friedrich G. Kupka y Tanya A. Ryabchikova por proporcionarme informaci\'on acerca de la selecci\'on de fuerzas de oscilador para el multiplete 42 de \ion{Fe}{ii}.

Durante este tiempo he podido compartir mis experiencias con una gran cantidad de amigos y compa\~neros de despacho y trabajo. Entre ellos mencionar\'e a David, Jaime, Juanjo, Marta, Natxo y Yago (despacho 512), Alberto, \'Alex, Alfredo, \'Alvaro, Andr\'es, Arantxa, Carlos Hoyos, Chiqui, Elena, Guillermo, Itziar, Javier, Marcelo, Marcos Jim\'e-nez, Mar\'{\i}a Jes\'us, Michael, V\'{\i}ctor (despacho 301), Enrique, H\'ector, Jose, Luis, Marcos L\'opez-Caniego, Mari\'angeles, Mercedes, Ra\'ul, Rub\'en G. Benito, Yago (grupo de astrof\'{\i}sica), \'Alex, Alfonso, \'Angel, Gast\'on, Guillermo, Enrique, Jaime, Nestor, Rub\'en Moreno, St\'ephane (expedici\'on del pabell\'on B), Daniela y Laura (Arcetri).

Nunca se tienen demasiados amigos.
Por eso quiero agradecer
a Juanjo por esa complicidad con la que siempre nos contamos nuestros proyectos e inquietudes,
a Diego G. Batanero por esa inagotable energ\'{\i}a al servicio de sus amigos,
a Ismael por su sinceridad y consecuencia con sus ideales,
a Salvador por esa manera de darlo todo por quien lo merece, a Lucas por haber madurado juntos,
a \'Alvaro por su amistad incondicional y sin fisuras,
a Diego Garc\'{\i}a por su alegre manera de vivir la vida,
a Juan Carlos por esa inagotable capacidad para sorprenderse y desarrollar nuevos proyectos,
a Alicia por sus denodados esfuerzos por conservar una amistad en la distancia y a \'Oscar por haber compartido intensamente la carrera y la tesis.
Tambi\'en quiero agradecer su amistad y todos los ratos que hemos pasado juntos a Ana (Madrid), Patricia (Instituto), Fernando (Bilbao), Aram, Borja, Joaqu\'{\i}n, Margarita, Miriam, Neni, Paco, Peco, Pili, Sergio (Huelva), Eduar\-do, Guillermo, Iv\'an, L\'azaro, Luis G. Prado, Marcos, Nico, Nuria, Rub\'en, Silvia (Colegio Mayor), Celia, David, Florencio, Luis Fern\'andez, Olga, Maite, Marimar (compa\~neros de F\'{\i}sica) Andrea, Circe, Jose, Juan Pedro, Teresa y Vicente (Tenerife).

Quiero agradecer al Departamento de F\'{\i}sica Te\'orica de la Universidad Aut\'onoma de Madrid el apoyo prestado durante la elaboraci\'on de esta tesis, as\'{\i} como el haberme podido integrar durante un a\~no a sus actividades docentes en calidad de profesor ayudante.

Durante esta tesis, el doctorando ha disfrutado de la beca AP98-29045605 de Formaci\'on de Profesorado Universitario (Ministerio de Educaci\'on Cultura y Deporte), as\'{\i} como de financiaci\'on parcial por parte de los proyectos ESP98-1339 del Plan Nacional del Espacio (Comisi\'on Interministerial de Ciencia y Tecnolog\'{\i}a) y AYA2001-1124 del Plan Nacional de Astronom\'{\i}a y Astrof\'{\i}sica (Direcci\'on General de Investigaci\'on, MCyT).

\chapter*{Abstract\markboth{Abstract}{Abstract}}
\pdfbookmark{Abstract}{abstract}

This thesis presents the results of a high spectral resolution ($\lambda / \Delta \lambda $ = 49000) study of the circumstellar (CS) gas around the intermediate mass, pre-main sequence UXOR stars BF~Ori, SV~Cep, UX~Ori, WW~Vul and XY~Per.
The results are based on a set of 38 \'echelle spectra covering the spectral range 3800-5900~\AA, monitoring the stars on time scales of months, days and  hours.

All spectra show a large number of Balmer and metallic lines with variable blue\-shif\-ted and redshifted absorption features superimposed to the photospheric stellar spectra.
Synthetic Kurucz models are used to estimate rotational velocities, effective temperatures and gravities of the stars.
The best photospheric models are subtracted from each observed spectrum to determine the variable absorption features due to the circumstellar gas; those features are characterized, via multigaussian fitting, in terms of their velocity, $v$, dispersion velocity, $\Delta v$, and residual absorption, $R_{\rm max}$.

The absorption components detected in each spectrum can be grouped by their similar radial velocities and are interpreted as the signature of the dynamical evolution of gaseous clumps.
Most of the events undergo accelerations/decelerations at a rate of tenths of \ms.
The typical timescale for the duration of the events is a few days.
The dispersion velocity and the relative absorption strength of the features do not show drastic changes during the lifetime of the events, which suggests that they are gaseous blobs preserving their geometrical and physical identity.

A comparison of the intensity ratios among the transient absorptions suggests a solar-like composition for most of the CS gas.
This confirms previous results and excludes a very metal-rich environment as the general cause of the transient features in UXOR stars.
These data are a very useful tool for constraining and validating theoretical models of the chemical and physical conditions of the CS gas around young stars; in particular, it is suggested that the simultaneous presence of infalling and outflowing gas should be investigated in the context of detailed magnetospheric accretion models, similar to those proposed for the lower mass T Tauri stars.  

WW Vul is unusual because, in addition to infalling and outflowing gas with properties similar to those observed in the other stars, it shows also transient absorption features in metallic lines with no obvious counterparts in the hydrogen lines.
This could, in principle, suggest the presence of CS gas clouds with enhanced metallicity around WW~Vul.
The existence of such a metal-rich gas component, however, needs to be confirmed by further observations and a more quantitative analysis.

All these results have been published by \citet[Chapter \ref{uxori}]{mora2002} and \citet[Chapter \ref{haebe}]{mora2004}.

Rotational velocities for a large sample of pre-main sequence and Vega-type stars have been determined from high resolution ($\lambda / \Delta \lambda $ = 49000) \'echelle spectra.
The first minimum of the Fourier transform of many photospheric line profiles has been used to estimate the velocities.
The resulting velocities have been published, along with spectral types determined by other co-authors, by \citet[see Chapter \ref{vsini}]{mora2001}.

\vspace{0.5cm}
\noindent {\bf Key words.} Stars: formation -- Stars: pre-main sequence -- Stars: circumstellar matter -- Accretion: accretion disks -- Lines: profiles -- Stars: individual: BF~Ori, SV~Cep, UX~Ori, WW~Vul, XY~Per -- Stars: rotation

\chapter*{Resumen\markboth{Resumen}{Resumen}}
\pdfbookmark{Resumen}{resumen}

En esta tesis se presentan los resultados de un estudio, basado en espectros de alta re\-so\-lu\-ci\'on ($\lambda / \Delta \lambda $ = 49000), del gas circunestelar (CircumStellar, CS) en las estrellas UXOR de masa intermedia BF~Ori, SV~Cep, UX~Ori, WW~Vul y XY~Per.
Las observaciones empleadas son un conjunto de 38 espectros \'echelle, obtenidos en el rango espectral 3800-5900~\AA, con los que se ha efectuado un seguimiento de las estrellas en escalas temporales de meses, d\'{\i}as y horas.

Todos los espectros muestran un gran n\'umero de componentes de absorci\'on des\-pla\-za\-das al rojo y al azul superpuestas sobre el espectro fotosf\'erico estelar, en l\'{\i}neas de Balmer y met\'alicas.
Se han utilizado espectros sint\'eticos, generados a partir de los programas y modelos de atm\'osfera de Kurucz, para estimar velocidades de rotaci\'on, temperaturas efectivas y gravedades estelares.
Se han sustra\'{\i}do los mejores modelos fotosf\'ericos de cada espectro observado para determinar las componentes de absorci\'on variables debidas al gas circunestelar.
Dichas componentes han sido caracterizadas, mediante ajustes gaussianos multicomponente, en t\'erminos de su velocidad, $v$, dispersi\'on de velocidades, \deltav, y absorci\'on residual, $R_{\rm max}$.

Las componentes de absorci\'on detectadas en cada espectro se pueden agrupar en eventos de acuerdo a la similitud de velocidades radiales, lo cual es interpretado como la huella de la evoluci\'on din\'amica de condensaciones de gas.
La mayor\'{\i}a de los eventos experimentan aceleraciones/desaceleraciones del orden de d\'ecimas de \ms.
El tiempo de vida t\'{\i}pico de estos eventos es de unos pocos d\'{\i}as.
Ni la dispersi\'on de velocidades ni la intensidad relativa de las absorciones de las componentes muestran cambios dr\'asticos durante el tiempo de vida de los eventos.
Esto sugiere que son originados por peque\~nas nubes de gas que mantienen su identidad geom\'etrica y f\'{\i}sica.

El estudio de las relaciones de intensidad entre distintas l\'{\i}neas para los distintos eventos sugiere una composici\'on similar a la solar para la mayor parte del gas CS.
Esto confirma algunos resultados previos y excluye un medio ambiente muy rico en metales como el origen com\'un de las componentes transitorias en estrellas UXOR.
Los datos obtenidos suponen restricciones o validaciones observacionales que pueden ser aplicadas en los modelos te\'oricos de condiciones f\'{\i}sico-qu\'{\i}micas del gas CS en estrellas j\'ovenes.
En particular, se sugiere que la presencia simult\'anea de gas en ca\'{\i}da y eyectado por la estrella deber\'{\i}a ser investigado en el contexto de modelos detallados de acreci\'on magnetosf\'erica, similares a los propuestos para las estrellas T~Tauri de baja masa.

Se ha descubierto que WW~Vul es una estrella peculiar porque, adem\'as de mostrar gas en ca\'{\i}da y eyecci\'on con propiedades similares a las observadas en las otras estrellas, presenta tambi\'en componentes de absorci\'on transitorias en l\'{\i}neas met\'alicas sin contrapartida obvia en las l\'{\i}neas de hidr\'ogeno.
Este hecho podr\'{\i}a, en principio, sugerir la presencia de nubes de gas CS de metalicidad elevada alrededor de WW~Vul.
La exis\-ten\-cia de una componente de gas rico en metales debe, sin embargo, ser confirmada mediante observaciones adicionales y un an\'alisis cuantitativo detallado.

Todos estos resultados han sido publicados por \citet[cap\'{\i}tulo \ref{uxori}]{mora2002} y \citet[cap\'{\i}tulo \ref{haebe}]{mora2004}.

Tambi\'en se han determinado velocidades de rotaci\'on para una gran muestra de es\-tre\-llas PMS y de tipo Vega mediante espectros \'echelle de alta resoluci\'on ($\lambda / \Delta \lambda $ = 49000).
Para determinar las velocidades se ha utilizado el primer m\'{\i}nimo de la transformada de Fourier de diversos perfiles de l\'{\i}nea fotosf\'ericos.
Las velocidades obtenidas han sido publicadas por \citet[ver cap\'{\i}tulo \ref{vsini}]{mora2001}, junto con determinaciones de tipos espectrales realizadas por otros coautores.

\vspace{0.5cm}
\noindent {\bf Palabras clave.} Estrellas: formaci\'on -- Estrellas: pre-secuencia principal -- Estrellas: materia circunestelar -- Acreci\'on: discos de acreci\'on -- L\'{\i}neas: perfiles -- Estrellas: individuales: BF~Ori, SV~Cep, UX~Ori, WW~Vul, XY~Per -- Estrellas: rotaci\'on

\cleardoublepage

\pdfbookmark{Table of contents}{tableofcontents}
\tableofcontents
\cleardoublepage

\pdfbookmark{List of Figures}{listoffigures}
\listoffigures
\cleardoublepage

\pdfbookmark{List of Tables}{listoftables}
\listoftables
\cleardoublepage

\pagenumbering{arabic} \setcounter{page}{1}

\chapter{Introduction}

This thesis, entitled ``Kinematics of the circumstellar gas around UXOR stars'', has been developed in the scientific framework of the evolution of circumstellar (CS) disks around Pre-main sequence (PMS) stars.
Its specific topic is the detection and characterization of Transient Absorption Components (TACs) in high resolution spectra of five intermediate mass PMS Herbig Ae/Be (HAeBe) stars.
These objects belong to the HAe subgroup of stars, which comprises HAeBe stars with masses lower than 5~$M_\odot$.
The stars in this thesis have photopolarimetric behaviour similar to the HAe star UX~Orionis (included in the sample), therefore they are called UXORs.
It is believed that the UXORs are surrounded by large CS protoplanetary disks seen edge-on.

The current status of the research projects most relevant for this thesis is reviewed in Sections~\ref{introduction_extrasolar_planets}~ to~\ref{introduction_magnetospheric_accretion}.
The evolution of the research carried out in this thesis is exposed in Sections~\ref{introduction_thesis_subject} and~\ref{introduction_objectives}.
Finally, the structure of this dissertation is shown in Section~\ref{introduction_thesis_structure}.

\section{Discovery of the first extrasolar planets}
\label{introduction_extrasolar_planets}

The search for extrasolar planets is a very interesting research field, both from the point of view of fundamental astrophysics (formation and evolution of stars and planetary systems) and the philosophical viewpoint (possible existence of extraterrestrial life).

The first detection of extrasolar planets was made by \citet{wolszczan1992}, who discovered three bodies orbiting around the pulsar PSR1257+12 via very precise measurements of the emission time of the pulses.
Two of these planets are similar to the Earth in mass and orbital period.
The third body is somewhat smaller.
However, \citet{mayor1995} were the first to discover an extrasolar planet around a Main Sequence (MS) star: 51~Peg.
Their method relies in precise measurements of the radial velocity of the stars with accuracies down to $\sim$10~m\,s$^{-1}$.
This work has become a fundamental milestone in the star and planetary formation field.
Many exoplanets around MS stars have been discovered so far with the radial velocities technique, at least 120\footnote{An updated list of all discovered exoplanets can be found in the ``Extrasolar Planets Encyclopaedia'', {\tt \href{http://www.obspm.fr/encycl/encycl.html}{http://www.obspm.fr/encycl/encycl.html}}} at the date of writing up (April 6$^{\rm th}$ 2004).

As soon as the number of discovered planets around MS stars was significant, it became apparent that the ``classic'' paradigm of planetary system formation was inadequate to describe the observations.
This should not be surprising, as the theory was developed from only one particular case: the Solar System.
The main lack of the theory is its inability to explain the common appearance of giant planets very close to the star, even nearer than the orbital distance of Mercury.
These bodies, which are called ``Hot-Jupiters'', are the easiest to detect via the radial velocities method.

\section{Evolution of circumstellar disks}

The theories of the formation of planetary systems have been very much improved since the discovery of the first extrasolar planets.
However, there is not yet a theory which explains, in a self-consistent way, the physical mechanisms involved in the generation of the observed large diversity of planetary systems.
It is also believed that the theories will be reformed again when methods capable of detecting telluric planets at distances $\sim$1~AU are developed.

However, there is agreement among the theoreticians that the planets are formed in the CS accretion disks \citep{ruden1999}.
These disks are originated during the gravitational collapse undergone by a molecular cloud core prior to the generation of a star.
The angular momentum accumulated by a cloud core is very large because of its very big initial dimensions ($R_{\rm core}\sim$~0.1~pc).
In this way, it is impossible for the cloud material to fall directly onto the centre of gravity.
The matter is accumulated in a disk, which rotates around the central condensation (protostar), where the stellar mass is being accreted \citep{hartmann1998}.
The evolution of a disk is regulated by the matter that arrives at its outer edge from the collapsing envelope and several internal processes (mainly viscous evolution and gravitational instabilities).
All these processes imply a parameter, namely mass accretion rate, $\dot{M}$, from the disk to the central protostar.

The evolution of protoplanetary disks in low mass T~Tauri stars is rather well understood \citep{hartmann1998}.
The observations suggest that, for a Classical T~Tauri Star (CTTS) with a typical mass of 1$M_\odot$, the accretion rate from the envelope to the disk is about  10$^{-6} M_\odot$/yr at the beginning of the core collapse.
This rate steadily decreases until, eventually, it vanishes after 0.1-0.2~Myr.
However, the observed accretion rates, $\dot{M}$, from the disk to the protostar are two orders of magnitude lower.
\citet{hartmann1998} suggests that most of the matter is transferred to the protostar during some FU~Orionis outbursts (he admits, however, that the number of detected FU~Ori objects is not large enough, compared to the number of identified PMS stars).
These phenomena are characterized by a huge increase in the stellar luminosity during brief lapses of time ($t \sim$~100~yr) and are generally associated with dramatic changes in the accretion rate $\dot{M}$.
According to \citet{hartmann1998}, if the accretion rate from the envelope to the disk is higher than that from the disk to the protostar, the disk mass grows steadily.
Disks may become gravitationally unstable if their mass reaches a certain threshold value.
These instabilities could allow large amounts of matter to fall quickly onto the star.
The potential energy released would give rise to the outburst, characterized by a large increase in brightness.

When the envelope eventually collapses, the result is a CTTS surrounded by a CS disk, whose evolution is essentially dominated by viscous friction processes \citep{hartmann1998}.
In the CTTS stage, the accretion rate $\dot{M}$ experiments a much smoother evolution: it decreases steadily as the disk is being depleted of matter (there is room for possible EXor outbursts, which are less violent than FU~Ori ones).
Even though the disk becomes less massive, its size increases due to angular momentum transport in the outward direction.
This scenario is only valid in isolated stars.
If the star is in a binary system, the companion would stop the disk expansion, and the disk would be consumed in a much shorter time.
It is thought \citep{calvet2000} that this is the origin of Weak T~Tauri Stars (WTTSs), which have similar ages than CTTSs but lack traces of accretion or CS disks.

\begin{figure}
\centerline{\includegraphics[width=0.75\textwidth,clip=true]
                            {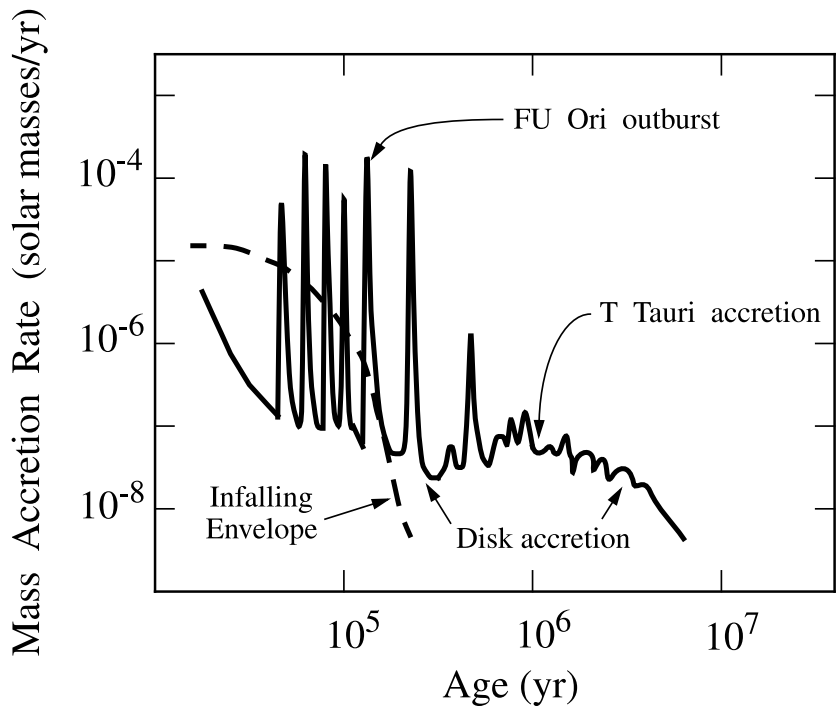}}
\caption
[Temporal evolution of the accretion rate in a typical CTTS]
{Temporal evolution of the accretion rate in a typical CTTS \citep[taken from][]{calvet2000}.
Three different regimes can be identified in the figure.
The first one extends from 0 to 0.1-0.2~Myr and corresponds to the gravitational collapse of the molecular cloud core.
It can be appreciated that the accretion rate from the envelope to the disk (dashed line) is higher than the accretion rate, $\dot{M}$, from the disk to the protostar.
This difference of material generates sporadic gravitational instabilities in the disk that, eventually, give rise to FU~Ori outbursts.
During the second stage, which spans from 0.1-0.2~Myr to a few Myr, the CS disk gives matter to the star and expands via viscous friction processes.
Finally, after a few Myr, the accretion disk dissipates and no more matter is transferred to the star.
}
\label{introduction_accretion_rate}
\end{figure}

A schematic representation of the temporal evolution of the mass accretion rate for a typical CTTS is shown in Figure~\ref{introduction_accretion_rate}, taken from \citet{calvet2000}.
Three different regimes can be identified in the figure: molecular cloud core collapse, viscous accretion of the disk and disk dissipation.

The protoplanetary disks around HAeBe stars are much less known.
Moreover, there are differences between the stars of spectral types B9 and later (HAe) and those of earlier types (HBe).
In general, most of the HAe stars have disks similar to those of CTTSs during the whole PMS phase and up to about 10~Myr \citep{nattappiv}, when second generation disks (see below) begin to be detected.
On the other hand, the HBe stars usually lack protoplanetary disks.
This has been interpreted by \citet{nattappiv} as a consequence of the intense radiation fields present in the neighbourhood of these objects.
The radiation could destroy the disk before the primordial molecular cloud remnants are dissipated and the star becomes detectable.

\section{Grain growth and planetesimals}

\begin{figure}
\centerline{\includegraphics[width=\hsize,clip=true]
                            {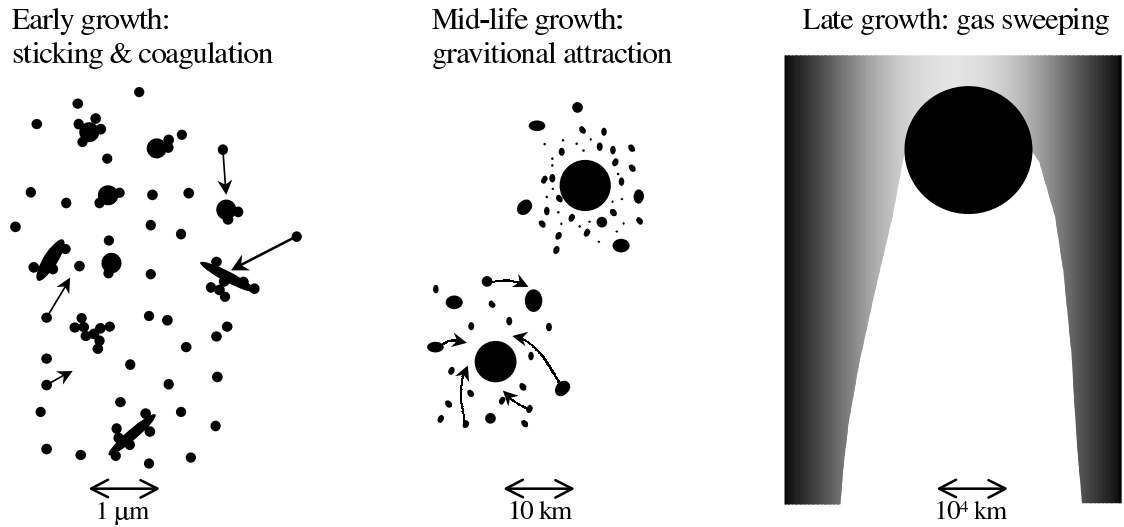}}
\caption[Different phases in the process of planet formation]
{Different phases in the process of planet formation \citep[taken from][]{beckwith2000}.
It is currently believed that the planets are formed via dust grain growth processes inside the CS disks.
At first (left), the grains, which have sizes of $d \sim$~1~$\mu$m typical of the IS medium, grow by inelastic collisions.
After about $t \sim$~10$^4$~yr, the grains reach macroscopic sizes of $d \sim$~1~m.
These processes continue until the grains grow up to $d \sim$~1~km after $t \sim$~10$^4$~yr.
These bodies, called planetesimals, have a non-negligible gravitational field that accelerates their growth (centre).
When a massive rocky core, $M \sim 1 M_\oplus$, is formed, it will accrete all the CS gas in its surroundings.
It will eventually become a gas giant planet if there is enough gas available.
The whole process is supposed to finish after $t \sim$~1~Myr.
}
\label{introduction_grain_growth}
\end{figure}

As we said in the previous section, it is believed that the formation of planets takes place in protoplanetary disks \citep{ruden1999}.
The whole process of planetary formation can be divided into three general stages \citep{beckwith2000}:
{\it 1.} The dust grains at the disk, which have initial sizes typical of the interstellar (IS) medium ($d \sim$~1~$\mu$m), grow up to sizes $d \sim$~1~km in about $t$~$\sim$~10$^4$~yr.
{\it 2.} The solid bodies of $d$~$\sim$~1~km, called planetesimals, grow in a very fast runaway process until the telluric planets and giant planet rocky cores are formed.
{\it 3.} The rocky cores accrete the gas located in nearby orbits by gravitational attraction.
The final mass of the planet depends on the remaining reservoir of gas available when the rocky core is created.
In this way, telluric, ice giant or gas giant planets can be formed.
It is expected that the whole planet formation process finishes after $t$~$\sim$~10$^6$~yr.
Figure~\ref{introduction_grain_growth}, \citep[taken from][]{beckwith2000}, shows an scheme of the three described phases of planet formation.
 
Many numerical simulations \citep[e.g.][]{weidenschilling1997,kokubo2002} show that, once the planetesimals have been formed, the formation of a planetary system is almost unavoidable.
The reason is that the gravitational field generated by the planetesimals (and not the dust grains) makes the captures of material much easier.
\citet{safronov1969} first demonstrated that the gravity makes the collisional cross section of a planetesimal to grow with the fourth power of the radius $\sigma \sim r^4$, which is a much higher increment than that of the geometrical cross section, which raises with the second power of radius $\sigma_{\rm g} \sim r^2$.
This effect favours the captures of matter by the more massive planetesimals and originates a runaway accretion phase of the biggest objects.
This stage will eventually finish when the planetesimals fuse to form the telluric planets and rocky cores of the planetary system.

Planetesimal formation is the most critical and worst known phase during the whole planetary formation process.
It was at first thought that these bodies are formed via gravitational instabilities in the disk, which could generate the gravitational collapse of a significant amount of mass \citep[e.g.][]{goldreich1973}.
However, this hypothesis has been discarded because the instabilities cannot be set up until the grains have grown to sizes $d \sim$~10-100~m \citep{cuzzi1993,weidenschilling1995}.
The only alternative mechanism for generating planetesimals is the gradual growth of dust grains by inelastic collisions \citep{beckwith2000}.
Initially, the dust grains have a size typical of those in the IS medium ($d \sim$~1~$\mu$m).
It is thought that the grains can grow by inelastic collisions until they reach macroscopic dimensions of $d \sim$~1~m in about $t \sim$~10$^4$~yr.
The settling of grains in the midplane of the disk is one of the most important factors in the growth, because it rises the frequency of collisions.
The bodies continue growing until they become planetesimals ($d \sim$~1~km) after $t \sim$~10$^4$~yr.

The main problem about the planetesimals is the time needed for their formation, because if this interval is longer than the lifetime of the gas disk, it would be impossible to generate giant planets.
This has motivated several scientists to study both the physics involved and the observational evidences of grain growth.
Many experiments have been performed to understand the physics of the process.
\citet{poppe2000} measured the adhesion efficiency between micrometric particles of different materials and sizes.
\citet{supulver1997} studied the influence of the outer frost layers in the adhesion, which is important for particles of sizes around $d \sim$~1-100~cm.
\citet{kouchi2002} analyzed the influence of the organic surface layers in particles of size $d \sim$~1~mm.
\citet{paraskov2003} have developed an experimental setup, which will allow them to study the role of erosion in high velocity impacts with grains of size $d \sim$~1~mm.
In addition, many theoretical studies have been carried out.
For example, \citet{dominik1997} calculated critical junction velocities between grains of several materials, while \citet{kornet2001} simulated numerically the growth of grains in the disks.

Currently, there are not undisputable proofs of grain growth in CS disks.
The most promising indications arise from the detailed study of the spectra and Spectral Energy Distributions (SEDs) in the millimetre and submillimetre ranges \citep{beckwith2000}.
The simplest theoretical models suggest that the accumulation of mass in big size opaque bodies lowers the extinction.
This phenomenon must be carefully modelled in order to make accurate predictions, because the vast diversity of molecules present in CS disks, the unknown spatial distribution of the millimetre emission and the complexities of light reprocessing in the disk do not allow a direct interpretation of the observations.
For example, \citet{dalessio2001} have predicted millimetre fluxes according to self-consistent models of protoplanetary disks.
It is hoped that the definitive proofs will be obtained as soon as the millimetre interferometer ALMA becomes operational.
This instrument will join a high sensitivity together with high spatial resolution \citep{beckwith2000}.

\section{Search for planetesimals: $\beta$~Pictoris}
\label{introduction_beta_pic}

It has been seen in the previous section that the computational modelling of planetesimals is feasible with the facilities already available.
However, the direct observation of planetesimals is extremely complicated.
The surface of one of such bodies is several orders of magnitude lower than the sum of the surface of all the dust grains in the disk.
In this way, both the thermal emission and the scattered light of a planetesimal are hidden by the dust.
Even though direct detections are impossible, \cite{lagrange1988} proposed the existence of planetesimals orbiting around the star $\beta$~Pictoris as the best explanation to its spectroscopic activity.

$\beta$~Pic is a MS star of spectral type A5V, approximately 20~$\pm$~10~Myr old \citep{barrado1999}.
$\beta$~Pic is one of the prototypical Vega stars.
Those stars are MS objects with prominent infrared excesses, named after Vega ($\alpha$~Lyr), the first identified star of the class.
The Vega phenomenon was discovered after some routine calibrations were performed with the satellite IRAS \citep{aumann1984}.
It has been interpreted as the signature of possible circumstellar disks.
This hypothesis was confirmed by \citet{smith1984}, who obtained a coronagraphic scattered light image of the $\beta$~Pic disk.
This was the first image of a CS disk, which was oriented edge-on.

It was soon apparent \citep{backman1993} that the $\beta$~Pic disk is formed primarily by dust grains of size $d \ga 1 \mu$m.
Due to dissipation by collisions, radiation pressure and Poynting-Robertson drag, the grains have a lifetime much shorter than the stellar age.
In this way, it should exist a source which replaces the grains as they evaporate, otherwise the disk could not survive for a long time.
Protoplanetary disks are very different from the $\beta$~Pic disk, because they are mainly composed of gas, they are accreted or dissipated in much earlier stages and they are unable to provide a continuous supply of dust grains.
In other words, the $\beta$~Pic disk is an object of different origin, formed after the gravitational collapse of the molecular cloud core, so it is thought to be of ``second generation''.

Many studies about $\beta$~Pic were conducted after the discovery of its CS disk.
A central gap clear of dust, up to a radius of about 25~AU was discovered \citep{lagage1994}, several asymmetries in the disk, including a warp, were detected \citep{burrows1995}.
Amorphous and crystalline silicates have been discovered in the disk \citep{telesco1991,knacke1993}.
Finally, small amounts of gas have been detected by means of Transient Absorption Components (TACs) and stable absorptions superimposed over several photospheric spectral lines \citep{hobbs1985}.

The stable components have the same radial velocity as the star and are relatively narrow, \deltav~$\sim$~2~\kms, \citep{lagrange1998}, so the gas responsible for the absorption is placed at a constant distance from the star.
This seems a paradox, because many of the ions showing stable components have a $\beta$ rate (quotient between the radiation pressure and the gravitational attraction) much greater than 1.
For example, $\beta_\ion{Ca}{ii} = 35$, $\beta_\ion{Fe}{ii} = 4.87$ \citep{lagrange1998}.
The same authors propose that the CS gas responsible for the absorption is hold by a neutral hydrogen torus (this element is little affected by the radiation pressure), so the radial velocity of the gas is zero.
The amount of \ion{H}{i} needed is small and compatible with the actual upper limits (see below).
If this model is true, the gas causing the absorptions needs to be continually replenished, because the torus delays the outward migration, but does not stop it.
This explanation has the additional property of being compatible with the theory the same authors have developed to explain the transient components (see below).

Most of the absorption components (TACs) observed in $\beta$~Pic are redshifted, so they are called RACs (Redshifted Absorption Components).
There are a few detections of blueshifted components \citep{bruhweiler1991,crawford1998}, called BACs (Blueshifted Absorption Components).
All the detections of CS gas in $\beta$~Pic have been made in metallic lines and never in hydrogen.
This surprising result motivated \citet{freudling1995} to search for neutral hydrogen in the $\beta$~Pic disk via the 21~cm \ion{H}{i} line.
The detection was negative, so they could establish an upper limit to the hydrogen column density of $N(\rm{H}) < 10^{19} \rm{cm}^{-2}$.

The RACs in $\beta$~Pic display a wide range of radial velocities, from $v \simeq$ 10~\kms\ to $v \ga$~300~\kms.
It has been observed a correlation between the radial velocity and the width of the RACs \citep{lagrange1996}.
In order to achieve a high radial velocity, the gas needs to get very close to the star (e.g. at distances of $r \sim$~0.1~AU for gas with velocity $v \sim$~20~\kms) in high ellipticity orbits.
However, the high radiation pressure exerted by $\beta$~Pic prevents much of the observed ions to approach so near the star.
In this way, a source that provides the gas at distances $r \sim$~0.1~AU with infall radial velocities is needed.

The model developed by \citet{lagrange1988} provides the best explanation of the RACs in $\beta$~Pic.
They assume that the RACs are generated by the evaporation of solid bodies (FEBs, Falling, Evaporating Bodies).
These bodies would be around 1~km in size (i.e. they are planetesimals) and would approach the star in highly excentrical orbits, as the Solar System comets do.
In this way, the stellar radiation would evaporate the outer layers of the planetesimals, thus generating a coma made by dust and gas in different ionization stages.
The dust would continuously replenish the CS disk and the evaporated gas would generate the transient absorptions and the stable ones when the outward gas migration would be stopped by the gas torus.
The tail would not have any influence in the RAC generation, because it would be highly collimated and would not be able to obscure a significant fraction of the stellar disk.

\begin{figure}
\centerline{\includegraphics[width=\hsize,clip=true]
                            {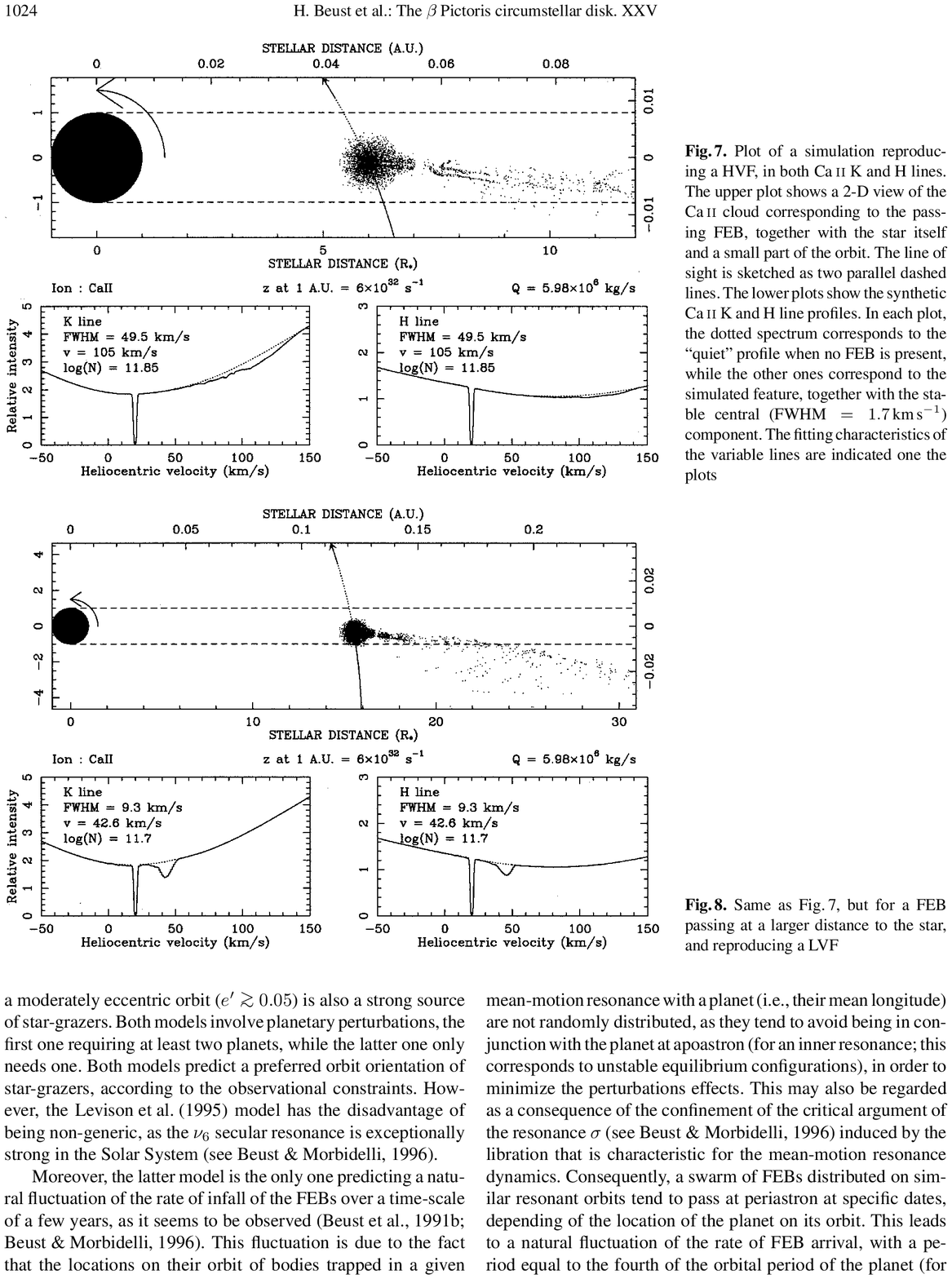}}
\caption[Simulation of a RAC via a FEB]
{Simulation of a RAC via a FEB \citep[taken from][]{beust1998}.
Top: a planetesimal approaches $\beta$~Pic ($r \sim$~0.1~AU) in an elliptical orbit.
Its outer layers evaporate and form a coma made of gas and dust which covers a significant fraction of the stellar disk.
The observer is located at the right and the line of sight is represented by two horizontal dashed lines.
Bottom: Two graphics show the computed absorption profiles for the K (left) and H (right) lines of \ion{Ca}{ii}.
$\beta$~Pic is a high speed rotator, so its photospheric lines are much broader than the RACs and correspond to the pseudocontinuum upon which the absorptions are formed.
Each line shows two CS components.
The first one is a very narrow, intense and saturated absorption which has the same heliocentric radial velocity of the star ($v \simeq$~20~\kms).
This is the stable absorption and cannot be reproduced by the FEB simulations, so it is introduced by hand in the spectra.
The second absorption is originated by the comet coma.
It is broader and shallower, but it is also saturated, as can be seen from the relative intensities of the two lines in the \ion{Ca}{ii} doublet.
This component appears at a radial velocity of $v \simeq$~45~\kms, which is the radial velocity of the star plus the velocity of the planetesimal projected on the line of sight.}
\label{introduction_feb}
\end{figure}

\citet{beust1990} performed a series of numerical simulations, later improved by \citet{beust1996} and \citet{beust1998}, which confirmed the ability of the FEB mechanism to produce RACs in $\beta$~Pic.
An illustration of such simulations can be seen in Figure~\ref{introduction_feb}, taken from \citet{beust1998}.
The FEB model has two fundamental problems: first, an efficient physical mechanism to increase the orbital excentricity of the planetesimals is needed, in order to explain the high number of RACs detected per year; second, there is a much large number of RACs than BACs but, if the system was symmetrical, the same number of comets approaching and moving away from the star should be observed.
\citet{beust_morbidelli_1996} possibly provided the best explanation assuming the existence of a Jupiter-like giant planet in a slightly elliptical orbit ($M \sim 1 M_J$, $r \sim$~20~AU, $e \ga 0.05$) and the presence of a certain amount of asteroids in the disk.
This planet would perturb the orbits of the planetesimals, by means of the (4:1) mean motion resonance, and would put them in high excentricity orbits, so they could become the comets required by the FEB model.
The excentricity of the planetary orbit would reproduce, for selected lines of sight, the observed statistical predominance of RACs over BACs.

Summarizing, the observation of RACs in $\beta$~Pic is nowadays the best indication of the possible existence of planetesimals in MS stars \citep[see the review paper by][]{lagrange2000}.
Furthermore, $\beta$~Pic was the strongest candidate star to harbour an extrasolar planet until the discovery of 51~Peg~B by \citet{mayor1995}.

\section{Search for planetesimals: UXORs}

The discovery of the $\beta$~Pic disk motivated many astronomers to look for other CS disks in PMS and MS stars.
The disk of $\beta$~Pic has some characteristics that make it quite easy to detect: it is nearby, large and seen edge-on.
The following detections of CS disks were conducted about a decade later by \citet{odell1993}, who identified CS disks using WFPC images taken with the Hubble Space Telescope (HST).
They found about ten PROtoPLanetarY DiskS (proplyds) in the Orion nebula.
The disks appeared as dark regions over the bright background of the \ion{H}{ii} region created by the strong radiation field of the Trapezium cluster.
This technique has been extended, using adaptive optics, to terrestrial telescopes.
Another method used to detect CS disks is optical and infrared coronagraphy.
It has been successfully performed with the HST and adaptive optics equipped terrestrial telescopes.
The coronagraph is no longer needed in the mid-infrared and longer wavelength regimes.
Some disks have been detected in direct images obtained in the submillimetre and mid-infrared ranges.
The highest spatial resolution has been reached with interferometric techniques, in the infrared and millimetre domains.

The improvements made in the available instrumentation has allowed the observation of many CS disks.
Examples of primordial disks are those of TW~Hya \citep{weinberger1999b,krist2000}, $\rho$~Oph \citep{brandner2000}, LkH$\alpha$~101 \citep{tuthill2001}, MBM~12 \citep{jayawardhana2002} and Carina \citep{smith2003}.
Several second generation disks have been found in main sequence stars, e.g. HR~4796 \citep{koerner1998,jayawardhana1998}, Vega and Fomalhaut \citep{holland1998} and $\epsilon$~Eri \citep{greaves1998}.
Some HAeBe stars near the Zero Age Main Sequence (ZAMS) also have second generation disks, e.g. AB~Aur \citep{marsh1995,nakajima1995}, HD~163296 \citep{mannings1997} and HD~141569 \citep{weinberger1999a,augereau1999,merin2004}.

It has been shown in the previous section that \citet{lagrange1988} proposed that both the disks and the RACs in $\beta$~Pic are originated by the evaporation of planetesimals.
This motivated some astronomers to look for CS gas in other stars as a complementary alternative to the direct detection of the disks (these searches were specially intense when no other detection of CS disks was achieved).
If the FEB hypothesis is applicable to other stars, the detection of TACs could suggest the existence of a CS disk edge-on or nearly edge-on.
\citet{hobbs1986} and \citet{lagrange1990} first conducted systematic searches of CS gas (TACs) in Vega, A-shell and F-shell MS stars.
They used high resolution optical spectra and identified a bunch of candidate stars (e.g. HR~10 y 51~Oph).
Some authors \citep[e.g.][]{grady1991,lecavelier1997} successfully extended the search for TACs to the ultraviolet (UV) range.

Later on, \citet{grinin1994} detected TACs in visible spectra of the HAe star UX~Ori, whereas \citet{grady1995} found spectral variability in some UV lines.
These results stimulated many authors \citep[e.g.][]{grinin1996,grady1996,dewinter1999} to extend the search to other HAeBe stars \citep[see the review paper by][]{grady2000}.
It was apparent that the HAe stars classified as UXORs were the most prone to present TACs in their spectra.
As it will be explained below, the UXORs can be considered the progenitors of $\beta$~Pic stars, due to the similarity in mass, spectral type and disk orientation with respect to the line of sight.
The study of the possible relation between TACs, CS disks and planetesimals in MS and PMS stars became a fundamental problem in the stellar formation and evolution field.

\begin{figure}
\begin{center}
\includegraphics[clip=true,width=0.6\hsize,angle=0.5]
                {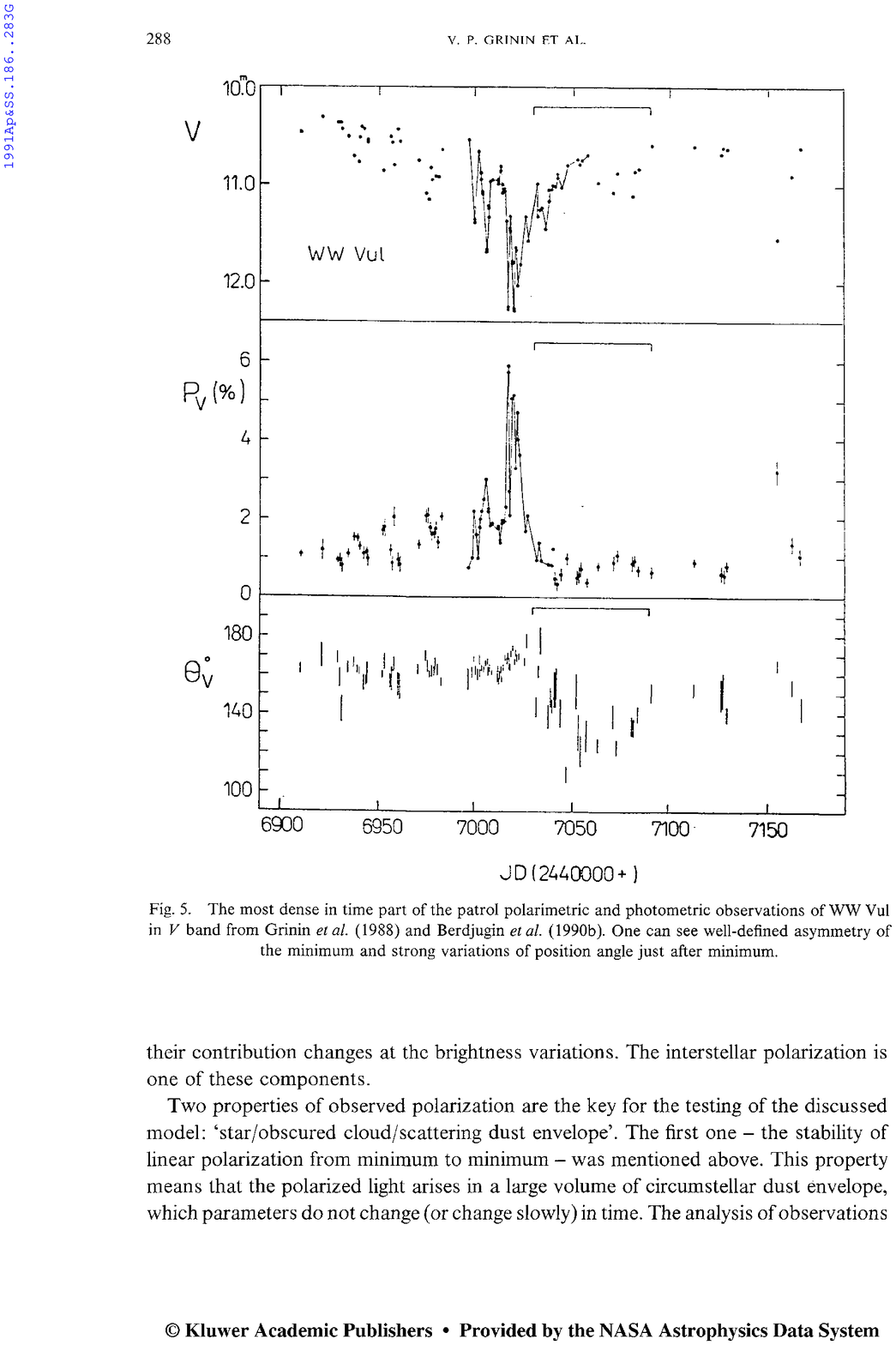}
\includegraphics[clip=true,width=0.6\hsize]
                {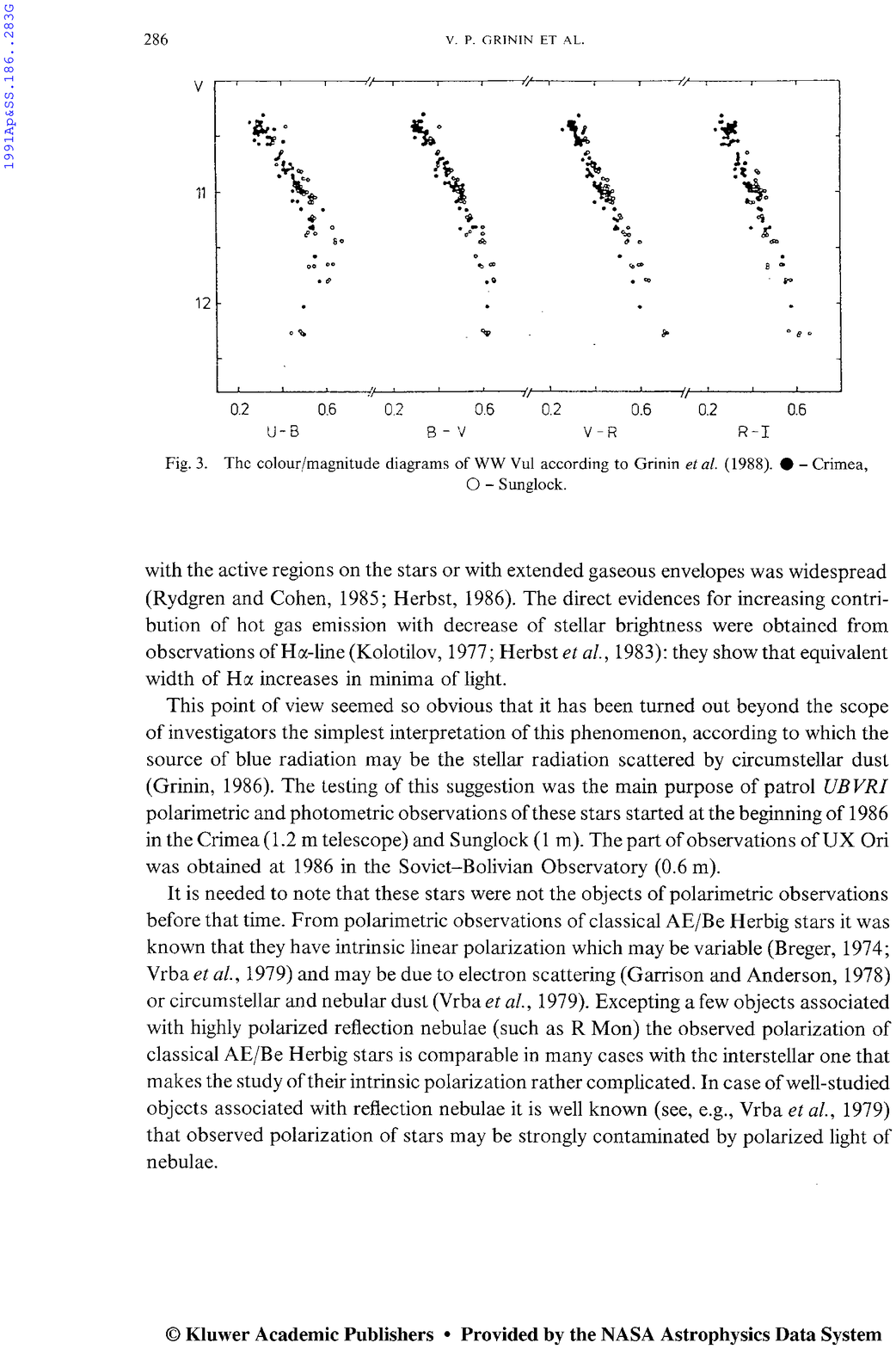}
\end{center}
\caption[Light curves for the UXOR star WW~Vul]
{Light curves for the UXOR star WW~Vul, taken from \citet{grinin1991}.
Top: Photopolarimetric data of WW~Vul obtained by \citet{grinin1991} during 300 nights are shown.
During this period, the star experienced an UXOR deep minimum $\Delta V \simeq 2$.
It can be appreciated that the drops in brightness are always matched by increases in polarization, up to values of $P_V \simeq 6\%$ in the photometric  minimum.
Bottom: Several colour-magnitude diagrams are shown for the star.
In a darkening event, the star departs from the top of the diagrams (maximum brightness), goes down to lower brightness and, eventually reaches the photometric minimum.
It then goes up until it recovers its original luminosity.
At first the brightness follows a linear extinction law, until the extinction reaches $\Delta V \simeq 1$.
For greater dimming, the reddening becomes increasingly lower than that given by the linear law, i.e. the slope of the curve grows or, even, changes its sign.
The biggest variations are found in the colour indices $U-B$ and $B-V$, for which the star becomes even bluer when the brightness decreases.}
\label{introduction_uxor}
\end{figure}

The UXORs are PMS stars, mainly HAe, which show photometric variations similar to the star Algol.
\citet{hoffmeister1949} first created a specific class to group all these objects together.
The initial name, RW~Aur stars, eventually became UXOR, because it was soon realized that the star UX~Ori presented the Algol behaviour in a much stronger way.
The UXORs alternate periods of low activity and almost constant brightness with severe dimming episodes ($\Delta V \sim 2.5$).
During the initial stages of darkening the stars redden, until there is a (turnaround) point after which the star, instead of getting redder, begins to turn bluer.
\citet{grinin1988} found out that the decreases in brightness were followed by increases in polarization up to values of $P_V \sim 6\%$ in the deepest photometric minima.
Figure~\ref{introduction_uxor}, taken from \citet{grinin1991}, shows sample light curves of the UXOR star WW~Vul, including a photometric minimum.

\citet{grinin1988} proposed the first model that explained successfully the photopolarimetric behaviour of UXORs.
According to \citet{grinin1988}, the light received by an observed is the sum of two components, namely the stellar radiation and the light scattered by an (almost) edge-on CS disk.
The photopolarimetric variations are attributed to the passage of an opaque dust cloud in front of the line of sight.
The cloud dims the stellar light according to an interstellar extinction law, i.e. the star becomes redder when fainter.
On the other hand, the disk luminosity is assumed to be constant.
As the disk is seen in scattered light, it is supposed to be bluer than the star and to have a non-zero polarization.

According to \cite{grinin1988}, the maximum brightness corresponds to instants where no cloud is crossing the line of sight, so the contribution of the disk to the total luminosity is negligible and only the stellar light is observable.
When a big cloud crosses the line of sight, the stellar brightness fades following the extinction law, so the star draws a straight line in the upper part of the colour-magnitude diagrams.
If the cloud is large enough, the amount of light scattered by the disk can be comparable to, or greater than, that emitted by the star.
This happens during deep minima, $\Delta V \ge 1$, and implies both a blueing of the observed light and an increase in the polarization.
On the other hand, the luminosity of the disk places a limit on the depth of the photometric minima ($\Delta V_{\rm max} \simeq 2.5$), because the disk, due to its large dimensions, remains visible even when the star has been fully obscured (i.e., the role of the cloud is similar to that of a coronagraph).

\citet{grinin1991} proposed that dust clouds are common in all HAe dense protoplanetary disks.
Therefore, the UXOR phenomenon would be observed in all PMS stars with edge-on disks, i.e. the UXOR phenomenon is purely geometrical.
This hypothesis receives additional support from the fact that, apart from the deep photometric minima, the UXORs are very similar to other non-UXOR HAe stars \citep{nattappiv}.

\citet{natta_whitney_2000} made some numerical simulations of the UXOR behaviour.
They used disk models more realistic than those utilized by \citet{grinin1988} (e.g. they considered flared disks).
\citet{natta_whitney_2000} found out that the UXOR phenomenon should be observable for those stars whose disks have an inclination respect to the line of sight of \mbox{$i \sim 45$-$68^\circ$}.
Lower inclinations do not allow a full occultation of the stellar light and larger angles cause too much extinction for the star to be optically detected.
A simple statistical calculation shows that about a half of the stars should present the UXOR phenomenon.
Currently, the best non-biased photometric sample of HAeBe stars is that of the Hipparcos catalogue.
\citet{vandenancker1998} carefully studied the variability presented by each of the HAeBe stars observed with the satellite.
It can be drawn from their study that about 30\% of the analysed HAe stars show a photometric variability greater than 0.5~mag.
This figure is roughly compatible with the prediction of 50\% given by \citet{natta_whitney_2000}.

The TACs in UXORs were initially interpreted as the undisputable trace of the presence of planetesimals \citep[see references therein]{grady2000}.
However, this interpretation has been recently questioned because the UXORs, contrary to $\beta$~Pic, present spectroscopic activity in the Balmer and \ion{Na}{i} lines.
\ion{Na}{i} atoms cannot survive more than a few seconds in the strong radiation field of an A type star.
This is hardly compatible with a cloud of a size comparable to the stellar disk, because the atoms should have been travelling about a day since their ejection from the comet.
\citet{sorelli1996} found that this would only be possible if the planetesimals were significantly larger than in $\beta$~Pic or if they fully evaporate in a single approach to the star.

On the other hand, the detection of TACs in the Balmer series is not consistent with a cometary composition for the CS gas.
\citet{natta2000} used a Non-Local Thermodynamical Equilibrium (NLTE) code to show that the abundances of the gas cloud which originated some TACs in UX~Ori are solar or nearly solar.
This is not compatible with the high metallicities ($m$~$\geq$~500) expected in highly hydrogen-depleted bodies such as planetesimals.
Besides, \citet{beust2001} showed that the FEB mechanism is not capable of generating TACs in HAeBe stars, because the strong stellar winds highly collimate the comets comae, so they could not cover a significant fraction of the stellar  disk.
The evaporation of comets would only be efficient if they approach to the star in wind free cavities.
Finally, \citet{hartmann1994} suggest that the RACs in CTTSs are formed in a completely different scenario: magnetospheric accretion of gas.
The latter possibility is explored in the next section.

Summarizing, it can be said that prior to this thesis the relation between TACs and planetesimals was not clear except for $\beta$~Pic.
This interesting question has been one of the main guidelines followed in the research conducted in the present work.

\section{Magnetospheric accretion}
\label{introduction_magnetospheric_accretion}

Classical T~Tauri stars present an intense photometric and spectroscopic activity.
Several of these phenomena can be explained in terms of the magnetospheric accretion of gas from the CS disk: brightness increments modulated by the rotation period (hot spots), excess of UV radiation (veiling), significant emission in hydrogen and metallic lines and presence of RACs (the simultaneous observation of extended line emission and a RAC is called an inverse P~Cygni profile).
In this Section, the fundamentals of the magnetospheric accretion theory and its relation with TACs in HAe stars will be reviewed.

\citet{hartmann1994} proposed the first self-consistent theoretical model capable of computing emission line profiles generated by magnetospheric accretion in CTTSs.
These calculations have been later improved \citep{muzerolle1998a,muzerolle2001} and the resulting profiles have been compared to real CTTSs spectra \citep{muzerolle1998b,muzerolle2001}.
The main hypothesis in the model is that the material accumulated in the circumstellar disk falls onto the star along the stellar magnetic field lines.
The kinetic energy acquired by the infalling matter is released, after the collision with the stellar surface, as UV radiation.
This model qualitatively explains the presence of hot spots (impact regions in the stellar surface), the veiling of the spectra and the strong emission present in some lines.
The material gains a significant velocity during the infall, so it could be expected the presence of RACs, due to the absorption of the continuum by the gas, for lines of sight corresponding to large inclinations of the disk.
Other phenomena, like the possible appearance of high velocity BACs, are out of the scope of the accretion models and can only be explained if high velocity winds are assumed.

\begin{figure}
\centerline{
\includegraphics[clip=true,width=0.75\hsize]
                {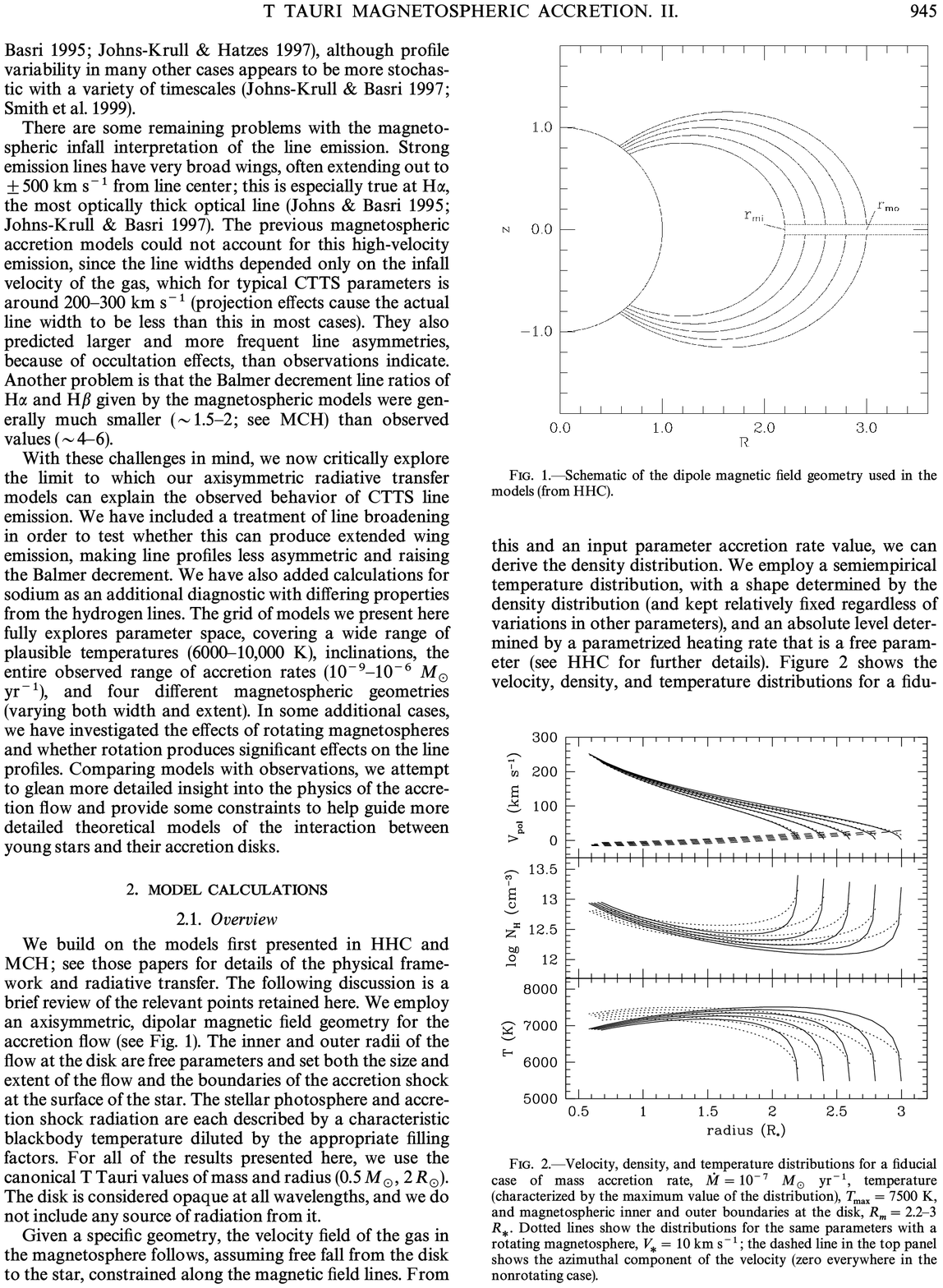}}
\caption[Geometry assumed in the magnetospheric accretion models]
{Geometry assumed in the magnetospheric accretion models \citep[taken from][]{hartmann1994}.
The figure represents a section in the radial direction, which includes the star (half circle in the left), the disk (bar in the right) and the magnetosphere (field lines).
The stellar magnetic field is assumed to be purely dipolar and its intersection with the disk is bound to the range $R_m = 2.2$-$3.3$~$R_\star$.
The CS gas located inside the magnetosphere experiences a free fall to the star along the field lines.
The kinetic energy acquired during the fall is released as UV and visible radiation when the material hits the stellar surface.
This is the origin of the line veiling.
A detailed study of the radiative transfer reproduces the emission profiles and RACs observed in many lines in CTTSs.}
\label{introduction_magnetosphere}
\end{figure}

\citet{hartmann1994} postulated the existence of a perfectly dipolar magnetic field whose force lines are bound to the disk and rotate in a rigid body-like picture.
It is assumed that the magnetic field destroys the structure of the disk and channels the CS matter to the stellar surface.
In this way, the gas undergoes a free fall following a trajectory given by the field lines.
The magnetosphere has axial symmetry and extends from an initial radius to an outer radius.
This symmetry forces the profiles to be non time-variable, because the accretion is evenly distributed along the whole disk.
The disk is assumed to be thin and completely opaque.
Figure~\ref{introduction_magnetosphere}, taken from \citet{hartmann1994}, shows the geometry assumed by the theory.

The gas is considerably heated during its free-fall ($T \sim 8000$~K).
Currently, this process is not fully understood.
This motivated \citet{hartmann1994} to use semiempirical heating laws that relate the gas temperature with the radial distance and the departure point from the disk.
These temperature distributions are relatively constant and are characterized by the highest temperature achieved by the gas ($T \sim$~6000-10000~K).
The density along each trajectory can be calculated if the accretion rate is known.
Once the density, temperature and velocity of the gas are calculated for every point in the magnetosphere, a detailed radiative transfer calculation can be performed in order to determine the contribution of each parcel of accreted material to the circumstellar profile of several lines.
This computation is complex and uses the extended Sobolev method \citep{muzerolle2001}.
In order to make the calculation easier, the star and the shock region in the stellar surface are replaced by black bodies of appropriate temperature and area.
The free parameters in the model are the stellar mass, the stellar radius, the rotation velocity, the inner and outer magnetosphere radii, the accretion rate, the maximum temperature achieved by the gas and the inclination of the rotation axis with respect to the line of sight.

\citet{muzerolle2001} have performed a thorough study of the parameter space, in order to find what regions are compatible with the observations.
They have used a typical CTTS with $M_\star = 0.5 M_\odot$, $R_\star = 2 R_\odot$.
The remaining parameters are in the following ranges: Accretion rate $\dot{M}: 10^{-9}$-$10^{-6} M_\odot / {\rm yr}$, maximum temperature of the gas $T_{\rm max}: 6000$-$10000$~K, inclination of the line of sight $i: 10$-$75^\circ$, rotation velocity $V_\star: 0$-$25$~\kms.
Four different prototypical magnetospheres were considered: small/wide ($R_m = 2.2$-$3$~$R_\star$), small/narrow ($R_m = 2.8$-$3$~$R_\star$), large/wide ($R_m = 5.2$-$6$~$R_\star$) and large/narrow ($R_m = 5.8$-$6$~$R_\star$).

In order to confirm the validity of the calculations, \citet{muzerolle2001} tried to reproduce the observed profiles and the veiling factors for H$\alpha$ and the \ion{Na}{i}~D doublet for several CTTSs.
They found a reasonable agreement for the stars with intermediate-low accretion rates.
However, the H$\alpha$ models fail to reproduce the profiles for the higher accretion rate objects (e.g. DR~Tau, $\dot{M} = 5 \times 10^{-7} M_\odot / {\rm yr}$), which have so strong winds that the lines display classical P~Cygni profiles that hide the effects of accretion.

\begin{figure}
\centerline{
\includegraphics[clip=true,width=0.75\hsize]
                {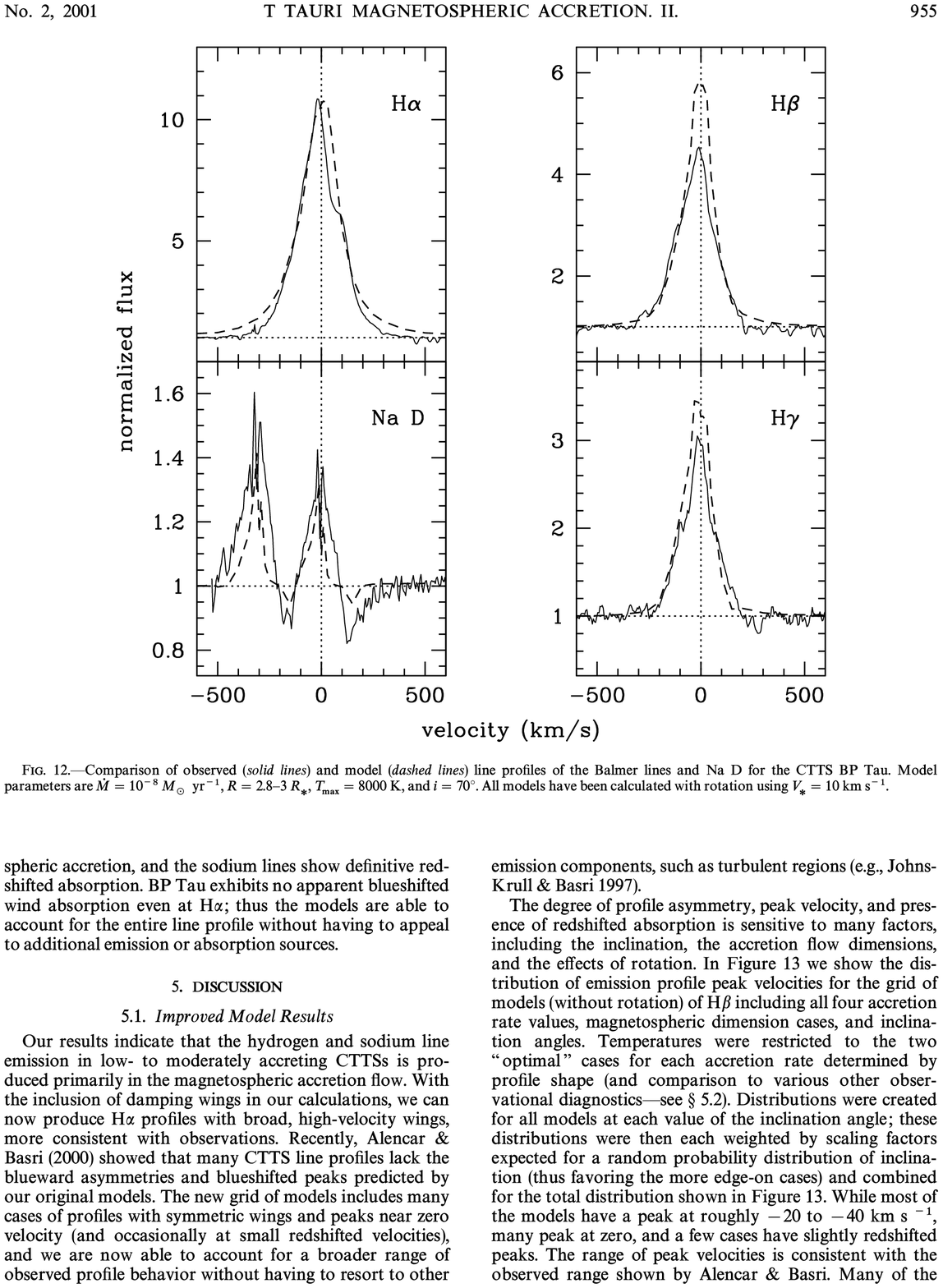}}
\caption[Comparison between synthetic and observed accretion profiles in the CTTS BP~Tau]
{Comparison between synthetic (dashed lines) and observed (solid lines) accretion profiles in the CTTS BP~Tau \citep[taken from][]{muzerolle2001}.
It can be appreciated a good agreement in the Balmer lines, both in the profile shape and the total emitted flux.
Inverse P~Cygni profiles, composed of extended emission and a RAC, are reproduced in the \ion{Na}{i}~D lines.
The parameters used in the calculation are $\dot{M} = 10^{-8} M_\odot / {\rm yr}$, $R_m = 2.8$-$3$~$R_\star$, $T_{\rm max} = 8000$~K, $i = 70^\circ$ and $V_\star = 10$~\kms.
The spectra of H$\beta$ and H$\gamma$ were obtained 4 years after those of H$\alpha$ and \ion{Na}{i}~D.
However, the agreement is good, which suggests that the accretion rate of the object has not substantially varied during that interval.}
\label{introduction_bptau_profiles}
\end{figure}

The best results are obtained for the star BP~Tau and are shown in Figure~\ref{introduction_bptau_profiles}.
It can be appreciated that both the shape of the profiles, including the large wings due to the Stark broadening, and the approximate emission fluxes have been reproduced.
The \ion{Na}{i} line models reproduce the observed inverse P~Cygni profiles.
The parameters used in the calculation are $\dot{M} = 10^{-8} M_\odot / {\rm yr}$, $R_m = 2.8$-$3$~$R_\star$, $T_{\rm max} = 8000$~K, $i = 70^\circ$ and $V_\star = 10$~\kms.
The H$\beta$ and H$\gamma$ spectra were obtained 4 years after those of H$\alpha$ and \ion{Na}{i}~D.
However, the agreement is good, which suggests that the accretion rate of the object has not substantially varied during that interval.

It has been shown that the magnetospheric accretion models reproduce properly the main features of the CTTSs spectra.
In particular, the RACs of the inverse P~Cygni profiles observed in many CTTSs are obtained both in hydrogen and metallic lines.
This achievement motivated \citet{sorelli1996} and \citet{natta2000} to claim that the presence of TACs in UXORs could be explained in terms of non axisymmetric magnetospheric accretion.
Therefore, the infall of material onto the star would be restricted to a kind of magnetic field tubes (funnels).
This hypothesis explains the spectral variability in terms of the creation, destruction and rotation of accretion funnels.

Despite its success in CTTSs, the profiles calculated by \citet{muzerolle2001} are not directly applicable to intermediate mass UXOR stars, i.e. the kind of objects studied in this thesis.
First, it is not clear that these stars can sustain a stable magnetosphere, with fields strong enough and rigidly anchored to the CS disk.
Second, the high rotation velocity of these objects, namely $V_\star \sim$~200~\kms, would imply a very quick revolution of the magnetosphere, which would substantially modify the synthetic line profiles.
Third, the line profiles studied in this thesis do not present much emission, which is restricted to a few lines in the Balmer series (H$\beta$, H$\gamma$ and H$\delta$).
That is, the profiles are composed of different overlapping CS absorptions without emission, so they are very different from the inverse P~Cygni profiles observed in typical CTTSs.
Finally, the TACs exhibit a fast evolution, in times of days and even hours without any apparent periodicity.
This temporal variability cannot be explained in terms of axisymmetric steady models.

\begin{figure}
\centerline{
\includegraphics[clip=true,width=\hsize]
                {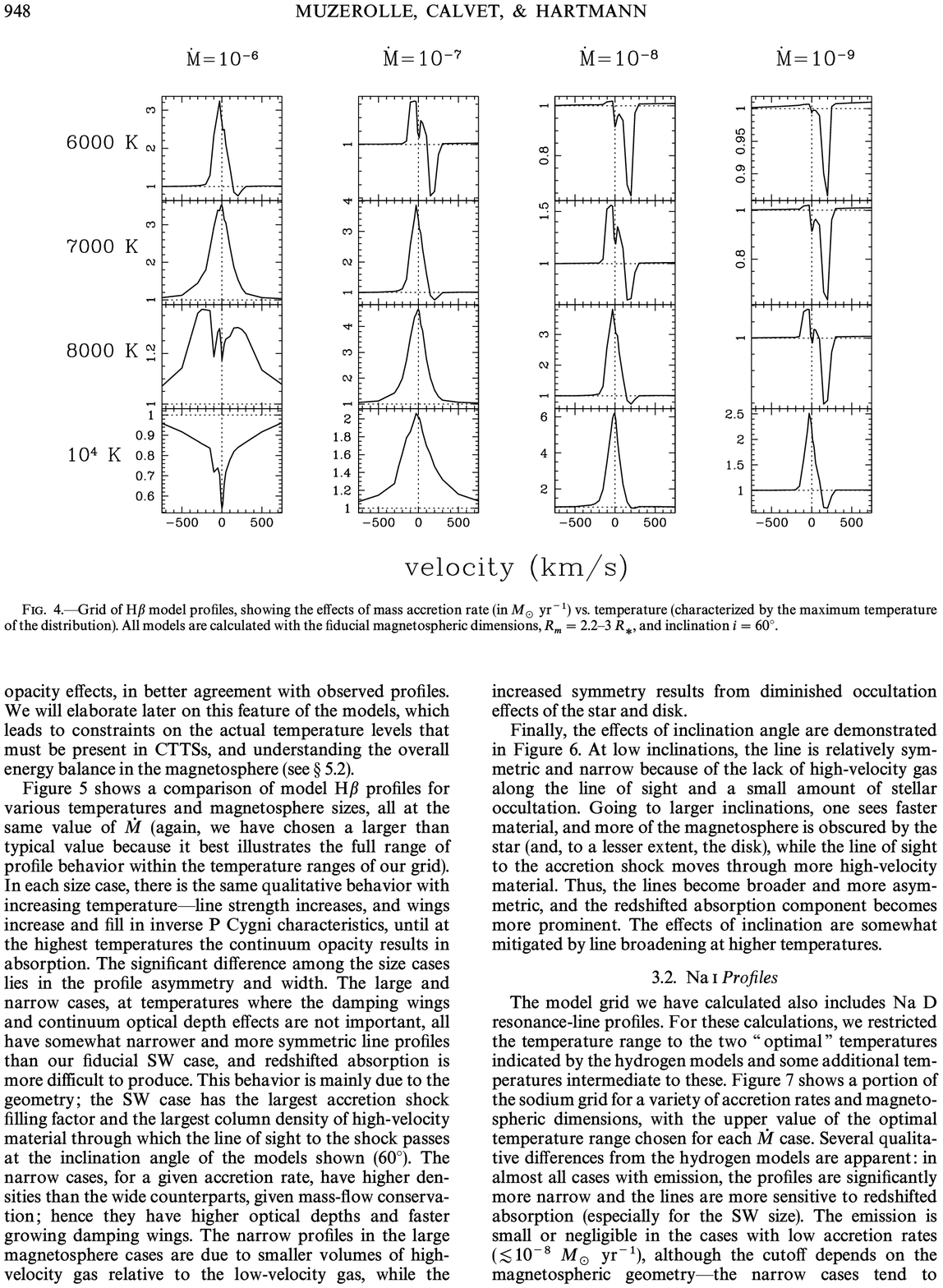}}
\caption[Emission and absorption H$\beta$ CS profiles generated by magnetospheric accretion]
{Emission and absorption H$\beta$ CS profiles generated by magnetospheric accretion, \citep[taken from][]{muzerolle2001}.
These profiles have been calculated for a typical CTTS with $M_\star = 0.5 M_\odot$, $R_\star = 2 R_\odot$, and a magnetosphere of type small/wide $R_m = 2.2$-$3$~$R_\star$.
The rotation axis is inclined 60$^\circ$, a typical value for UXOR stars.
It can be seen that it is necessary to combine a high (low) accretion rate with a low (high) maximum temperature in order to achieve the strong emissions observed in T~Tauri stars.
The profiles with greater resemblance to the ones observed in UXOR stars are those obtained with the lowest temperatures and accretion rates, which are characterized by an almost absence of emission combined with narrow absorption components.
}
\label{introduction_accretion_profiles}
\end{figure}

In Figure~\ref{introduction_accretion_profiles}, taken from \citet{muzerolle2001}, the models with the closest relationship to the UXOR stars are shown.
The displayed profiles correspond to the H$\beta$ line and have been computed for a typical CTTS with $M_\star = 0.5 M_\odot$, $R_\star = 2 R_\odot$ and $V_\star = 0$~\kms.
The magnetosphere is of type small/wide ($R_m = 2.2$-$3 R_\star$).
The line of sight is inclined 60$^\circ$ respect to the rotation axis.
This is a typical value for HAe stars with UXOR activity \citep{natta_whitney_2000}.
The free parameters in the simulation are the accretion rate, $\dot{M}$, and the maximum temperature, $T_{\rm max}$.
The profiles shown only represent the CS contribution to the lines, excluding the stellar photosphere and the veiling continuum.

The parameter space can be divided into three loose regions with different types of line profiles: high accretion rate and temperature, low accretion rate and temperature and, finally, high (low) accretion rate and low (high) temperature.
The inverse P~Cygni profiles are obtained when one parameter is high and the other one is kept low.
This implies some restrictions to the models that have been extensively studied by \citet{muzerolle2001}.
If a high accretion rate and a high temperature are combined, the material flux produces a greater level of continuum veiling than that generated in the shock region.
This produces strong absorptions, including an extremely broad component (\deltav~$\sim$~500~\kms).
Finally, for low accretion rates and temperatures, low emission profiles with redshifted narrow absorption components (\deltav~$\sim$~100~\kms) are obtained.
Only the later case is similar to the observed profiles in UXOR stars, which will be thoroughly studied in Chapters~\ref{uxori} and~\ref{haebe}.
If the models explained in this Section were applicable to HAe stars, it would be deduced from the observations that the physics of accretion is less violent in UXORs than in CTTSs.

\section{Evolution of the subject of the present PhD thesis}
\label{introduction_thesis_subject}

The structure and contents of the present PhD thesis considerably differ from the initial approach followed by the author and his advisor.
This a natural consequence of the research process, because only after the analysis of the data began, it was realized that the initial approach had to be modified and the real power of the observational data could be established.
In this Section the evolution that the thesis has followed until it reached its actual structure and contents will be exposed, as well as the reasons that motivated those changes.

The initial objective was the study of the evolution of the properties of planetesimals in CS disks, from the beginning of the PMS stage until the MS.
It was expected to find new observational physical restrictions that could be included in the theoretical models of star formation.
It was initially assumed, as a working hypothesis, that the TACs are the footprints of the presence of cometesimals, both in PMS and MS stars.
In this way, the thesis was defined as the search and study of TACs in a significative sample of stars (broad range of ages and masses) with CS disks.

The adopted observational strategy was to obtain a large amount of high resolution \'echelle spectra for a big sample of stars.
12 observing nights were used for this program in the framework of the EXPORT collaboration (see details in Chapter~\ref{observations}).
The WHT telescope (4.2m, La Palma) equipped with the UES spectrograph was used.
After the spectra were reduced, a database composed by 198 \'echelle spectra of 49 stars with possible CS disks was available, both primordial (PMS stars) and second generation (MS stars).
The PMS sample of stars includes UXORs, HAeBe with observed disks, HAeBe with spectroscopic behaviour similar to $\beta$~Pic, CTTSs, WTTSs and Early T~Tauri Stars (ETTSs).
The observed MS stars are of types Vega, A-shell with possible $\beta$~Pic activity and Post T~Tauri Stars (PTTSs) taken from the catalog of \citet{lindroos1986}.

The initial EXPORT sample of objetcs included in the observing proposal was biased towards intermediate mass objects: only a 30\% of the stars were T~Tauri or PTTSs.
On the other hand, the sensitivity of the instrument together with the low brightness of the late spectral type stars made the observation of low mass stars very difficult: only about 20\% of the total number of spectra obtained correspond to T~Tauri stars or PTTSs, and many of them have a low Signal to Noise Ratio (SNR).
These restrictions, imposed by the sample studied and the available instrumentation, have limited the research in this thesis to the intermediate mass stars, mainly of spectral type A.

A preliminary analysis, made after the reduction of the data, revealed that most of the MS stars do not present any TAC, with the noteworthy exception of HR~10, an A-shell star similar to $\beta$~Pic (see Section~\ref{introduction_beta_pic}).
However, there were only two spectra of HR~10 available, taken within an interval of three months.
These two spectra were not enough to make a thorough study of the properties and interpretation of TACs in MS stars, so this kind of objects was excluded from this thesis because, on the one hand, the sample would not be significant and, on the other hand, the time interval between the two spectra is excessively large.
It will be seen in Chapters~\ref{uxori} and~\ref{haebe} that the TACs vary in timescales of days and even hours.

Summarizing, after the preliminary analysis was conducted, three unavoidable requirements were identified in order to study the presence of planetesimals in CS disks: presence of TACs in the spectra, moderate to high SNR and good temporal coverage (many spectra per star with an approximate time delay of one day, during many nights).
This restricted the list of possible objects of study basically to a few HAe stars, most of them of UXOR type.
Finally, the five objects with the best temporal monitoring and larger spectral variability of the sample were selected: BF~Orionis, SV~Cephei, UX~Orionis, WW~Vulpeculae and XY~Persei.
The star HD~163296 was also studied for a while.
However, its strong stellar winds produce considerable alterations of the line profiles, so it was not possible to make an adequate characterization of the TACs present in its spectra.

BF~Ori, UX~Ori, WW~Vul \citep[e.g. see][]{grinin1991} and SV~Cep \citep{rostopchina2000} have been classified as UXORs.
\citet{oudmaijer2001} found a strong photopolarimetric variability in XY~Per, but they could not undoubtedly establish its membership to the UXOR class.
Since the spectral variability of XY~Per is, according to the TACs, identical to that of the remaining objects studied, from now on it will be considered an UXOR for all practical purposes.

Once the initial expectations of this study were delimited to the detection and characterization of planetesimals in the EXPORT subsample of UXOR stars, the question of the relation between TACs and planetesimals already remained.
As it has been said before, \citet{natta2000} demonstrated that, for a series of strong TACs detected in UX~Ori, the chemical abundances were approximately solar and not compatible with the evaporation of planetesimals.
The simulations made by \citet{beust2001} also led to the same conclusion, so the link between TACs and planetesimals became doubtful.
Due to the high relevance these results have for the present thesis, it was initiated a collaboration with Dr.~Natta (Astrophysical Observatory of Arcetri, Italy) to thoroughly study the connection between TACs and planetesimals.
We put in common the extensive experience with theoretical models of Dr.~Natta and our collection of excellent quality \'echelle spectra and our capability of synthesize photospheric spectra.

A substantial part of the collaborative efforts, devoted to an exhaustive study of the TACs in UX~Ori, were made during two short stays in the Astrophysical Observatory of Arcetri in 2000 (2 months) and 2001 (3 months)
It was verified that the TACs are not generally related to planetesimals but to CS gas clumps of solar metallicity, probably related to magnetospheric accretion phenomena.
It was also discovered that the extraordinary temporal coverage of the EXPORT spectra allowed to study, with an accuracy without precedents, the kinematics of the gas originating the TACs in UX~Ori.
These results have been published by \citet[from now on Paper I]{mora2002}.

Once it was realized the success of the analysis of UX~Ori, a similar study for the remaining selected stars was started: BF~Ori, SV~Cep, WW~Vul and XY~Per.
This new work has been successfully finished and has been published by \citet[from now on Paper II]{mora2004}.
The conclusions obtained in \citetalias{mora2002} and \citetalias{mora2004} do not substantially support the link between TACs and planetesimals, so it was needed to change again the subject and title of the thesis to its final form: ``Kinematics of the circumstellar gas around UXOR stars''.

\section{Objectives and carried out work}
\label{introduction_objectives}

The research carried out in this PhD thesis has not been developed in a linear way (see previous Section).
That is, there existed some initial objectives, according to the relevant part in the EXPORT observational proposal, but they were modified many times as a consequence of the partial results obtained during the research. 
Therefore, in this Section we do not describe the (obsolete) primitive objectives, but the significant milestones achieved during the whole PhD research process.

\begin{itemize}

\item
The first duty accomplished in this thesis was the obtention of the required observational data: a set of \'echelle spectra.
The observations were carried out in the framework of the EXPORT collaboration.
The student joined the collaboration when the observational proposal was written, the observing time allocated and many of the observations (2 out of 12 nights) performed (see Chapter~\ref{observations}).
The raw data were available from the very beginning of the research.

\item
Thanks to the good weather during the spectroscopic observations (100\% of clear nights), a total amount of 198 \'echelle spectra of many objects was obtained.
The reduction of the spectra required, along with the observing runs, the first year and a half of this thesis, and is described in Chapter~\ref{reduction}.

\item
Once the data were reduced, a preliminary inspection of the spectra, in order to classify the observed objects and estimate their interest was started.
The first results of this study were presented by \citet{mora2000a}.
The selection process of interesting objects has been long, because of the different changes in the thesis subject, due to the partial results that were obtained.

\item
In order to characterize properly the TACs in the reduced spectra, it was necessary to determine and subtract the photospheric spectrum of the stars studied.
It was decided to use synthetic spectra to estimate the photospheric spectra.
The codes SYNTHE and SYNSPEC were used for this purpose, so they had to be properly installed and, sometimes, modified.
The spectral synthesis process is described in detail in Sections~\ref{uxori_the_photospheric_spectrum} and~\ref{haebe_the_photospheric_spectra}.

\item
A fundamental parameter in the computation of stellar spectra is, along with the effective temperature and the gravity, the rotation velocity projected on the line of sight, \vsini.
A working group by Dr.~Montesinos, Dr.~Solano (who was the coordinator) and the student was constituted with the only purpose of systematically determine \vsini\ for all the stars in the EXPORT sample.
In this way, besides the obtention of fundamental parameters for the ulterior analysis of the data, a research of a great scientific value on its own was carried out.
The results have been published by \citet{mora2001}, along with the spectral type determinations for all the stars in the sample made by Dr.~Mer\'{\i}n.
In Chapter~\ref{vsini} a detailed exposition of the method employed and the results obtained for the stars in this thesis can be found.

\item
After the best photospheric spectrum is obtained for each star, it is subtracted from the observed spectrum.
The study of the subtraction residual $R$ has been done in several metallic and hydrogen lines and revealed the presence on many TACs in the spectra.
In general, there appeared several overlapped TACs in each spectrum.
In order to characterize simultaneously and self-consistently all the TACs multicomponent gaussian fits were used.
Estimations of the radial velocity $v$, velocity dispersion \deltav\ and intensity $R_{\rm max}$ were obtained for each TAC.
This process is described in Section~\ref{uxori_multigaussian_fitting}

\item
The study of the the multigaussian deconvolution data revealed that absorption components detected simultaneously in different lines can be grouped according to their radial velocity.
The origin and variability of these absorptions have been attributed to the dynamical evolution (acceleration or deceleration) of gas clouds present in the CS disks.
The kinematics of CS gas clumps in UXOR stars has been traced with a detail without precedents.
The method used in the identification of these clouds is exposed in Sections~\ref{uxori_results} and~\ref{haebe_circumstellar_contribution}.

\item
Apart from the infall and outflow velocities of the CS gas, the dispersion velocity and the intensity of the absorptions have also been studied in detail.
The possible cross-relationships between different parameters have also been analysed.
Highly significant correlations between $v$~vs.~\deltav\ and \deltav~vs.~$R_{\rm max}$ have been found (see Sections~\ref{uxori_kinematics} and~\ref{haebe_kinematics}).

\item
The study of the $R_{\rm max}$ values allowed us to examine what line multiplets are saturated for each gas blob and each star.
These results impose restrictions to the physical parameters of the cloud, remarkably the density of the CS medium, which should be included in any realistic model explaining the existence of TACs in UXOR stars (e.g. FEBs or magnetospheric accretion).
In some stars (UX~Ori and XY~Per) the intensity ratios in the Balmer lines have been related to the possible emission of radiation by the CS cloud.
The most relevant results can be found in Sections~\ref{uxori_r_parameter}, \ref{uxori_origin_cs_gas} and~\ref{haebe_r_ratios}.

\item
Finally, a metallicity analysis similar to that of \citet{natta2000} has been performed for the whole sample of stars in this thesis.
The results show that, in general, the CS gas clouds in UXOR stars have a solar or nearly solar metallicity.
Otherwise, WW~Vul seems to be a different object because, in addition to solar metallicity TACs, it shows metallic absorption components without an evident counterpart in hydrogen lines.
The significance of these results is discussed in Sections~\ref{uxori_origin_cs_gas}, \ref{uxori_gas_dynamics}, \ref{haebe_origin_cs_gas} and~\ref{haebe_wwvul}.

\end{itemize}

\section{Brief description of the structure of this thesis}
\label{introduction_thesis_structure}

Chapter~\ref{observations} explains all the information related to the observations used in this thesis: The initial observing proposal, the observations themselves and the final set of spectra.
Chapter~\ref{reduction} shows the details of the reduction procedures used in this thesis.
Chapter~\ref{vsini} explains the method used by \citet{mora2001} in the measurement of the projected rotation velocities for all the EXPORT stars.
Chapter~\ref{uxori} includes the complete study of UX~Ori, performed by \citet{mora2002} and originally published in Astronomy \& Astrophysics.
Chapter~\ref{haebe} consists of the the work carried out by \citet{mora2004} about BF~Ori, SV~Cep, WW~Vul and XY~Per, which was originally published in Astronomy \& Astrophysics.
Finally, Chapter~\ref{conclusions} contains the conclusions obtained in this PhD thesis.

\chapter{Observations}
\label{observations}

In this Chapter, the observational campaigns carried out to obtain the spectra used in this thesis will be described.
Section~\ref{observations_proposal} explains the scientific case of the observing proposal, composed of three different subprograms.
The data used in the present research are only a fraction of those obtained in the subproposal devoted to the study of CS disks, whose main features are discussed in Section~\ref{observations_subprogramme_disks}.
Section~\ref{observations_export_echelle} contains some general remarks about the observations and the database generated with the reduced spectra.
The detailed information relative to the observations of the stars studied in this thesis (number and distribution of the spectra, SNR, etc.) are not discussed here but in Chapters~\ref{uxori} and~\ref{haebe} (Sections~\ref{uxori_observations} and~\ref{haebe_observations}).

\section{The EXPORT observational proposal}
\label{observations_proposal}

The spectra used in this thesis were obtained during the observing runs awarded to the EXPORT collaboration\footnote{The web page of the EXPORT collaboration, {\tt \href{http://laeff.esa.es/EXPORT/}{http://laeff.esa.es/EXPORT/}}, contains detailed information about its objectives, proposal, observations and results.} during the International Time (see below) in 1998 and 1999.
One of the main objectives of the EXPORT proposal was to perform a precursor scientific study for the DARWIN mission of the European Space Agency (ESA).
DARWIN is one of the most ambitious projects currently managed by ESA.
The mission involves the development of an infrared interferometer in space capable of detecting and characterizing extrasolar telluric planets.
The DARWIN project, apart from being extremely expensive, requires the development of several technological precursor missions, which would address the viability of the new technologies to be implemented in the real mission.
Therefore, there is not currently a realistic schedule to carry out the  project.
The EXPORT proposal \citep{eiroa2000a}, titled ``Planetary systems: their formation and properties'', tried to contribute to the scientific case for DARWIN, conducting astronomical research of high scientific value on their own.

The observing proposal was submitted in 1997.
At that time, the detections of extrasolar planets that came after the discovery of 51~Peg~B by \citet{mayor1995} were questioning the validity of the accepted star formation paradigm.
The EXPORT collaboration, composed of about twenty European and American astronomers, was constituted with the main objective of applying for the ``International Time''\footnote{The ``International Time'' amounts the 5\% of the total observing time available} available at the Canary Islands telescopes in 1998.
The proposal, which was awarded with the total International Time, consisted of the following research lines:

\begin{itemize}

\item
Formation and evolution of planetary systems via observational studies of protoplanetary disks around PMS and MS Vega-type stars.
The allocated telescope time was:
12 nights with the WHT (4.2~m, UES high resolution \'echelle spectrograph),
16 nights with the INT (2.5~m, IDS intermediate resolution spectrograph),
16 nights with the NOT (2.5~m, TURPOL photopolarimeter) and
15 nights with the CST (1.5~m, infrared photometer and CAIN infrared camera).

\item
Search for planetary atmosphere signatures in the spectra of $\tau$~Boo and 51~Peg.
This subprogramme was assigned 4 nights with the WHT telescope equipped with the UES spectrograph.

\item
Search for new exoplanets by means of photometric transits in stellar clusters and microlensing techniques: 16 nights with the JKT telescope (1~m, CCD camera) were allocated for the quest for planets via transits.
A certain amount of alert time for the microlensing searches for planets was allocated in the IAC-80 telescope (0.8~m, CCD camera).

\end{itemize}

Table~\ref{observations_schedule} shows a summary of all the observing time granted to the EXPORT collaboration (excluding the microlensing alert time with the IAC-80 telescope).
The weather was quite good: 100\% of the nights were adequate for the spectroscopy programmes and more than half of the nights were photometric.

\begin{table}
\caption[EXPORT observations schedule]
{EXPORT observations schedule, excluded the microlensing alert time available in the IAC-80 telescope.
WHT (4.2~m): high resolution spectroscopy.
INT (2.5~m): intermediate resolution spectroscopy.
NOT (2.5~m): $UBVRI$ photopolarimetry.
CST (1.5~m): $JHK$ photometry.
JKT (1.0~m): CCD images.}
\label{observations_schedule}
\begin{center}
\begin{tabular}{lcccc}
\hline
\hline
Telescope & May   1998 & July  1998 & October 1998 & January 1999 \\
\hline
WHT       & 14-17      & 28-31      & 23-26        & 28-31        \\
INT       & 14-17      & 28-31      & 24-28        & 29-31        \\
NOT       & 14-17      & 28-31      & 23-27        & 29-31        \\
CST       & 15-17      & 28-31      & 23-26        & 28-31        \\
JKT       &  --        & 25-31      & 24-1         &  --          \\
\hline
\end{tabular}
\end{center}
\end{table}

\section{Formation and evolution of circumstellar disks subprogramme}
\label{observations_subprogramme_disks}

The objectives of this subprogramme were the following:

\begin{enumerate}

\item
Study of the evolution of CS disks, from the PMS (dense protoplanetary disks) to the MS stages (second generation dust disks).

\item
Characterization of the planetesimals during the whole disk lifetime: appearance-disappearance, absorption profiles and chemical abundances.
As a working hypothesis it was assumed that the TACs are the signature of the evaporation of planetesimals (as it probably happens in $\beta$~Pic) both in MS and PMS stars.

\end{enumerate}

The selected sample consists of stars with possible CS disks: UXOR with detected TACs, PMS with possible UXOR photopolarimetric behaviour, HAe near the ZAMS, Vega-type, A-shell with detected TACs and PTTSs.
In order to be as more representative as possible, the sample studied includes objects with a broad range of ages and stellar masses.
However, the sample is biased towards intermediate mass stars.

It has been shown in the previous Section that the observations allocated to this subprogramme consist of optical high and intermediate resolution spectroscopy, optical photopolarimetry and infrared photometry.
Many of the objects studied have also been observed with the infrared space telescope ISO, whose data archive was released in 1999.
Because of their great interest, the available mid-infrared ISO spectra, obtained with the SWS spectrograph, were studied.

The study of the disks required the photopolarimetric data and the ISO spectra.
The initial objective was to combine the construction of excellent quality SEDs (due to the simultaneity of the photometric data) with the study of the solid state features present in the ISO spectra.
The SED construction for all the stars in the sample is one of the most interesting issues of the PhD thesis of Dr~Mer\'{\i}n, which has been recently defended \citep{merin_thesis}.
The study of the ISO spectra has been restricted to the characterization of the $\lambda \sim 10 \mu$m silicate emission band \citep{palacios2000}.

The characterization of the TACs for the EXPORT sample is exactly the subject of this thesis.
In principle, all the available observations should have been used.
The intermediate resolution spectra were expected to be the most important tool, because they should have allowed to perform the majority of the detections and statistics of TACs.
The high resolution spectra would have been dedicated to the detailed characterization (kinematics, profiles, abundances) of the most interesting TACs.
The visible photopolarimetry and the infrared photometry should have been used to study the relation between the photometric variability (e.g. UXOR minima) and the spectroscopic activity.
Finally, the study of the ISO spectra should have revealed traces of the CS solid material.

It has been shown in Section~\ref{introduction_thesis_subject} that, after an initial inspection of the data, it was decided to study the TACs only in UXOR stars.
The variability showed by the stars studied was small, with the remarkable absence of deep photometric minima.
No correlation between the photopolarimetric and spectroscopic variabilities was detected.
It was also realized that, in order to characterize properly TACs (number, velocity, width and intensity), high resolution was absolutely needed.
Therefore, the intermediate resolution spectra were excluded from the study.
Finally, when it was confirmed that the TACs in UXOR stars are not generally related to the evaporation of planetesimals but to gas accretion, it was decided to defer the study of the ISO spectra for a future research.
In summary, the only observations used in the detailed analysis of the TACs are the high resolution \'echelle spectra.

\section{EXPORT observations of \'echelle spectra}
\label{observations_export_echelle}

It has been said in Section~\ref{observations_proposal} that 12 nights of telescope were allocated for the \'echelle observations devoted to the evolution of CS disks.
The nights were scheduled in 4 blocks, separated by 3 months intervals: 2 nights in May 1998, 4 in July 1998, 2 in October 1998 and 4 in January 1999 \citep{mora2000a}.
This strategy permitted to perform long-term, short-term and very-short-term monitoring (months, days and hours, respectively) of the most interesting objects.

The instrument used was the ``Utrecht Echelle Spectrograph'' (UES), which was permanently located in a Nasmyth focus of the ``William Herschel Telescope'' (WHT, 4.2~m) during all its operational life.
The selected configuration of UES provided spectra in the range 3800-5900~\AA\, with a spectral resolution $R = \lambda / \Delta\lambda = 49\,000$.
A CCD detector of $2048 \times 2048$ pixels was used to record the spectra.
The overlap between contiguous orders rendered a continuous coverage in the mentioned spectral range.
The projection on the sky of the selected entrance slit was 1.15 arcseconds.

The reduced spectra consist of a set of 198 spectra of 49 astronomical objects.
Table~\ref{observations_objects} shows a detailed list of them.
The classification in the table corresponds to that made by \citet{mora2001}, according to the spectral type, SED and photopolarimetric behaviour of each star.
The integration times range from 10 minutes for the brightest objects (e.g. HR~10, $V = 6.23$, SNR~$\sim$~2400) to 45 minutes for the faintest stars (e.g. VV~Ser, $V \sim 11.9$, SNR~$\sim$~14).
The observed limiting magnitude is about $V \simeq 12$ and corresponds to 45~minutes exposures.
Larger integrations have not been performed, because of the cosmic rays ($\sim$2\% of affected pixels for 45~minutes exposures) and the need to obtain spectra for a large sample of stars.

\begin{table}
\caption[List of high resolution \'echelle spectra obtained by the EXPORT collaboration with the WHT telescope (4.2~m, La Palma)]
{List of high resolution \'echelle spectra obtained by the EXPORT collaboration with the WHT telescope (4.2~m, La Palma).
The classification shown corresponds to that made by \citet{mora2001}, according to the spectral type, SED and photopolarimetric behaviour of each star.
Column~3 includes the total number of exposures obtained per star.}
\label{observations_objects}
\begin{center}
\begin{tabular}[t]{p{5.5em}lc}
\hline
\hline
Name            & Type          &\# exp.\\
\hline
HD 23362        & CTTS          & 1     \\
HD 31293        & HAeBe         & 1     \\
HD 31648        & HAeBe         & 4     \\
HD 34282        & HAeBe         & 2     \\
HD 34700        & ETTS          & 2     \\
HD 58647        & HAeBe         & 3     \\
HD 109085       & Vega          & 4     \\
HD 123160       & CTTS          & 6     \\
HD 141569       & HAeBe         & 11    \\
HD 142666       & HAeBe         & 6     \\
HD 142764       & Vega          & 2     \\
HD 144432       & HAeBe         & 4     \\
HD 163296       & HAeBe         & 9     \\
HD 190073       & HAeBe         & 2     \\
HD 199143       & MS            & 2     \\
HD 233517       & CTTS          & 4     \\
HR 10           & A-shell       & 2     \\
HR 26   A       & MS            & 2     \\
HR 419          & PTTS          & 1     \\
HR 2174         & Vega          & 1     \\
HR 4757 A       & MS            & 3     \\
HR 4757 B       & PTTS          & 2     \\
HR 5422 A       & MS            & 2     \\
HR 9043         & Vega          & 2     \\
BD+31 643       & Vega          & 5     \\
\hline
\end{tabular}
\hspace{0.7cm}
\begin{tabular}[t]{p{5.5em}lc}
\hline
\hline
Name            & Type          &\# exp.\\
\hline
$\lambda$ Boo   & Vega          & 13    \\
VX Cas          & HAeBe         & 3     \\
BH Cep          & ETTS          & 8     \\
SV Cep          & HAeBe         & 7     \\
49 Cet          & Vega          & 2     \\
Z  CMa          & HAeBe         & 2     \\
24 CVn          & A-shell       & 4     \\
51 Oph          & HAeBe         & 8     \\
KK Oph          & HAeBe         & 5     \\
T  Ori          & HAeBe         & 1     \\
BF Ori          & HAeBe         & 4     \\
CO Ori          & ETTS          & 3     \\
HK Ori          & ETTS          & 1     \\
NV Ori          & ETTS          & 1     \\
RY Ori          & ETTS          & 1     \\
UX Ori          & HAeBe         & 10    \\
XY Per          & HAeBe         & 9     \\
VV Ser          & HAeBe         & 11    \\
17 Sex          & A-shell       & 4     \\
CQ Tau          & ETTS          & 1     \\
DR Tau          & CTTS          & 2     \\
RR Tau          & HAeBe         & 3     \\
RY Tau          & ETTS          & 4     \\
WW Vul          & HAeBe         & 8     \\
\hline
\end{tabular}
\end{center}
\end{table}

\begin{figure}
\centerline{\includegraphics[width=0.80\textwidth,clip=true]
           {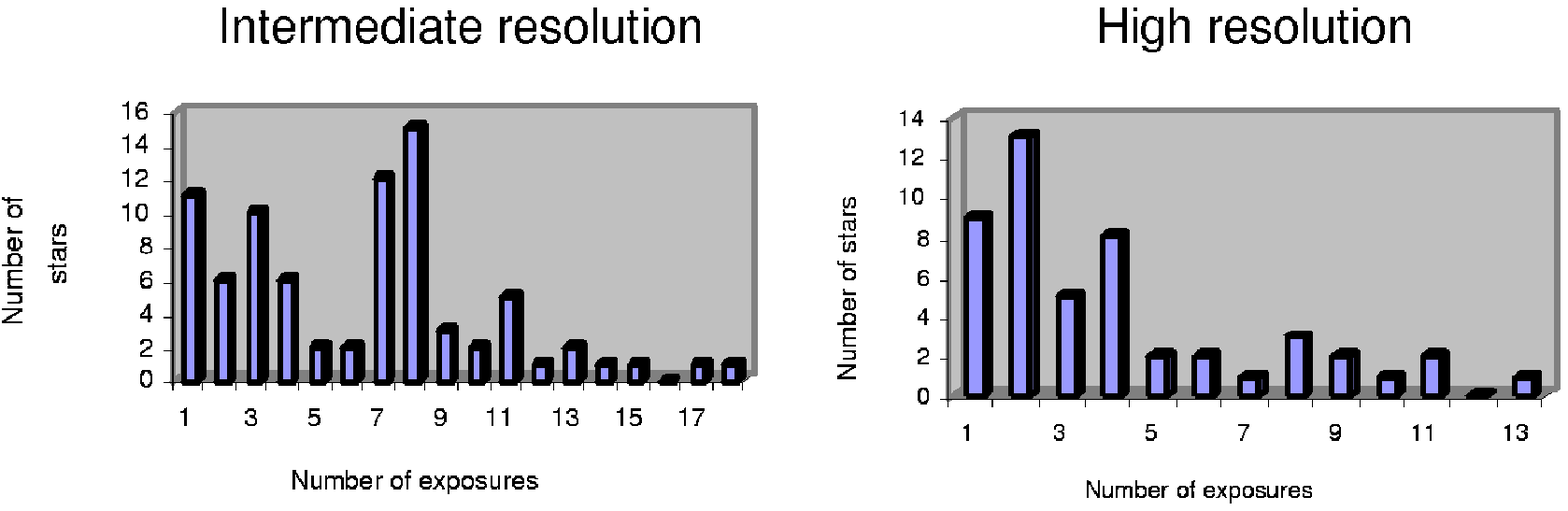}}
\caption
[Histogram of the number of spectra obtained per star]
{Histogram of the number of spectra obtained per star.
A significant number of stars was observed only once, because of their apparent lack of CS material (evaluated in real-time from the inspection of the intermediate resolution INT spectra) or their low brightness.
Thanks to the information gathered from the intermediate resolution spectra, the observations could be optimized and the most interesting objects were more frequently observed.}
\label{observations_histogram}
\end{figure}

Figure~\ref{observations_histogram} shows a histogram of the number of UES spectra obtained per star.
It can be appreciated that a significant number of objects was observed only once.
This was due to two reasons.
The first one is that the intermediate resolution INT spectra obtained simultaneously were real-time analysed to evaluate the relative interest of the objects.
In this way, those stars not showing significant traces of CS material in the INT spectra were rejected for subsequent observations.
This approach has been generally valid.
However there are exceptions, e.g. RY~Ori \citep{eiroa2000b} presents a normal intermediate resolution spectra, but shows a large amount of TACs in its only high resolution spectrum.
The second reason is that some stars were too faint to obtain enough high SNR UES spectra.
Using the real-time information provided by the INT and observing with the highest possible frequency those objects of greater apparent interest allowed us to optimize the observational strategy.

Currently, all the data obtained by the EXPORT collaboration are available to the scientific community.
All the photopolarimetric data have been published and analysed \citep{oudmaijer2001,eiroa2001,eiroa2002}.
Many studies have been done to exploit the spectroscopic data obtained by the collaboration \citep{mora2001,mora2002,mora2004,merin2004,baines2004}.
Because of the large amount of data involved and the peculiarities of the reduction, the spectra are only available under request.

\chapter{Reduction of the observational data}
\label{reduction}

In this chapter, the method used in the reduction of the high resolution EXPORT spectra is described.
The EXPORT database is composed of 198 \'echelle spectra (see Chapter~\ref{observations}).
38 of them have been used in this thesis (4 of BF~Ori, 7 of SV~Cep, 10 of UX~Ori, 8 of WW~Vul and 9 of XY~Per).
The reduction of \'echelle spectra is mostly a standardized process, which is described from Section~\ref{reduction_software} to Section~\ref{reduction_calibration}.
However, the normalization of the extracted spectra is not trivial.
In Section~\ref{reduction_normalization}, the normalization procedure applied to the present set of data is reviewed.
In Section~\ref{reduction_validity}, the quality of the reduced spectra and the limitations shown by the data, in order to be used in other researches different from this thesis, are discussed.

\section{Characteristics of an \'echelle spectrum}
\label{reduction_characteristics}

In general, \'echelle diffraction gratings are used to obtain high resolution spectra $R = \lambda / \Delta \lambda \ga 50\,000$ \citep{schroeder2000}.
These gratings are designed to operate in high interference orders (m~$\sim$~100).
Thus high dispersion is obtained, but the free spectral range severely diminishes for each order ($\lambda_{\rm max} - \lambda_{\rm min} \sim 100 \, \AA$) and a spatial overlap occurs between contiguous orders.
Special filters that only transmit the light of a spectral order can be used if only a small wavelength range, completely contained in an spectral order, is required.
However, if a large spectral coverage is wanted, a low dispersive power optical element, in addition to the \'echelle grating, can be used.
Thus the different orders can be spatially separated and a large portion of the spectrum ($\lambda_{\rm max} - \lambda_{\rm min} \ga 2000 \, \AA$) can be recorded in a single CCD detector.

\begin{figure}
\centerline{\includegraphics[width=0.75\textwidth]{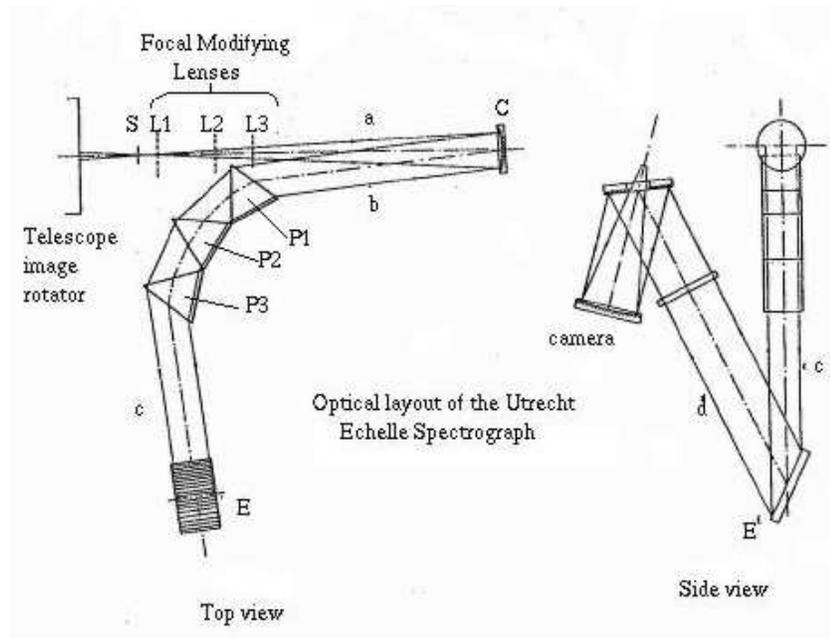}}
\caption[Optical diagram of the UES \'echelle spectrograph]
{Optical diagram of the UES \'echelle spectrograph, taken from \citet{manual_ues}.
The high dispersive power element (\'echelle diffraction grating: E) and the low dispersive power component (set of three different prisms: P1, P2 and P3) can be easily identified.}
\label{reduction_ues}
\end{figure}

The spectra used in this thesis were obtained with the UES \'echelle spectrograph placed on the WHT telescope (see Chapter~\ref{observations}).
An optical diagram of UES is shown in Figure~\ref{reduction_ues}, taken from \citet{manual_ues}.
The \'echelle grating and the cross disperser element, composed of three different prisms, can be clearly identified.
The instrument was set up to obtain spectra in the wavelength range 3800-5900~\AA\ with a resolution $\lambda / \Delta \lambda = 49\,000$.
49 diffraction orders (m~=~95-153) of the grating were used.
Each order covers an approximate range of 100~\AA.
The spectra were recorded on a square CCD detector of $2048 \times 2048$ pixels.

\begin{figure}
\centerline{\includegraphics[width=0.5\textwidth,trim=0 219 0 0]
                            {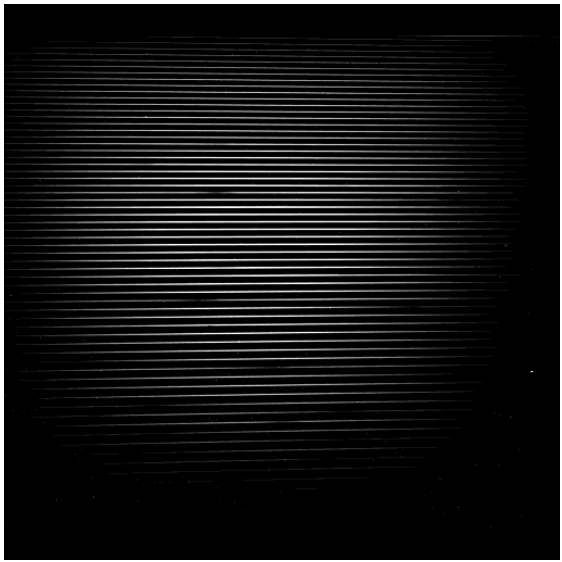}}
\caption[Raw \'echelle spectrum prior to the reduction.]
{Raw \'echelle spectrum of the star UX~Ori.
The spectrum is divided into several approximately horizontal strips.
Each band corresponds to an interference order of the diffraction grating.
The interference order increases from top (m~=~95) to bottom (m~=~153).
For each individual order, the wavelength increases from left to the right.}
\label{reduction_raw_spectrum}
\end{figure}

The UES raw spectrum of a hot star is shown in Figure~\ref{reduction_raw_spectrum}.
The different spectral orders are shown as narrow approximately horizontal strips.
The wavelength increases from left to the right.
The spectral direction is parallel to the $x$ axis whereas the spatial direction runs through the $y$ axis.
The interference order increases from top to bottom.
The orders with the highest interference number have the lowest wavelength, the highest width and the greatest inter-order spatial separation.

The whole process of \'echelle reduction processes can be classified into two categories.
On the one hand, those corrections derived from the use of CCD detectors: bias, flat-field and cosmic rays.
On the other hand, those specific tasks of the \'echelle reduction: Extraction of the orders, scattered light subtraction, wavelength calibration, and continuum normalization.

\section{Software}
\label{reduction_software}

The astronomical environment MIDAS has been used both for the reduction and the analysis of the data.
This software package contains basic image manipulation commands and high level \'echelle reduction procedures \citep{midas1999a,midas1999b}.
The IRAF routine {\tt continuum} was also used during the normalization of the extracted spectra.
In addition, many programs have been developed by the author in the MIDAS programming language.
They provided a certain degree of automation in the reduction (i.e. they constitute a pipeline) and rendered the tools to normalize the continuum.

\section{Bias correction}

The first step in the whole reduction procedure is the bias subtraction of the raw images.
The bias is an electrical potential, which is added to that created by the electrons stored in each pixel of a CCD prior to their passage through the analog to digital converter.
Thus it is assured that all the pixels will supply a positive number of counts, because the read-out noise may randomly generate an incorrect negative number of counts in pixels with very little or no stored photoelectrons.
Two beneficial effects arise if negative counts are avoided.
On the one hand, one encoding bit per pixel is saved.
On the other hand, the possibility that a pixel with a large number of stored electrons displays wrongly a negative number of counts is prevented \citep{howell2000}.

If the temperature of the CCD is kept fixed during the whole night, the bias value should be constant.
It is enough to take zero seconds exposures with the aperture closed to estimate the level of counts introduced by the bias.
The bias frames are almost constant and their median is the bias value.
This number must be subtracted from all the spectra before any other correction is performed, because the counts do not correspond to photons captured by the CCD.

It was verified during the reduction that the bias does vary during the night.
Perhaps it was due to problems in the thermal isolation of the CCD, or to the use of the maximum read-out speed available by the camera electronics.
This effect is unimportant, since the dispersed light subtraction corrects these variations afterwards.

The bias values were around 800 counts.
It was observed that this value strongly depends on the CCD read-out speed.

\section{Detection of \'echelle orders}
\label{reduction_detection}

\begin{figure}
\centerline{
\includegraphics[width=0.50\textwidth]
                {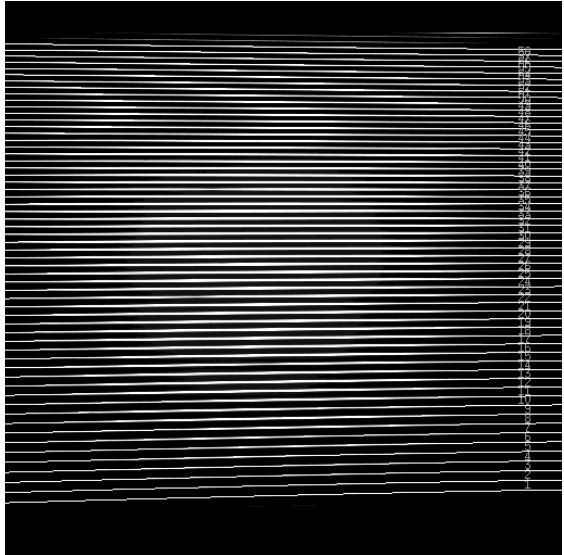}}
\caption[Detection of \'echelle orders]
{Detection of \'echelle orders.
A raw spectrum, similar to that of Figure~\ref{reduction_raw_spectrum} is shown.
The detected orders polynomial function, obtained from the fit of a standard star, is overplotted as a set of white lines.
Each order is identified by a number, which is not equal to the interference order of the \'echelle grating.
It can be seen that the orders detection is excellent, because it is almost impossible to distinguish the observed spectrum from the fit.}
\label{reduction_orders_detection}
\end{figure}

Each of the \'echelle orders is an independent spectrum which has to be reduced.
Each order has a different curvature, width and position in the CCD, as it can be seen in Figure~\ref{reduction_raw_spectrum}.
The first step in the \'echelle reduction procedure is the characterization of all these parameters for each order.
As they are nearly horizontal, the shape of the orders can be expressed via the following equation.

\begin{equation}
y = f(x,m)
\end{equation}

\noindent where $y$ is the position of the order centroid in the ordinate axis for the abscissa $x$ value.
It has been verified that $f$ can be properly approximated by low order polynomials \citep{midas1999b}.
The polynomials used for the UES spectra are of 3$^{\rm rd}$ degree in $x$ and 4$^{\rm th}$ degree in $m$.
It was verified that the position of the orders did not show significant variations during each night.
Therefore, it was enough to perform a single order detection for each observing night.
The detection was made using very bright standard stars, which gave a good SNR in the whole set of orders.
In Figure~\ref{reduction_orders_detection} a raw stellar spectrum is shown.
The polynomial function obtained from the fit to a standard star is also displayed.
It can be appreciated that the agreement is excellent.

The width of the orders in the spatial direction is needed to extract correctly the spectrum.
Vertical cuts were made in the observed spectra.
It was verified that the orders have spatial profiles approximately gaussian in shape.
In this way, the total width of an order is defined as 6\,$\sigma$, being $\sigma$ the dispersion value obtained from a gaussian fit to the spatial profile of the order.
This value maximizes the SNR of the spectrum, because all the stellar light is extracted and the less possible amount of noise is added.

As MIDAS uses the same extraction slit for all the spectral orders, the value adopted for the slit is the width of the bluest orders, which are the broadest ones.
The width of the spatial profile strongly depends on the seeing, so it was necessary to compute (using a program) a different slit size for each spectrum.

\section{Scattered light subtraction}

\begin{figure}
\includegraphics[angle=-90,width=0.49\textwidth]
                {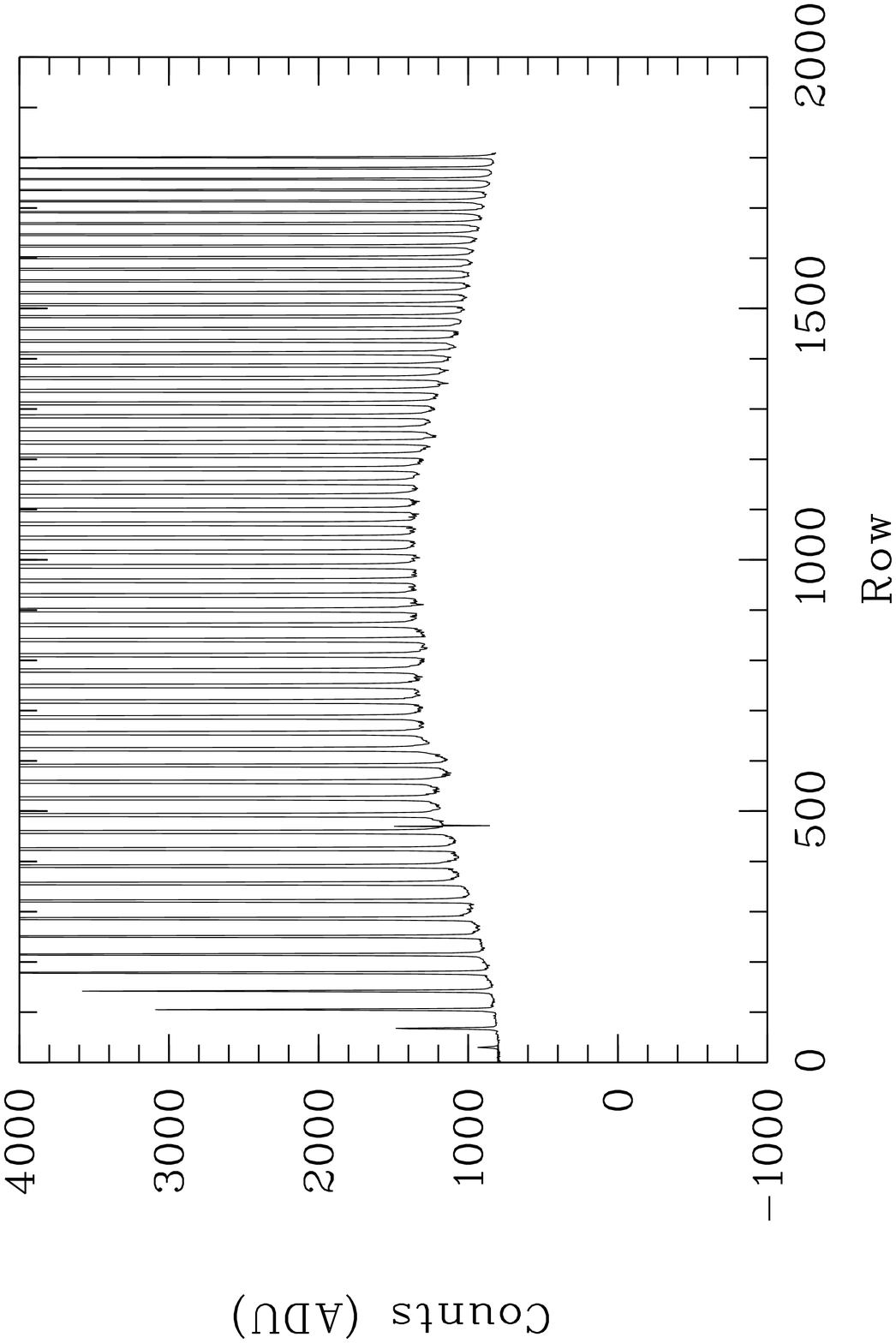}
\hspace{0.01\textwidth}
\includegraphics[angle=-90,width=0.49\textwidth]
                {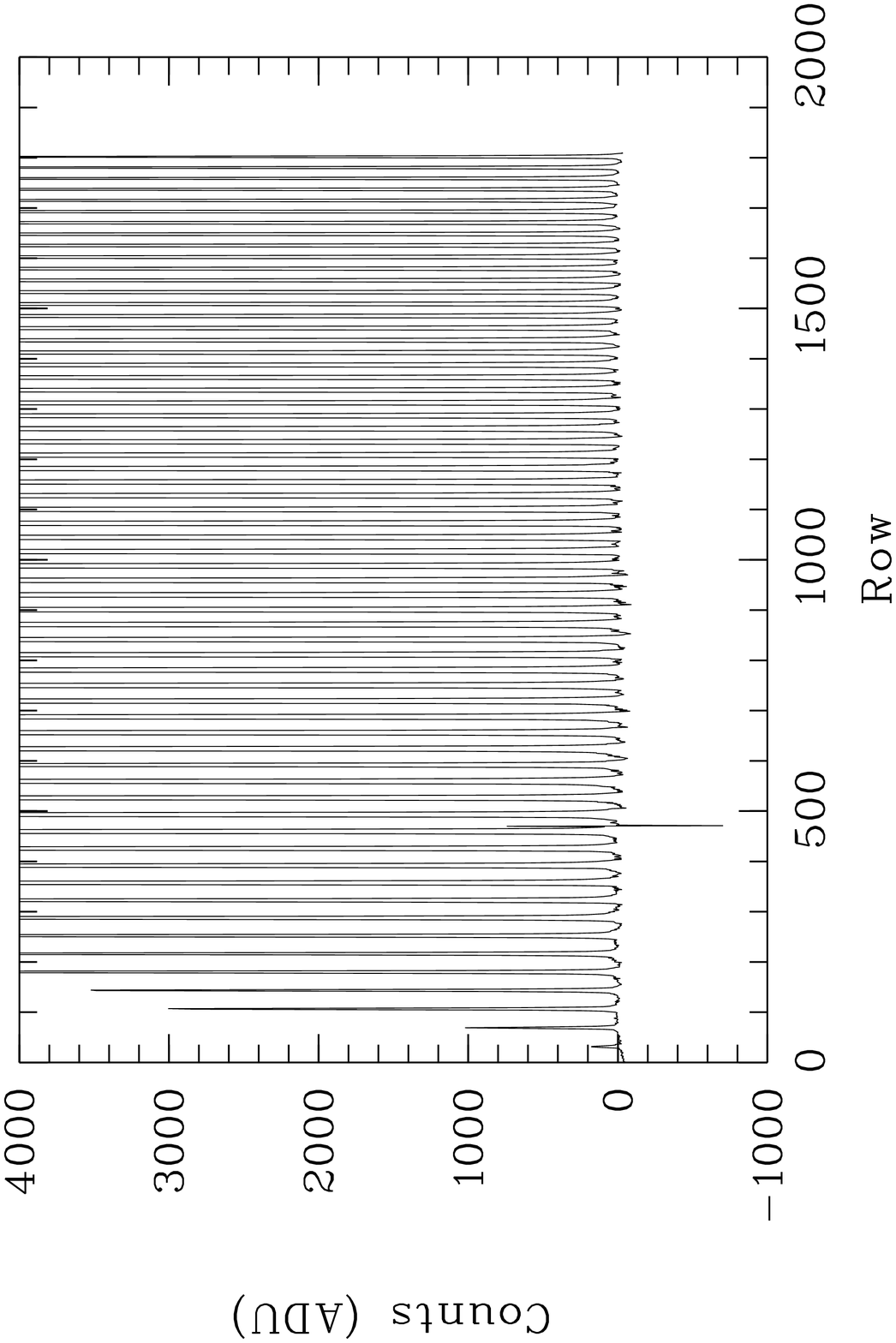}
\caption[Dispersed light subtraction]
{Dispersed light subtraction.
Left: the figure shows a cross section of an \'echelle spectrum in the spatial direction.
The orders can be identified as narrow profiles, approximately gaussian in shape, superimposed over a light continuum of smooth spatial variability.
In order to remove the dispersed light, the continuum is approximated by a bidimensional cubic spline polynomial.
This function is created from a suitable sampling of the inter-order space.
Right: a cross section of the same \'echelle spectrum in the spatial direction after the dispersed light has been subtracted.}
\label{reduction_background}
\end{figure}

The spectrographs have many imperfections that originate additional sources of illumination on the CCD.
Therefore, the light collected by the CCD is the sum of the astronomical spectrum and a scattered light continuum.
Figure~\ref{reduction_background} shows a cross section of an \'echelle spectrum in the spatial direction.
In the figure, the orders appear as narrow profiles superimposed over a light continuum of smooth spatial variability.
The dispersed light is not proportional to the intensity of the stellar spectrum, so it must be removed prior to the extraction of the spectrum.
Otherwise, the equivalent width of all the spectral lines would be underestimated.

The dispersed light presents a smooth variation across the CCD, so its intensity can be estimated for all the pixels using the values obtained from the inter-order space.
This space is sampled at regular intervals, in order to create a dense network of data points.
A point for every 100 pixels was used in the spectral direction of the inter-order space for the UES spectra.
These points were used to generate a bidimensional cubic spline polynomial that reproduces accurately the dispersed light continuum.
Once the dispersed light is estimated for every pixel, its subtraction is immediate.
An example of this procedure is displayed in Figure~\ref{reduction_background}.

\section{Removal of cosmic rays}

\begin{figure}
\includegraphics[height=0.49\textwidth,angle=-90]
                {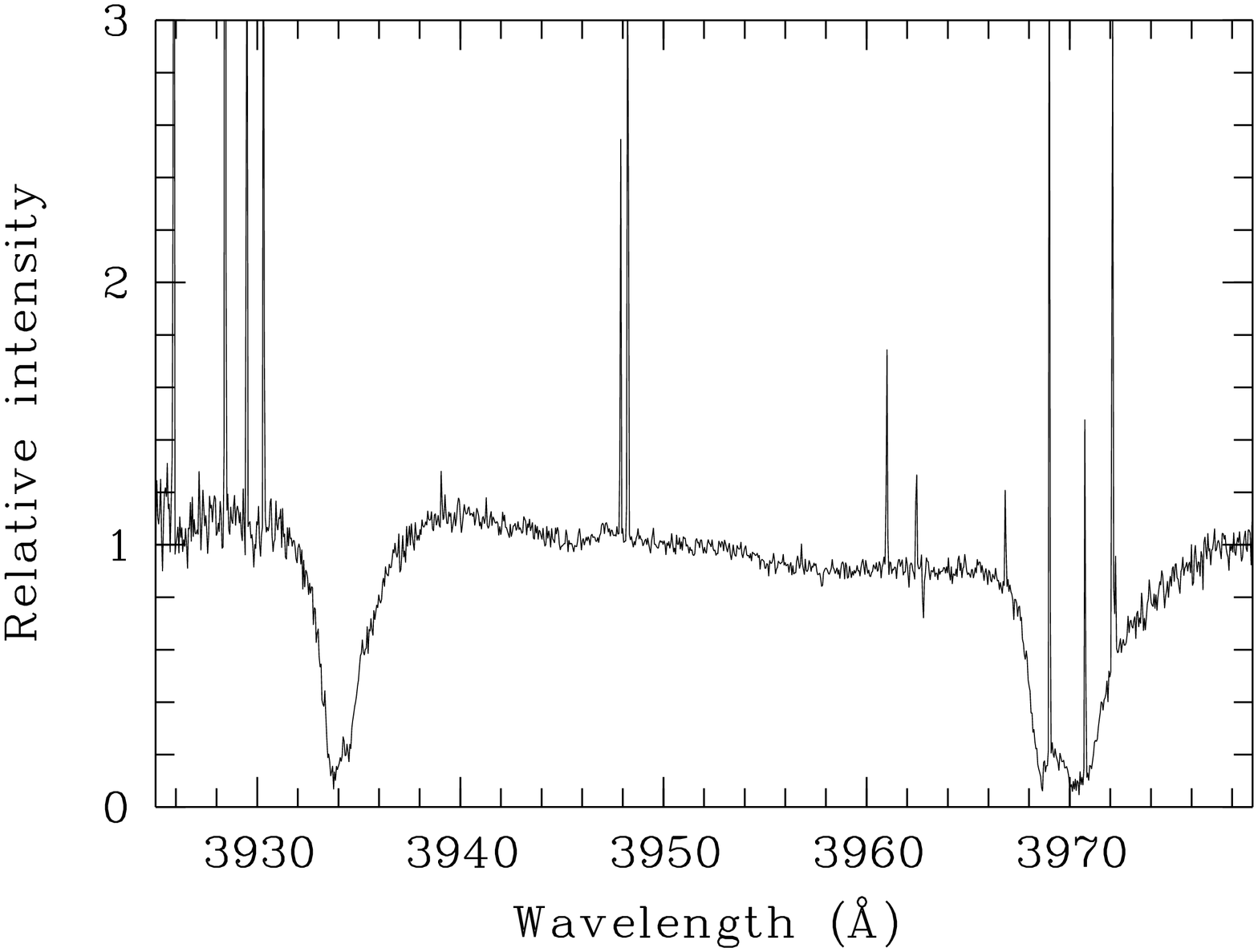}
\hspace{0.01\textwidth}
\includegraphics[height=0.49\textwidth,angle=-90]
                {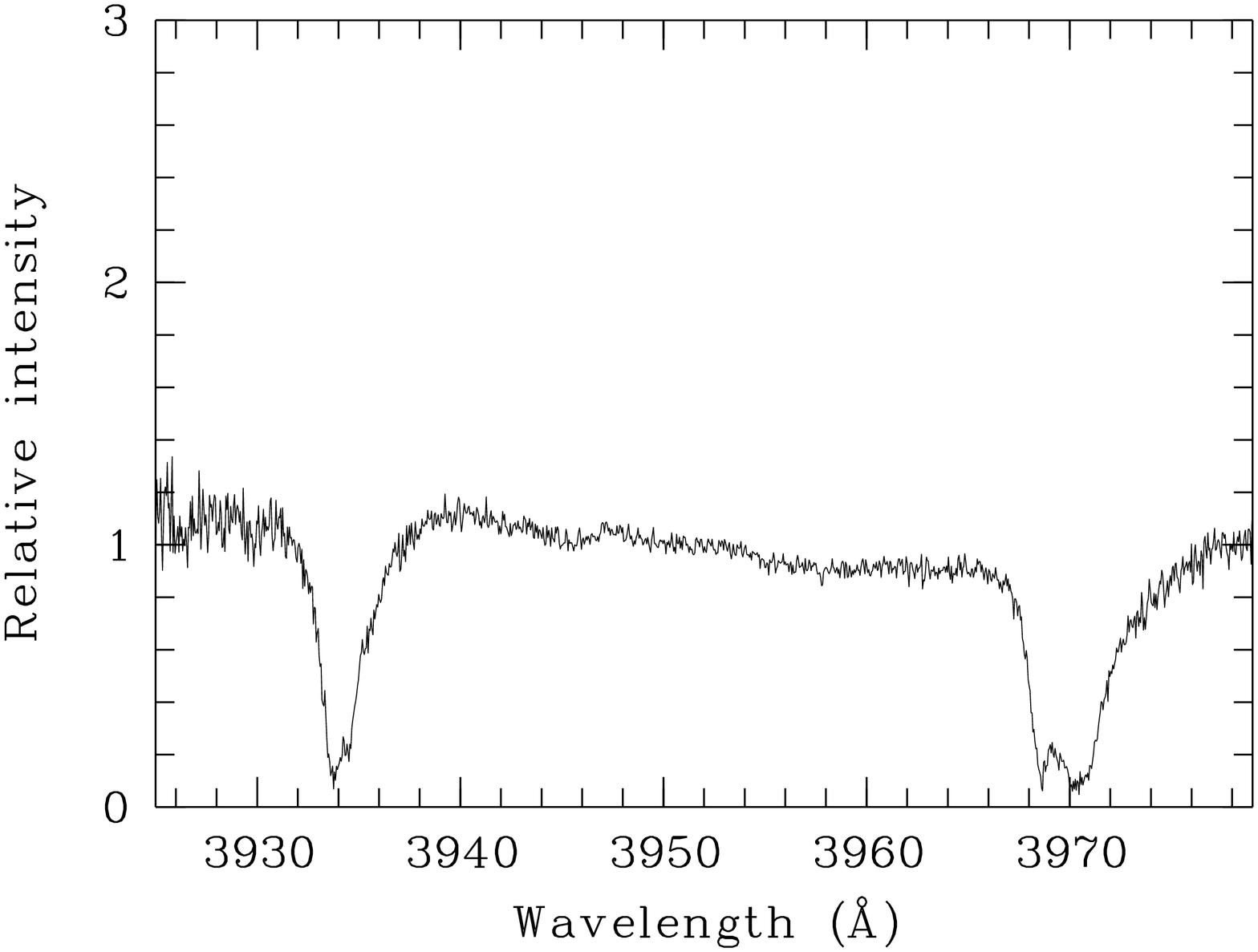}
\caption[Removal of cosmic rays]
{Removal of cosmic rays.
Left: the figure shows the spectral region of the \ion{Ca}{ii} doublet for WW~Vul.
The spectrum has been reduced without applying the subtraction of cosmic rays process.
Right: the same spectrum is shown, but the algorithm for identification and removal of cosmic rays has been applied during the reduction procedure.
It can be verified that the process has subtracted all the cosmic rays but the absorption lines remain unaltered.}
\label{reduction_cosmic_rays}
\end{figure}

The cosmic rays interact with the CCDs during the observations.
When a cosmic ray hits a pixel, many photoelectrons are generated, because the particle is charged and it has a large amount of kinetic energy.
Thus, the information stored in the pixels that have interacted with the cosmic ray (usually $n \sim$~1-5) is lost.
If a spectrum polluted by cosmic rays is extracted, there randomly appear very narrow peaks (1-3 pixel wide), apparently in ``emission'' (see Figure~\ref{reduction_cosmic_rays}).

The high resolution of the \'echelle spectra makes the spatial profile of each order to slowly vary along the spectral direction.
This effect can be used to identify the affected pixels and to reconstruct the stellar spectrum, to a certain extent.
Thus, the spatial profiles along all the pixels in the spectral direction of each order are extracted, and an ``observed'' profile is obtained for each pixel along each order.
For each pixel and each order, the median of the 30 nearest observed profiles along the spectral direction is computed, so two sets of ``median'' and ``RMS error'' (Root Mean Square, RMS) profiles can be obtained.
The method for detecting and rejecting cosmic rays is a simple sigma-clipping peak rejection algorithm.
If for any  pixel in the observed profile, the difference between the observed and the median values is greater than 4 times the RMS error (4\,$\sigma$), then the observed value is replaced by the median value.

This method fails when the fraction of pixels affected is so high that it is not possible to remove them from the median profiles.
Thus a practical limit to the duration of the CCD integrations is imposed, because the number of cosmic rays grows linearly with time.
The longest exposures taken lasted for 45~minutes and yielded fractions of bad pixels of about $\sim$2\%.

It has been verified a posteriori that this method generally produces very good results.
An example is shown in Figure~\ref{reduction_cosmic_rays}.
However, the algorithm has sometimes ruined the continuum of the spectrum, because periodic spurious absorptions have been introduced in the reddest orders.
The width of these artificial absorptions is about $\sim$5~\AA\ and the depth is around $\sim$5\% of the continuum.

\section{Flat-field correction}

The flat-field correction has to be taken into account because of the variations of sensitivity between different pixels in a CCD.
These differences are generally small.
However, dangerous defects can be present, for example cool or hot pixels, which present a substantially lower or higher quantum efficiency, respectively.
Those bad pixels can be isolated or may belong to defective columns.
These columns are specially damaging because they can destroy the information contained in a whole order.

\begin{figure}
\centerline{
\includegraphics[angle=-90,width=0.5\textwidth]
                {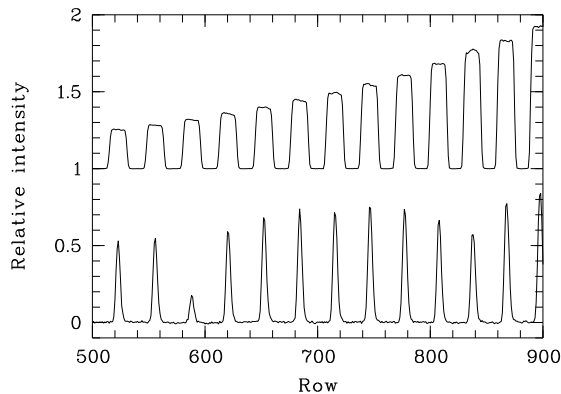}}
\caption[Spatial profile of a flat-field and an astronomical object]
{Spatial profile of a flat-field (top) and an astronomical object (bottom).
It can be appreciated that the orders of the astronomical object have an spatial profile approximately gaussian in shape and they are contained inside those of the flat-field, whose shape is similar to a square wave.
The origin of the difference is that the observed object is a point source (a star), while the flat-field is similar to the spectrum of an extended object.
The flat-field is originated when the spectrograph flat-field entrance slit (which is larger than that of the astronomical objects) is uniformly illuminated with a tungsten lamp.}
\label{reduction_flat_field}
\end{figure}

Flat-field exposures are used to identify and correct this undesirable effect.
The name comes from the techniques developed for the observation and reduction of astronomical images.
For this kind of observations it is possible to obtain images of fields uniformly illuminated (flat-fields), so any deviation from flatness can be attributed to imperfections in the CCD or the instrument \citep{howell2000}.
However, it is not possible to illuminate uniformly all the pixels in the detector while doing spectroscopy.
The spectroscopic flat-field is the spectrum obtained when the spectrograph entrance slit is illuminated with an incandescent filament lamp (a tungsten lamp in our case).
The entrance slit used for the flat-fields is larger than that used for the astronomical objects, the reason being to assure that the spatial profile of any order in any astronomical object is contained inside the spatial profile of the flat-fields (see Figure~\ref{reduction_flat_field}).

The flat-field correction involves the division of the spectra of astronomical objects between the median of all the flat-fields obtained during a night.
Thus, the pixel-to-pixel differences in sensitivity are eliminated
However, the shape of the extracted orders is distorted, because the observed spectrum has been divided by the curvature function of the flat-field.
This effect can be removed if the stellar spectra are multiplied by the blaze function of the flat-field after the extraction of the spectrum.
The curvature function of the flat-field can be obtained by extracting the median flat-field as if it were an astronomical object.

The MIDAS \'echelle routines perform the flat-field correction after the dispersed light has been subtracted.
It has been confirmed a posteriori that this approach is wrong, because the photoelectrons generated by the scattered light in a pixel are proportional to its sensitivity.
Because of this error in the software, the reduction was more difficult than expected.
On the one hand, the detection of the orders was done twice (objects and flat-fields).
On the other hand, the dispersed light had to be subtracted from the median flat-field, this being a CPU time consuming procedure.

The flat-field correction could not be used for two observational campaigns, namely October 1998 and January 1999, because the slits used (objects and flat-fields) were not well aligned, so the spatial profiles of the stellar objects were not contained inside the flat-field spatial profiles.
Despite this fact, it has been afterwards verified that the absence of flat-field correction has not significantly degraded the SNR of those spectra.

\section{Extraction of the spectra}

\begin{figure}
\includegraphics[width=0.49\textwidth]{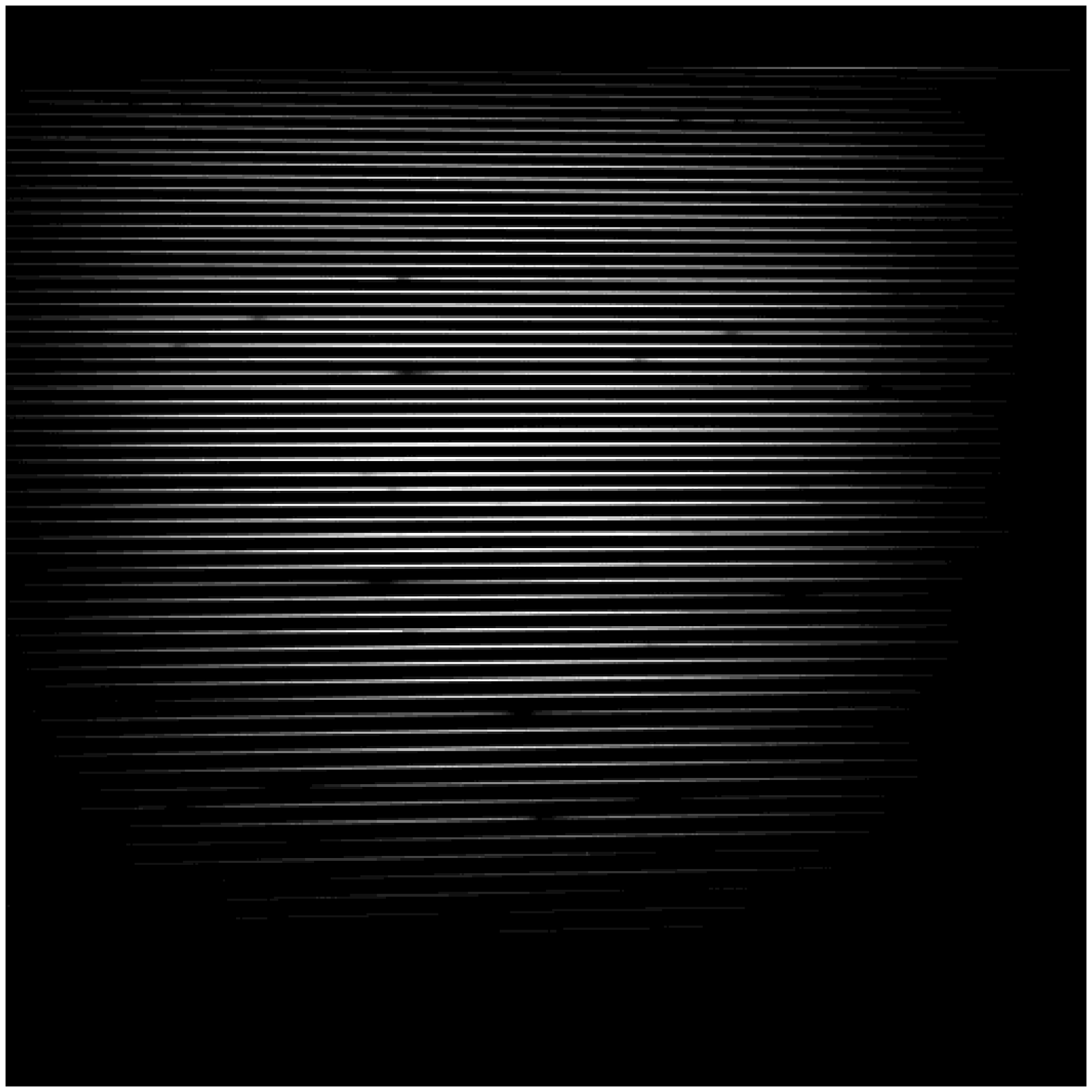}
\hspace{0.01\textwidth}
\includegraphics[width=0.49\textwidth]{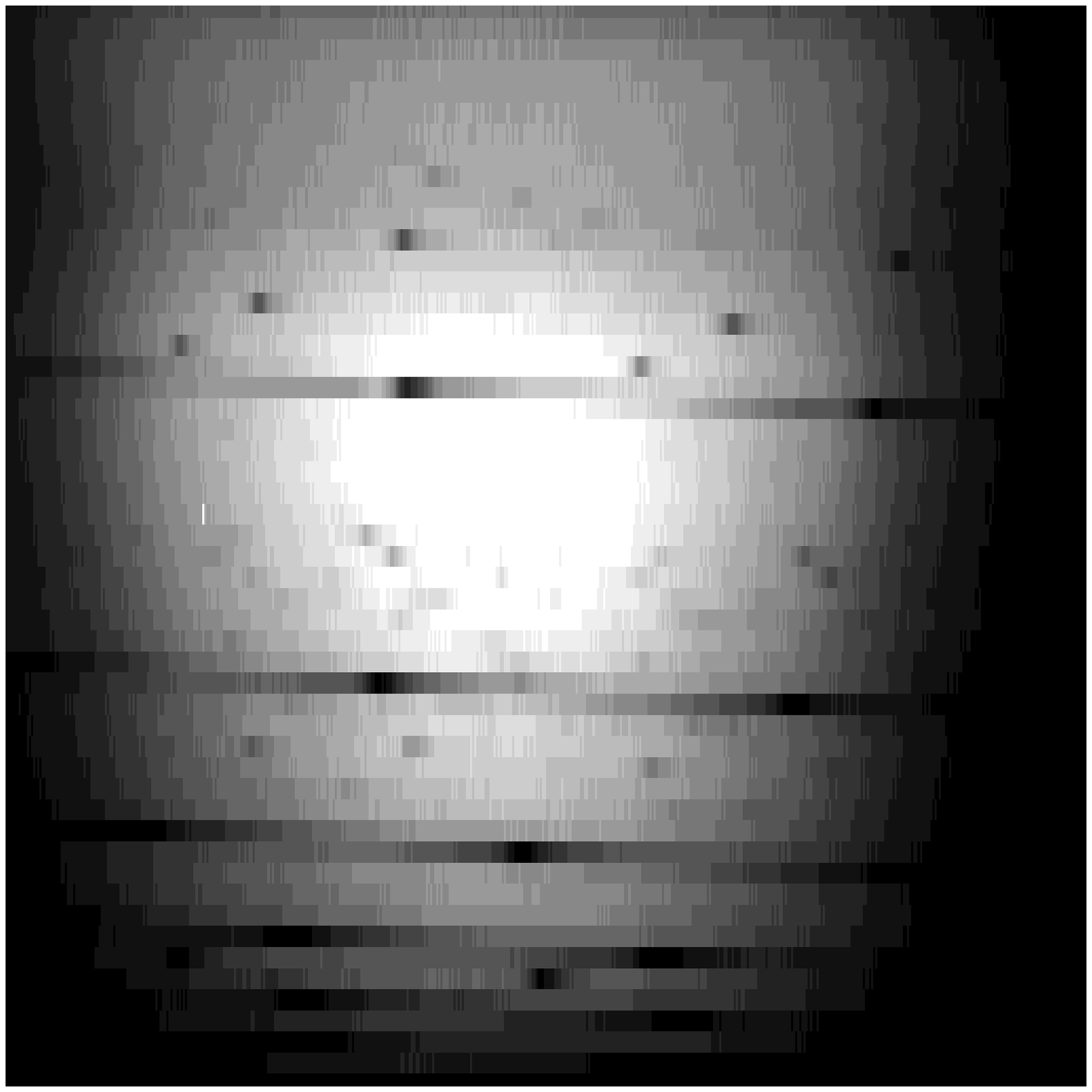}
\caption[Extraction of a spectrum]
{Extraction of a spectrum.
If the extraction procedure is applied to the bidimensional spectrum on the left, the spectrum on the right is obtained.
The extracted spectrum is composed of 59 rows, corresponding to each of the orders of the spectrum on the left.}
\label{reduction_extraction}
\end{figure}

Once the corrections for bias, scattered light and cosmic rays have been applied, the spectra are ready for being extracted.
The extraction is the obtention of a one-dimensional spectrum for each order of a bidimensional \'echelle spectrum recorded on the CCD.
For any order and any position (pixel) in its spectral direction, the stellar light is distributed across the whole corresponding spatial profile.
The extraction is the assignment to the pixel in that order of a value equal to the sum of all the counts stored across the spatial profile.
The size of the profile is defined by a suitable extraction slit (see Section~\ref{reduction_detection}).
It needs to be long enough to gather all the light from the object, but it does not have to be exceedingly large in order to minimize the noise.
An example of spectrum extraction can be seen in Figure~\ref{reduction_extraction}.

\section{Wavelength calibration}
\label{reduction_calibration}

\begin{figure}
\centerline{\includegraphics[width=0.5\textwidth]{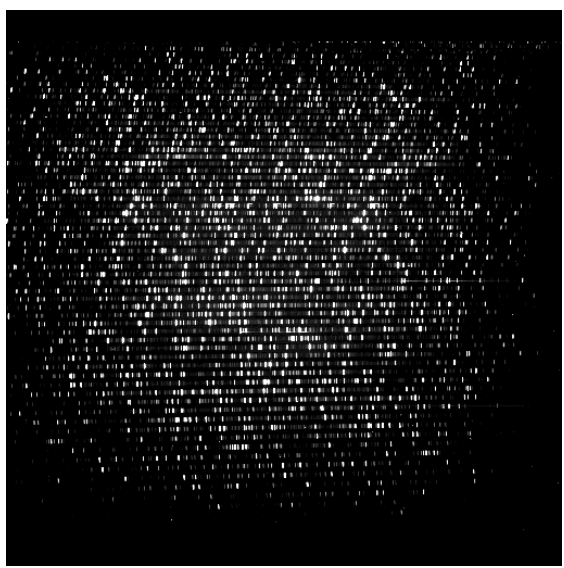}}
\caption[Th-Ar arc lamp spectrum]
{Th-Ar arc lamp spectrum.
It can be seen that it is a discrete spectrum, composed of a huge amount of emission lines whose wavelengths and relative intensities are tabulated for the strongest transitions.}
\label{reduction_arc}
\end{figure}

The wavelength calibration is a procedure that allows the extracted spectra to be expressed in terms of the wavelength and not of the pixel number along the spectral direction.
The spectra produced by arc lamps filled with vapour of diverse atomic elements is used.
These lamps generate a discrete emission spectrum composed of the lines corresponding to the electron transitions of the atoms in the gas.
A thorium-argon (Th-Ar) lamp was used for the reduction of the UES spectra.
An example of arc lamp spectrum is displayed in Figure~\ref{reduction_arc}.
The wavelength and the relative intensity of the strongest lines is tabulated for these lamps.
Thus, if a sufficient number of lines is identified in each order ($\sim$20), the relation between the pixel numbers and the wavelengths can be found.
In general, it is enough to use low order polynomial fits (n~$\leq$~4 in our case).

The wavelength calibration is fully integrated in MIDAS.
Thus it is only needed to perform interactively a few identifications and the program is then capable to identify iteratively a sufficient number of lines for each order without any further input.
In general, the calibrations are excellent.
The typical RMS errors are $\sim$20~m{\AA} (0.39~pixels), i.e. 5 times lower than the spectral resolution ($\sim$100~m{\AA}).

The Th-Ar lamp got damaged during one observing night (January 29$^{\rm th}$ 1999) and the reserve iron-argon lamp was used instead.
The available identification tables for the Fe-Ar lamps did not have a sufficiently large number of identifications.
A new table was created from the existing tables of different lamps, included in the MIDAS and IRAF packages.
The calibration results were acceptable, but worst than those obtained with the Th-Ar lamp.

\section{Continuum normalization}
\label{reduction_normalization}

The wavelength calibrated spectra have a continuum defined by the curvature function of the orders, which is the product of the spectrograph response function by the spectral distribution of the star.
If the curvature function is obtained, the observed spectrum can be normalized dividing it by the aforementioned function.
The continuum normalization is the last step in the reduction procedure carried out in this thesis, because a spectrophotometric calibration was not needed.

The continuum of a high SNR spectrum obtained from a low rotation velocity (\vsini\ $\le$ 50~\kms) F spectral type (or earlier) star is quite easy to determine.
The higher the temperature, the fewer and shallower are the photospheric lines.
Moreover, low rotation velocities generate less overlaps between neighbouring lines.
If the previous conditions are given, the characterization of the envelope function of each extracted order can be easily performed via a moderate order polynomial ($\sim$4-7).
However, it is impossible to fit properly the orders containing the Balmer series lines, because these lines are as broad as the orders.
The Balmer series is essential for this thesis, so it was made the approximation that the curvature function in an order containing a Balmer line is equal to the logarithmic average of the curvatures in the contiguous orders.
This approximation, being inexact, produces reasonable results, because it has been verified that the curvature changes slowly from order to order.

Only a few of the stars in the whole sample strictly comply with all the stated requirements, specially that about the spectral type, because the EXPORT observations cover the full range B9-K7.
However, it is enough to determine the curvature function for a single (normalization standard) star each night.
The curvature function obtained is used to normalize the remaining stars.
If this procedure is performed to a (target) star of the same spectral type as the normalization standard star and the seeing of both observations is the same, then a target star spectrum normalized to the quotient between the fluxes of the target and the normalization standard stars is obtained.
If the target star has a different spectral type or it has been observed under different seeing conditions, orders without curvature will be obtained, but they will not be normalized to any fixed quantity, because the quotient of the fluxes depends now on the wavelength.

Once the spectrum of a single non-variable star has been normalized to unity, it can be used to determine the curvature function of the orders of any other spectrum of the same star.
It is enough to take into account the changes in the radial velocity of the normalized spectrum before dividing it by the observed spectrum.
Thus the curvature function calculation can be ``propagated'' towards other observing nights and the number of polynomial fits needed is minimized.
Moreover, once the spectral orders of the target stars are normalized by means of the curvature function of the normalization standard star, they can be merged.
Then the complete normalized spectrum of any object can be generated and new normalization standard star spectra can be generated.

\section{Validity of the reduction method}
\label{reduction_validity}

It has been verified that the extraction and wavelength calibration of the spectra are excellent.
It has also been verified that the subtraction of cosmic rays has degraded the signal to noise ratio of a few stars to levels about $\leq$20.
Fortunately, the stars in this thesis have not been significantly affected by this effect.

The reduction method used in this thesis has been compared with those used by two astronomers experienced in the use of \'echelle spectra: Dr.~Cameron (University of St. Andrews, United Kingdom) and Dr.~Grinin (Crimean Astrophysical Observatory, Ukraine and Pulkovo Astronomical Observatory, Russia).
The spectra reduced for this thesis are similar in quality to those obtained by Dr.~Cameron for his research activities.
However, it became apparent that the normalization of our spectra is worst than that achieved by the group of Dr.~Grinin for a series of \'echelle spectra of UXOR stars \citep{grinin2001}.
Two are the reasons of this difference.
On the one hand, this group has a collection of, properly normalized, intermediate resolution spectra obtained in the same wavelength range than the \'echelle spectra for all the stars studied.
On the other hand, they heavily rely on synthetic spectra to adequately normalize the Balmer lines by means of an iterative fit process.
Even though the use of artificial spectra from the very moment of the reduction may introduce an additional error source, the goodness of the results obtained justify the validity of the procedure.

The reduced spectra can be used without reservation for any purpose which requires small spectral ranges ($\Delta \lambda \la \pm$~10~\AA,  $\Delta v \la \pm$~600~\kms).
However, the continuum is not accurately defined if wider wavelength ranges are used.
This was made clear in an attempt made by Dr.~Montesinos to use the UES spectra to determine stellar effective temperatures and surface gravities using the Balmer lines ($\Delta\lambda \ga \pm$~100~\AA).
It was verified that such large intervals led to erroneous results.
In this thesis no TAC with radial velocity greater than 500~\kms\ has been analysed, so the validity of the spectral analysis, performed in Chapters ~\ref{uxori} and~\ref{haebe}, is guaranteed.

\chapter[Calculation of \vsini\ via the Fourier transform of line profiles]
{Calculation of projected rotational velocities via the Fourier transform of line profiles}
\label{vsini}

The fit of stellar parameters is an iterative process which will be explained thoroughly in Sections~\ref{uxori_the_photospheric_spectrum} and~\ref{haebe_the_photospheric_spectra}.
This procedure needs some initial estimates for the following stellar parameters: effective temperature (\teff), surface gravity (\logg) and rotational velocity projected on the line of sight (\vsini).
These values have been obtained from the spectral types and \vsini\ rotational velocities derived by \citet{mora2001} for the whole EXPORT sample.
The initial values of \vsini\ can be directly used in the fit of stellar parameters.
However, in order to obtain \teff\ and \logg\ from the spectral type, some empirical calibrations are needed.
In this thesis, the relations obtained by \citet{dejager1987} and \citet{habets1981} have been used.
In this Chapter, the method employed by \citet{mora2001} for the measurements of \vsini\ will be described in detail.

\section{Theoretical basis}

Many methods for the determination of stellar rotational velocities have been developed so far.
The most important procedures are briefly described by \citet{gray1992}.
Some are based in spectroscopic measurements.
Two of them are particularly suited for their application to \'echelle spectra: simultaneous fitting of \vsini\ and other stellar parameters via synthetic spectra and the use of the Fourier Transform of isolated line profiles.
The first method is extensively used in Section~\ref{haebe_the_photospheric_spectra}, whereas the second one has been used by \citet{mora2001} (strictly speaking, the procedure used is somewhere in between both generic methods, because a small amount of synthetic spectra is also utilized).

As it can be seen in \citet{gray1992}, the first spectroscopic determinations of rotational velocities were performed by visual comparison of the spectra of target stars with some template spectra of rotational standard stars, which were grouped in catalogues \citep[e.g.][]{slettebak1975}.
A more advanced technique is based on the use of empirical relations between the Full Width at Half Maximum (FWHM) of some lines and \vsini.
An example of such relations is given by \citet{gray1992} from the data of \citet{slettebak1975}.
These calibrations give precisions in the measurements around 10-20\%.

More advanced procedures came from the use of the whole line absorption profile ($D(\lambda)=F_\nu/F_c$), and not only the FWHM.
In this way, the Fourier transform of the profile ($d(\sigma)$) can be computed and the spectrum changes from the wavelength space ($\lambda$, \AA) to the spatial frequency domain ($\sigma$, \AA$^{-1}$).
In what follows, the fundamentals of the simplest, although very efficient, method of measurement of \vsini\ from the Fourier Transform will be described.
This method is very precise.
Errors as low as 5\% can be achieved in favourable cases.

The main hypothesis of the method is that the star rotates as a rigid body and the emitted spectrum $H(\lambda)$ is the same for all points in the stellar disk.
Therefore, the observed integrated spectrum is found to be the spectrum emitted by any part of the disk, $H(\lambda)$, convolved with a certain rotational profile, $G(\lambda)$:

\begin{equation}
D(\lambda) =\int_{-\infty}^\infty
H(\lambda - \Delta\lambda) \, G(\Delta\lambda) \, d\Delta\lambda
\end{equation}

A linear law of limb darkening is also postulated.
This law relates the intensity emitted from the disk surface element exactly located under the line of sight ($I_c^0$) with the light emitted from any other element ($I_c$) whose radius vector $\vec{R}$ subtends an angle $\theta$ with the observer's direction.

\begin{equation}
I_c/I_c^0 = 1 - \epsilon (1 - \cos \theta)
\end{equation}

Thus, an analytical formula for the rotational profile: $G(\lambda_{line} + \Delta\lambda)$ can be found:

\begin{equation}
G(\Delta \lambda) =
\frac
{2(1 - \epsilon) [1 - (\Delta\lambda / \Delta\lambda_L)^2]^\frac{1}{2} 
+ \frac{1}{2}\pi\epsilon [1 - (\Delta\lambda / \Delta\lambda_L)^2]}
{\pi\Delta\lambda_L (1 - \epsilon/3)}
\end{equation}

\noindent where $\Delta\lambda_L = \lambda_{\rm line} \times v\sin i / c$.
\citet{carroll1933} found the following approximate expression for the line profile Fourier transform $g(\sigma)$, assuming a limb darkening coefficient $\epsilon$~=~0.6.

\begin{equation}
g(\sigma) = \frac{J_1(x)}{x} - \frac{3 \cos x}{2 x^2} + \frac{3 \sin x}{2 x^3}
\end{equation}

\noindent where $x = 2 \pi \sigma \lambda_{\rm line} v \sin i / c$.
This function is the sum of three oscillatory components.
The first component contains a first-order Bessel function.
Being an oscillating fuction, the profile function has several zeroes.
The first one ($\sigma_1$) is located at:

\begin{equation}
\sigma_1 = \frac{0.660 \, c}{\lambda_{\rm line} v\sin i}
\label{vsini_ecuacion}
\end{equation}

It is known that the transform of the observed spectrum $d(\sigma)$ is the product of the transform of the profile $g(\sigma)$ by the transform of the spectrum emitted by every infinitesimal region of the stellar disk $h(\sigma)$:

\begin{equation}
d(\sigma) = h(\sigma) \times g(\sigma)
\end{equation}

So every frequency $\sigma$ which makes the rotational profile $g(\sigma)$ zero will also make any line profile transform $d(\sigma)$ to vanish.
If the zeroes of $h(\sigma)$ correspond to higher frequencies than those of $g(\sigma)$, it would be enough to measure the first zero of $d(\sigma)$ to obtain a value of \vsini.
If equation~\ref{vsini_ecuacion} is inverted:

\begin{equation}
v\sin i = \frac{0.660 \, c}{\lambda_{\rm line} \sigma_1}
\end{equation}

\section{Operational procedure}

According to the previous Section, the rotational velocity can be immediately computed if the first zero of the Fourier transform of a suitable line profile is known.
Actually, the line profiles are not only affected by rotation but by many other factors, such as the presence of nearby faint lines, low SNR, defective reduction, etc.
Therefore, the transform does not completely vanish, so instead of zeroes we will have minima in the power spectrum (the modulus of the transform).
The higher the quality of the selected lines and the spectra, the deeper the minima will be.
The method consists of three steps: search for strong isolated lines whose main broadening mechanism is rotation, normalization of the line profiles and computation of the Fourier transforms.
Synthetic spectra have been used to perform properly the first two stages in the method, as it will be seen below.

It is impossible to find true isolated lines in a stellar spectrum, due to the huge amount of actual optical transitions.
It will be assumed that an isolated line is that with ``higher intensity'' than those in the wavelength range where the line exists, i.e. the neighbourhood of the line absorption profile, which will be rotationally broadened.
It can be easily determined from photospheric zero rotational velocity spectra if any line is of ``higher intensity''.
First, a synthetic spectrum with a temperature and gravity similar to those in the star should be generated.
The spectra have been calculated using the stellar model atmospheres given by \citet{kuruczCD13} and the code SYNTHE \citep{kuruczCD18}.
Since the depth of the lines does not vary quickly with changes in the stellar parameters, very accurate initial values for \teff\ and \logg\ are not needed.
It is enough to estimate them from the spectral types.
Initial guesses of the spectral types were obtained from the astrophysical database Simbad ({\tt \href{http://simbad.u-strasbg.fr}{http://simbad.u-strasbg.fr}}).

\begin{figure}
\begin{center}
\includegraphics[clip=true,angle=-90,width=0.49\textwidth]
{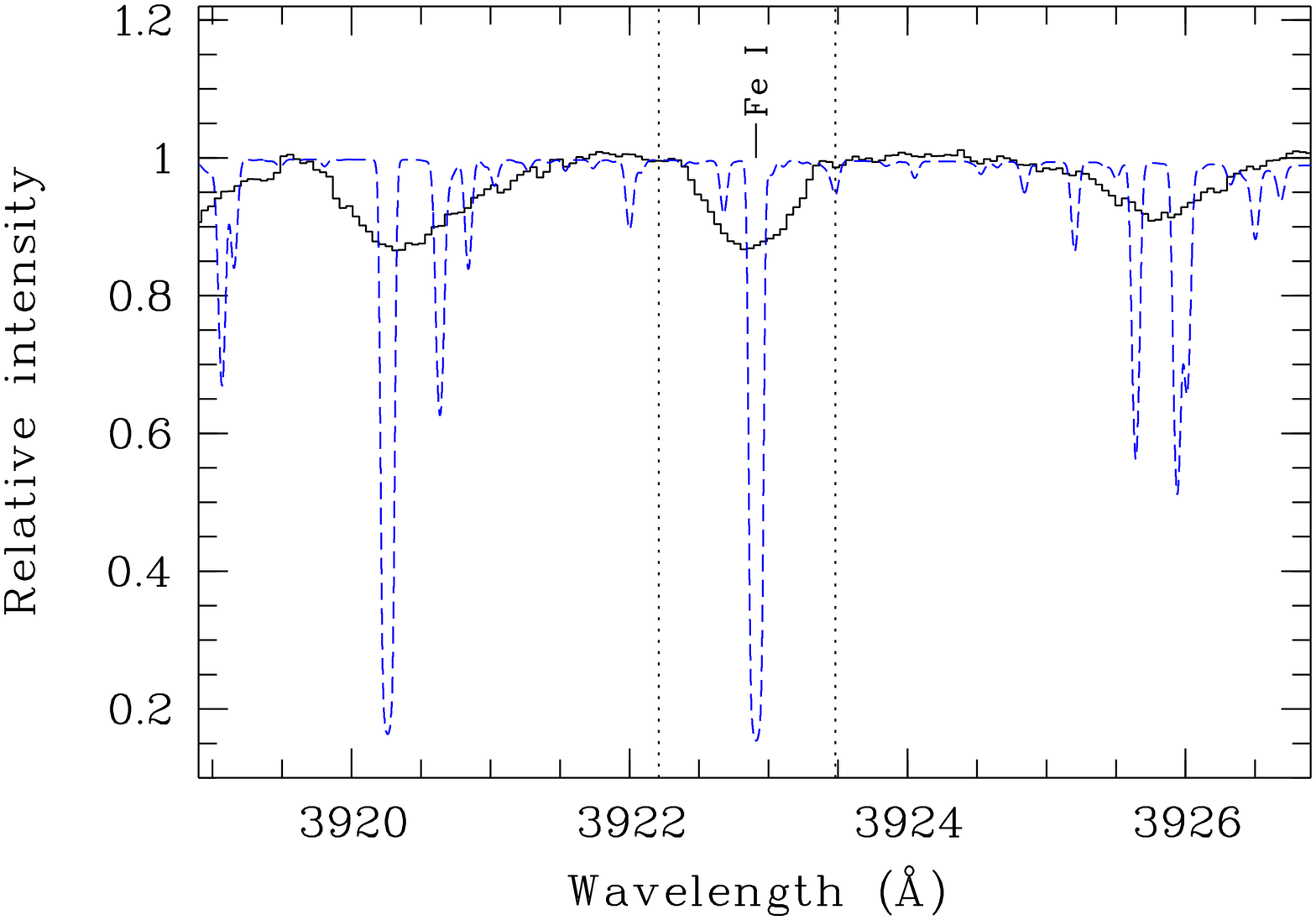}
\includegraphics[clip=true,angle=-90,width=0.49\textwidth]
{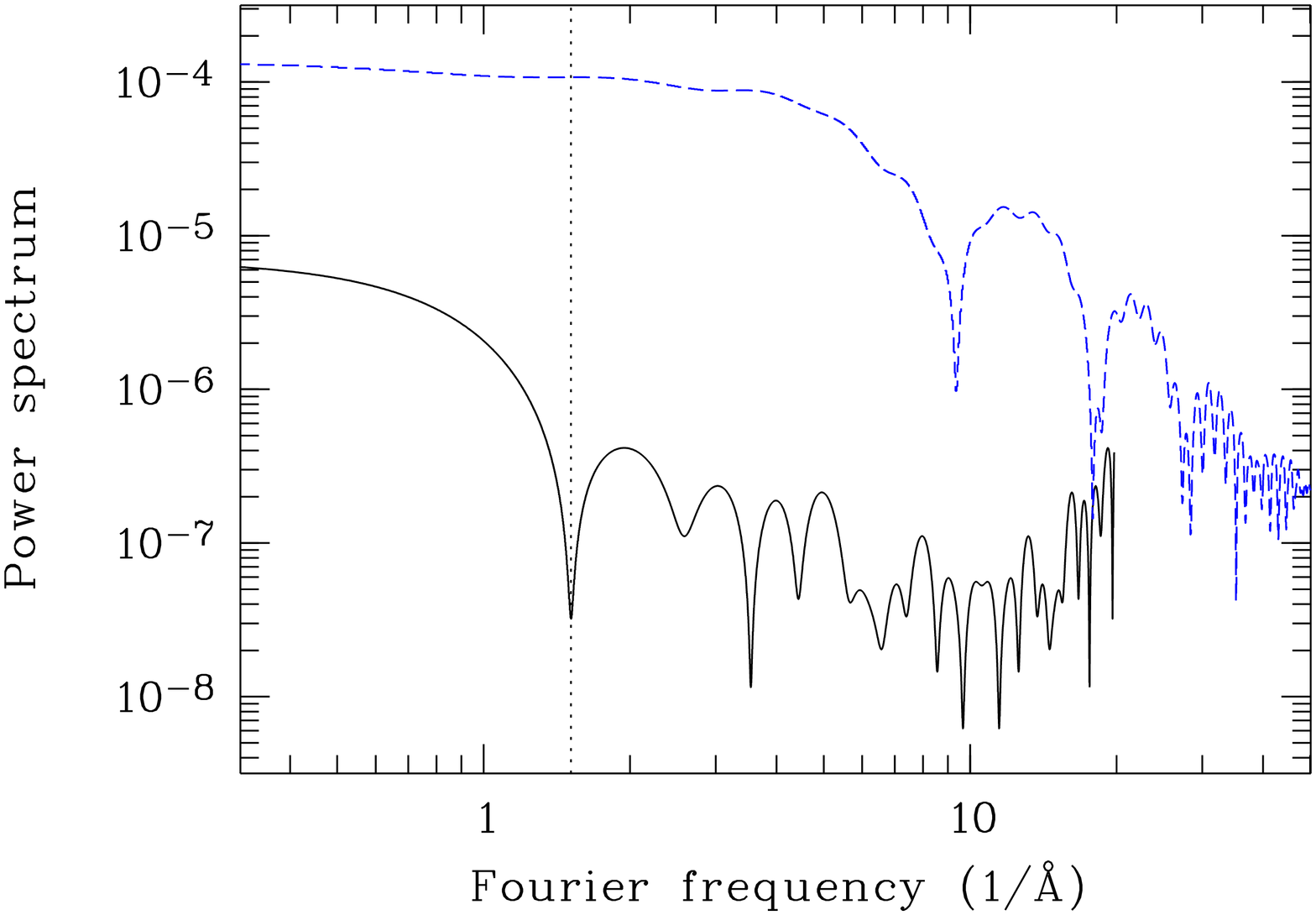}
\includegraphics[clip=true,angle=-90,width=0.49\textwidth]
{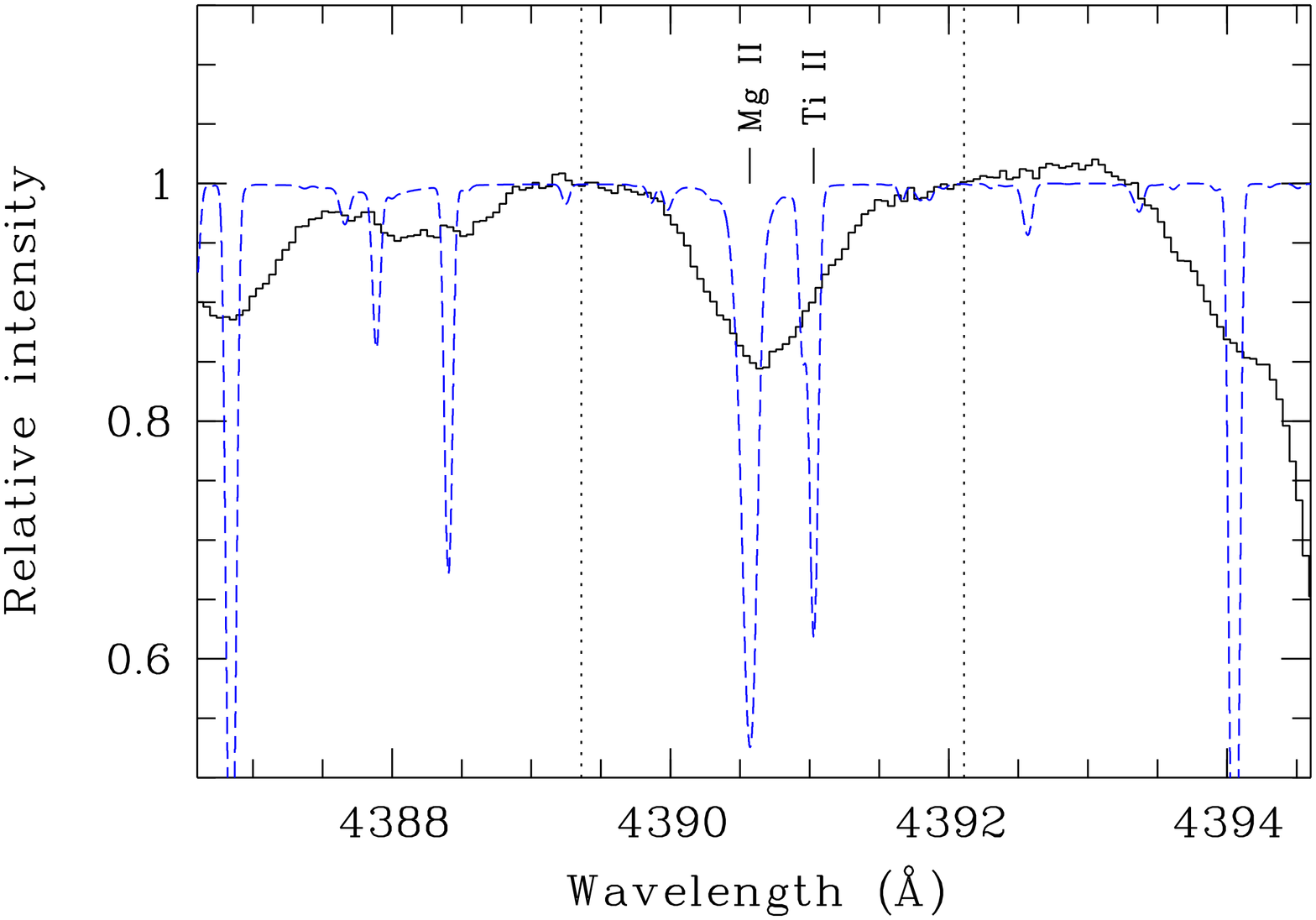}
\includegraphics[clip=true,angle=-90,width=0.49\textwidth]
{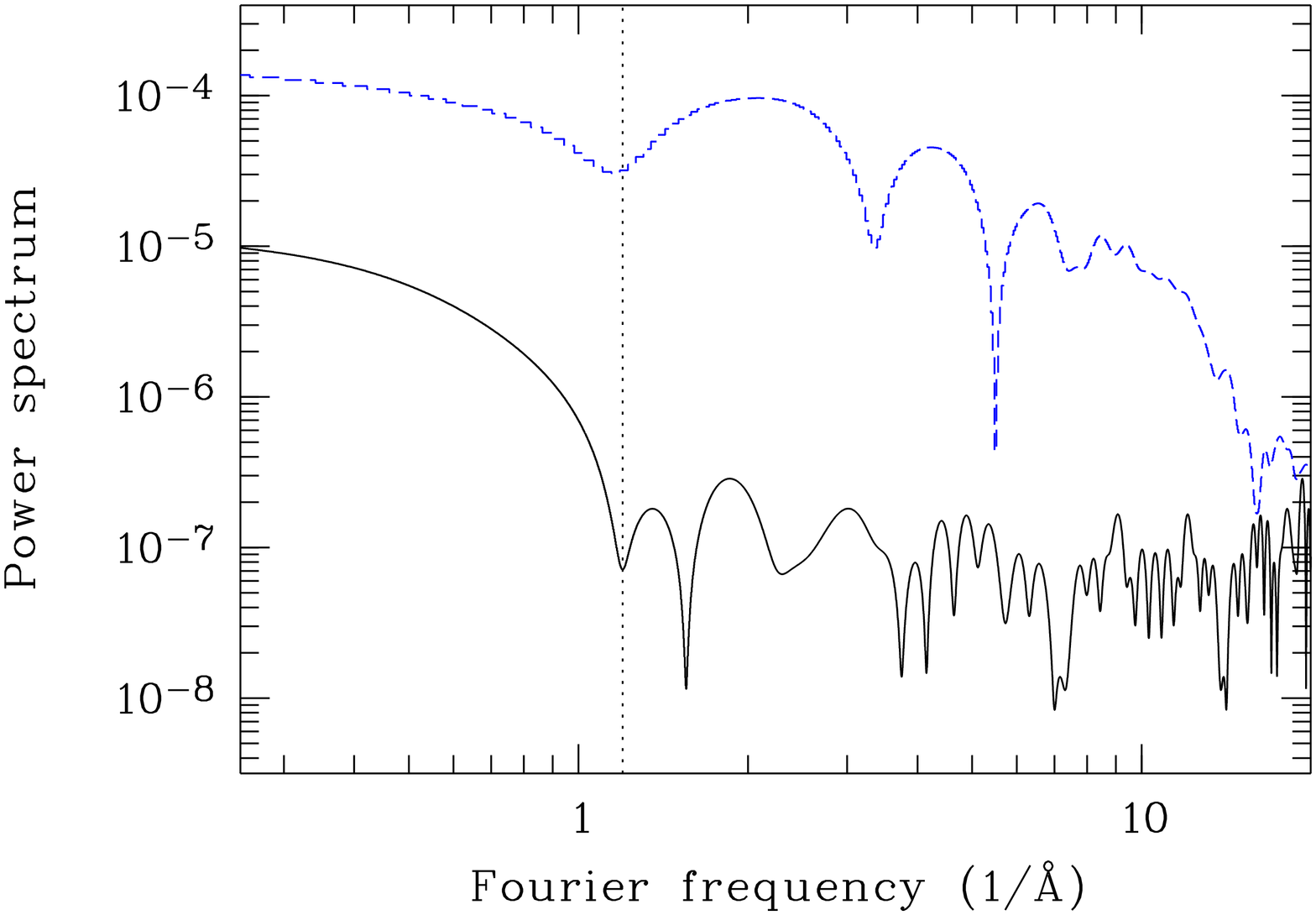}
\end{center}
\caption[Line selection for \vsini\ estimations]
{Line selection for \vsini\ estimations.
Left:
two portions of the spectrum of BF~Ori are shown (solid black line).
The top spectrum was obtained in JD~1209.55 and the bottom one in JD~1112.63.
Synthetic spectra generated using \teff~=~8750~K, \logg~=~3.5 and \vsini~=~0 are overplotted (blue dashed line).
The upper and lower bounds of the interval used for the Fourier transform computations are shown by two dotted lines.
The ions responsible for the most important electron transitions in the interval are also shown in each graph.
Right:
next to each line selection graph, the power spectrum of both the observed profile (black solid line) and the synthetic profile (blue dashed line) are displayed.
The placement of the first minimum in the observed profile is shown in each graph as a dotted vertical line.
It can be appreciated that the selection made in the upper left panel is correct, because the first minimum of the Fourier transform of the synthetic spectrum is located at a higher frequency.
However, the selection made in the lower left graphic is wrong because both minima match in frequency.}
\label{vsini_selection}
\end{figure}

The Fourier transform of both the observed absorption profile (in order to measure \vsini) and the non-broadened synthetic spectrum were evaluated in the same wavelength range.
If the first minimum of both transforms match, then the value obtained for \vsini\ is wrong, because the frequency used $\sigma_1$ is not related to rotation but to the distance between the line studied and another one of significant intensity.
On the other hand, if the first zero of the observed profile happens at a lower frequency than that of the synthetic spectrum, it can be undoubtedly affirmed that the target line is of ``higher intensity'' than those in its neighbourhood.
Two examples of line selection can be seen in Figure~\ref{vsini_selection}.
One of the examples is correct, while the another one is wrong.

\begin{figure}
\centerline{
\includegraphics[clip=true,angle=-90,width=0.60\textwidth]
                {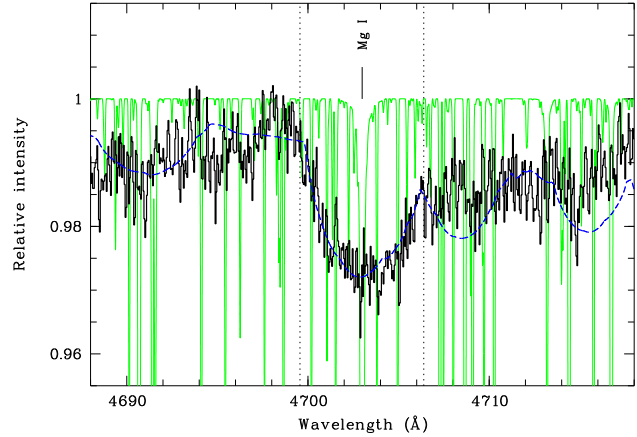}}
\caption[Continuum placement for \vsini\ calculations]
{Continuum placement for \vsini\ calculations.
A portion of the spectrum of XY~Per, obtained in JD~1113.55, is shown as a black solid line.
The blue dashed line is a synthetic spectrum used to place the continuum of the observed spectrum.
The same spectrum but without broadening is shown in green.
It can be seen that most of the observed broad absorption component, which will be used for the \vsini\ determination, is mainly composed by a strong broadened line of \ion{Mg}{i}.
The dotted lines indicate the interval used to calculate the power spectrum.
If the apparent edges of the observed line had been used to normalize the line profile, the equivalent width of the line would have been underestimated by about 50\%.}
\label{vsini_continuum}
\end{figure}

Once the right lines are selected, the profiles must be properly normalized.
In principle it should be enough to fix the left and right edges to unity.
However, if the rotation velocity is high (\vsini~$\ga$~100~\kms), the lines will generally be broad and shallow.
This favours neighbouring lines to blend, so the continuum will not be given by the apparent edges of the lines.
If the spectrum is normalized according to the procedure described, the equivalent width of the line will be underestimated and the determination of the rotation velocity will be wrong.
Synthetic spectra can also be used to minimize this problem.
If the borders of the lines are not set to unity but fitted to a continuum given by an artificial spectrum of the star, then a more reasonable value for the equivalent width of the line will be used.
The artificial spectra required are those used in the line selection process.
The value of \vsini\ is needed to broaden the synthetic spectra, but this magnitude is exactly what is being looked for.
Once again, very precise values are not needed, so \vsini\ data previously published by other authors or, when not available, the typical values relative to the spectral types, have been used.
The procedure described is explained in Figure~\ref{vsini_continuum}.

Fourier transforms were computed by means of the MIDAS routine {\tt FFT/POWER}, which provides the real part, the imaginary part and the power spectrum of the transform.
Once the power spectrum is obtained, the determination of its first minimum location and, therefore, \vsini\ are trivial.

The rotational velocity has been measured for the largest possible number of available lines in the spectra of each star.
Thus the average values and statistical errors finally presented by \citet{mora2001} have been calculated.
In general, the statistical errors obtained are about 10-20~\kms.

\section{Uncertainties in the method}

There are more precise and complete methods for the determination of the rotational velocity based on the Fourier transform \citep{gray1992}.
These alternative methods do not only use the position of the first minimum, but the whole shape of the function.
Therefore, additional information can be obtained, as the value of $\epsilon$, the identification of rotation or macroturbulence as the dominant line broadening mechanism, and even a measurement of the differential rotation in the star.
The great advantage of using only the position of the first minimum for the determination of \vsini\, relies in its absolute independence from theoretical models (e.g. the variation of $\epsilon$ with the temperature and gravity) and experimental measurements (e.g. the use of rotational standard stars or instrumental and thermal broadening profiles).

The assistance of synthetic spectra for the selection of lines and continuum fitting has allowed the method to be applied to very high rotational velocity (\vsini\ $\ga$~200~\kms) stars.
This is undoubtedly an advantage (4 out of 5 stars in this thesis have \vsini~$\ge$ 200~\kms).
However, the results obtained are less model-independent, because errors on the computation of the artificial spectra have been introduced in the final values obtained.

The resolution of the observations places a lower limit to the range of measurable velocities.
This can be expressed in terms of the Fourier transform.
The 6~\kms\ resolution of the UES spectra can be represented by a narrow instrumental profile which is convolved with the stellar spectrum.
This instrumental profile is a function, approximately gaussian in shape, with a FWHM of 6~\kms\ and a minimum in its power spectrum located at the frequency corresponding to this velocity.
The observed line profile transform is the product of the three individual transforms (rotational profile, instrumental profile and spectrum emitted by every region of the stellar disk).
Thus, another minimum in the power spectrum will be found exclusively due to the instrumental profile, so, if \vsini\ is lower than the resolution, the first minimum of the transform will not be generated by rotation but by the instrumental profile.

The limb darkening law used is another source of systematic errors.
The method relies in the use of a linear function with $\epsilon$~=~0.6.
This approximation is valid for the Sun and stars of similar spectral types.
Many authors have questioned the validity of this empirical relation.
\citet{diaz-cordoves1992} demonstrated that the difference between a linear law and other models of limb darkening is very small for the wavelength ranges explored in this thesis, at most $\sim$5\% in the less favourable conditions.
However, the specific value of $\epsilon$ used in the \vsini\ measurements does matter.

\begin{figure}
\centerline{\includegraphics[angle=-90,width=0.65\textwidth,clip=true]
           {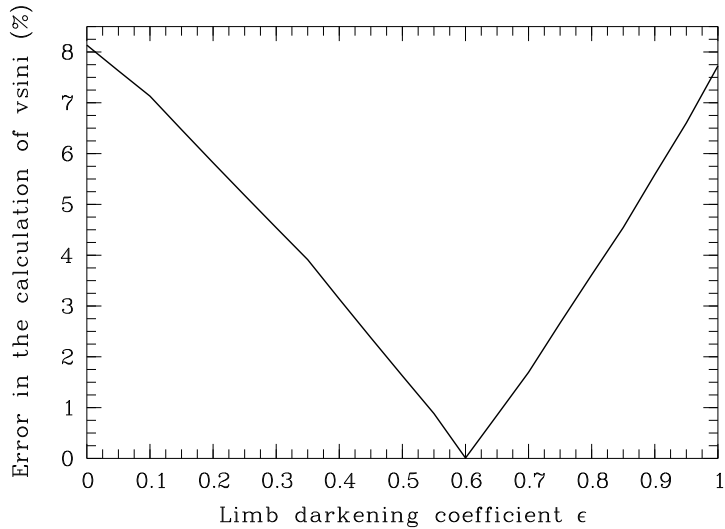}}
\caption[Error made in the the calculation of \vsini\ according to the linear limb darkening coefficient of the star.]
{Error made in the the calculation of \vsini\ according to the linear limb darkening coefficient of the star.
In the figure, the maximum error made when \vsini\ is measured for stars with limb darkening coefficient $\epsilon \ne$~0.6 is shown.
This error is negligible ($\le$~1.6\%, $\epsilon$~=~0.50-0.65) for the stars studied in this thesis, but could be significant ($\le$~7~\%) for the coolest objects in the EXPORT sample, which present the highest $\epsilon$ values.}
\label{vsini_limb_darkening}
\end{figure}

The influence of $\epsilon$ will be maximum when the rotational profile fully dominates the observed line shape.
This happens when \vsini\ is much greater than both the resolution of the observations (6~\kms) and the macroturbulence ($\sim$2~\kms).
\citet{solano1997} estimated the error associated with the selection of a wrong limb darkening coefficient.
We have improved this method as follows.
A grid of rotational profiles with \vsini~=~100~\kms\ and $\epsilon$ values ranging from 0.0 to 1.0 with a step of 0.05 has been generated.
A set of artificial spectral lines has been generated from these profiles with an arbitrary central wavelength of 5000~\AA.
The line shape is exclusively regulated by the rotational profile, \vsini\ has been calculated for these artificial lines with the method described in this section (i.e. assuming $\epsilon$~=~0.6).
Thus the errors in the measurement of \vsini\ arising from a wrong value of $\epsilon$ have been estimated and are shown in Figure~\ref{vsini_limb_darkening}.
\citet{diaz-cordoves1995} provide theoretical predictions for the limb darkening coefficients.
It can be verified that the maximum range of variation for the whole EXPORT sample is $\epsilon$~=~0.35-0.95.
The maximum error will be of 7\% for the cooler stars (K spectral type).
If only the five stars studied in this thesis are considered, the limb darkening coefficients will be restricted to the range $\epsilon$~=~0.50-0.65, which implies a maximum error of 1.6\%.
It can be concluded that the limb darkening law used is not a significant source of error, except for, possibly, the cooler stars in the EXPORT sample.

\section{Comparison with values obtained by spectral synthesis}

Table~\ref{vsini_comparison_table} shows the values of \vsini\ calculated via the Fourier transform and those obtained in Section~\ref{haebe_the_photospheric_spectra} by means of synthetic spectral fits to photospheric lines.
The values relative to the star HD~163296 have also been included, because its stellar parameters were also calculated by the spectral synthesis method.
That star was initially included in this thesis, but its strong stellar winds rendered the TACs deconvolution values unusable.
It can be seen that the agreement between both completely independent methods is remarkable for the whole set of stars.

\begin{table}
\caption[Comparison between both \vsini\ measurement methods]
{Comparison between two \vsini\ measurement methods.
The values of \vsini\ for the stars in this thesis are shown in the table.
Column~2 includes the results obtained searching for the first minimum in the power spectrum.
Column~3 gives the values retrieved from the simultaneous fit of \teff, \logg\ and \vsini\ by means of spectral synthesis.
The errors of these measurements are discussed in Section~\ref{haebe_the_photospheric_spectra}.
The rotational velocity was not determined by both methods for UX~Ori, because the value obtained from the Fourier transform provided an excellent agreement between the observed and synthetic spectra.

\dag The values obtained for HD~163296, not included in this thesis, are given for comparison purposes only.}
\label{vsini_comparison_table}
\begin{center}
\begin{tabular}{lcc}
\hline
\hline
Star             & Fourier transform & Spectral synthesis \\
\hline
BF Ori           & 37  $\pm$ 2       & 37                 \\
SV Cep           & 206 $\pm$ 13      & 225                \\
UX Ori           & 215 $\pm$ 15      & --                 \\
WW Vul           & 220 $\pm$ 22      & 210                \\
XY Per           & 217 $\pm$ 13      & 200                \\
HD 163296$^\dag$ & 133 $\pm$ 6       & 125                \\
\hline
\end{tabular}
\end{center}
\end{table}

\chapter{A dynamical study of the circumstellar gas in UX Orionis}
\label{uxori}

\begin{center}
{\small
\noindent
  A.~Mora$^{1}$,
  A.~Natta$^{2}$,
  C.~Eiroa$^{1}$,
  C.A.~Grady$^{3}$,
  D.~de Winter$^{4}$,
  J.K.~Davies$^{5}$,
  R.~Ferlet$^{6}$,
  A.W.~Harris$^{7}$,
  B.~Montesinos$^{8,9}$,
  R.D.~Oudmaijer$^{10}$,
  J.~Palacios$^{1}$,
  A.~Quirrenbach$^{11}$,
  H.~Rauer$^{7}$,
  A.~Alberdi$^{8}$,
  A.~Cameron$^{12}$,
  H.J.~Deeg$^{13}$,
  F.~Garz\'on$^{13}$,
  K.~Horne$^{12}$,
  B.~Mer\'{\i}n$^{9}$,
  A.~Penny$^{14}$,
  J.~Schneider$^{15}$,
  E.~Solano$^{9}$,
  Y.~Tsapras$^{12}$, and
  P.R.~Wesselius$^{16}$
}
\end{center}

\newcounter{institutes}
{\scriptsize
   \setlength{\rightmargin}{1.3cm} \setlength{\leftmargin}{1.3cm}
\begin{list}
  {$^{\arabic{institutes}}$}
  {\usecounter{institutes}
    \setlength{\partopsep}{0cm} \setlength{\topsep}{0.2cm}
    \setlength{\parsep}{0cm}    \setlength{\itemsep}{0cm}}
\item Departamento de F\'{\i}sica Te\'orica C-XI, Universidad Aut\'onoma de
      Madrid, Cantoblanco 28049 Madrid, Spain
\item Osservatorio Astrofisico di Arcetri, Largo Fermi 5, I-50125 Firenze,
      Italy
\item NOAO/STIS, Goddard Space Flight Center, Code 681, NASA/GSFC, Greenbelt,
      MD 20771, USA
\item TNO/TPD-Space Instrumentation, Stieltjesweg 1, PO Box 155, 2600 AD Delft,
      The Netherlands
\item Astronomy Technology Centre, Royal Observatory, Blackford Hill,
      Edinburgh, EH9 3HJ, UK
\item CNRS, Institute d'Astrophysique de Paris, 98bis Bd. Arago, 75014 Paris,
      France 
\item DLR Department of Planetary Exploration, Rutherfordstrasse 2, 12489
      Berlin, Germany
\item Instituto de Astrof\'{\i}sica de Andaluc\'{\i}a, Apartado de Correos
      3004, 18080 Granada, Spain
\item LAEFF, VILSPA, Apartado de Correos 50727, 28080 Madrid, Spain
\item Department of Physics and Astronomy, University of Leeds, Leeds LS2 9JT,
      UK
\item Department of Physics, Center for Astrophysics and Space Sciences,
      University of California San Diego, Mail Code 0424, La Jolla,
      CA 92093-0424, USA
\item Physics \& Astronomy, University of St. Andrews, North Haugh, St. Andrews
      KY16 9SS, Scotland, UK
\item Instituto de Astrof\'{\i}sica de Canarias, La Laguna 38200 Tenerife,
      Spain
\item Rutherford Appleton Laboratory, Didcot, Oxfordshire OX11 0QX, UK
\item Observatoire de Paris, 92195 Meudon, France
\item SRON, Universiteitscomplex ``Zernike'', Landleven 12, P.O. Box 800, 9700
      AV Groningen, The Netherlands
\end{list}

Originally published in Astronomy and Astrophysics {\bf 393}, 259-271 (2002)

Received 14 May 2002 / Accepted 11 July 2002
}

\section*{Abstract}
We present the results of a high spectral resolution ($\lambda / \Delta \lambda $ = 49000) study of the circumstellar (CS) gas around the intermediate mass, pre-main sequence star UX~Ori.
The results are based on a set of 10 \'echelle spectra covering the spectral range 3800 -- 5900 \AA, monitoring the star on time scales of months, days and hours.
A large number of transient blueshifted and redshifted absorption features are detected in the Balmer and in many  metallic lines.
A multigaussian fit is applied to determine for each transient absorption the velocity, $v$, dispersion velocity, $\Delta v$, and the parameter $R$, which provides a measure of the absorption strength of the CS gas.
The time evolution of those parameters is presented and discussed.
A comparison of intensity ratios among the transient absorptions suggests a solar-like composition of the CS gas.
This confirms previous results and excludes a very metal-rich environment as the cause of the transient features in UX Ori.
The features can be grouped by their similar velocities into 24 groups, of which 17 are redshifted and 7 blueshifted.
An analysis of the velocity of the groups allows us to identify them as signatures of the dynamical evolution of 7 clumps of gas, of which 4 represent accretion events and 3 outflow events.
Most of the events decelerate at a rate of tenths of m\,s$^{-2}$, while 2 events accelerate at approximately the same rate; one event is seen experiencing both an acceleration and a deceleration phase and lasts for a period of few days.
This time scale seems to be the typical duration of outflowing and infalling events in UX Ori. 
The dispersion velocity and the relative absorption strength of the features do not show drastic changes during the lifetime of the events, which suggests they are gaseous blobs preserving their geometrical and physical identity.
These data are a very useful tool for constraining and validating theoretical models of the chemical and physical conditions of CS gas around young stars; in particular, we suggest that the simultaneous presence of infalling and outflowing gas should be investigated in the context of detailed magnetospheric accretion models, similar to those proposed for the lower mass T Tauri stars.  

\vspace{0.5cm}
\noindent {\bf Key words.} Stars: formation -- Stars: pre-main sequence -- Circumstellar matter -- Accretion, accretion disks -- Lines: profiles -- Stars: individual: UX Ori

\section{Introduction}

The detection of planetesimals is highly relevant for the study of the formation and evolution of planetary systems, since it is nowadays accepted that planets form from CS disks via the formation of planetesimals \citep{beckwith2000}.
Several lines of evidence suggest that the young main sequence A5V star $\beta$~Pic \citep[20 Myr, ][]{barrado1999} hosts planetesimals inside its large CS disk.
The main argument is the presence of transient Redshifted Absorption Components (RACs) in high resolution spectra of strong metallic lines, like \ion{Ca}{ii}~K 3934~\AA.
These spectroscopic events have been interpreted as being caused by the evaporation of comet-like highly hydrogen-depleted bodies.
The interpretation is known as the Falling Evaporating Bodies (FEB) scenario \citep[ and references therein]{lagrange2000}.
However, the presence or absence of planetesimals during the pre-main sequence (PMS) phase is a controversial observational topic.
The time scale for the formation of planetesimals \citep[$\sim$10$^4$~yr, ][]{beckwith2000} is shorter than the duration of the PMS phase ($\sim$1--10~Myr), which suggests that they should exist during PMS stellar evolution.

UX~Ori-like PMS objects (UXORs) are characterized by a peculiar photo-polarimetric variability, which has been interpreted as the signature of massive, almost edge-on, CS disks \citep{grinin1991}.
Most UXORs have A spectral types and therefore are the PMS evolutionary precursors of $\beta$~Pic.  
UXORs have been reported to show RACs \citep{grinin1994,dewinter1999}; they also show Blueshifted Absorption Components (BACs) in their spectra. 
In this paper, the acronym TAC (Transient Absorption Component) will be used to denote both RACs and BACs.
In analogy to $\beta$~Pic, RACs observed in UXORs have been interpreted in terms of the FEB scenario \citep{grady2000}.
However, this interpretation has recently been questioned by \citet{natta2000}, who used the spectra obtained by \citet{grinin2001} to analyze the dynamics and chemical composition of a very strong, redshifted event in UX Ori itself, an A4 IV star \citep{mora2001} $\sim 2\times 10^6$ year old \citep{natta1999}.
Gas accretion from a CS disk was suggested in \citet{natta2000} as an alternative to the FEB scenario to explain the observed RACs in UX Ori.
In addition, \citet{beust2001} have found that the FEB hypothesis cannot produce detectable transient absorptions in typical HAe CS conditions, unless the stars are relatively old ($\ge 10^7$ yr).

A detailed observational study of the kinematics and chemistry of TACs, e.g. by means of \'echelle spectra which simultaneously record many metallic and hydrogen lines, can discriminate between the two scenarios. 
For instance, in a FEB event large metallicities are expected, while gas accreted from a CS disk would have approximately solar abundances.
A strong observational requirement is set by the time scale of monitoring of the TACs.
In \citet{natta2000} spectra were taken 3 days apart.
Since UX Ori is a highly variable star, there is some ambiguity in identifying transient spectral components of different velocities detected over this time interval as having the same physical origin.
A better time resolution is needed in order to ensure that the TACs observed at different velocities  are due to the dynamical evolution of the same gas.
The EXPORT collaboration \citep{eiroa2000a} obtained high resolution \'echelle spectra of a large sample of PMS stars \citep{mora2001}.
About 10 PMS stars showed TACs which were intensively monitored ($\Delta$t~$\leq$~1~day).
The study of the kinematics and chemistry of these events provides important tools for identifying their origin, or at least to put severe constraints on it.   

This paper presents an analysis of the TACs observed in UX Ori by EXPORT and shows that the results are not compatible with a FEB scenario.
The layout of the paper is as follows:
Section~\ref{uxori_observations} presents a brief description of the EXPORT observations.
Section~\ref{uxori_analysis} presents the procedures followed in the analysis of the spectra.
Section~\ref{uxori_results} presents the kinematic and chemical results of the detected TACs.
Section~\ref{uxori_discussion} gives a discussion of the dynamics and nature of the gas.
In Section~\ref{uxori_conclusions}, we present our conclusions.

\section{Observations}
\label{uxori_observations}

High resolution  spectra of UX~Ori were taken in October 1998 and January 1999 using the Utrecht Echelle Spectrograph (UES) at the 4.2m WHT (La Palma Observatory).
We collected 10 \'echelle spectra in the wavelength range 3800~--~5900~\AA, with resolution $\lambda / \Delta\lambda$~=~49000 (6~km/s); exposure times range between 20 and 30 minutes.
The observing log is given in Table~\ref{uxori_photometry}, in which the long-term (months) and short-term (days, hours) monitoring is evident.
Standard MIDAS and IRAF procedures have been used for the spectroscopic reduction \citep{mora2001}.
Final typical signal to noise ratio (SNR) values are $\sim$150.

\begin{table}
\caption[EXPORT UES/WHT UX~Ori observing log]
{EXPORT UES/WHT UX~Ori observing log. The Julian date (-2450000) of each  exposure is given in column 1.
Columns 2 to 5 give simultaneous $V$, $H$, $K$ and \%P$_V$ photopolarimetric data, when available, taken from \citet{oudmaijer2001} and \citet{eiroa2001}.}
\label{uxori_photometry}
\centerline{
\begin{tabular}{lllll}
\hline
\hline
Julian date & $V$    & \%P$_V$ & $H$  & $K$  \\
\hline
1112.5800   &  9.94  &  1.33   & 8.19 & 7.37 \\
1113.6034   &  9.82  &  1.32   & 8.20 & 7.37 \\
1113.7194   &  --    &  --     & --   & --   \\
\hline
1207.5268   &  --    &  --     & --   & --   \\
1208.5072   &  --    &  --     & 8.41 & 7.56 \\
1209.4204   &  9.80  &  1.15   & 8.31 & 7.62 \\
1210.3317   &  9.89  &  1.26   & --   & --   \\
1210.3568   &  9.86  &  1.34   & --   & --   \\
1210.4237   &  9.88  &  1.30   & 8.34 & --   \\
1210.5156   &  9.90  &  1.31   & 8.34 & --   \\
\hline
\end{tabular}}
\end{table}

EXPORT also carried out simultaneous intermediate resolution spectroscopy, optical photo-polarimetry and near-IR photometry of UX Ori during the nights when the UES spectra were taken \citep{mora2001,oudmaijer2001,eiroa2001}.
Table~\ref{uxori_photometry} also gives photo-polarimetric data taken simultaneously with the UES spectra; they show that UX Ori was always in its bright state with little variation, but significant \citep{oudmaijer2001}.

\begin{figure}
\centerline{\includegraphics[height=0.65\hsize,angle=-90,clip=true]
                            {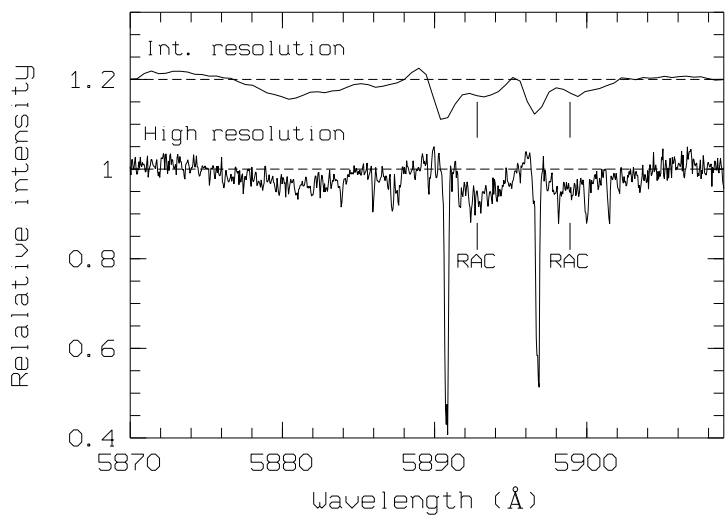}}
\caption[Comparison between intermediate and high resolution spectra]
{Comparison between intermediate (top, J.D. 2451210.3098, 10~min. exposure time) and high (bottom, J.D. 2451210.3317, 30~min. exposure time) resolution spectra.
The \ion{Na}{i}~D spectral region is shown.
A sharp IS zero velocity component and a much broader RAC (indicated in the figure) can be seen in both lines of the doublet in the high resolution spectrum.
They are also detected in the intermediate resolution spectrum, but the big difference in the width of the components is not clear.
For the sake of clarity  an artificial vertical displacement of 0.2 units has been applied to the intermediate resolution spectrum.}
\label{uxori_wht_vs_int}
\end{figure}

\section{Analysis of the spectra}
\label{uxori_analysis}

A detailed analysis of TACs requires high  SNR ($>$~25) and high spectral resolution ($<$~10~km/s) in order to resolve the kinematic components observed simultaneously in different lines, as is shown in Fig. 1.
This figure presents the \ion{Na}{i}~D lines in  one of the UES spectra together with the simultaneous intermediate resolution spectrum (R~$\sim$~6000) in the same lines. 
The UES spectrum clearly shows a sharp, IS, narrow zero velocity component and a RAC in both lines of the doublet while both components, though visible, cannot be cleanly separated in the lower resolution spectrum.
This is why we restrict our further analysis  of the absorptions to the UES spectra.

All the UX~Ori UES spectra show CS spectral features in a variety of lines.
The features are seen in absorption and can be either blueshifted or redshifted; some underlying emission is also present (see below).
Their profiles are complex and blended components are directly seen in many cases.  
Thus, the analysis of the CS contribution to the observed spectra requires a careful subtraction of the stellar photospheric spectrum and a method to characterize the blended kinematical components.
In this section we describe the procedure we have followed, which is the one used in \citet{natta2000}, although we have improved it by considering all detected TACs simultaneously.

\subsection{Subtraction of the UX Ori photospheric spectrum}
\label{uxori_the_photospheric_spectrum}

Firstly we estimate the radial velocity of the star.
The heliocentric correction of each observed spectrum is computed using MIDAS.
Then, the radial velocity of the star is estimated by fitting the  Na~I~D sharp IS absorption.
This absorption  is  stable in velocity (velocity differences in our spectra are less than 1.0~km/s) and is related to the  radial velocity of the star \citep{finkenzeller1984}.
A value of 18.3~$\pm$~1.0~km/s is obtained, in good agreement with the 18~km/s estimate by \citet{grinin1994}.   

Kurucz (1993) model atmospheres assuming solar metallicity and turbulence velocity of 2~km/s have been used to synthesize the photospheric spectra, which are later used for a comparison with the observed ones.
Atomic line data have been obtained from the VALD online database \citep{kupka1999}, and the codes SYNTHE \citep{kurucz1979,jeffery1996} and SYNSPEC \citep{hubeny1995} are used to synthesize the metallic lines and Balmer hydrogen lines, respectively. 
In this way we compute a large grid of synthetic spectra with the effective temperature, gravity and rotation velocity as free variable parameters. 

\begin{figure}
\centerline{\includegraphics[height=0.75\hsize,angle=-90,clip=true]
                            {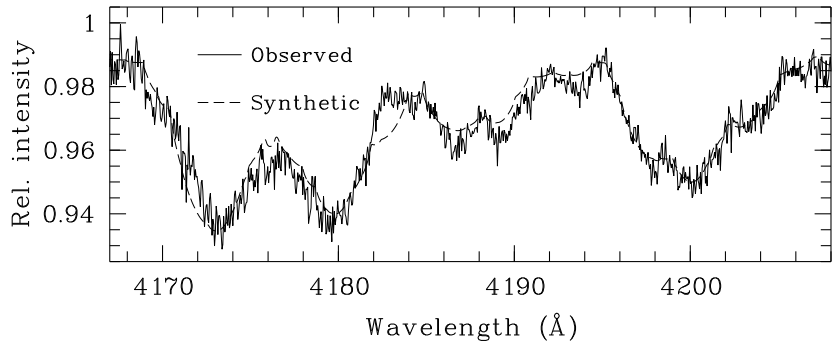}}
\caption
[Comparison of the observed average spectrum of UX Ori to the synthetic spectrum with stellar parameters T$_{\rm eff}$~=~9250~K, \logg~=~4.0 and ${\it v\sin i}$~=~215~km/s]
{Comparison of the observed average spectrum of UX Ori (solid line) to the synthetic spectrum with stellar parameters T$_{\rm eff}$~=~9250~K, \logg~=~4.0 and ${\it v\sin i}$~=~215~km/s (dashed line).  
The spectral region shown includes the blends at 4172~--~4179~{\AA} (mainly \ion{Fe}{ii} and \ion{Ti}{ii}) and 4203~{\AA} (\ion{Fe}{i}).}
\label{uxori_obs_vs_syn}
\end{figure}

The best set of free parameters is estimated by comparing a number of lines among the observed spectra and the synthetic ones.
This is not an easy  task since most UX Ori lines are variable to some extent, and the choice of pure photospheric lines is not trivial.
Our choice is to consider a number of weak lines that show a minimum degree of variability (less that 1\%) in the UES spectra.
The blends located at 4172~--~4179~{\AA} (mainly \ion{Fe}{ii} and \ion{Ti}{ii}) and 4203~\AA\ (\ion{Fe}{i}) have been used to estimate T$_{\rm eff}$ (effective temperature) and \logg\ (logarithm of the surface gravity), since they are sensitive to changes in T$_{\rm eff}$ and \logg\ for early type A stars \citep{gray1987,gray1989b}.
The synthetic spectra have also been broadened allowing for stellar rotation.
The best set of free parameters we find is T$_{\rm eff}$~=~9250~K, \logg~=~4.0, {\it v $\sin i$}~=~215~km/s.  
Fig.~\ref{uxori_obs_vs_syn} shows the comparison between the synthetic and the average observed spectrum in the above-mentioned lines; the agreement is excellent.
The stellar parameters of the synthetic spectrum are in very good agreement with the spectral type and rotational velocity derived elsewhere \citep{mora2001}.

\begin{figure}
\centerline{\includegraphics[width=0.65\hsize,clip=true]
                            {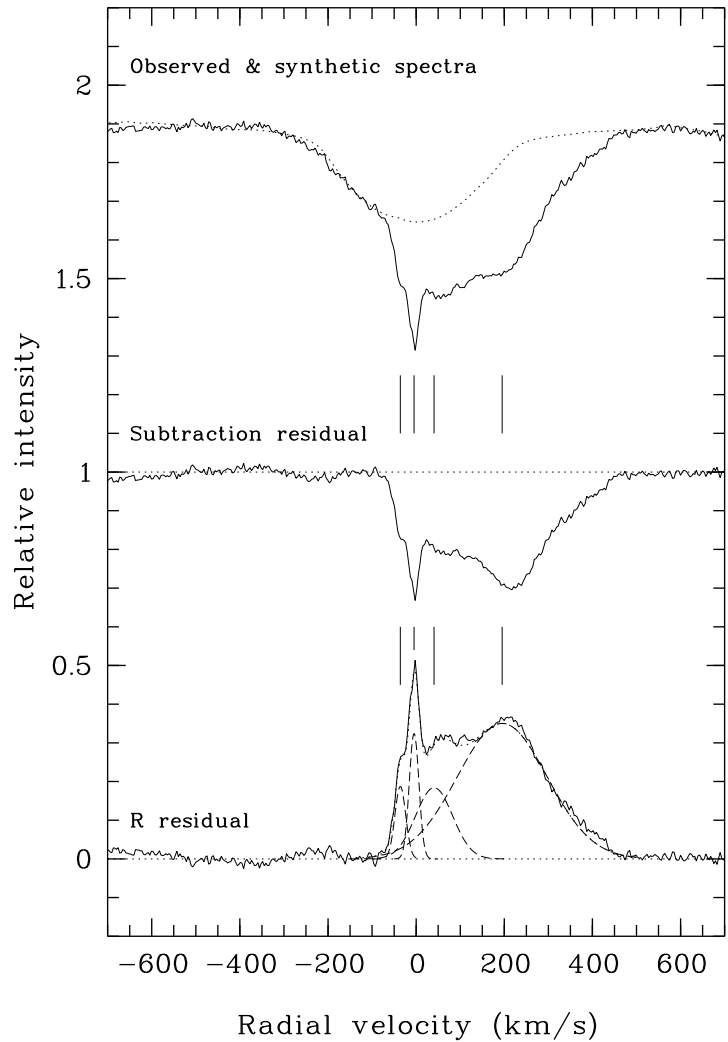}}
\caption[Subtraction of the photospheric spectrum]
{Subtraction of the photospheric spectrum. 
The spectral range corresponds to the \ion{Ca}{ii}~K line. 
Top: observed (solid line) and synthetic (dotted line) spectra are displayed.
Middle: Residual due to the CS contribution (observed -- photospheric).
Bottom: Normalized residual absorption, $R$ = (1 -- observed~/~photospheric).
Vertical tick marks indicate the 4 absorption components of the \ion{Ca}{ii}~K which can be identified in this spectrum, including the narrow IS absorption.
In the bottom we show the 4 gaussians (dashed lines), whose sum (dotted line) excellently reproduces the $R$ profile (solid line).} 
\label{uxori_subtraction}
\end{figure}

The next step consists of the subtraction of the best synthetic photospheric spectrum from each observed spectrum. 
This has been carried out by fitting the continuum with a linear law between two points selected at both sides of those lines showing TACs.
As an example Fig.~\ref{uxori_subtraction} shows the \ion{Ca}{ii}~K line results, where the synthetic spectrum and the J.D. 2451209.4204 UX Ori observed one are plotted together with the residual after the subtraction.
The  residual corresponds to the CS contribution to the observed spectra.
As a more convenient way to represent this contribution we define the $R$ parameter (normalized residual absorption):

\centerline{$R = 1 - F_{\rm obs} / F_{\rm syn}$} 

$R$~=~0 means no CS absorption, and $R$~=~1 denotes complete stellar light occultation, i.e. $R$ quantifies the CS absorption strength and could be directly compared to the models in 
\citet{natta2000}.
$R$ is also plotted in Fig.~\ref{uxori_subtraction}. 
  
\subsection{$R$ multicomponent gaussian fitting}
\label{uxori_multigaussian_fitting}

The CS contribution presents a very complex line profile caused by the blending of different transient features. 
As an example, Fig.~\ref{uxori_subtraction} shows the components of the \ion{Ca}{ii}~K line which are clearly present in that particular observation. 
For the sake of analysing the profile we assume that it is caused by gas with different kinematics and that each component can be represented by means of a gaussian function.
This choice provides a direct interpretation of the fit parameters in terms of a physical characterization of the gaseous transient absorption components: the gaussian center gives the central velocity ($v$) of each component, the gaussian  FWHM (Full Width at Half Maximum) provides the velocity dispersion ($\Delta v$) and finally the gaussian peak gives the value of the residual absorption $R$.
Since the features are blended a multigaussian fit is needed, which has been carried out by means of the IRAF {\tt ngaussfit} routine. 
Fig.~\ref{uxori_subtraction} shows the results of the multigaussian fit for the \ion{Ca}{ii}~K line.
4 gaussians, i.e. 4 kinematical components, each with their corresponding parameters, provide an excellent agreement with the residual CS contribution profile.

This analysis is applied to the lines with the strongest variability, which are  good tracers of the CS gas properties: H$\beta$ 4861~\AA, H$\gamma$ 4340~\AA, H$\delta$ 4102~\AA, H$\epsilon$ 3970~\AA, H$\zeta$ 3889~\AA, \ion{Ca}{ii}~K 3934~\AA, \ion{Ca}{ii}~H 3968~\AA, \ion{Na}{i}~D2 5890~{\AA} and \ion{Na}{i}~D1 5896~\AA. 
All Balmer lines with good SNR  present transient absorption features (even the noisy H$\kappa$~3750~\AA, 10$^{\rm th}$ line in the Balmer series). 
They are also observed in many other lines (e.g. \ion{He}{i}, \ion{Fe}{i} and \ion{Fe}{ii}), as already pointed out by \citet{grinin2001}, but a corresponding discussion is deferred to a future work.

\begin{figure}
\centerline{\includegraphics[width=0.65\hsize,clip=true]{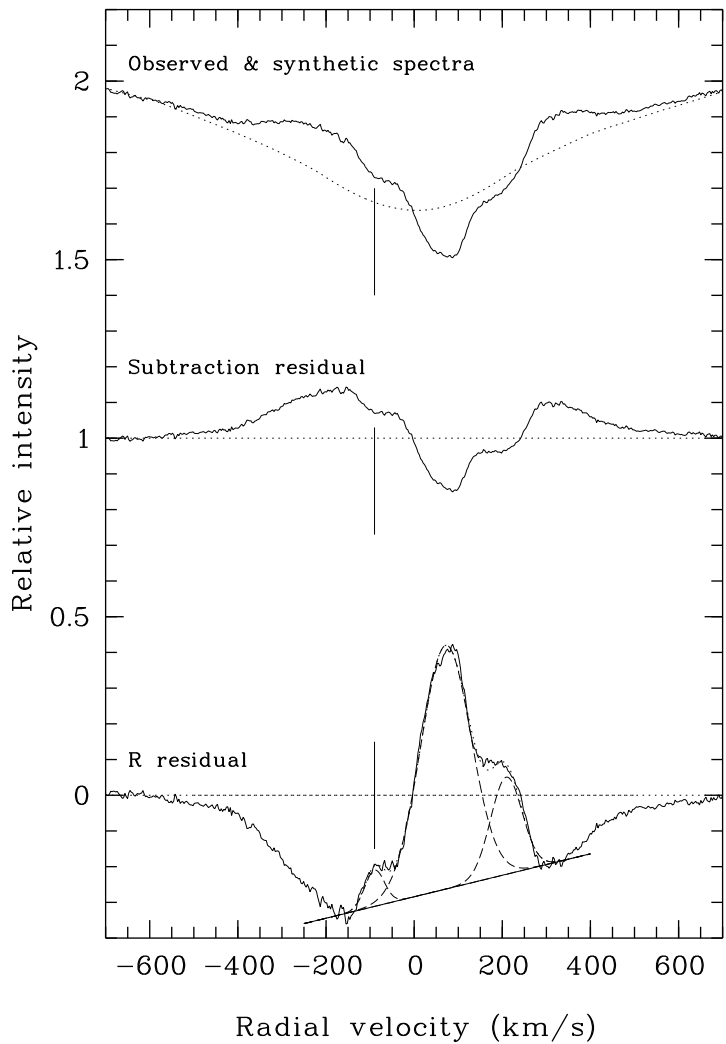}}
\caption[Underlying emission in the H$\beta$ absorption line]
{Underlying emission in the H$\beta$ absorption line.
Top: Normalized observed spectrum (solid line) and photospheric model (dotted line). 
The line profile is clearly modulated by an emission contribution at $\simeq\pm$300~km/s in the line wings.
Middle: Subtracted spectrum (observed -- synthetic).
Bottom: $R$ curve. 
The vertical mark points to an absorption component which would be undetectable if the continuum level were not redefined.
This has been done by means of a linear fit to the peaks of the emission, shown as throughs in the $R$ plot.
This fit (solid line at the bottom) is the new zero level for the gaussians.
3 gaussians (dashed lines) whose sum (dotted line) provides the optimal fit to the $R$ curve are also plotted at the bottom.}
\label{uxori_hbemission}
\end{figure}

Underlying line emission is often significant in the Balmer lines, particularly in H$\beta$, H$\gamma$ and H$\delta$; in fact, H$\alpha$ is seen in emission (equivalent width $\sim$~10~\AA) in the simultaneous INT spectra.
\ion{Ca}{ii}~K and H also show little emission.
In these cases $R$ is underestimated and negative values can artificially arise when the synthetic photospheric spectrum is subtracted.
This is especially bad for faint absorption components near the emission peaks since the latter appear as minima in the $R$ plot.  
In order to avoid contamination by these emission components, we define a new ``zero level", estimated by a linear fit to the minima observed in the $R$ plot. Fig.~\ref{uxori_hbemission} illustrates this approach.
The consistency of the results obtained with different lines confirms its validity.  
Further improvements should be based on theoretical modelling or higher order fitting of the emission.

\begin{table}
\caption
[Identified transient absorption components in the Balmer and \ion{Ca}{ii} and \ion{Na}{i} lines of UX Ori]
{\footnotesize Identified transient absorption components in the Balmer, \ion{Ca}{ii} and \ion{Na}{i} lines of UX Ori.
JD is the Julian Date ($-$2450000).
Numbers in column 3 represent the event assigned to the particular absorptions (see text section 4).
A ``0'' in that column corresponds to the narrow IS absorptions, while a ``--" means that the absorption is not associated with a particular event. 
Columns 4 to 6 give: $v$, the radial velocity of the transient absorption, $\Delta v$, the FWHM and $R$,  the normalized residual absorption.}
\label{uxori_master_table}
\centerline{
\tiny
\begin{tabular}{lllrrl}
\hline
\hline
Line & JD & Event & $v$ (km/s) & $\Delta v$ (km/s) & $R$ \\
\hline
H$\beta$      & 1112.5800 & 1 & 209 &  84 & 0.27 \\
H$\gamma$     & 1112.5800 & 1 & 207 &  78 & 0.21 \\
H$\delta$     & 1112.5800 & 1 & 202 &  73 & 0.16 \\
H$\epsilon$   & 1112.5800 & 1 & 206 &  68 & 0.19 \\
\ion{Ca}{ii} K& 1112.5800 & 1 & 199 &  91 & 0.26 \\
\ion{Na}{i} D2& 1112.5800 & 1 & 169 & 132 & 0.10 \\
\ion{Na}{i} D1& 1112.5800 & 1 & 183 & 123 & 0.09 \\
H$\beta$      & 1112.5800 & 2 &  77 & 135 & 0.68 \\
H$\gamma$     & 1112.5800 & 2 &  63 & 172 & 0.59 \\
H$\delta$     & 1112.5800 & 2 &  72 & 183 & 0.43 \\
H$\epsilon$   & 1112.5800 & 2 &  56 & 199 & 0.55 \\
\ion{Ca}{ii} K& 1112.5800 & 2 &  62 & 155 & 0.49 \\
H$\beta$      & 1112.5800 & 3 &--90 &  52 & 0.10 \\
H$\gamma$     & 1112.5800 & 3 &--91 &  51 & 0.10 \\
H$\delta$     & 1112.5800 & 3 &--88 &  41 & 0.06 \\
H$\epsilon$   & 1112.5800 & 3 &--85 &  47 & 0.18 \\
\ion{Ca}{ii} K& 1112.5800 & 3 &--92 &  61 & 0.11 \\
\ion{Ca}{ii} K& 1112.5800 & 0 &   5 &  17 & 0.23 \\
\ion{Ca}{ii} H& 1112.5800 & 0 &   6 &  12 & 0.23 \\
\ion{Na}{i} D2& 1112.5800 & 0 &   6 &  11 & 0.65 \\
\ion{Na}{i} D1& 1112.5800 & 0 &   6 &  11 & 0.50 \\
H$\zeta$      & 1112.5800 & --& 111 & 225 & 0.25 \\
H$\beta$      & 1113.6034 & 1 & 132 & 238 & 0.41 \\
H$\gamma$     & 1113.6034 & 1 & 109 & 217 & 0.33 \\
H$\delta$     & 1113.6034 & 1 & 175 & 136 & 0.16 \\
H$\epsilon$   & 1113.6034 & 1 & 157 & 100 & 0.22 \\
H$\zeta$      & 1113.6034 & 1 & 104 & 258 & 0.18 \\
\ion{Ca}{ii} K& 1113.6034 & 1 & 114 & 246 & 0.31 \\
H$\beta$      & 1113.6034 & 2 &  55 &  75 & 0.58 \\
H$\gamma$     & 1113.6034 & 2 &  57 &  64 & 0.50 \\
H$\delta$     & 1113.6034 & 2 &  55 & 109 & 0.53 \\
H$\epsilon$   & 1113.6034 & 2 &  50 & 105 & 0.56 \\
H$\zeta$      & 1113.6034 & 2 &  56 &  71 & 0.23 \\
\ion{Ca}{ii} K& 1113.6034 & 2 &  53 &  59 & 0.41 \\
\ion{Ca}{ii} H& 1113.6034 & 2 &  60 &  58 & 0.63 \\
\ion{Na}{i} D2& 1113.6034 & 2 &  61 &  38 & 0.09 \\
H$\beta$      & 1113.6034 & 3 &--51 &  42 & 0.07 \\
H$\gamma$     & 1113.6034 & 3 &--30 &  69 & 0.07 \\
\ion{Ca}{ii} K& 1113.6034 & 3 &--56 &  45 & 0.06 \\
\ion{Ca}{ii} K& 1113.6034 & 0 &   2 &  17 & 0.30 \\
\ion{Ca}{ii} H& 1113.6034 & 0 &   5 &  15 & 0.26 \\
\ion{Na}{i} D2& 1113.6034 & 0 &   5 &  12 & 0.68 \\
\ion{Na}{i} D1& 1113.6034 & 0 &   5 &  12 & 0.53 \\
H$\beta$      & 1113.7194 & 1 & 114 & 188 & 0.36 \\
H$\gamma$     & 1113.7194 & 1 &  90 & 169 & 0.36 \\
H$\delta$     & 1113.7194 & 1 & 182 &  89 & 0.11 \\
H$\epsilon$   & 1113.7194 & 1 & 143 & 121 & 0.20 \\
H$\zeta$      & 1113.7194 & 1 & 165 &  84 & 0.09 \\
\ion{Ca}{ii} K& 1113.7194 & 1 &  84 & 205 & 0.35 \\
H$\beta$      & 1113.7194 & 2 &  52 &  75 & 0.62 \\
H$\gamma$     & 1113.7194 & 2 &  57 &  58 & 0.44 \\
H$\delta$     & 1113.7194 & 2 &  59 & 121 & 0.56 \\
H$\epsilon$   & 1113.7194 & 2 &  50 &  96 & 0.51 \\
H$\zeta$      & 1113.7194 & 2 &  57 & 108 & 0.38 \\
\ion{Ca}{ii} K& 1113.7194 & 2 &  56 &  50 & 0.35 \\
\ion{Ca}{ii} H& 1113.7194 & 2 &  60 &  60 & 0.60 \\
\ion{Na}{i} D2& 1113.7194 & 2 &  61 &  44 & 0.10 \\
\ion{Na}{i} D1& 1113.7194 & 2 &  60 &  28 & 0.06 \\
H$\beta$      & 1113.7194 & 3 &--44 &  72 & 0.13 \\
H$\gamma$     & 1113.7194 & 3 &--22 &  99 & 0.13 \\
\ion{Ca}{ii} K& 1113.7194 & 3 &--58 &  52 & 0.08 \\
\ion{Ca}{ii} K& 1113.7194 & 0 &   6 &  19 & 0.31 \\
\ion{Ca}{ii} H& 1113.7194 & 0 &   5 &  19 & 0.20 \\
\ion{Na}{i} D2& 1113.7194 & 0 &   5 &  13 & 0.64 \\
\ion{Na}{i} D1& 1113.7194 & 0 &   5 &  12 & 0.49 \\
H$\beta$      & 1207.5268 & 4 &  78 & 221 & 0.26 \\
H$\gamma$     & 1207.5268 & 4 &  48 & 211 & 0.29 \\
H$\delta$     & 1207.5268 & 4 &  52 & 223 & 0.24 \\
H$\epsilon$   & 1207.5268 & 4 &  68 & 207 & 0.17 \\
H$\zeta$      & 1207.5268 & 4 &  83 & 128 & 0.16 \\
\ion{Ca}{ii} K& 1207.5268 & 4 &  51 & 269 & 0.32 \\
H$\beta$      & 1207.5268 & 5 &--60 &  88 & 0.61 \\
H$\gamma$     & 1207.5268 & 5 &--53 &  90 & 0.46 \\
H$\delta$     & 1207.5268 & 5 &--49 &  82 & 0.39 \\
H$\epsilon$   & 1207.5268 & 5 &--50 &  90 & 0.19 \\
H$\zeta$      & 1207.5268 & 5 &--43 &  92 & 0.30 \\
\ion{Ca}{ii} K& 1207.5268 & 5 &--59 &  70 & 0.32 \\
\ion{Ca}{ii} H& 1207.5268 & 5 &--56 &  63 & 0.38 \\
\ion{Na}{i} D2& 1207.5268 & 5 &--54 &  22 & 0.10 \\
\ion{Ca}{ii} K& 1207.5268 & 0 &   6 &  33 & 0.17 \\
\ion{Ca}{ii} H& 1207.5268 & 0 &   1 &  21 & 0.15 \\
\ion{Na}{i} D2& 1207.5268 & 0 &   3 &  13 & 0.59 \\
\ion{Na}{i} D1& 1207.5268 & 0 &   3 &  12 & 0.49 \\
H$\beta$      & 1207.5268 & --&  20 &  62 & 0.20 \\
H$\gamma$     & 1208.5072 & 4 &  89 & 285 & 0.20 \\
H$\delta$     & 1208.5072 & 4 &  84 & 268 & 0.18 \\
H$\epsilon$   & 1208.5072 & 4 & 109 & 294 & 0.14 \\
H$\zeta$      & 1208.5072 & 4 &  91 & 331 & 0.15 \\
\ion{Ca}{ii} K& 1208.5072 & 4 &  77 & 279 & 0.25 \\
H$\beta$      & 1208.5072 & 5 &--35 & 107 & 0.60 \\
\hline
\end{tabular}
\hspace{1.5cm}
\begin{tabular}{lllrrl}
\hline
\hline
Line & JD & Event & $v$ (km/s) & $\Delta v$ (km/s) & $R$ \\
\hline
H$\gamma$     & 1208.5072 & 5 &--34 &  85 & 0.47 \\
H$\delta$     & 1208.5072 & 5 &--44 &  81 & 0.36 \\
H$\epsilon$   & 1208.5072 & 5 &--37 &  52 & 0.15 \\
H$\zeta$      & 1208.5072 & 5 &--40 &  76 & 0.20 \\
\ion{Ca}{ii} K& 1208.5072 & 5 &--38 &  42 & 0.33 \\
\ion{Ca}{ii} H& 1208.5072 & 5 &--39 &  34 & 0.28 \\
\ion{Na}{i} D2& 1208.5072 & 5 &--34 &  18 & 0.18 \\
\ion{Na}{i} D1& 1208.5072 & 5 &--36 &  22 & 0.10 \\
H$\beta$      & 1208.5072 & 6 &--95 &  30 & 0.10 \\
H$\gamma$     & 1208.5072 & 6 &--91 &  29 & 0.12 \\
\ion{Ca}{ii} K& 1208.5072 & 6 &--82 &  41 & 0.22 \\
\ion{Ca}{ii} H& 1208.5072 & 6 &--83 &  42 & 0.19 \\
\ion{Ca}{ii} K& 1208.5072 & 0 &   1 &  26 & 0.26 \\
\ion{Ca}{ii} H& 1208.5072 & 0 & --1 &   8 & 0.14 \\
\ion{Na}{i} D2& 1208.5072 & 0 &   3 &  12 & 0.61 \\
\ion{Na}{i} D1& 1208.5072 & 0 &   3 &  13 & 0.47 \\
H$\beta$      & 1209.4204 & 4 & 207 & 309 & 0.43 \\
H$\gamma$     & 1209.4204 & 4 & 200 & 267 & 0.36 \\
H$\delta$     & 1209.4204 & 4 & 201 & 263 & 0.28 \\
H$\epsilon$   & 1209.4204 & 4 & 135 & 278 & 0.49 \\
H$\zeta$      & 1209.4204 & 4 & 191 & 313 & 0.21 \\
\ion{Ca}{ii} K& 1209.4204 & 4 & 195 & 240 & 0.35 \\
\ion{Na}{i} D2& 1209.4204 & 4 & 143 & 111 & 0.04 \\
H$\beta$      & 1209.4204 & 5 &--11 &  76 & 0.48 \\
H$\gamma$     & 1209.4204 & 5 &--22 &  53 & 0.30 \\
H$\delta$     & 1209.4204 & 5 &--14 &  54 & 0.16 \\
\ion{Ca}{ii} K& 1209.4204 & 5 &--36 &  29 & 0.19 \\
\ion{Ca}{ii} H& 1209.4204 & 5 &--32 &  26 & 0.12 \\
\ion{Ca}{ii} K& 1209.4204 & 0 & --5 &  26 & 0.32 \\
\ion{Ca}{ii} H& 1209.4204 & 0 & --4 &  17 & 0.20 \\
\ion{Na}{i} D2& 1209.4204 & 0 &   3 &  12 & 0.62 \\
\ion{Na}{i} D1& 1209.4204 & 0 &   4 &  12 & 0.50 \\
H$\gamma$     & 1209.4204 & --&  35 &  85 & 0.15 \\
\ion{Ca}{ii} K& 1209.4204 & --&  41 &  99 & 0.18 \\
H$\beta$      & 1210.3317 & 4 & 138 & 287 & 0.66 \\
H$\gamma$     & 1210.3317 & 4 & 138 & 255 & 0.54 \\
H$\delta$     & 1210.3317 & 4 & 142 & 270 & 0.43 \\
H$\epsilon$   & 1210.3317 & 4 & 110 & 277 & 0.64 \\
H$\zeta$      & 1210.3317 & 4 & 151 & 265 & 0.31 \\
\ion{Ca}{ii} K& 1210.3317 & 4 & 124 & 265 & 0.54 \\
\ion{Na}{i} D2& 1210.3317 & 4 & 115 & 137 & 0.06 \\
\ion{Na}{i} D1& 1210.3317 & 4 & 133 & 150 & 0.06 \\
H$\beta$      & 1210.3317 & 7 &   3 &  46 & 0.19 \\
H$\gamma$     & 1210.3317 & 7 &   9 &  39 & 0.10 \\
\ion{Ca}{ii} K& 1210.3317 & 0 & --3 &  19 & 0.20 \\
\ion{Ca}{ii} H& 1210.3317 & 0 &   1 &   8 & 0.16 \\
\ion{Na}{i} D2& 1210.3317 & 0 &   1 &  12 & 0.57 \\
\ion{Na}{i} D1& 1210.3317 & 0 &   2 &  13 & 0.46 \\
H$\beta$      & 1210.3568 & 4 & 132 & 287 & 0.64 \\
H$\gamma$     & 1210.3568 & 4 & 132 & 260 & 0.53 \\
H$\delta$     & 1210.3568 & 4 & 128 & 238 & 0.43 \\
H$\epsilon$   & 1210.3568 & 4 & 107 & 280 & 0.63 \\
H$\zeta$      & 1210.3568 & 4 & 140 & 249 & 0.29 \\
\ion{Ca}{ii} K& 1210.3568 & 4 & 119 & 296 & 0.52 \\
\ion{Na}{i} D2& 1210.3568 & 4 & 113 & 109 & 0.04 \\
\ion{Na}{i} D1& 1210.3568 & 4 & 142 & 127 & 0.04 \\
H$\beta$      & 1210.3568 & 7 &   6 &  41 & 0.18 \\
H$\gamma$     & 1210.3568 & 7 &  17 &  37 & 0.08 \\
\ion{Ca}{ii} K& 1210.3568 & 0 &   0 &  15 & 0.14 \\
\ion{Ca}{ii} H& 1210.3568 & 0 & --2 &  12 & 0.14 \\
\ion{Na}{i} D2& 1210.3568 & 0 &   2 &  13 & 0.60 \\
\ion{Na}{i} D1& 1210.3568 & 0 &   3 &  13 & 0.48 \\
H$\beta$      & 1210.4237 & 4 & 134 & 240 & 0.66 \\
H$\gamma$     & 1210.4237 & 4 & 122 & 242 & 0.55 \\
H$\delta$     & 1210.4237 & 4 & 125 & 229 & 0.44 \\
H$\epsilon$   & 1210.4237 & 4 &  89 & 253 & 0.64 \\
H$\zeta$      & 1210.4237 & 4 & 126 & 230 & 0.30 \\
\ion{Ca}{ii} K& 1210.4237 & 4 & 117 & 272 & 0.51 \\
\ion{Na}{i} D2& 1210.4237 & 4 & 120 &  99 & 0.05 \\
H$\beta$      & 1210.4237 & 7 &  10 &  53 & 0.22 \\
H$\gamma$     & 1210.4237 & 7 &  20 &  26 & 0.07 \\
\ion{Ca}{ii} K& 1210.4237 & 0 & --1 &  18 & 0.15 \\
\ion{Ca}{ii} H& 1210.4237 & 0 & --2 &  12 & 0.13 \\
\ion{Na}{i} D2& 1210.4237 & 0 &   2 &  13 & 0.61 \\
\ion{Na}{i} D1& 1210.4237 & 0 &   3 &  13 & 0.50 \\
H$\beta$      & 1210.5156 & 4 & 124 & 217 & 0.62 \\
H$\gamma$     & 1210.5156 & 4 & 117 & 218 & 0.51 \\
H$\delta$     & 1210.5156 & 4 & 111 & 223 & 0.40 \\
H$\epsilon$   & 1210.5156 & 4 &  76 & 262 & 0.60 \\
H$\zeta$      & 1210.5156 & 4 & 111 & 237 & 0.25 \\
\ion{Ca}{ii} K& 1210.5156 & 4 & 111 & 265 & 0.44 \\
\ion{Na}{i} D2& 1210.5156 & 4 & 132 & 111 & 0.04 \\
H$\beta$      & 1210.5156 & 7 &  10 &  53 & 0.23 \\
H$\gamma$     & 1210.5156 & 7 &  13 &  49 & 0.15 \\
\ion{Ca}{ii} K& 1210.5156 & 7 &  22 &  17 & 0.11 \\
\ion{Ca}{ii} H& 1210.5156 & 7 &  17 &  15 & 0.07 \\
\ion{Ca}{ii} K& 1210.5156 & 0 & --1 &  22 & 0.20 \\
\ion{Ca}{ii} H& 1210.5156 & 0 & --3 &  15 & 0.20 \\
\ion{Na}{i} D2& 1210.5156 & 0 &   3 &  13 & 0.61 \\
\ion{Na}{i} D1& 1210.5156 & 0 &   3 &  13 & 0.50 \\
\hline
\end{tabular}}
\end{table}

\section{Results}
\label{uxori_results}

Table~\ref{uxori_master_table}  gives the radial velocity shift $v$, the velocity dispersion $\Delta v$ and the  absorption strength $R$ of each identified transient absorption, computed according to the procedure described in the previous section.
Uncertainties affecting the values of Table~\ref{uxori_master_table} are difficult to quantify, especially in the case of blended components.
We are confident, however, that the values are a good representation of the gas properties in the case of sharp, isolated features.
The results allow us to identify trends, based on the similar and consistent behaviour of many different lines.

\begin{figure}
\centerline{ \
\includegraphics[angle=180,width=0.85\hsize,clip=true]{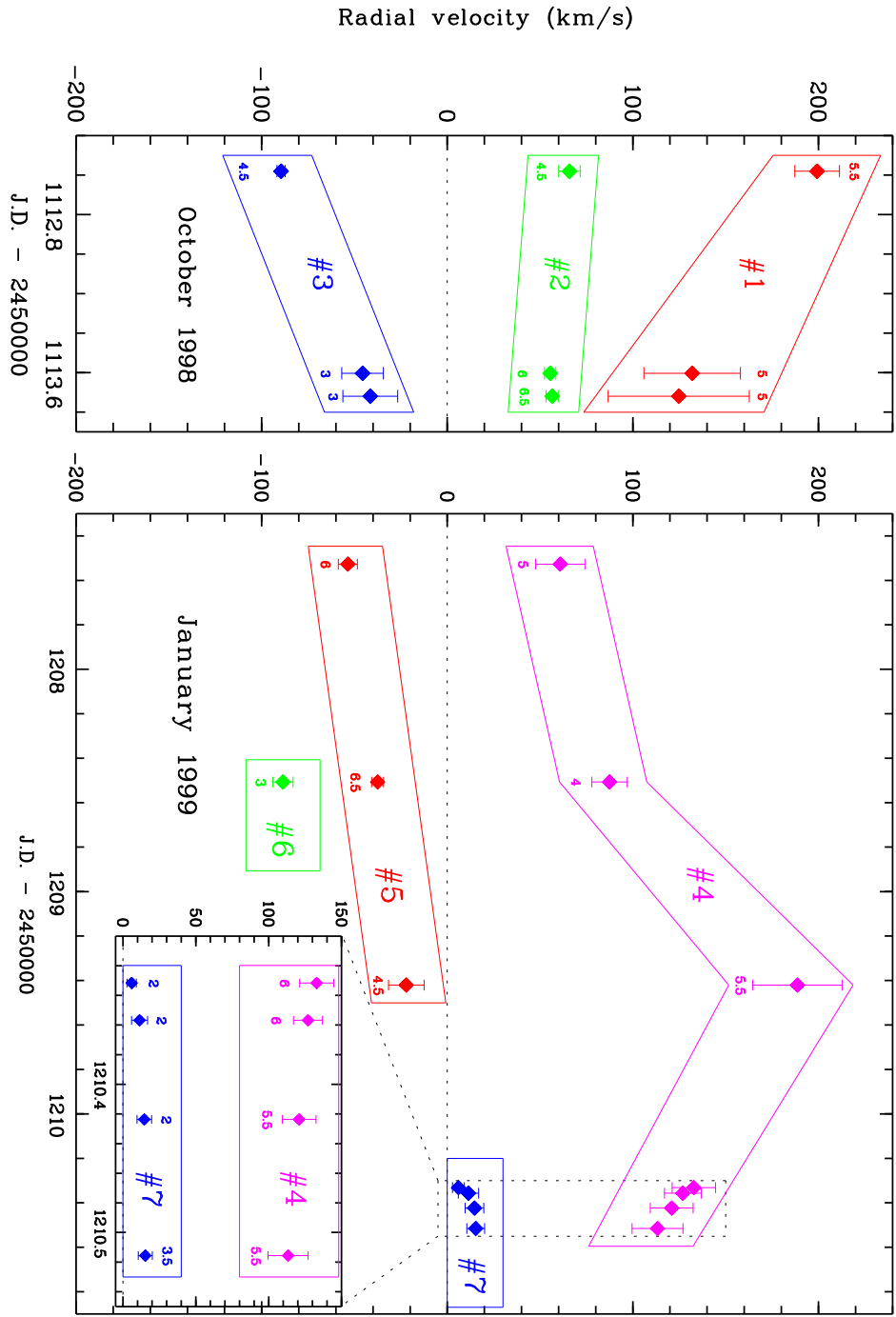}}
\caption
[Observed transient absorption components in each of the UES spectra (characterized by the Julian date of the observations)]
{\small
Observed transient absorption components in each of the UES spectra (characterized by the Julian date of the observations).
Each point corresponds to one ``TAC'' and represents the mean velocity of the features with similar radial velocities detected in different lines. 
The radial velocity behaviour allows us to follow the temporal evolution of the gas causing the ``TAC''.
7 different dynamic events are identified; each event is enclosed in a box with an identification number (\#1, \#2,....).
Error bars show the rms error of the average velocity and the number of lines used to estimate the average velocity is indicated.
Fractional numbers come from the weight attributed to individual lines (see text).
For the sake of clarity, the spectra taken during January 31$^{\rm st}$ 1999 (JD 2451210.4) are expanded along the x--axis
(This figure is available in color in electronic form).}
\label{uxori_master}
\end{figure}

\begin{figure}
\includegraphics[height=0.5\hsize,angle=-90,clip=true]{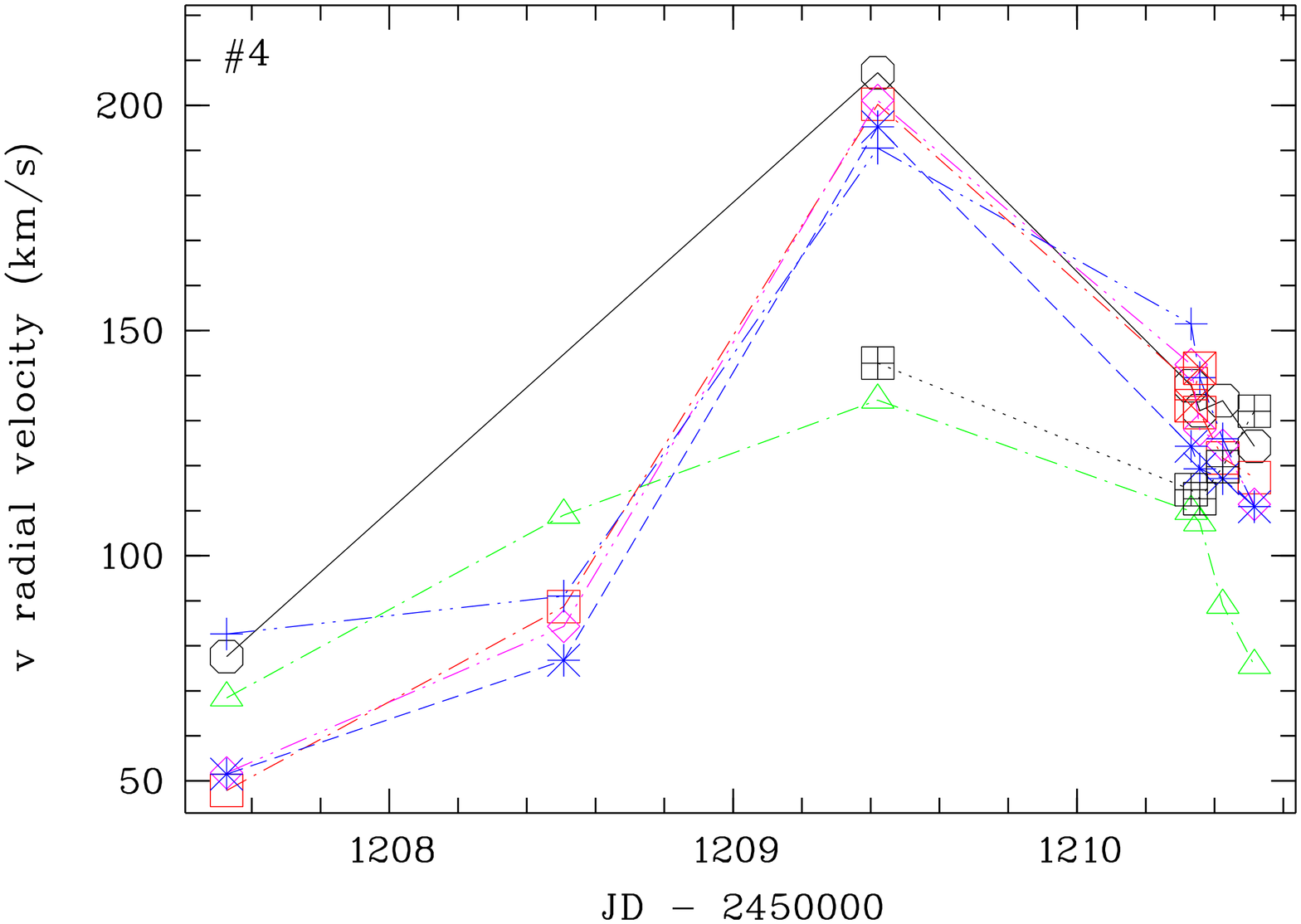}
\includegraphics[height=0.5\hsize,angle=-90,clip=true]{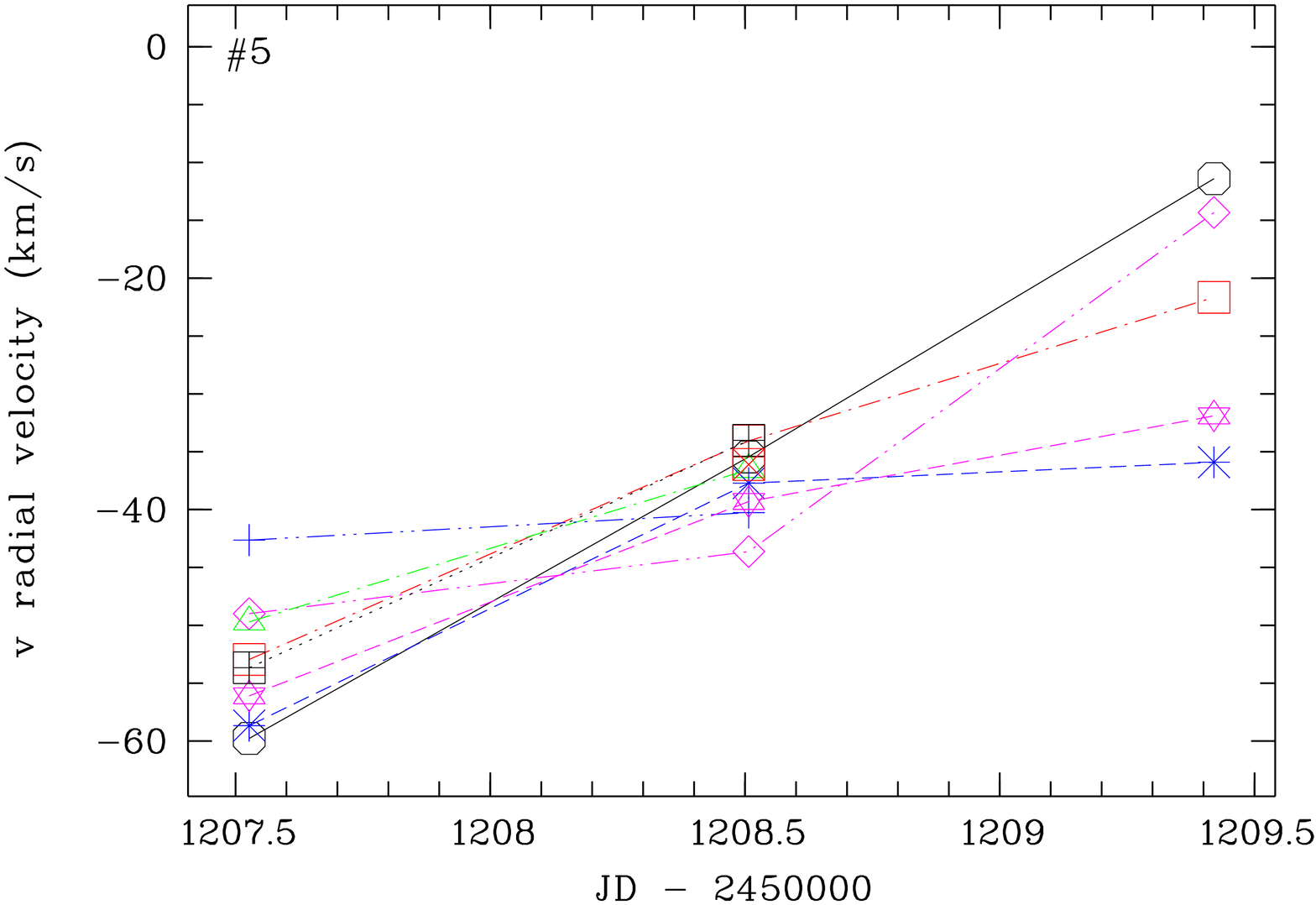}
\includegraphics[height=0.5\hsize,angle=-90,clip=true]{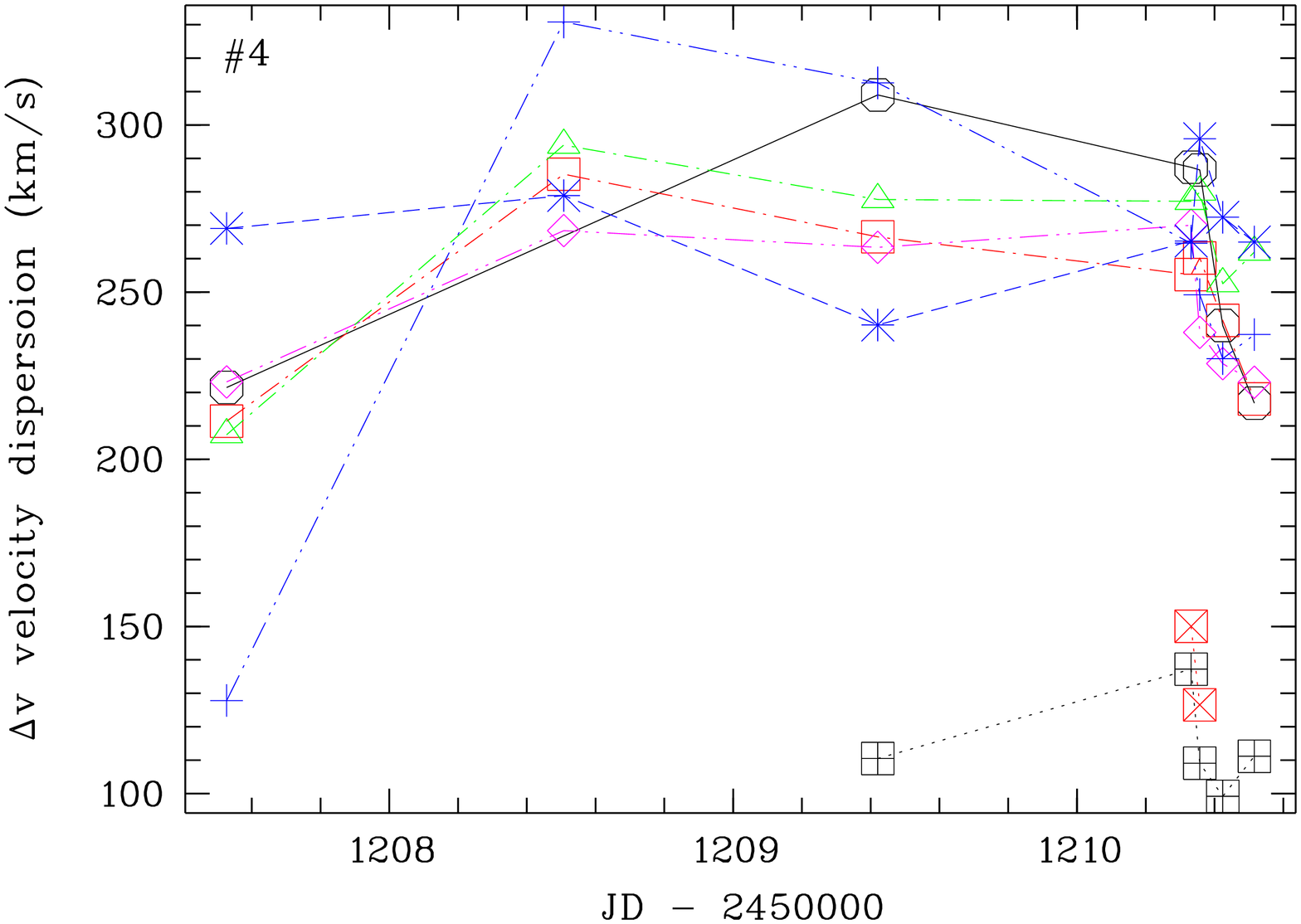}
\includegraphics[height=0.5\hsize,angle=-90,clip=true]{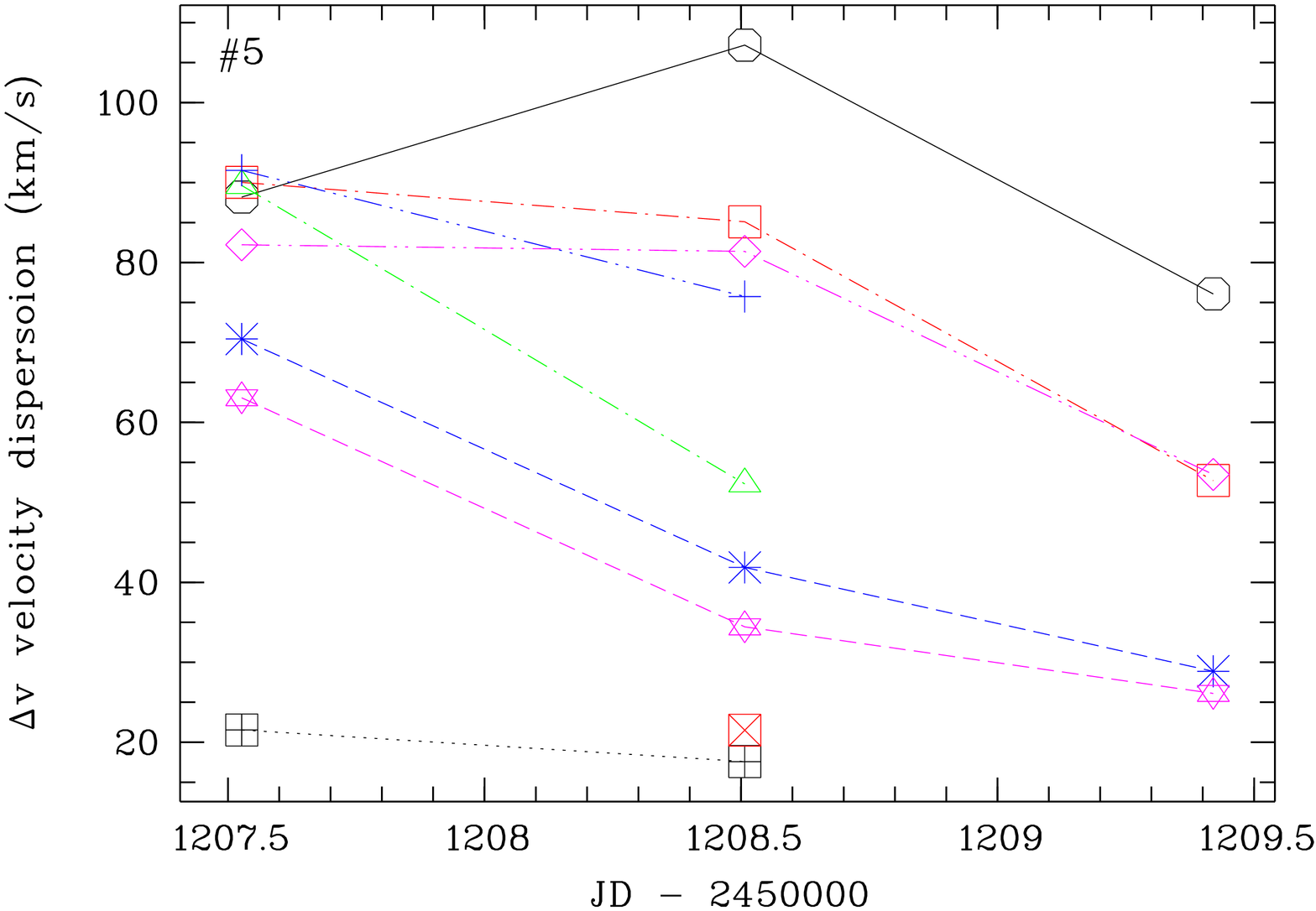}
\includegraphics[height=0.5\hsize,angle=-90,clip=true]{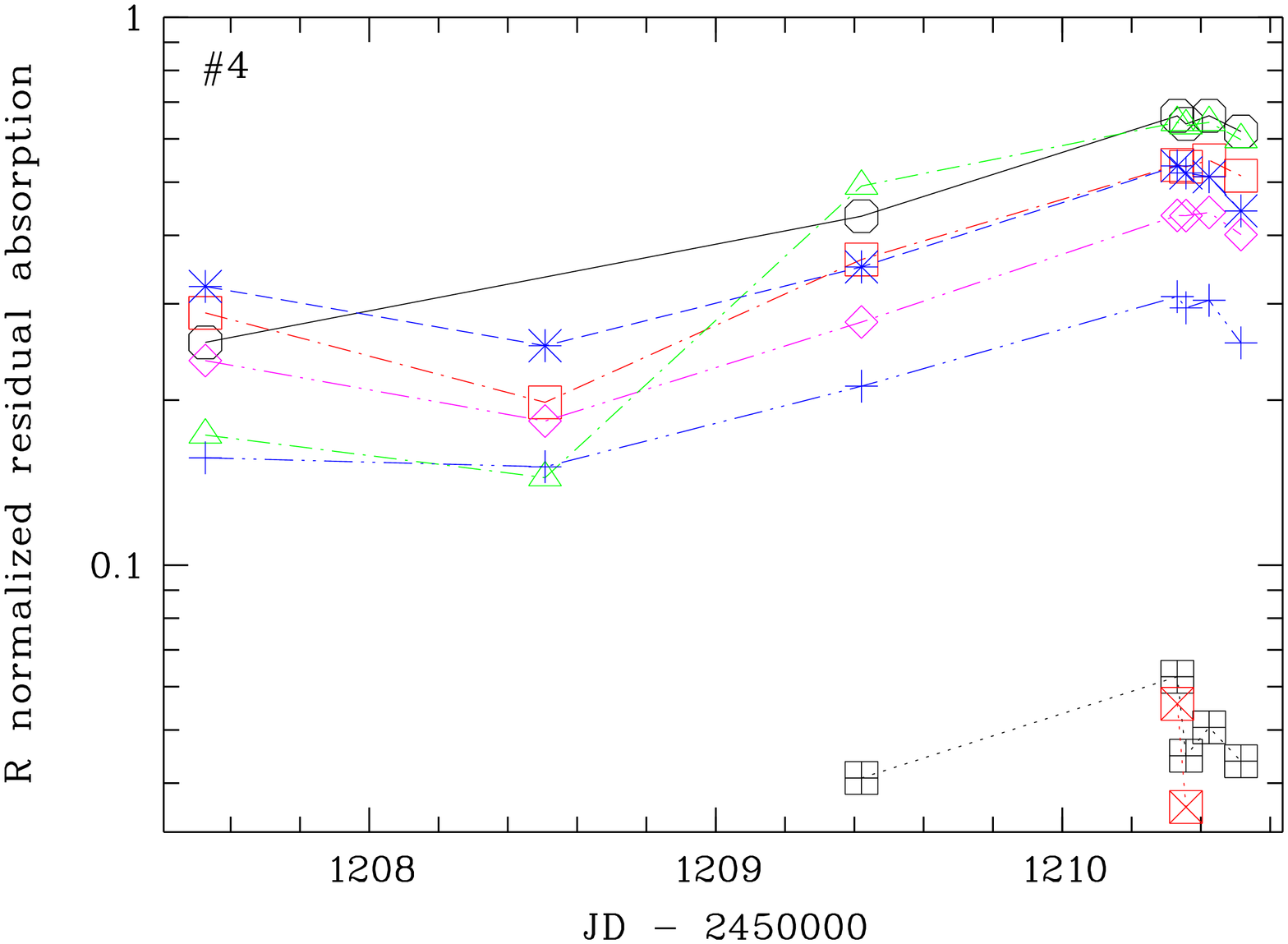}
\includegraphics[height=0.5\hsize,angle=-90,clip=true]{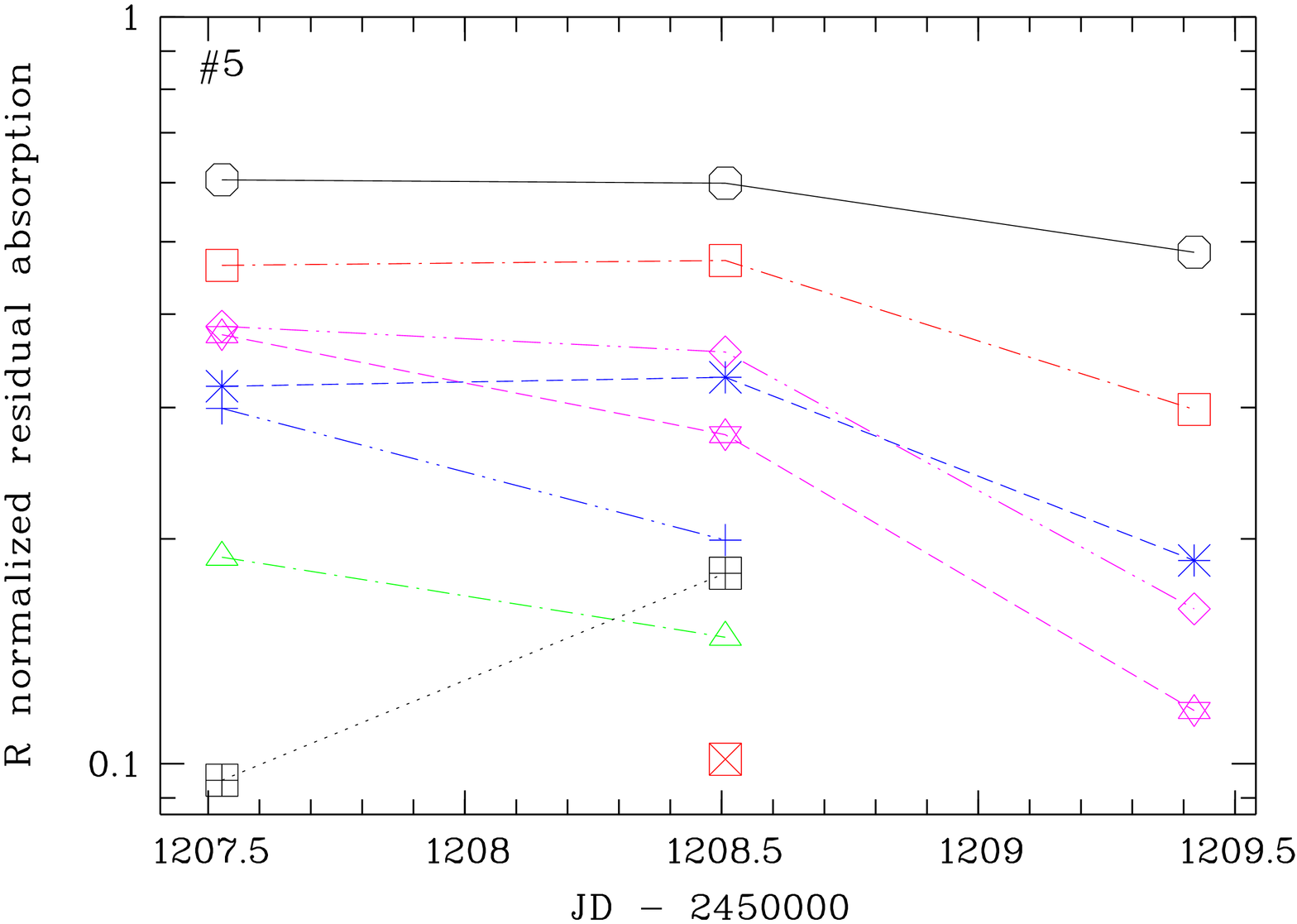}
\includegraphics[height=1.0\hsize,angle=-90,clip=true]{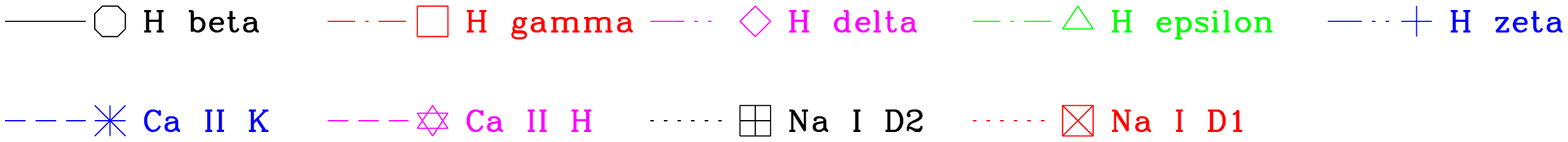}
\caption
[Events \#4 and \#5.
The temporal evolution of $v$, $\Delta v$ and $R$ is shown for every line as a function of the JD.]
{Events \#4 and \#5.
The temporal evolution of $v$, $\Delta v$ and $R$ is shown for every line as a function of the JD.
From left to right: events \#4 and \#5.
From top to bottom: time evolution of $v$ (km/s), $\Delta v$ (km/s) and $R$
(This figure is available in color in electronic form).}
\label{uxori_events_4_5}
\end{figure}

Absorption components at different radial velocities - either redshifted or blueshifted - are found in the Balmer and metallic lines. 
\ion{Ca}{ii} and \ion{Na}{i} components with the stellar radial velocity are also detected.
Features in different lines with similar radial velocities are detected within each spectrum; thus,  it is reasonable to assume that they form in approximately the same region.
The absorptions with similar velocities appearing in different lines are called a TAC, which can be characterized by the average of the radial velocity of the individual lines.
A total of 24 TACs can be identified in our spectra - 17 RACs and 7 BACs.
Fig.~\ref{uxori_master} plots the average radial velocity of each detected TAC as a function of the observing time.
Error bars show the rms error and the number of individual lines used to estimate the average velocity is indicated.
Fractional numbers come from the fact that a weight of 1/2 is assigned to those absorption line features that are less certain (the blended lines H$\epsilon$ and \ion{Ca}{ii}~H, the weak line H$\zeta$ and the \ion{Na}{i} doublet affected by telluric lines), and weight 1 to the rest.
Fig.~\ref{uxori_master} clearly shows that more than one TAC is present in each observed spectrum and that there is a radial velocity shift when comparing TACs in different spectra. 
The shift does not seem to be accidental; there appears to be a systematic temporal evolution of the TACs from one spectrum to the following one when their radial velocities are analysed.
This is clearly evident in the behaviour of the TACs observed in the four spectra taken during the night J.D.  2451210.5 with a time interval of around 4 hours from the first to the last one.  
Thus, the data reveal groups of TACs representing a dynamical evolution of the gas with which they are associated.
We call each of these groups an "event". 
In total, 7 events are identified, of which 4 are redshifted and 3 blueshifted.
The detailed temporal evolution of $v$, \deltav\ and $R$ for each individual line absorption component of 2 of these events (1 blueshifted and 1 redshifted) is shown in Fig.\ref{uxori_events_4_5}.  

\subsection{Kinematics}
\label{uxori_kinematics}

In this section we describe the kinematic behaviour of the circumstellar gas as revealed by our observations.
Since they cover only a limited amount of time, it is difficult to estimate if the trends we have observed are always present, or if they are specific to our observations, and a different choice of times, or a longer coverage of the star, would give a different picture.
Only additional observations, and longer time coverage, can tell us what is the case.
With this caveat, this database is nevertheless a good starting point for the discussion of the TAC phenomenon in UX~Ori.

Among the detected events, the infalling gas (RACs) shows the largest velocities.
For instance, event \#1 shows a maximum average radial velocity of about 200 km/s which is roughly a half of the value of the stellar escape velocity, $\sim$410 km/s, while the maximum average velocity of the outflowing gas (BACs) is about 100 km/s.
This observational result could be biased since we begin detecting outflowing gas at maximum velocities, which is not always the case for the infalling gas (e.g. events \#4 and \#7). 
We also note that blueshifted gas is detected when there are redshifted absorptions, but the opposite is not true; we do not know if there is a physical meaning behind this result. 

The dynamics of events \#1 and \#2 (RACs) and \#3 and \#5 (BACs) denotes a deceleration of the gas.
In event \#4 the gas first accelerates and then decelerates; the data of \#7 suggest an acceleration, while event \#6 is only present in one spectrum.
An estimate of the acceleration of the events over the time interval covered by the data is given in Table~\ref{uxori_acceleration}.
Positive values mean acceleration while negative ones mean deceleration.
Table~\ref{uxori_acceleration} also gives the duration of each event, which is defined as the time it takes to reach zero velocity (for decelerating gas) or to go from zero to the maximum observed velocity for accelerating gas.
The estimates suggest that the events typically last for a few days and that accelerations are a fraction of \ms.
It is interesting to point out that the acceleration and deceleration phases of event \#4 are of the same order of magnitude.
We also note that our data do not allow us to conclude that this event is unique, since we may have missed the accelerating part of the other events.
In this respect, it would have been of great interest to follow the evolution of event \#7, which represents infalling gas with a significant acceleration, departing from a velocity very close to the stellar velocity (though it is only well detected in H$\beta$ and H$\gamma$).

\begin{figure}
\centerline{\includegraphics[clip=true,angle=-90,width=0.80\hsize]
                            {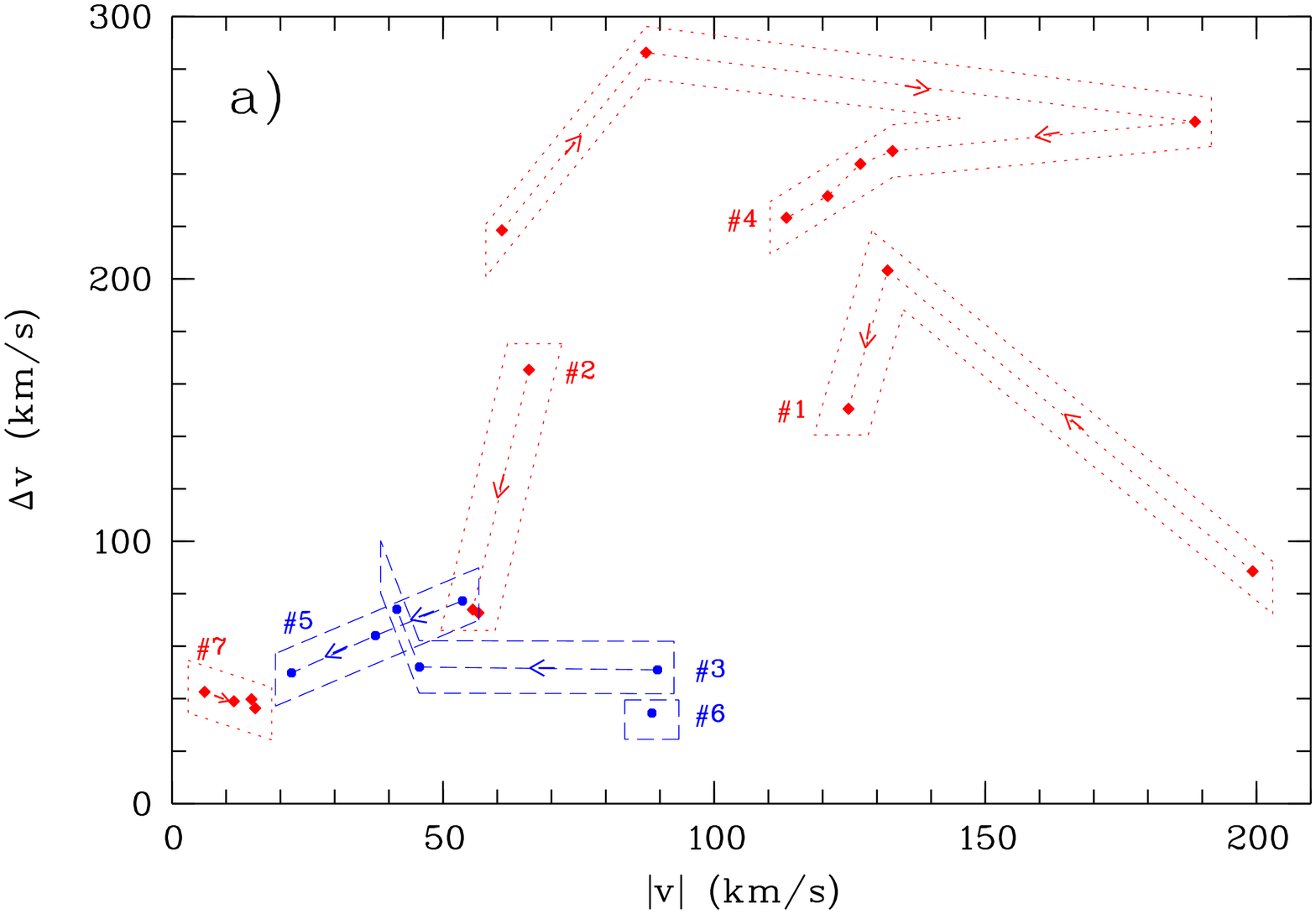}}
\centerline{\includegraphics[clip=true,angle=-90,width=0.80\hsize]
                            {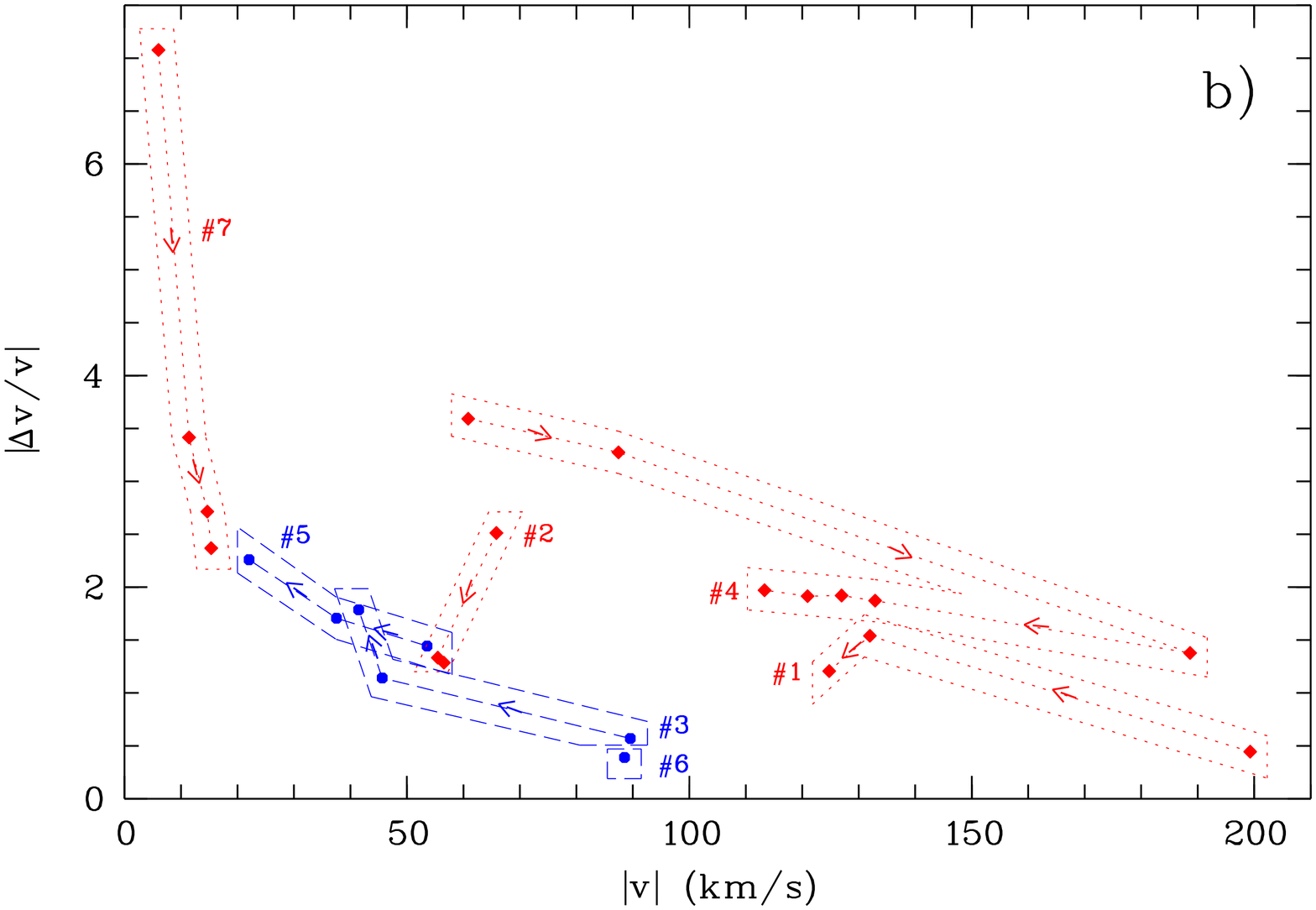}}
\caption
[\deltav\ versus $v$ and the complementary $|\Delta v/v|$ versus $v$ diagrams of all detected TACs]
{\deltav\ versus $v$ (top) and the complementary $|\Delta v/v|$ versus $v$ diagrams of all detected TACs.
Events enclosed in boxes with dotted lines are infalls and these in dashed lines are outflows.
The arrows show the dynamical evolution (acceleration/deceleration) of the events.
Redshifted gas appears to be broader than blueshifted gas.
The location of each event in the diagrams suggests that its velocity dispersion does not change substantially throughout the dynamical lifetime
(This figure is available in color in electronic form).}
\label{uxori_deltav_vs_v}
\end{figure}

All the transient line absorption features are very broad (Table~\ref{uxori_master_table}), as is evident in the raw spectra. 
We have computed for each TAC of each event the average value of \deltav, weighted as in the case of $v$ (the rms dispersion of \deltav\ among the different lines in a TAC is typically $\sim$25\%).
There is a tendency for \ion{Na}{i} lines to be narrower than hydrogen and \ion{Ca}{ii} lines ($\Delta v_{\rm \ion{Na}{i}}~\simeq 0.5~\Delta v_{\rm H,\ion{Ca}{ii}}$).
This will not be investigated further in this paper, but it could give interesting clues to the physical conditions of the gas.
Fig.~\ref{uxori_deltav_vs_v} shows  \deltav\  versus $v$ and \deltav/$v$ versus $v$ for each detected TAC, which are two complementary views.
Each event with the corresponding TACs are enclosed in boxes.
The diagrams show that events are well separated and that the velocity dispersion does not change drastically along their temporal evolution (perhaps with the exception of event \#2). 
This behaviour is remarkable in event  \#4 which has both acceleration and deceleration phases.
Thus, these results suggest that events could be characterized by a kind of ``intrinsic'' velocity dispersion throughout their lifetimes.
 
\begin{table}
\caption[Estimates of the acceleration and duration of each transient event]
{Estimates of the acceleration and duration of each transient event.
The time scale for the event's lifetime is estimated as $\tau = |v_{\rm max}/a|$.
The acceleration and deceleration phases of event \#4 are shown separately.
$\tau$ for event \#7  means the time needed to achieve $v_{\rm max}$ departing from zero velocity. 
}
\label{uxori_acceleration}
\centerline{
\begin{tabular}{llll}
\hline
\hline
Event & Type     & $a$(\ms) &$\tau$ (days)\\
\hline
1     & infall   & -0.8  &  3.0           \\
2     & infall   & -0.1  &  8.1           \\
3     & outflow  & -0.5  &  2.1           \\
4$+$  & infall   & +0.8  &  2.6           \\
4$-$  & infall   & -0.8  &  2.7           \\
5     & outflow  & -0.2  &  3.2           \\
7     & infall   & +0.6  &  0.3           \\
\hline
\end{tabular}
}
\end{table}

\subsection{$R$ parameter}
\label{uxori_r_parameter}

Table~\ref{uxori_master_table} gives the parameter $R$ for each absorption component.
The ratio of the $R$ values among different lines hardly varies for all 24 identified TACs.
This is shown in Table~\ref{uxori_ratios_table}, where we give the ratio of H$\beta$, H$\delta$, H$\zeta$, \ion{Ca}{ii}~K and \ion{Na}{i}~D2 to H$\gamma$.
H$\gamma$ has been chosen as the reference line because it is clearly present in all events.
To estimate the errors we have applied a sigma-clipping algorithm to reject bad points; the corresponding values are given in Table~\ref{uxori_ratios_table}. 
The remaining lines do not have a well defined ratio to H$\gamma$; we think this has  no physical meaning, since H$\epsilon$ and \ion{Ca}{ii}~H are blended and the relatively weak \ion{Na}{i}~D1 is more affected than \ion{Na}{i}~D2 by the strong telluric absorptions in this wavelength interval.

\begin{figure}
\centerline{\includegraphics[clip=true,angle=-90,width=0.80\hsize]
                            {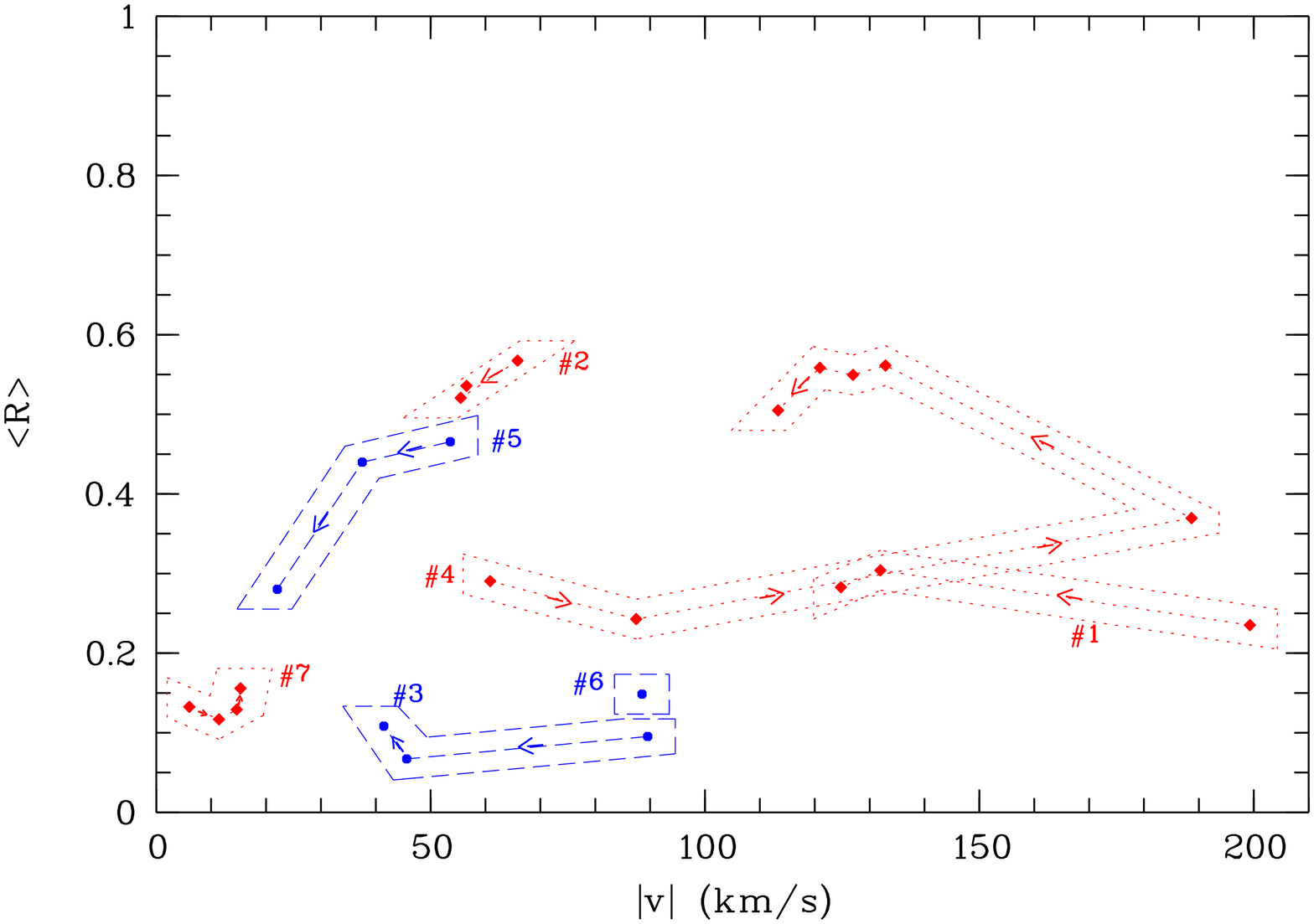}}
\caption[``Intensity'' $<R>$ versus radial velocity $v$ of each TAC]
{``Intensity'' $<R>$ versus radial velocity $v$ of each TAC.
Events and their dynamical evolution are marked as in Fig~\ref{uxori_deltav_vs_v}. 
$<R>$ is different for each event, but there seems to be no large changes among the TACs which form an event
(This figure is available in color in electronic form).}
\label{uxori_r_vs_v}
\end{figure}

The fact that the ratio of $R$ between different lines is rather constant allows us to characterize each TAC by an ``intensity" ($<R>$) in the following manner.
The ratios $R_{\rm line}$/$R_{\rm H\gamma}$ of each TAC are used to compute an ``equivalent average $R_{\rm H\gamma}$ $<R>$". $<R>$ is estimated from H$\beta$, H$\delta$, \ion{Ca}{ii}~K (weight 1.0) and H$\zeta$ (weight 0.5).
Fig.~\ref{uxori_r_vs_v} shows $<R>$ as a function of the TAC velocities and events are enclosed in boxes.
No correlation is seen. 
However, as in the case of \deltav, the estimated $<R>$ values do not seem to change drastically from TAC to TAC within individual events.     

\begin{table}
\caption[Ratios of the average $R$ parameter of several lines to H$\gamma$]
{Ratios of the average $R$ parameter of several lines to H$\gamma$.
The ratios (column 2) are estimated using all TACs in which the corresponding line absorption is present.
A sigma-clipping algorithm has been applied in order to reject bad $R$ values. 
The statistical error, the sigma-clipping value adopted and the fraction of the rejected values are given in columns 3 to 5.
The theoretical value for the Balmer lines is shown for comparison (gf$_{\rm line}$/gf$_{\rm H\gamma}$
).}
\label{uxori_ratios_table}
\centerline{
\begin{tabular}{llllll}
\hline
\hline
Line           & Ratio  & Error & clip. $\sigma$& \% Rej. & Theor. \\
\hline
H$\beta$       & 1.2    & 0.2   & 2.5           &  13     & 2.67   \\ 
H$\delta$      & 0.8    & 0.2   & 2.5           &  0      & 0.49   \\ 
H$\zeta$       & 0.6    & 0.15  & 2.5           &  0      & 0.18   \\ 
\ion{Ca}{ii} K & 0.9    & 0.2   & 2.5           &  5      &  --    \\
\ion{Na}{i} D2 & 0.14   & 0.06  & 2.0           &  20     &  --    \\
\hline
\end{tabular}
}
\end{table}

\section{Discussion}
\label{uxori_discussion}

The intensity and dynamics of the transient components are revealed by the spectra in such detail that a careful analysis in the context of  specific models (magnetospheric accretion models, to mention one obvious example) could be used to validate the theoretical scenarios.
This is in itself a huge effort, well beyond the scope of this paper. There are, however, several points that we think are worth mentioning.
Firstly, the data indicate that the events are the signatures of the dynamical evolution of gaseous clumps in the UX Ori CS disk.
There do not seem to be large changes in the velocity dispersion and the relative line absorption strengths through the lifetime of each clump.
This suggest that they are `blobs' which basically preserve their geometrical and physical identity.
In this respect, event \#4 is remarkable since during its estimated life of $\sim$6 days it shows both an acceleration and a deceleration.
In addition, the data reveal other aspects concerning the nature and dynamics of the events which are indicated in the following subsections.  
 
\subsection{Origin of the circumstellar gas}
\label{uxori_origin_cs_gas}

Based on spectra qualitatively similar to those discussed here, \citet{natta2000} analysed the chemical composition of one infalling event in UX Ori.
Non-LTE models were used to estimate the ionization and excitation of Balmer (up to  H$\delta$), \ion{Ca}{ii} and \ion{Na}{i} lines.
Roughly solar abundances were found for the gas causing the redshifted features of that event.
The present data suggest a similar nature for the gas causing the identified blobs.
We detect evidence of infalling gas in the Balmer lines and in a number of metallic lines with similar velocities, which also display roughly constant strength ratios to the Balmer lines; these ratios are $< 1$ (see Table~\ref{uxori_ratios_table}).
Similar result are found for the outflowing gas (blueshifted absorptions).
Thus, we can conclude that the CS gas in UX Ori is not very metal rich.
In addition, the data suggest that the physical conditions of the outflowing and infalling gas are rather similar and they likely co-exist in space. 

A second point is that in both RACs and BACs the CS gas has a significant underlying emission, at least in the Balmer lines we have examined.
This can be seen easily if we compare the values of the observed average ratio for the Balmer lines to the ratio of the opacity in the lines (see Table~\ref{uxori_ratios_table}, columns 2 and 6).
If the CS gas was just absorbing the photospheric flux, then the ratio of any two lines originating from the same lower level should be equal to the ratio of their opacities.
This is clearly not the case (see also similar results in \citet{natta2000}).
There are a number of reasons why this can happen.
The first is that the observed absorption is due to a very optically thick cloud, whose projected size is smaller than the stellar surface.
In this case, however, $R$ should be the same for all the lines, roughly equal to the occulted fraction of the stellar surface, and this does not seem to be the case.
In particular, the \ion{Na}{i}~D1 absorption is always weaker than the \ion{Na}{i}~D2 (R$_{\rm \ion{Na}{i}D2}$~/~R$_{\rm \ion{Na}{i}D1}$~=~1.4~$\pm$~0.4).
A second possibility is that there is significant emission in the lines.  

R, for any given component, can be estimated from the ratio $F_{\rm obs} / F_{\rm syn}$.
This can be done following eq.~5 of \citet{rodgers2002}, which provides this ratio under the assumption of a stellar and a circumstellar contribution to the observed flux; the circumstellar contribution is caused by a spherical occulting cloud and a more extended envelope.
In our zero order approach we neglect the extended envelope, and we assume a small optical depth and black-body emission for the spherical occulting cloud.
In this case, R can be written as:

\begin{equation}
R= \tau \>\>\Big[1- {{B_\nu(T_{\rm ex})}\over{B_\nu({T_\star})}} \> 
\big({{R_{\rm cloud}}\over{R_\star}}\big)^{2}\Big]
\end{equation}

where $\tau$ is the optical depth in the line, and $T_{\rm ex}$ is the black-body excitation temperature of the gas in the cloud. $R_{\rm cloud}$ is the projected radius of the CS cloud.
Some "filling-in" of the absorption features ($R < \tau$) may occur if $T_{\rm ex} \la T_\star$ and $R_{\rm cloud} \ga R_\star$ (e.g. $T_{\rm ex} \simeq 7000 K$ and $R_{\rm cloud}/R_{\rm \star} = 1.6$, which is the corotation radius for UX~Ori, reproduce the observed ratios).

\subsection{Gas Dynamics}
\label{uxori_gas_dynamics}

As already pointed out, the discussion in the previous subsection together with the results of \citet{natta2000} indicates that the TACs in UX Ori do not arise from the evaporation of solid bodies, but in a substantially different scenario.  
The simultaneous presence of infalling and outflowing gas in the immediate vicinity of UX~Ori is reminiscent of the predictions of magnetospheric accretion models, as developed by a number of authors in the last several years (see \citet{hartmann1998} for a basic account of this theory).
All these models assume stationary conditions, and it is difficult to derive quantitative predictions on the time dependent behaviour of the CS lines.
However, some simple points can be made to constrain the different models.

Let us start with the infall motions we detect. 
The simplest case we can consider for comparison is that of a blob of gas moving along the field lines of a stellar dipole magnetic field, seen in absorption against the stellar photosphere.
The infalling gas is practically contained in a region of the size of the corotation radius, which for UX~Ori is $R_{\rm co} \sim 1.6$ $R_\star$. 
This is in agreement with the discussion in subsection 5.1.
The small value of the corotation radius implies that the maximum poloidal velocity the blob will reach is of order $\sim V_{\rm esc} (1-R_{\rm star}/R_{\rm co}) \sim 220$ \kms.
 Since  the UX~Ori disk is seen almost edge-on \citep{voshchinnikov1988}, the  observed velocity shift will be of the same order, which is in agreement with what we observe.
A simple calculation shows that the blob will be seen to accelerate from very low velocity to this maximum value in a time interval of about 1 day, until it hits the star.
The acceleration seen in event \#4 is reminiscent of such a behaviour, although the time scale is longer.
On the contrary, the decelerating infall motions we  see (events \#1,2,4) require a modification of this simple model.
A significant distortion of the velocity pattern can be caused by the rapid rotation of the star, which can create a sort of centrifugal barrier for the infalling gas.
Using the expression of the poloidal velocity given by eq. (8.76) of \citet{hartmann1998} for the UX~Ori parameters, we find a  decrease of the maximum velocity (to about 180 \kms), but a similar acceleration pattern and time scale.
Still, the interaction of the fast rotating magnetosphere with the accreting disk material should be considered in detail.
\citet{muzerolle2001} have included rotation in a study of T~Tauri Stars (TTS), where, however, it is much smaller than the escape velocity at the stellar surface and the change of the poloidal velocity is negligible.
Similar calculations for cases where the stellar rotation is a significant fraction of the escape velocity  would be very valuable.

Outflowing gas is seen in events \#3 and \#5.
In both cases we start detecting the outflow at negative velocities of 100--80 \kms, and we see deceleration on time scale of 2--3 days.
Stellar-field-driven wind models \citep[X-wind, see ][ and references therein]{shu2000} predict that the wind leaves the disk surface close to the corotation radius on almost radial trajectories, accelerating rapidly to a speed of the order of $\Omega_\star R_{\rm co} \sim  350$ \kms.
This is significantly more than the maximum outflow velocity we observe, although the time coverage of our spectra can bias this result since we always find the maximum velocity of the outflow in the first spectrum taken in both the October 98 and January 99 observing runs.
Nevertheless, the deceleration we observe is not predicted by the models, unless collimation of the wind into the direction of the magnetic axis (i.e., away from the line of sight to the observer) occurs before the maximum speed is achieved and takes place on time scales of few days.
Winds driven by disk magnetic field \citep[see ][ and references therein]{konigl2000} tend to produce slower outflows with large velocity dispersions from a broad region of the disk.
\citet{goodson1997} discuss models where the interaction between the rotating stellar field and a disk field results in a time-dependent launching of a two component outflow, of which one (the disk wind) has some similarity to what we observe.
However, also in these models it is not clear what could cause the observed deceleration.
It is noteworthy that, once the acceleration phase is over, a wind can be decelerated by the effect of the stellar gravity \citep{mitskevich1993}. 
In UX~Ori, the deceleration caused by the central star is $\sim 0.2-0.5$ m\,s$^{-2}$ at distances of $\sim 14-20$ $R_\star$, where the local escape velocity is $\sim$ 130 \kms.
An episode of outflow consistent with being gravitationally decelerated was observed in the TTS SU~Aur by \citet{petrov1996}.

The predictions of simple magnetospheric accretion models have been compared to the observed line variability in a number of TTS, with moderate success.
The best case is that of SU~Aur, a rather massive TTS with a rotation period of about 3 days.
This star shows evidence of simultaneous infall and outflow motions in various lines, similar to those observed in UX~Ori \citep{johns1995,petrov1996,oliveira2000}.
The infalling velocity is clearly modulated by the rotation period of the star, and a model with magnetically channelled accretion in a dipole field inclined by a few degrees to the rotation axis of the star may account for most of the observed infall properties.
As in UX~Ori, the outflow is  very variable, with maximum velocity similar to that of the infalling gas and a tendency to decelerate on a time scale of days, with no obvious rotational modulation \citep{johns1995}.
Note that the timing strategy of our UX~Ori observations is not suited to the detection of modulations on the time scale of the stellar rotation period, which is about 17 hr.

UX~Ori may provide a useful test-case.
On one hand, it is likely that the structure of the stellar magnetic field is more complex than in TTS, since UX~Ori (a A-type star of 2--3 $M_\odot$) lacks the surface convective layer that may create a simple magnetic field in TTS. 
On the other hand, one may take advantage of the fact that the average accretion rate of UX~Ori is rather low \citep{tambovtseva2001}.
The CS gas is  mostly  seen in absorption against the continuum, when a sporadic increase of the accretion rate in the disk creates gas ``blobs" which move along the magnetic field lines, tracing their pattern more clearly than in objects with higher accretion rates, where broad emission dominates the CS lines.
We would like to mention that \citet{pontefract2000} also suggest magnetospheric accretion as a promising model to explain the H$\alpha$ spectropolarimetry data of the Herbig Ae star AB~Aur.

One final point to keep in mind in this context is that variations along the flow (both for infall and outflow) of the source function in the various lines may have an important effect on the comparison between observations and models, so that detailed calculations are required for validation of the models. 

\section{Conclusions}
\label{uxori_conclusions}

The data presented in this paper allow us to analyse the spectroscopic behaviour of the CS gas disk around UX Ori on time scales of months, days and hours.
Significant activity in the CS disk is always present, which manifests itself in the continuous appearance and disappearance of absorption components detected in hydrogen and in many metallic lines.
This activity is not related to substantial variations of the stellar photosphere.
Blobs of gas experiencing infalling and outflowing motions are the likely origin of the transient features.
Blobs undergo accelerations/decelerations of the order of tenths of m\,s$^{-2}$ and last for a few days.
Detectable changes in the gas dynamics occur on a time scale of hours, but the intrinsic velocity dispersion of the blobs appears to remain rather constant.
No noticeable differences are seen in the properties of the infalling and outflowing gas, although infalls might have larger velocity dispersion.   
The relative absorption strength of the transient absorptions suggests gas abundances similar to the solar metallicity, ruling out the evaporation of solid bodies as the physical origin of the spectroscopic features. 
We suggest that the data should be analysed in the context of detailed magnetospheric accretion models, similar to those used for T Tauri stars.

\chapter[Dynamics of the CS gas in BF~Ori, SV~Cep, WW~Vul and XY~Per]
{Dynamics of the circumstellar gas in the Herbig Ae stars BF~Orionis, SV~Cephei, WW~Vulpeculae and XY~Persei}
\label{haebe}

\begin{center}
{\small
\noindent
  A.~Mora$^{1}$,
  C.~Eiroa$^{1,2}$,
  A.~Natta$^{3}$,
  C.A.~Grady$^{4}$,
  D.~de Winter$^{5}$,
  J.K.~Davies$^{6}$,
  R.~Ferlet$^{7}$,
  A.W.~Harris$^{8}$,
  L.F.~Miranda$^{9}$
  B.~Montesinos$^{10,9}$,
  R.D.~Oudmaijer$^{11}$,
  J.~Palacios$^{1}$,
  A.~Quirrenbach$^{12}$,
  H.~Rauer$^{8}$,
  A.~Alberdi$^{9}$,
  A.~Cameron$^{13}$,
  H.J.~Deeg$^{14}$,
  F.~Garz\'on$^{14}$,
  K.~Horne$^{13}$,
  B.~Mer\'{\i}n$^{10}$,
  A.~Penny$^{15}$,
  J.~Schneider$^{16}$,
  E.~Solano$^{10}$,
  Y.~Tsapras$^{13}$, and
  P.R.~Wesselius$^{17}$
}
\end{center}

{\scriptsize
   \setlength{\rightmargin}{1.3cm} \setlength{\leftmargin}{1.3cm}
\begin{list}
  {$^{\arabic{institutes}}$}
  {\usecounter{institutes}
    \setlength{\partopsep}{0cm} \setlength{\topsep}{0.2cm}
    \setlength{\parsep}{0cm}    \setlength{\itemsep}{0cm}}
\item Departamento de F\'{\i}sica Te\'orica C-XI, Universidad Aut\'onoma de
      Madrid, Cantoblanco 28049 Madrid, Spain
\item Visiting Scientist at ESA/ESTEC and Leiden University, The Netherlands
\item Osservatorio Astrofisico di Arcetri, Largo Fermi 5, I-50125 Firenze,
      Italy
\item NOAO/STIS, Goddard Space Flight Center, Code 681, NASA/GSFC, Greenbelt,
      MD 20771, USA
\item TNO/TPD-Space Instrumentation, Stieltjesweg 1, PO Box 155, 2600 AD Delft,
      The Netherlands
\item Astronomy Technology Centre, Royal Observatory, Blackford Hill,
      Edinburgh, UK
\item CNRS, Institute d'Astrophysique de Paris, 98bis Bd. Arago, 75014 Paris,
      France 
\item DLR Department of Planetary Exploration, Rutherfordstrasse 2, 12489
      Berlin, Germany
\item Instituto de Astrof\'{\i}sica de Andaluc\'{\i}a, Apartado de Correos
      3004, 18080 Granada, Spain
\item LAEFF, VILSPA, Apartado de Correos 50727, 28080 Madrid, Spain
\item Department of Physics and Astronomy, University of Leeds, Leeds LS2 9JT,
      UK
\item Department of Physics, Center for Astrophysics and Space Sciences,
      University of California San Diego, Mail Code 0424, La Jolla,
      CA 92093-0424, USA
\item Physics \& Astronomy, University of St. Andrews, North Haugh, St. Andrews
      KY16 9SS, Scotland, UK
\item Instituto de Astrof\'{\i}sica de Canarias, La Laguna 38200 Tenerife,
      Spain
\item Rutherford Appleton Laboratory, Didcot, Oxfordshire OX11 0QX, UK
\item Observatoire de Paris, 92195 Meudon, France
\item SRON, Universiteitscomplex ``Zernike'', Landleven 12, P.O. Box 800, 9700 AV Groningen, The Netherlands
\end{list}

Originally published in Astronomy and Astrophysics {\bf 419}, 225-240 (2004)

Received 3 June 2003 / Accepted 12 February 2004
}

\section*{Abstract}
We present high resolution ($\lambda$/$\Delta \lambda$~=~49\,000) \'echelle spectra of the intermediate mass, pre-main sequence stars BF Ori, SV Cep, WW Wul and XY Per.
The spectra cover the range 3800~--~5900~\AA\ and monitor the stars on time scales of months and days.
All spectra show a large number of Balmer and metallic lines with variable blueshifted and redshifted absorption features superimposed to the photospheric stellar spectra.
Synthetic Kurucz models are used to estimate rotational velocities, effective temperatures and gravities of the stars.
The best photospheric models are subtracted from each observed spectrum to determine the variable absorption features due to the circumstellar gas; those features are characterized in terms of their velocity, $v$, dispersion velocity, $\Delta v$, and residual absorption, $R_{\rm max}$.
The absorption components detected in each spectrum can be grouped by their similar radial velocities and are interpreted as the signature of the dynamical evolution of gaseous clumps with, in most cases, solar-like chemical composition.
This infalling and outflowing gas has similar properties to the circumstellar gas observed in UX~Ori, emphasizing the need for detailed theoretical models, probably in the framework of the magnetospheric accretion scenario, to understand the complex environment in Herbig Ae (HAe) stars.
WW Vul is unusual because, in addition to infalling and outflowing gas with properties similar to those observed in the other stars, it shows also transient absorption features in metallic lines with no obvious counterparts in the hydrogen lines.
This could, in principle, suggest the presence of CS gas clouds with enhanced metallicity around WW~Vul.
The existence of such a metal-rich gas component, however, needs to be confirmed by further observations and a more quantitative analysis.

\vspace{0.5cm}
\noindent {\bf Key words.} Stars: pre-main sequence -- Stars: Circumstellar matter -- Stars:individual: BF Ori, SV Cep, WW Vul, XY Per

\section{Introduction}

Observations reveal that the dynamics of the circumstellar (CS) gaseous disks around intermediate mass, young main sequence (MS) and pre-main sequence (PMS) stars is extremely complex.
Variable absorption components detected in many lines of different elements and ions constitute good examples of such complexity.
The kinematics and intensity strength of the absorption components contain relevant information on the physical properties of the gas.
Further, their detailed characterization and analysis provide clues and constraints on plausible formation mechanisms as well as on theoretical scenarios describing the structure and nature of the CS gaseous disks.

The presence of metal-rich planetesimals in the young MS  $\beta$~Pic system has been inferred both observationally and theoretically in a long series of papers \citep[e.g.][and references therein]{lagrange2000}.
Summarizing, there are two main arguments on which this inference is based.
Firstly, dust causing the far-IR excess and also seen in the $\beta$~Pic disk images may be second generation material continuously replenished by collisions of large solid bodies or slow evaporation. 
Secondly, transient spectral line absorptions, usually redshifted, of different chemical species in a wide range of ionization states can be modelled in terms of the evaporation of km-sized bodies on star-grazing orbits.
Star-grazing planetesimals have also been suggested to exist in the 51~Oph system, a star with an uncertain evolutionary status \citep{roberge2002,vandenancker2001}; also, blueshifted absorption in excited fine structure lines of \ion{C}{ii}$^*$ at 1037~\AA\ and \ion{N}{ii}$^*$ at 1085 and 1086~\AA\ have been used to infer the presence of $\sim$1~m bodies in the Vega-type binary MS system $\sigma$~Her \citep{chen2003}.  

Absorption features similar to those observed in $\beta$~Pic have been observed towards many HAe stars \citep[see][for the description of these stars as a subgroup of HAeBe stars]{nattappiv}, particularly in the UXOR-subclass \citep[e.g.][]{grady1996} and, by analogy, they have been interpreted as indicative of the presence of large solid bodies in the CS disks  around these PMS stars \citep[e.g.][and references therein]{grady2000}.
In principle, this interpretation is not in conflict with the accepted time scale for the formation of planetesimals of $\sim$10$^4$~yr \citep{beckwith2000}, which suggests that planetesimals should already be present during the PMS phase of stars ($\sim$1-10~Myr).
This explanation for the variable absorption features observed in HAe stars is, however, controversial and, in fact, \citet{grinin1994} pointed out other alternatives, more concretely dissipation of dust clouds and the simultaneous infall of cool gas onto the star.
\citet{natta2000} have analyzed the chemical composition of a strong redshifted event in UX~Ori and shown it to have a solar-like composition.
Instead of the planetesimal origin for the transient components, those authors suggested gas accretion from a CS disk.
This result is supported by \citet[from now on Paper~I]{mora2002} who presented the analysis of a large series of high resolution optical spectra of UX~Ori.
Many variable absorption events in hydrogen and metallic lines were attributed to the dynamical evolution of gaseous clumps with non metal-rich, roughly solar chemical compositions.
In addition, \citet{beust2001} have shown that the  $\beta$~Pic infalling planetesimal model would not produce detectable absorptions in typical PMS HAe CS conditions.
We also note that dust disks around HAe stars are primordial and can be explained in the context of irradiated PMS CS disk models \citep{natta2001}.
 
In this paper we present high resolution spectra of the HAe stars BF~Ori, SV~Cep, WW~Vul and XY~Per and perform an analysis similar to that carried out for UX~Ori \citepalias{mora2002}.
The spectra show very active and complex CS gas in these objects; many transient absorption features in hydrogen and metallic lines are detected, indicating similar properties of the gas around these stars to those of UX~Ori CS gas.
In addition, the spectra of WW Vul show metallic features without obvious hydrogen counterparts; in this sense, this star presents a peculiar behaviour.
The layout of the paper is as follows: Section~\ref{haebe_observations} presents a brief description of the observations.
Section~\ref{haebe_analysis} presents the results and an analysis of the photospheric spectra and the CS contribution.
Section~\ref{haebe_discussion} presents a short discussion on the kinematics and strength of the variable features, and on the metallic events detected in WW~Vul.
Finally, Sect.~\ref{haebe_conclusions} gives some concluding remarks.

\section{Observations}
\label{haebe_observations}

High resolution \'echelle spectra were collected with the Utrecht Echelle Spectrograph (UES) at the 4.2~m WHT (La~Palma observatory) during four observing runs in May, July and October 1998 and January 1999. 
28 spectra were obtained: 4 of BF~Ori, 7 of SV~Cep, 8 of WW~Vul and 9 of XY~Per (XY~Per is a visual binary system with a Herbig Ae primary and a B6Ve secondary separated by 1.4'', the good seeing during the observations allowed us to fully separate both components, in this paper only the HAe primary star has been studied). 
The wavelength range was 3800-5900~{\AA} and the spectral resolution, $\lambda / \Delta\lambda$,~=~49\,000 (6~\kms).
Wavelength calibration was performed using Th-Ar arc lamp spectra.
Typical errors of the wavelength calibration are $\sim$5 times smaller than the spectral resolution.
The observing log, exposure times and signal to noise ratio (SNR) values, measured at $\lambda \simeq 4680$~\AA, are given in Table~\ref{haebe_observing_log}.
Further details of the observations and reduction procedure are given by \citet{mora2001}.
For some spectra there are simultaneous optical photo-polarimetric and near-IR photometric observations \citep{oudmaijer2001,eiroa2001}.
Table~\ref{haebe_observing_log} also presents these simultaneous data.
At the time of the observations the stars were close to their brightest state, BF~Ori and XY~Per, or at average brightness, SV~Cep and WW~Vul \citep{herbst1999,eiroa2002}.

\begin{table}
\caption[EXPORT UES/WHT observing logs of BF~Ori, SV~Cep, WW~Vul and XY~Per]
{EXPORT UES/WHT observing logs of BF~Ori, SV~Cep, WW~Vul and XY~Per.
The Julian date ($-$2\,450\,000) of each spectrum is given in Col. 1.
Column 2 shows the exposure time in seconds.
Column 3 gives the SNR at $\lambda \simeq 4680$~\AA.
Columns 4 to 6 give simultaneous $V$, \%P$_V$ and $K$ photopolarimetric data, where available.
Typical errors are 0.10 in $V$, 0.05 in $K$ and 0.1\% in \%P$_V$.}
\label{haebe_observing_log}
\centerline{
\begin{tabular}{lrrlll}
\hline
\hline
{\bf BF Ori} &                 &     &        &         &      \\
\hline
Julian date  &t$_{\rm exp}$ (s)& SNR & $V$    & \%P$_V$ & $K$  \\
\hline
1112.6324    & 1800            & 280 &  9.65  &  0.56   & 7.91 \\
1113.6515    & 2700            & 190 &  9.79  &  0.75   & 7.85 \\
\hline
1209.5542    & 2700            & 210 &  --    &  --     & 7.76 \\
1210.4571    & 2700            & 280 &  --    &  0.14   & --   \\
\hline
             &                 &     &        &         &      \\
\hline
\hline
{\bf SV Cep} &                 &     &        &         &      \\
\hline
Julian date  &t$_{\rm exp}$ (s)& SNR & $V$    & \%P$_V$ & $K$  \\
\hline
950.6668     & 1800            & 100 & --     & --      & --   \\
950.6893     & 1800            & 120 & --     & --      & --   \\
\hline
1025.6260    & 2700            & 140 & --     & 0.96    & --   \\
1025.6595    & 1800            & 130 & --     & --      & --   \\
1026.6684    & 2700            & 140 & --     & --      & --   \\
\hline
1113.4730    & 2700            & 70  & 11.01  & 1.05    & --   \\
\hline
1209.3372    & 2700            & 170 & --     & --      & --   \\
\hline
             &                 &     &        &         &      \\
\hline
\hline
{\bf WW Vul} &                 &     &        &         &      \\
\hline
Julian date  &t$_{\rm exp}$ (s)& SNR & $V$    & \%P$_V$ & $K$  \\
\hline
950.6176     & 1800            & 130 &  --    &  --     & --   \\
950.6413     & 1800            & 130 &  10.89 &  0.69   & --   \\
951.6232     & 1800            & 120 &  --    &  --     & --   \\
951.6465     & 1800            & 150 &  --    &  --     & --   \\
\hline
1023.5186    & 1800            & 140 &  --    &  --     & 7.37 \\
1023.5423    & 2700            & 190 &  --    &  0.40   & --   \\
\hline
1112.3689    & 1800            & 110 &  10.77 &  0.37   & 7.44 \\
1113.3958    & 2700            & 120 &  11.03 &  0.65   & 7.50 \\
\hline
             &                 &     &        &         &      \\
\hline
\hline
{\bf XY Per} &                 &     &        &         &      \\
\hline
Julian date  &t$_{\rm exp}$ (s)& SNR & $V$    & \%P$_V$ & $K$  \\
\hline
1024.6728    & 1800            & 200 & --     & --      & --   \\
1024.6967    & 1800            & 170 & 9.04   & 1.49    & --   \\
1025.6948    & 1800            & 230 & --     & 1.55    & --   \\
1025.7171    & 1800            & 270 & --     & --      & --   \\
1026.7065    & 1800            & 220 & --     & --      & --   \\
\hline
1112.4978    & 1800            & 270 & 9.12   & 1.65    & 5.97 \\
1113.5483    & 2700            & 230 & 9.05   & 1.53    & 5.99 \\
\hline
1207.3786    & 2700            & 190 & --     & --      & --   \\
1209.3844    & 1800            & 260 & 9.51   & 1.58    & 6.18 \\
\hline
\end{tabular}}
\end{table}

\section{Analysis of the spectra and results}
\label{haebe_analysis}

\subsection{The photospheric spectra}
\label{haebe_the_photospheric_spectra}

Circumstellar absorptions with complex profiles and blended components are detected in hydrogen and metallic lines in all UES spectra of the four stars. 
The analysis of those spectral features requires the subtraction of the underlying photospheric spectra.
Such subtraction is carried out following the method outlined in detail in \citetalias{mora2002}.
Briefly, \citet{kuruczCD13} model atmospheres are used to synthesize photospheric spectra.
Four parameters are estimated: the heliocentric radial velocity (\vrad), the rotation velocity (\vsini), the effective temperature (\teff) and the surface gravity (\logg).
It has been shown by \citet{grinin2001} that the photospheric lines of BF Ori and WW Vul can be well reproduced using solar metallicity synthetic spectra.
On the other hand, the presence of CS components in most of the lines makes it very difficult to carry out a detailed abundance analysis.
We have thus decided to assume that the four stars have solar metallicities.
The atomic line data have been obtained from the VALD database \citep{kupka1999}.
The `best' synthetic model, defined by the parameters listed before, is selected by comparing some appropriate faint photospheric absorption lines among the observed spectra and the synthetic ones.
This is not straightforward because each star behaves differently, and the choice of pure photospheric lines in the spectra of such highly variable objects is not trivial.
Therefore, slightly different, ad-hoc approaches for each star are needed.
These approaches are discussed below and the stellar parameters giving the best synthetic spectra are given in Table~\ref{haebe_parameters}.
They are compatible within the uncertainties with the spectral types and rotational velocities quoted by \citet{mora2001} and with the results by \citet{grinin2001}, who studied BF~Ori and WW~Vul.   

\begin{table}
\caption
[Stellar parameters defining the `best' synthetic Kurucz models for each star]
{Stellar parameters defining the `best' synthetic Kurucz models for each star.
See Sect.~\ref{haebe_the_photospheric_spectra} for a discussion on the uncertainties of \teff, \logg\ and \vsini.}
\label{haebe_stellar_parameters}
\centerline{
\begin{tabular}{lcccc}
\hline
\hline
Star   & 
\parbox{0.8cm}{\centering {\teff} \\ (K)} & 
\logg\ & 
\parbox{1cm}{\centering \vsini\ \\ (\kms)} & 
\parbox{1cm}{\centering \vrad\  \\ (\kms)} \\
\hline
BF~Ori & 8750  & 3.5 & 37  &  23.1 $\pm$ 1.9 \\
SV~Cep & 10000 & 4.0 & 225 & -11.9 $\pm$ 0.8 \\
WW~Vul & 9000  & 4.0 & 210 & -10.4 $\pm$ 1.2 \\
XY~Per & 8500  & 3.5 & 200 &   8.3 $\pm$ 0.6 \\
\hline
\end{tabular}}
\label{haebe_parameters}
\end{table}

\begin{figure*}
\centerline{
\includegraphics[clip=true,angle=-90,width=\hsize]
                {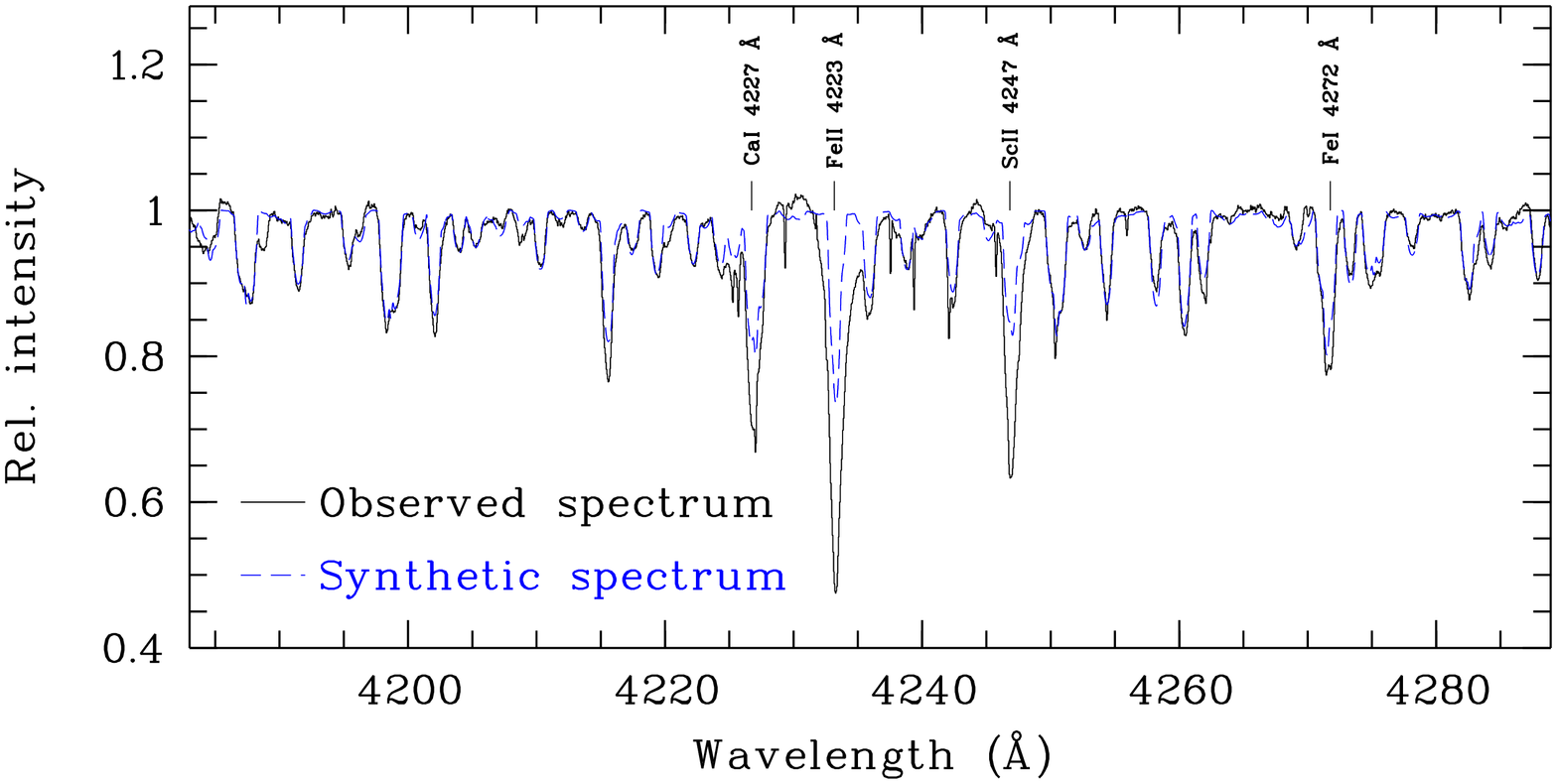}}
\caption[Synthetic vs observed spectrum for BF~Ori]
{Synthetic (dashed line) vs observed spectrum (solid line) for BF~Ori.
The agreement between the photospheric and the observed median spectrum is remarkable over a large wavelength range.
The extra absorption in the stronger lines (identified in the figure) is due to the circumstellar contribution.
This example illustrates the need to use very faint lines to estimate the stellar parameters.
(This figure is available in color in electronic form)}
\label{haebe_bfori_obs_vs_syn}
\end{figure*}

{\em BF~Ori:}
The photospheric lines are narrow and do not show a noticeable variability.
\vrad\ is estimated using faint photospheric lines and its rms error is low.
Estimated errors are  250~K  for \teff\ (the step in the synthetic spectrum grid created using Kuruzc's models), and 6~\kms\ for \vsini\ (the spectral resolution); the \logg\ values considered have been restricted to 3.5 and 4.0.
Fig.~\ref{haebe_bfori_obs_vs_syn} shows the excellent agreement between the synthetic  model and the observed median spectrum.
The extra absorption seen in the stronger lines is due to the circumstellar contribution.  

\begin{figure*}
\centerline{
\includegraphics[clip=true,angle=-90,width=0.5\hsize]
                {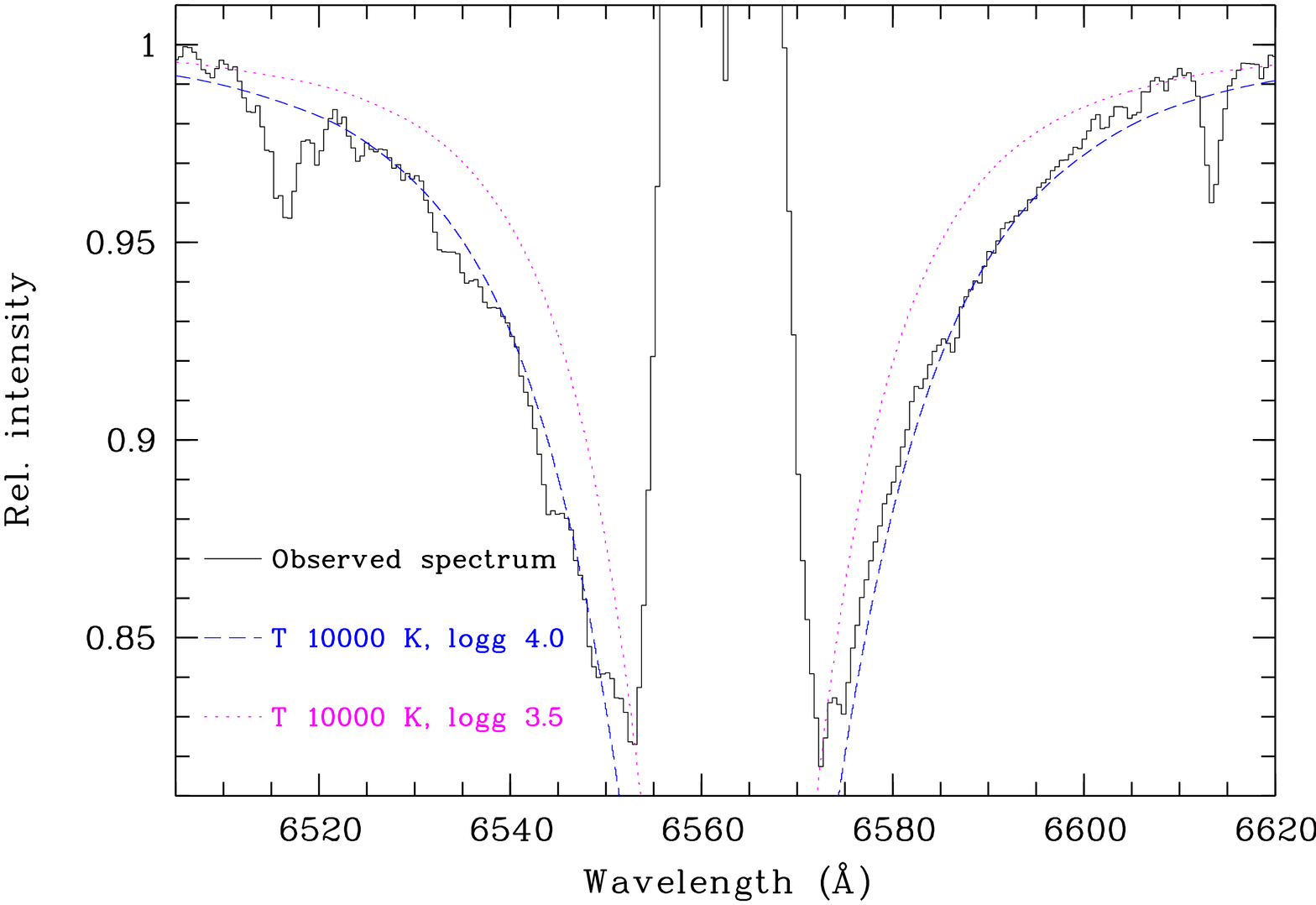}
\includegraphics[clip=true,angle=-90,width=0.5\hsize]
                {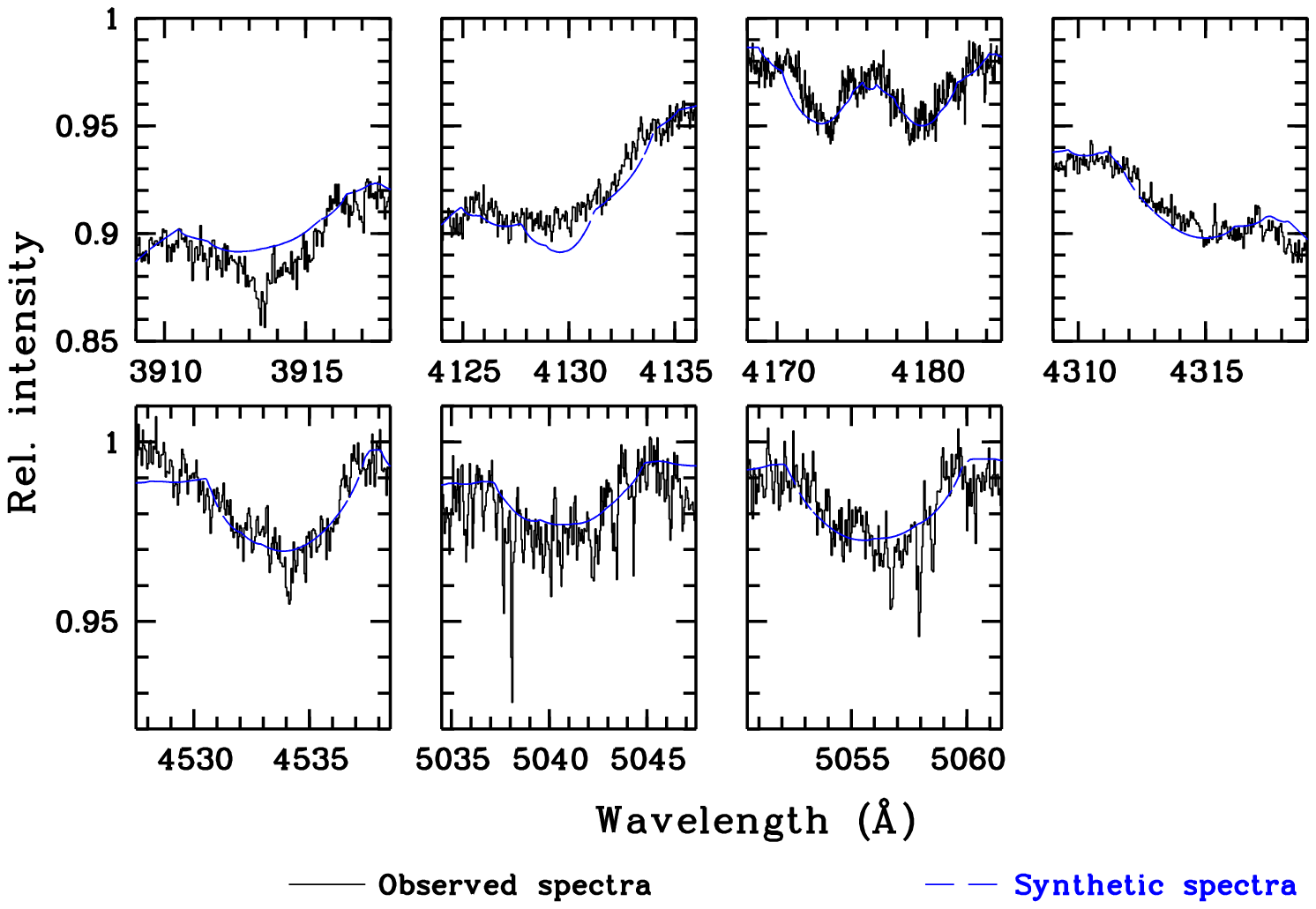}}
\caption[Synthetic vs observed spectra for SV~Cep]
{Synthetic vs observed spectra for SV~Cep.
Left: comparison of the intermediate resolution H$\alpha$ spectrum and two synthetic H$\alpha$ profiles.
The best result is obtained for \teff~=~10\,000~K and \logg~=~4.0 (blue dashed line). 
It was assumed that \vsini~=~206~\kms\ \citep{mora2001}.
Right: Median UES spectra (solid lines) of several blends compared with the best synthetic spectra, with \teff~=~10\,000~K, \vsini~=~225~\kms (blue dashed lines).
(This figure is available in color in electronic form)}
\label{haebe_svcep_obs_vs_syn}
\end{figure*}

{\em SV~Cep:}
According to \citet{finkenzeller1984} the radial velocity of HAe stars coincides within a few \kms with the radial velocity of interstellar (IS) lines.
The strong \ion{Na}{i}~D IS components in the line of sight of SV~Cep allow us to determine a precise value of the radial velocity of this star.
Its high temperature, rotation velocity and variability make it extremely difficult to identify photospheric lines in order to derive the Kurucz stellar parameters.
We have circumvented this problem by using the EXPORT intermediate resolution spectra \citep{mora2001}, which have very high quality H$\alpha$ profiles and are of great help in making a first estimate of \teff\ and \logg.
From the H$\alpha$ wings and the almost complete absence of weak photospheric lines in the high resolution spectra, \teff~$\simeq$~10\,000~K and \logg~=~4.0 are estimated.
It was assumed that \vsini~=~206~\kms \citep{mora2001}.
The adequacy of this choice can be seen in Fig.~\ref{haebe_svcep_obs_vs_syn}, where the 
observed median SV~Cep H$\alpha$ line and 2 synthetic Balmer profiles (\teff~=~10\,000~K, \logg~=~4.0, \vsini~=~206~\kms\ and \teff~=~10\,000~K, \logg~=~3.5, \vsini~=~206~\kms) are shown.
Using those values we have identified 7 blends of weak photospheric lines (absorption intensities $\leq$~4\% of the continuum) with very low variability.
The blends are at $\sim$~3913, 4129, 4176, 4314, 4534, 5041 and 5056~\AA.
A grid of photospheric spectra has been generated with different values of \teff\ and \vsini\ (\logg\ was assumed to be 4.0 from the $H\alpha$ analysis) and the rms differences between the synthetic and observed spectra have been estimated.
The lowest rms difference is obtained for \teff~$=$~10\,000~K and \vsini~$=$~225~\kms.
The differences between the H$\alpha$ wing profiles broadened to \vsini~$=$~225~\kms\ and \vsini~$=$~206~\kms\ are negligible, so it was not needed to compute \logg\ again.
Fig.~\ref{haebe_svcep_obs_vs_syn} also shows the best synthetic and the observed median spectra for the selected blends. 
Uncertainties of $\sim$500~K (the step of the Kurucz's models) and $\sim$10\% for \teff\ and \vsini\,, respectively, are estimated. 

\begin{figure}
\centerline{
\includegraphics[clip=true,angle=-90,width=0.75\hsize]
                {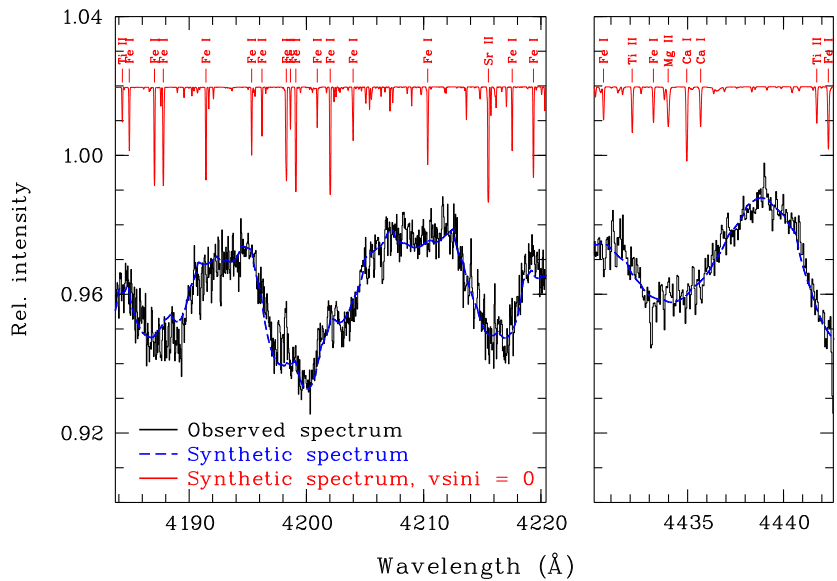}}
\caption
[The observed median UES spectrum of WW~Vul compared to the broadened synthetic one in two different spectral regions]
{The observed median UES spectrum of WW~Vul (black continuous line) compared to the broadened synthetic one (blue dashed lines) in two different spectral regions.
At the top of the figure, the unbroadened Kurucz model is shown with some line identifications (red solid line).  
(This figure is available in color in electronic form)}
\label{haebe_wwvul_obs_vs_syn}
\end{figure}

{\em WW~Vul:}
\vrad ~is estimated from the sharp \ion{Na}{i}~D IS lines.
The star was very active during the observing runs and its spectra show a large number of broad and variable absorption features, e.g. redshifted features are superimposed on practically each photospheric line in the JD 1113.40 spectrum, which pose severe difficulties for the selection of appropriate photospheric lines.
Nevertheless \vsini\ has been estimated with the \ion{Mg}{ii} 4481~\AA\ blended doublet. 
Many pairs of \teff~--~\logg\ values can reproduce the observed spectra, though the agreement between the observed and synthetic spectra is generally poor.
We have selected \teff~=~9000~K, \logg~=~4.0 as representative of the WW~Vul photosphere because of its compatibility with the results by \citet{mora2001}, though large errors are likely, but not larger than around 10\% in both \teff\ and \vsini.
Fig.~\ref{haebe_wwvul_obs_vs_syn} shows a comparison between the observed median spectrum and the broadened synthetic one in two spectral regions where a good fit is achieved; this figure also shows the unbroadened Kurucz's model with some line identifications.
 
\begin{figure*}
\centerline{
\includegraphics[clip=true,angle=-90,width=\hsize]
                {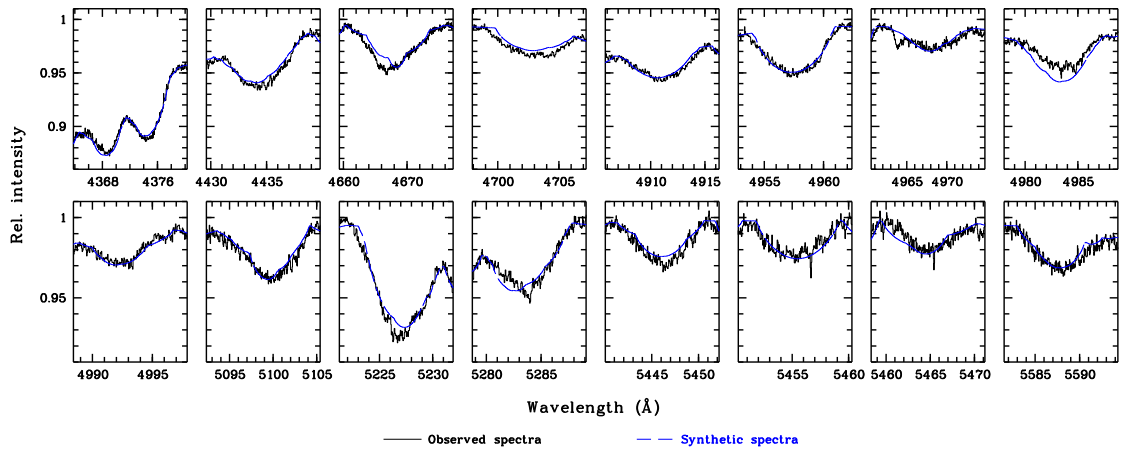}}
\caption
[Synthetic vs observed median spectra of XY~Per for 16 spectral features with very low CS contribution]
{Synthetic (dashed lines) vs observed median (continuous lines) spectra of XY~Per for 16 spectral features with very low CS contribution.
The stellar parameters of the synthetic spectrum are given in Table~\ref{haebe_stellar_parameters}.
(This figure is available in color in electronic form)}
\label{haebe_xyper_obs_vs_syn}
\end{figure*}

{\em XY~Per:}
\vrad ~is estimated from the \ion{Na}{i}~D IS components.
The photospheric lines are very broad but the high SNR of the spectra and the relatively low \teff ~allow us to identify a large number of faint line blends in order to perform a precise estimate of \teff, \logg\ and \vsini ~(16 faint blends with absorption intensities $\leq$~8\% of the continuum with very little CS activity could be identified).    
Errors of \teff\ and \logg\ are of the order of the step in the Kurucz models, 250~K and 0.5 respectively, while the error in  \vsini\ is very low, $<$10\%.
The comparison between the synthetic and the observed median spectra of the 16 blends is shown in Fig.~\ref{haebe_xyper_obs_vs_syn}.

\subsection{The circumstellar transient absorption contribution}
\label{haebe_circumstellar_contribution}

Once the best photospheric spectrum of each star is determined, the circumstellar contribution to each observed spectrum can be estimated by subtracting the synthetic one.
The residual spectra show transient absorption features, which can be characterized by means of the normalized residual absorption, defined as \mbox{$R = 1 - F_{\rm obs} / F_{\rm syn}$} \citep{natta2000}.
The $R$ profile of each line reflects the blending of several components.
A multigaussian fit providing the radial velocity, velocity dispersion and absorption strength is used to identify the individual components \citepalias[see][for details]{mora2002}.
Broad redshifted and blueshifted absorptions at different radial velocities are found in the Balmer and metallic lines for all 4 stars analyzed.
We apply the multigaussian fit to Balmer lines (H$\beta$ 4861~\AA, H$\gamma$ 4340~\AA, H$\delta$ 4102~\AA, H$\epsilon$ 3970~\AA, H$\zeta$ 3889~\AA), \ion{Ca}{ii}~K 3934~\AA, \ion{Ca}{ii}~H 3968~\AA, \ion{Na}{i}~D2 5890~{\AA} and \ion{Na}{i}~D1 5896~\AA, as well as to fainter metallic lines \ion{Fe}{ii}~42 multiplet (\mbox{a6S-z6Po} triplet: 4924~\AA, 5018~\AA\ and  5169~\AA), \ion{Ti}{ii} 4444~\AA, \ion{Ti}{ii} 4572~\AA, \ion{Fe}{i} 4046~\AA, \ion{Sc}{ii} 4247~\AA\ and \ion{Ca}{i} 4227~\AA.
We have chosen these ionic lines because they show significant CS variability and are relatively strong and isolated.
Narrow IS components (mainly \ion{Na}{i}, \ion{Ca}{ii} and \ion{Fe}{ii}) with the stellar radial velocity are also detected for all the stars. 

Consecutive spectra with a time delay of $\sim$1~hour of SV~Cep, XY~Per and WW~Vul were taken on several nights (see Table 1).
These spectra were quite similar and  the gaussian deconvolution of the $R$ profiles essentially provides the same values for the fit parameters; thus, any significant variation of the phenomena causing the transient absorptions is excluded on this time scale, at least during these observing periods.
This result gives us confidence in the identification of the components and allows us to add the spectra taken during the same night in order to increase the SNR.
Tables~\ref{haebe_master_table_bfori},~\ref{haebe_master_table_svcep},~\ref{haebe_master_table_wwvul}~and~\ref{haebe_master_table_xyper} give the radial velocity shift $v$, the velocity dispersion \deltav\ and the absorption strength $R_{\rm max}$, the peak of the $R$ profile, of each identified broad transient absorption component of the lines listed above for BF~Ori, SV~Cep, WW~Vul and XY~Per, respectively.
Column~1 gives the the corresponding Balmer or metallic line, Col.~2 gives the Julian Date, Col.~3 represents the event assigned to the particular absorptions, Col.~4 gives the radial velocity shift $v$, Col.~5 lists the velocity dispersion \deltav\ and Col.~6 gives the absorption strength $R_{\rm max}$.
JD values in Tables~\ref{haebe_master_table_svcep},~\ref{haebe_master_table_wwvul}~and~\ref{haebe_master_table_xyper} correspond to the starting time of the first spectrum of each night. In the following, whenever a JD is given, 2\,450\,000 is subtracted.

\begin{figure*}
\centerline{
\includegraphics[clip=true,angle=-90,width=\hsize]
                {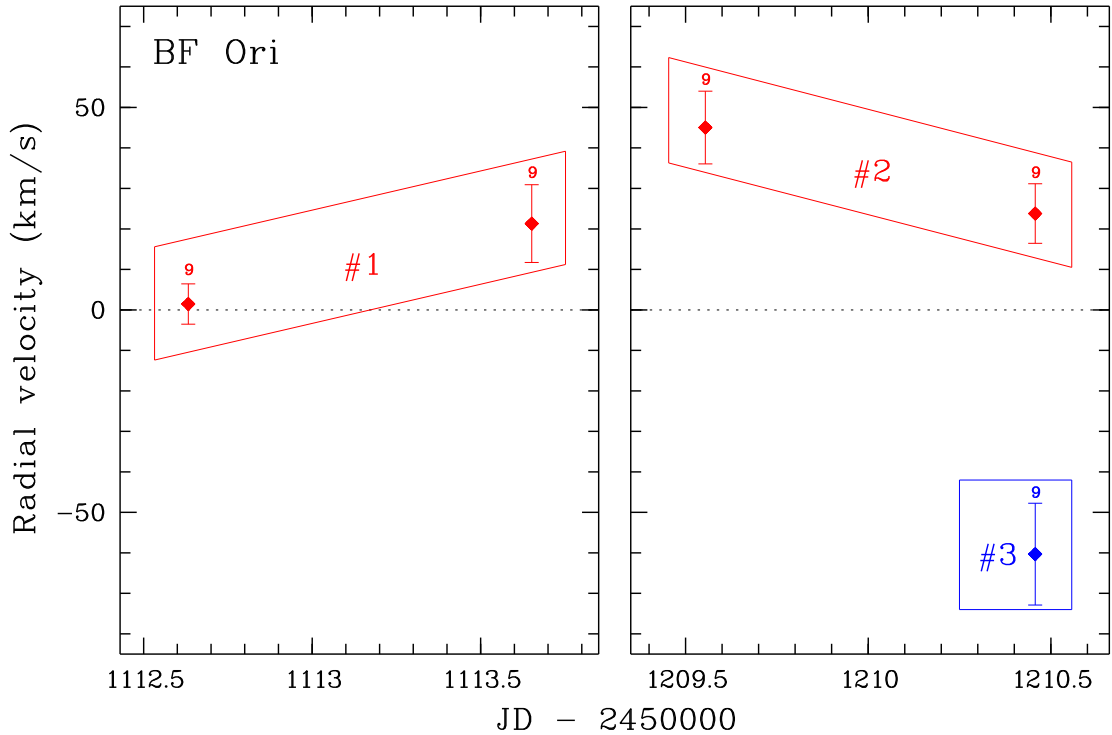}}
\caption[Events in BF~Ori]
{Events in BF~Ori.
Each point corresponds to the radial velocity of one TAC and represents the average velocity $<$$v$$>$ of the absorptions with similar radial velocities detected in different lines.
$<$$v$$>$ is a weighted average in which the higher intensity non-blended lines are assigned a weight of 1 and the rest 1/2.
Error bars show the rms error of the average velocity; the numbers above the data points indicate the weighted number of lines used to estimate the average.
Fractional numbers arise from the 1/2 weight attributed to some lines (see text).
Redshifted events (infalling gas) are printed in red colour, while blueshifted events (outflowing gas) are in blue.
(This figure is available in color in electronic form)}
\label{haebe_master_bfori}
\end{figure*}

\begin{figure*}
\centerline{
\includegraphics[clip=true,angle=-90,width=\hsize]
                {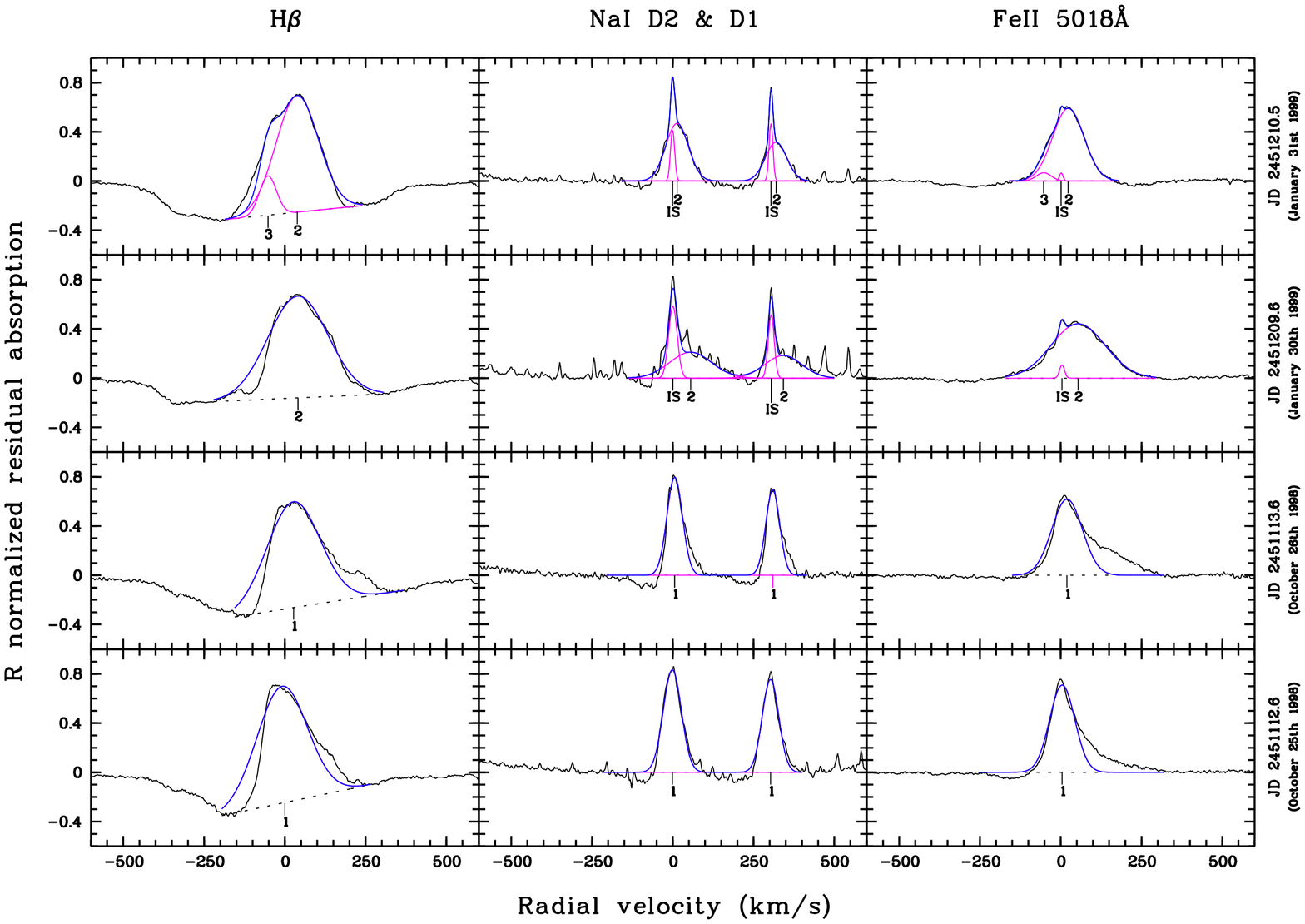}}
\caption[BF~Ori $R$ profiles]
{BF~Ori $R$ profiles.
The normalized residual absorption profiles ($R$ = 1 - $F_{\rm obs}$ / $F_{\rm syn}$) of H$\beta$ (left), \ion{Na}{i} D2 and D1 (middle) and \ion{Fe}{ii}~5018~\AA\ (right) are shown in the figure (black colour).
The corresponding spectra are indicated in the vertical right axis (Julian and civilian epochs).
The identified TACs (gaussian components, pink colour) and the reconstructed $R$ profile fit (blue colour) are displayed.
Event numbers are  shown under the gaussians.
The zero velocity interstellar components are marked as ``IS''.
(This figure is available in color in electronic form)}
\label{haebe_bfori_spectra}
\end{figure*}

\begin{figure*}
\centerline{
\includegraphics[clip=true,angle=-90,width=\hsize]
                {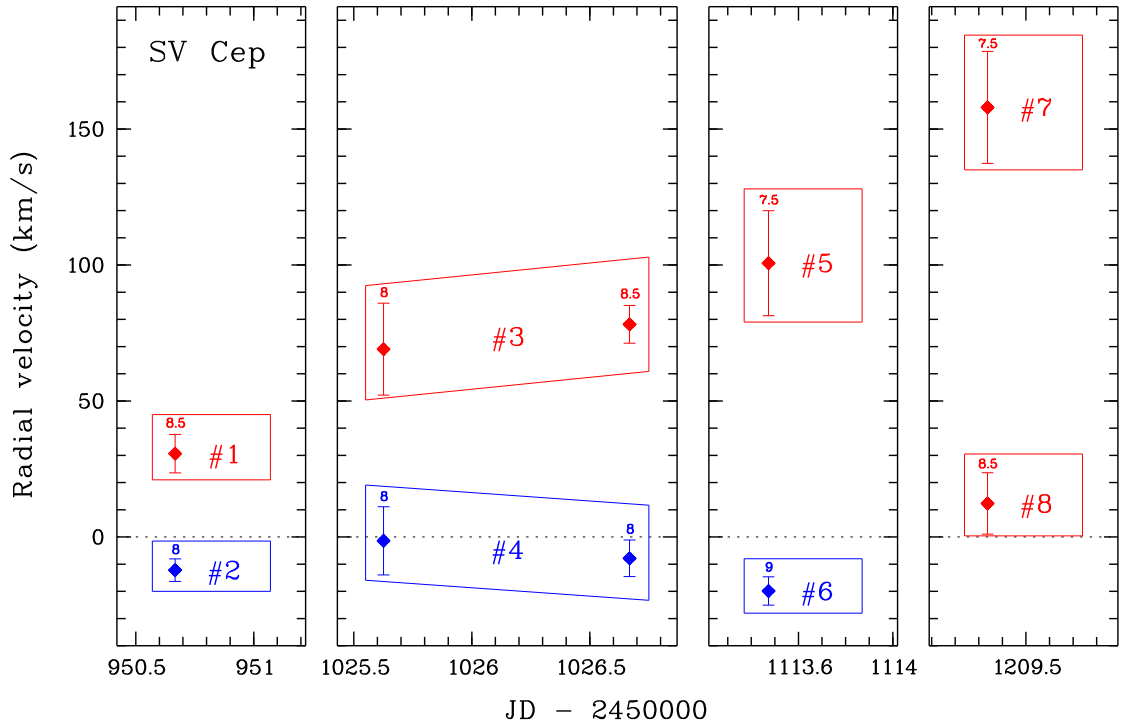}}
\caption[Events in SV~Cep]
{Events in SV~Cep.
Details as for Fig.~\ref{haebe_master_bfori}
(This figure is available in color in electronic form)}
\label{haebe_master_svcep}
\end{figure*}

\begin{figure*}
\centerline{
\includegraphics[clip=true,angle=-90,width=\hsize]
                {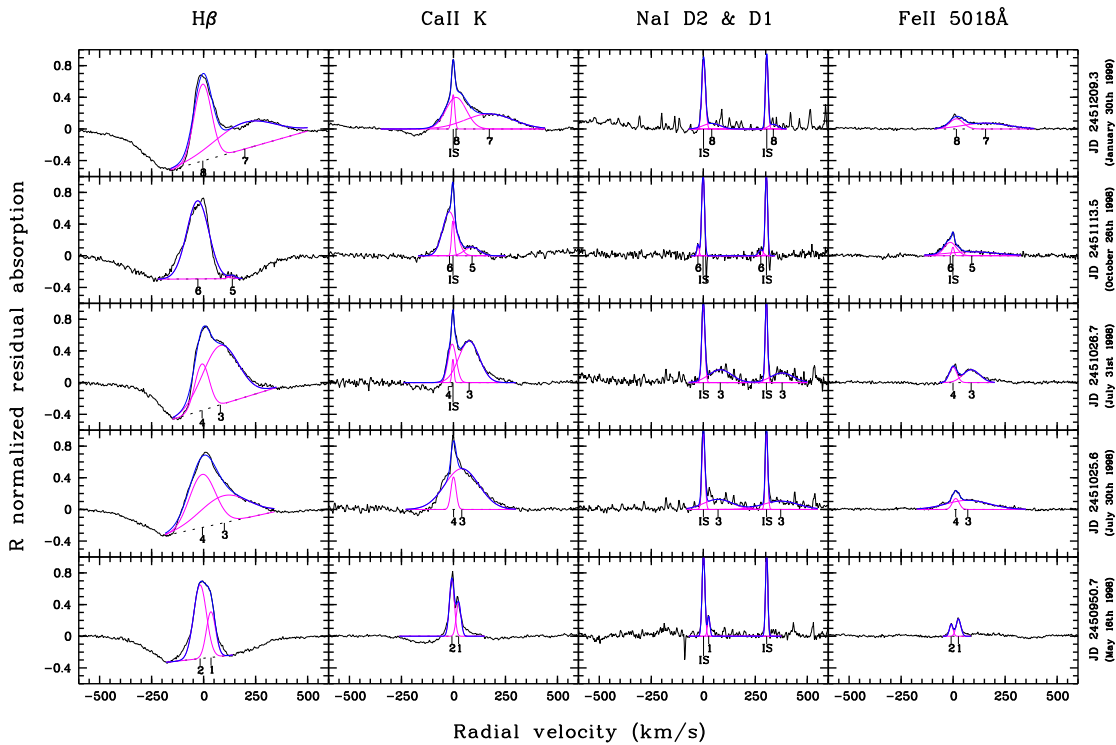}}
\caption[SV~Cep $R$ profiles]
{SV~Cep $R$ profiles.
Details as for Fig~\ref{haebe_bfori_spectra}.
(This figure is available in color in electronic form)}
\label{haebe_svcep_spectra}
\end{figure*}

\begin{figure*}
\centerline{
\includegraphics[clip=true,angle=-90,width=\hsize]
                {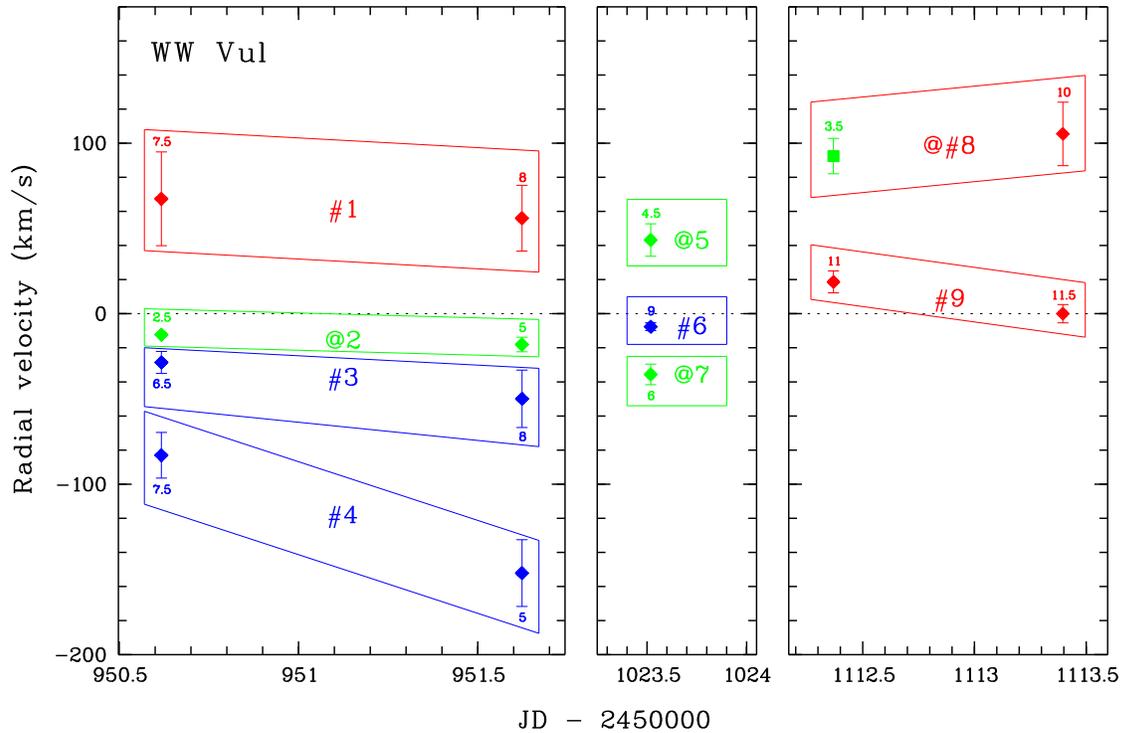}}
\caption[Events in WW~Vul]
{Events in WW~Vul.
Events marked with a ``\#'' (1, 3, 4, 6, 8 and 9; red for redshifted events and blue for blueshifted events) are detected both in hydrogen and metallic lines.
Events denoted with an ``@'' (2, 5 and 7; green) are only seen in metallic lines.
The square point of \#8 in JD~1112.37 correspond to TACs only observed in \ion{Fe}{ii} and \ion{Ti}{ii}.
Further details as for Fig~\ref{haebe_master_bfori}.
(This figure is available in color in electronic form).}
\label{haebe_master_wwvul}
\end{figure*}

\begin{figure*}
\centerline{
\includegraphics[clip=true,angle=-90,width=\hsize]
                {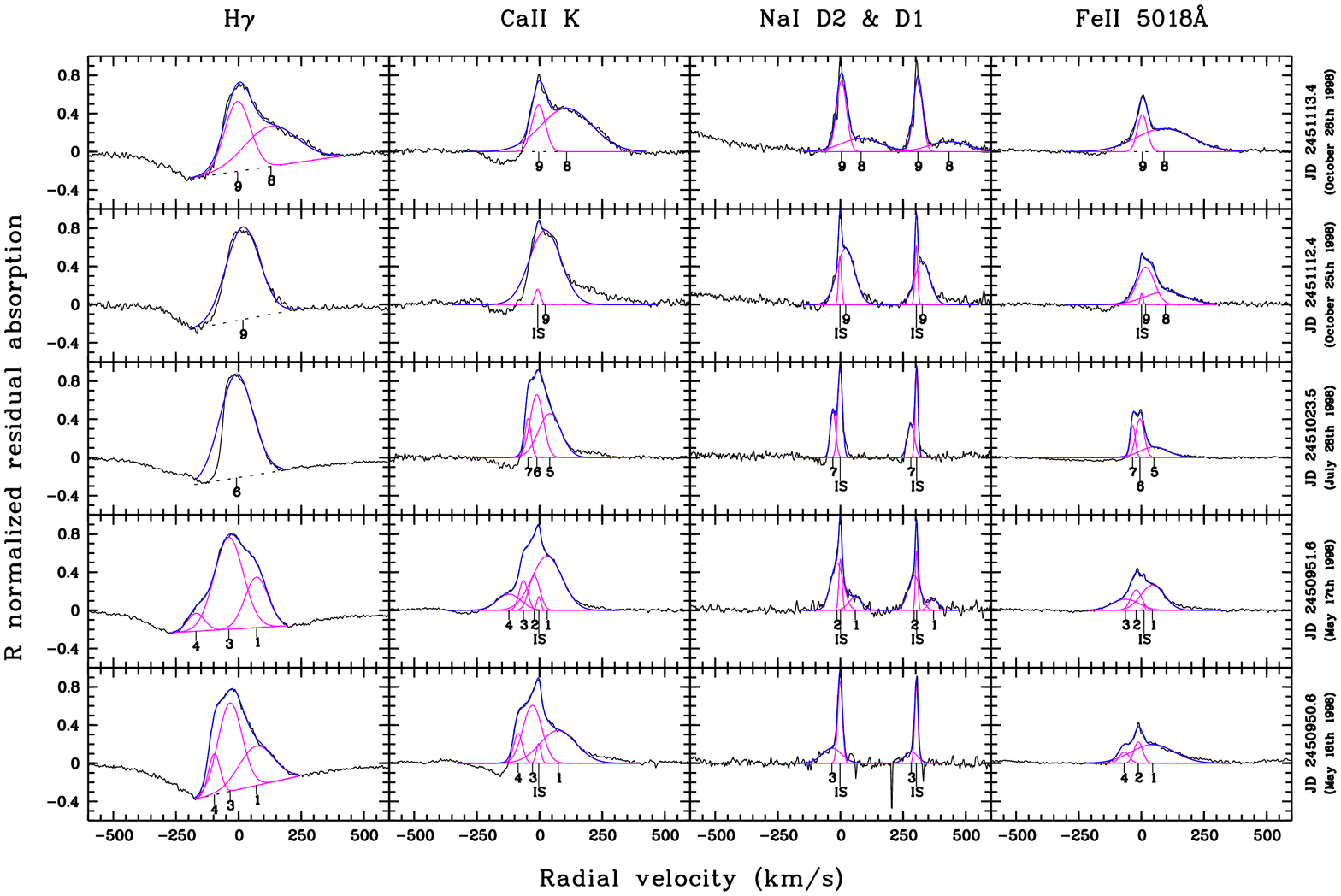}}
\caption[WW~Vul $R$ profiles]
{WW~Vul $R$ profiles.
Details as for Fig~\ref{haebe_bfori_spectra}.
(This figure is available in color in electronic form).}
\label{haebe_wwvul_spectra}
\end{figure*}

\begin{figure*}
\centerline{
\includegraphics[clip=true,angle=-90,width=\hsize]
                {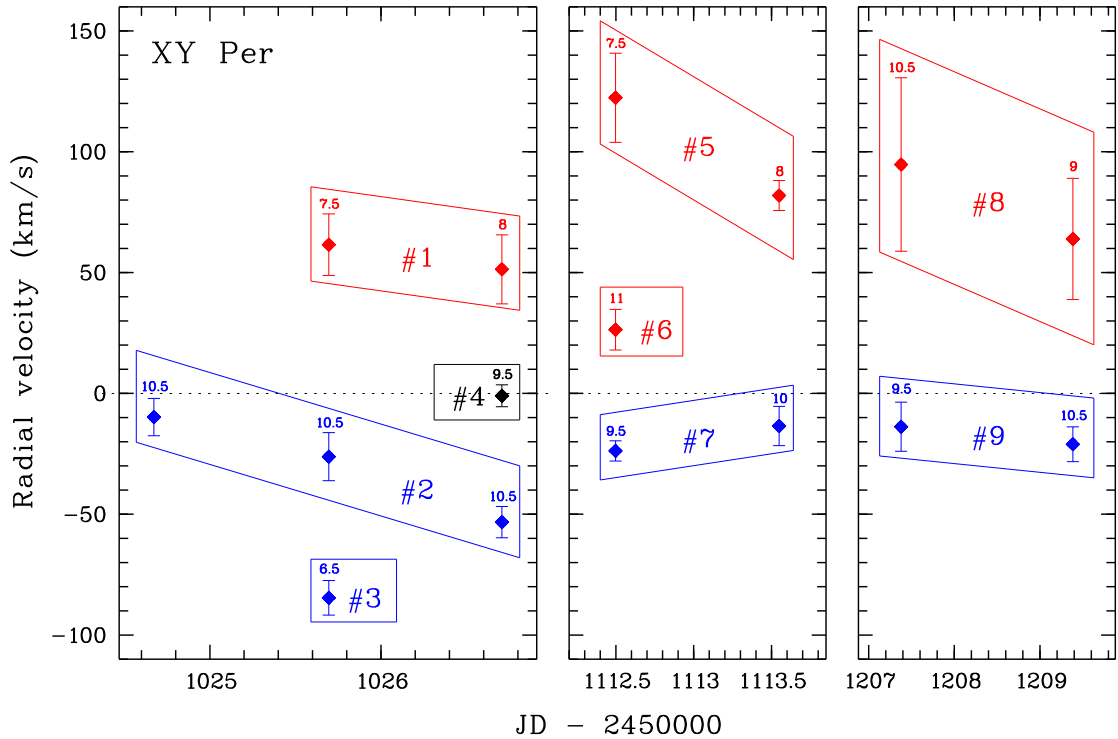}}
\caption[Events in XY~Per]
{Events in XY~Per.
Details as for Fig~\ref{haebe_master_bfori}.
(This figure is available in color in electronic form).}
\label{haebe_master_xyper}
\end{figure*}

\begin{figure*}
\centerline{
\includegraphics[clip=true,angle=-90,width=\hsize]
                {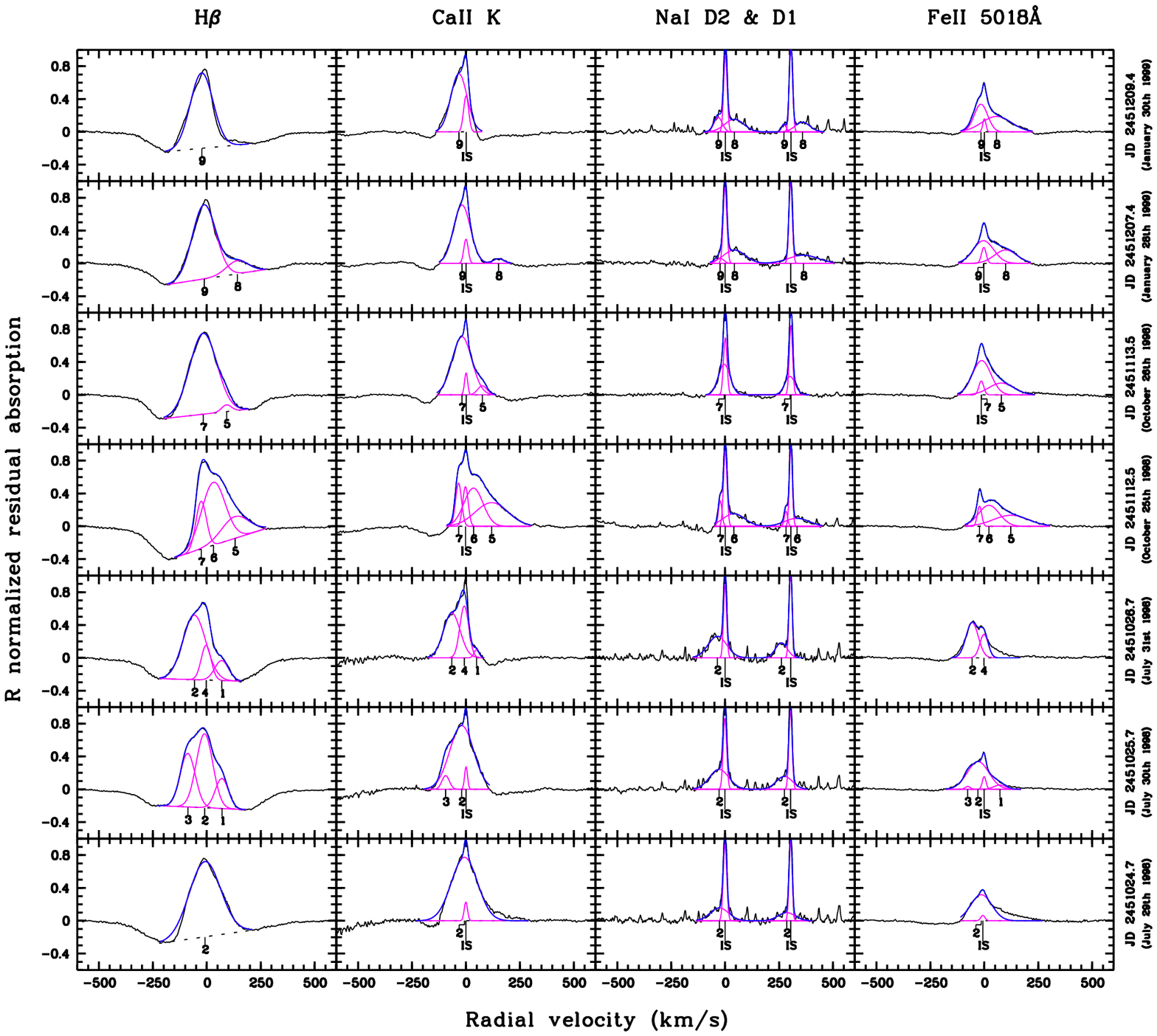}}
\caption[XY~Per $R$ profiles]
{XY~Per $R$ profiles.
Details as for Fig~\ref{haebe_bfori_spectra}.
(This figure is available in color in electronic form).}
\label{haebe_xyper_spectra}
\end{figure*}

Tables~\ref{haebe_master_table_bfori},~\ref{haebe_master_table_svcep},~\ref{haebe_master_table_wwvul}~and~\ref{haebe_master_table_xyper} show that absorption components with similar radial velocities appear/disappear simultaneously in different lines, as is observed in UX~Ori (\citetalias{mora2002}).
We assume that absorptions with similar velocities come from the same gaseous clump, which can be characterized by an average radial velocity $<$$v$$>$ (a Transient Absorption Component or TAC).
The time evolution of the TACs' velocity is referred to as an event and represents the dynamical evolution of the gaseous clumps.
We point out that there is an uncertainty in identifying TACs detected on different nights with the same gaseous clump; our assumption relies on the UX~Ori results \citepalias[see also below]{mora2002}.
Figs.~\ref{haebe_master_bfori},~\ref{haebe_master_svcep},~\ref{haebe_master_wwvul}~and~\ref{haebe_master_xyper} plot $<$$v$$>$  of the identified TAC versus JD.
$<$$v$$>$ is a weighted average in which the lines H$\beta$, H$\gamma$, H$\delta$, \ion{Ca}{ii}~K, \ion{Fe}{ii}~4924~\AA\ and \ion{Fe}{ii}~5018~\AA\ have an arbitrarily assigned weight of 1 because of their higher intensity and non-blended nature.
A weight of 1/2 is also arbitrarily assigned to the remaining lines in order to reflect that they are either blended (H$\epsilon$, \ion{Ca}{ii}~H and \ion{Fe}{ii}~5169\AA), weak (\ion{Sc}{ii}~4247\AA, \ion{Fe}{i}~4046\AA\ and \ion{Ca}{i}~4227\AA) or affected by telluric lines (\ion{Na}{i} D2 \& D1).
The weighted number of lines used in each average, which can be fractional because of the 1/2 weights, is plotted next to each point.
The TACs have been grouped according to the event they represent; thus, the figures show the dynamical evolution of the gaseous clumps.
Note that some events were only detected once (only one TAC).
 Figs.~\ref{haebe_bfori_spectra},~\ref{haebe_svcep_spectra},~\ref{haebe_wwvul_spectra}~and~\ref{haebe_xyper_spectra} show the $R$ profiles of some selected lines for the four stars.
The line absorption components and the corresponding  event identifications are indicated.
The lines are H$\beta$, \ion{Ca}{ii}~K (except for BF~Ori), \ion{Na}{i}~D2, \ion{Na}{i}~D1 and \ion{Fe}{ii}~5018~\AA.
H$\gamma$ is shown for WW~Vul (Fig.~\ref{haebe_wwvul_spectra}) since H$\beta$ has a large underlying emission.
The results for each star are presented in the following.

{\it BF~Ori:}
5 TACs grouped in 3 different events were detected in the spectra of BF~Ori (Fig.~\ref{haebe_master_bfori}).
Event \#1 is an accelerating redshifted event, first detected in the JD~1112.63 spectrum at approximately the stellar radial velocity.
Absorbing gas is seen in Balmer and metallic lines of \ion{Na}{i}, \ion{Fe}{ii}, \ion{Ti}{ii}, \ion{Sc}{ii}, and \ion{Ca}{i}.
The strongest \ion{H}{i} lines (H$\beta$, H$\gamma$ and H$\delta$) appear to be saturated, i.e. $R_{\rm max}$ is very close to unity, and the metallic lines are also very strong.
The Balmer lines are broader than the metallic ones.
The parameters of at least some of these lines might be influenced by IS gas absorption (a careful look at the \ion{Na}{i}~D lines shows the presence of two peaks in the JD~1113.65 spectrum, Fig.~\ref{haebe_bfori_spectra}).
The velocity dispersion of the lines tends to be larger when the event increases its velocity, while $R_{\rm max}$ values tend to decrease, though changes are modest.
Event \#2 is a strong redshifted decelerating event; the behaviour of its TACs  is in general similar to those of \#1.
\#3 represents blueshifted gas only detected in the last night of January~99 and is fainter than the redshifted ones.
The $R_{\rm max}$ values of the Balmer lines are low, but they might be saturated, because $R_{\rm max}$ does not appear to decrease as we follow the series in what will be called the ``expected Balmer decreasing trend'' for optically thin gas ($R_{\rm max}($H$\beta) > R_{\rm max}($H$\gamma) > R_{\rm max}($H$\delta) > R_{\rm max}($H$\zeta)$); also the relative intensity of the metallic lines with respect to the \ion{H}{i} ones seems to be lower.

The \ion{Ca}{ii}~K line has non-photospheric profiles with the simultaneous presence of redshifted and blueshifted components, but their radial velocities do not match  the absorptions observed in other lines (except the blueshifted TAC in JD~1210.45).

{\it SV Cep:}
Broad absorptions of \ion{H}{i}, \ion{Ca}{ii}, \ion{Na}{i} and \ion{Fe}{ii} are detected in the spectra of SV~Cep, but no variability is found in \ion{Ti}{ii}, \ion{Sc}{ii}, \ion{Fe}{i} and \ion{Ca}{i} (unlike the other stars in the paper).
The broad absorptions represent 10 TACs grouped in 8 different events: 3 of them correspond to outflowing gas and the remaining 5 to infalling gas (Figs.~\ref{haebe_master_svcep}~and~\ref{haebe_svcep_spectra}).
Blueshifted gas shows small radial velocities, on average $ \leq 20$~\kms , while redshifted components display velocities as high as 160~\kms.
Only one spectrum was taken in May~98, October~98 and January~99, i.e. the events of these periods are composed of 1 TAC only: these data represent isolated snapshots of the CS gas around SV~Cep and no temporal evolution can be inferred from them.
The 4 TACs detected in July~98 can be grouped in two events: \#3 corresponds to redshifted gas with practically constant radial velocity, and \#4 is gas observed at a velocity close to the stellar radial velocity or slightly blueshifted.
In general, the $R_{\rm max}$ values of the Balmer lines show the expected Balmer decreasing trend and are broader and much stronger than the metallic lines, $R_{\rm max}$(H$\delta$)\,/\,$R_{\rm max}$$(\ion{Fe}{ii})>3$.
There are, however, some exceptions.
\#1 shows relatively strong \ion{Fe}{ii} and it is not clear that the Balmer lines show the expected decreasing trend.
In \#5 the \ion{H}{i} lines are not broader than the \ion{Fe}{ii} ones, and the Balmer lines are probably saturated (but note that this event is very faint, and it could be an overinterpretation of the fit procedure).
The strongest \ion{H}{i} lines are saturated in \#7 and also in the JD~1026.66 spectrum of \#3.
There are anticorrelated changes in the \deltav\ and $R_{\rm max}$ values of \#3, but they show the same trend in \#4.

{\it WW Vul:}
15 TACs grouped in 9 different events are identified in the 5 spectra of WW~Vul (Figs.~\ref{haebe_master_wwvul}~and~\ref{haebe_wwvul_spectra}). 
4 events are seen in May 98. \#1 is redshifted gas with saturated Balmer lines and strong \ion{Fe}{ii} lines.
Both \ion{H}{i} and \ion{Fe}{ii} lines have similar \deltav\ and from one TAC to another $R_{\rm max}$ and \deltav\  show opposite trends.
\#2 corresponds to low velocity blueshifted gas clearly detected in the metallic lines but no counterpart in the hydrogen lines is apparent (see the \ion{Fe}{ii} 5018~\AA\ line in Fig.~\ref{haebe_wwvul_spectra}).
\#3 and \#4 are blueshifted accelerating events.
The only spectrum of July 98 (JD~1023.51) reveals 3 TACs: \#5 is a broad, $\Delta v > 100$~\kms  , redshifted event only detected in metallic lines (the broad wing of the \ion{Fe}{ii} 5018~\AA\ line profile in Fig.~\ref{haebe_wwvul_spectra}).
\#6 is a very low velocity blueshifted component \mbox{($v \simeq -5$~\kms)}; this event is significantly broader in the \ion{H}{i} lines (which are  saturared) than in the metallic ones (the IS contribution cannot be separated from this low velocity event).
\#7 is a relatively narrow blueshifted event only detected in metallic lines and is clearly distinguished as a peak in the line profiles.
Again, metallic redshifted absorptions at \mbox{$v \simeq 90 $ \kms} without \ion{H}{i} counterpart are detected on JD~1112.36 (October 98).
Similar metallic absorptions are also detected on JD~1113.39, but on this date saturated hydrogen components with basically the same kinematic parameters (including  the velocity dispersion) are present.
We tentatively identify both TACs with the same event, \#8, although we cannot exclude the possibility that the metallic absorptions detected on each date could be due to different gas. 
Finally, \#9 is a strong, redshifted, decelerating event, identified in \ion{H}{i} and in many metallic lines.
In this case, the hydrogen lines are considerably broader than the metallic ones.
The \ion{H}{i} lines seem saturated on the first night, while on the second one the expected Balmer decreasing trend is observable and on both nights the \ion{Fe}{ii} lines are relatively strong.

{\it XY Per:}
16 TACs grouped in 9 different events are detected in the 7 XY~Per spectra.
\#1 is a faint redshifted event detected on the last two (out of three) July~98 nights.
\#2 corresponds to blueshifted accelerating gas; all three TACs of this event are strong and broad.
\#3 corresponds to a relatively narrow and faint blueshifted  event only detected on JD~1025.69. 
\#4 is also a narrow and faint event, centered at approximately the stellar radial velocity, only detected on JD~1026.71; IS absorption could contribute to the metallic $R_{\rm max}$ values.
\#5 represents  decelerating redshifted gas. 
\#6 is a strong redshifted event only detected on JD~1112.49.
\#7 corresponds to decelerating blueshifted gas; the expected Balmer decreasing trend is clearly present in both TACs and metallic lines are very strong.
Both \deltav\ and $R_{\rm max}$ increase very significantly from JD~1112.49 to JD~1113.55.
\#8 is  decelerating redshifted gas detected on January 98 (note that the time interval between the two TACS grouped in this event is 48 hours); in both TACs the metallic lines are very broad and strong: \ion{Fe}{ii} lines are even broader and stronger than the Balmer ones which are faint and seem saturated.
Finally, \#9 represents accelerating blueshifted gas.    

\section{Discussion}
\label{haebe_discussion}

The objects studied in this work and UX~Ori are very much alike.
All are PMS HAe stars. UX~Ori, BF~Ori and WW~Vul are bonafide UXOR-type objects \citep[e.g.][]{grinin2000}, i.e. their light curves show high-amplitude variability ($\Delta m > 2.0$~mag), Algol-like minima, a blueing effect and an increase of the polarization when the object brightness decreases. 
SV~Cep also shows UXOR characteristics \citep{rostopchina2000,oudmaijer2001}.
Further data are required before XY~Per can be confidently classed as a UXOR-type object \citep{oudmaijer2001}, although it does share the same complex and variable spectroscopic behaviour of the other stars.

The time covered by the UX~Ori spectra allowed us to analyze and identify the TACs as due to the dynamical evolution of gaseous moving clumps; this identification and dynamical evolution was particularly convincing in the case of a clump detected in four spectra taken within a time interval of four hours \citepalias{mora2002}.
We rely on those results to ascribe to the same gaseous clump groups of \ion{H}{i} and metallic absorption components described in the previous Sect.
We are aware, however, that this identification is doubtful in some cases, since the number of spectra for the stars in this paper is smaller and the time coverage is poorer than for UX~Ori.
With this caveat in mind, we will discuss these spectra as we have done in \citetalias{mora2002} for UX~Ori.
We are quite confident that the main conclusions of this paper are not affected by the  uncertainties with which some specific event can be identified.

\subsection{Kinematics}
\label{haebe_kinematics}

Accelerating/decelerating blueshifted and redshifted events were detected in UX~Ori, the events seemed to last for a few days and their acceleration rates were a fraction of a \ms.
Within the present time coverage limitations, the same is observed in the outflowing and infalling gas of BF~Ori, WW~Vul, SV~Cep and XY~Per.
All 5 stars share the trend that infalling gas shows the largest velocities (the exception is event \#4 of WW~Vul in the JD~951.62 spectrum) and that blueshifted absorptions are detected when redshifted absorptions are present (the exception might be the JD~1024.67 spectrum of XY~Per which shows a low velocity outflow but no infalling gas). 
Similar results are also present in the spectra of \citet{grinin2001}. 

Infalling gas appears to have larger dispersion velocities than outflowing gas.
In addition, the \ion{H}{i} lines are  broader than the metallic ones in approximately 40~\% of the identified TACs, as indicated, for example, by the fact that $\Delta v_\ion{Fe}{ii} < 0.66 \Delta v_\ion{H}{i}$ (in UX~Ori this trend was noticed in the \ion{Na}{i} lines).
The fraction of TACs with broader \ion{H}{i} lines varies from star to star.
For example, in 12 out of 16 detected TACs in XY~Per \ion{H}{i} and \ion{Fe}{ii} have similar velocity dispersion, and only 3 out of 10 TACs in SV Sep follow this trend.
On  the other hand, the \ion{Fe}{ii} lines are broader than the \ion{H}{i} lines in only two events, \#5 of SV~Cep and \#8 of XY~Per. 
It is worth noting that the event of XY~Per has metallic $R_{\rm max}$ values at least as large as those of the Balmer lines and that the SV~Cep event is very weak (see above for this event).

There seems to be a correlation between the dispersion velocity and the velocity of the TACs in the sense that TACs appear to be broader when the velocity increases.
If we perform a linear regression ($\Delta v$~=~A~+~B~$\times \: |v|$) to the whole set of data, we find a correlation coefficient of 0.68.
There is also a suggestion of an anticorrelation between $<$$\Delta v_{\ion{Fe}{ii}}$$>$ and the $<R_{\rm max}>$ values of the TACs during their evolution, i.e. events become fainter when they increase their dispersion velocity.
This was also suggested in the case of UX~Ori.
For every event with more than 1 observation (14 in total), $<$$\Delta v_{\ion{Fe}{ii}}$$>$ and $<R_{\rm max}>$ have been normalized to the first observed values.
If 2 `anomalous' events, which are the most uncertain identifications in the whole sample, are removed  (WW~Vul \#8 -- one TAC is observed only in the 
metallic lines while the second TAC is detected in both metallic and \ion{H}{i} lines --, and XY~Per \#7 --  the $\Delta v$ and $<R_{\rm max}>$ variations 
are much more extreme than those of all other detected events) we find a linear correlation coefficient of -0.66.
Both correlations are highly significant, i.e., the probability of randomly obtaining such coefficients from 2 unrelated variables is $<$1\% for the data sets considered. We have to point out, however, that the  $<$$\Delta v_{\ion{Fe}{ii}}$$>$ vs $<R_{\rm max}>$ anticorrelation dissapears if 
the two `anomalous' events are included in the statistics.

\subsection{Line intensity ratios}
\label{haebe_r_ratios}

Many absorption components in Tables~\ref{haebe_master_table_bfori}~to~\ref{haebe_master_table_xyper} have $R_{\rm max}$ values close to unity, which suggests that they are saturated.
This is not the case even for the strongest events of UX~Ori \citepalias{mora2002} (in fact, the ratio among the $R_{\rm max}$ line values of the 24 UX~Ori TACs does not vary much, which allowed us to estimate a line residual absorption average). 
Nevertheless, we have followed the procedure of \citetalias{mora2002} to investigate whether a ``fixed'' ratio among the line absorption strengths might be present in those TACs which are most likely unsaturated.
Thus, we have excluded from this exercise lines wich show signs of being saturated, i.e. those with $R_{\rm max} >$~0.8 and TACs with Balmer lines of similar strength.
The line \ion{Fe}{ii}~4924~\AA ~has been taken as a reference as it shows the lowest statistical errors (other lines which have been considered are H$\zeta$, \ion{Ca}{ii}~K and \ion{Fe}{ii}~5018~\AA).
The ratio $R_{\rm max,\,line}/R_{\rm max,\,ref.line}$ (e.g. $R_{\rm max,\,H\delta}/R_{\rm max,\,\ion{Fe}{ii} 4924 \AA}$) has been computed for each line of every TAC, and later the mean $<$$R_{\rm max,\,line} / R_{\rm max,\,ref.line}$$>$ has been estimated using a sigma-clipping algorithm to reject bad points.
Table~\ref{haebe_ratios} gives $<$$R_{\rm max,\,line} / R_{\rm max,\,ref.line}$$>$ for each star, together with statistical errors and line rejections (\%).

\begin{table*}
\caption
[Ratios of the average $R_{\rm max}$ parameter of several lines to \ion{Fe}{ii}~4924~\AA\ ($<R_{\rm max,\,line} / R_{\rm max,\,\ion{Fe}{ii} 4924 \AA}>$) for each star]
{Ratios of the average $R_{\rm max}$ parameter of several lines to \ion{Fe}{ii}~4924~\AA\ ($<R_{\rm max,\,line} / R_{\rm max,\,\ion{Fe}{ii} 4924 \AA}>$) for each star.
The values correspond to lines in TACs which are most likely not saturated (see text).
Values with no error mean that only one TAC is available.
The sigma-clipping threshold adopted is 2.0~$\sigma$, except for the values followed by the symbol $\dagger$, in which 1.5~$\sigma$ has been used.
The percentage of rejected lines is given in brackets.}
\label{haebe_ratios}
\small
\centerline{
\begin{tabular}{lllll}
\hline
\hline
Line                 & BF Ori                 & SV Cep                 &
                       WW Vul                 & XY Per                 \\
\hline
H$\beta$             &  --                    & 5.23 $\pm$ 1.82        &
                       6.59                   & 3.40 $\pm$ 0.83 (14\%) \\
H$\gamma$            &  --                    & 5.38 $\pm$ 2.29        &
                       4.53                   & 2.68 $\pm$ 0.82 (11\%) \\
H$\delta$            &  --                    & 4.96 $\pm$ 1.85        &
                       4.83 $\pm$ 0.97        & 2.86 $\pm$ 1.15        \\
H$\zeta$             & 1.50 $\pm$ 0.30        & 3.71 $\pm$ 1.27        &
                       2.17 $\pm$ 0.69        & 2.09 $\pm$ 0.66        \\
\ion{Ca}{ii} K       & 5.79                   & 4.05 $\pm$ 0.97 (10\%) &
                       2.37 $\pm$ 0.65        & 2.20 $\pm$ 1.16 (8\%)  \\
\ion{Na}{i} D2       & 1.06 $\pm$ 0.40        & 1.27 $\pm$ 0.40        &
                      1.45 $\pm$ 0.64 (17\%)$^\dagger$&0.76 $\pm$ 0.12 (20\%)\\
\ion{Na}{i} D1       & 0.89 $\pm$ 0.39        & 0.85 $\pm$ 0.33        &
                       1.50 $\pm$ 0.92        & 0.48 $\pm$ 0.09 (11\%) \\
\ion{Fe}{ii} 5018\AA & 1.08 $\pm$ 0.15        & 1.41 $\pm$ 0.14        &
                       1.28 $\pm$ 0.11 (15\%) & 1.16 $\pm$ 0.17 (14\%) \\
\ion{Fe}{ii} 5169\AA & 1.18 $\pm$ 0.15        & 1.52 $\pm$ 0.30        &
                       1.37 $\pm$ 0.13 (15\%) & 1.30 $\pm$ 0.25 (7\%)  \\
\ion{Ti}{ii} 4444\AA & 0.57 $\pm$ 0.29        & --                     &
                       0.25 $\pm$ 0.08        & 0.30 $\pm$ 0.14 (7\%)  \\
\ion{Ti}{ii} 4572\AA & 0.45 $\pm$ 0.07        & --                     &
                       0.25 $\pm$ 0.08        & 0.32 $\pm$ 0.10 (7\%)  \\
\ion{Sc}{ii} 4247\AA & 0.47 $\pm$ 0.09        & --                     &
                       0.31 $\pm$ 0.06        & 0.33 $\pm$ 0.23        \\
\ion{Ca}{i}  4227\AA & 0.32 $\pm$ 0.06        & --                     &
                       0.18 $\pm$ 0.11        & 0.26 $\pm$ 0.08        \\
\ion{Fe}{i}  4046\AA &  --                    & --                     &
                       0.16 $\pm$ 0.06        & 0.18 $\pm$ 0.05        \\
\hline
\end{tabular}}
\end{table*}

Values quoted in Table~\ref{haebe_ratios} can be used to assess whether the gas causing the variable absorptions is optically thin or not, simply by comparing the $R_{\rm max}$ ratios of lines belonging to the same element multiplet with their gf ratios (we recall that the TACs suspected of being saturated have already been excluded).
We have analysed the Balmer, the \ion{Na}{i}~D doublet and the \ion{Fe}{ii} 42 lines, for which H$\delta$, \ion{Na}{i}~D2 and \ion{Fe}{ii}~5018~\AA\ have been taken as reference, respectively.
$R_{\rm max}$ ratios have been computed following the above procedure and gf values have been taken from the VALD database \citep{kupka1999} for \ion{H}{i} and \ion{Na}{i}.
\ion{Fe}{ii} lines do not have reliable experimental gf values, partly because \ion{Fe}{ii}~5169~\AA\ is blended with \ion{Mg}{i}~5167~\AA, \ion{Fe}{i}~5167~\AA\ and \ion{Mg}{i}~5173~\AA. Following the suggestion made by T.A.~Ryabchikova and F.G.~Kupka (private communication), we have used the semiempirical values computed according to \citet{raassen1998} and available at {\it \href{http://www.science.uva.nl/research/atom/levels/levtext.html}{http://www.science.uva.nl/research/atom/levels/levtext.html}}.
The comparison between the estimated ratios and the theoretical gf shows that the \ion{Fe}{ii} 42 triplet is most likely saturated in BF~Ori but not in the rest of the stars, and that the \ion{Na}{i}~D doublet is also probably saturated in BF~Ori and WW~Vul (Table~\ref{haebe_multiplets}).
Concerning the Balmer lines (note we are referring to those TACs apparently unsaturated) their ratios are very different from the  gf ratios.
This is, in principle, similar to the case of UX~Ori where the lines do not seem to be saturated in any of its events.
\citetalias{mora2002} suggested that the UX~Ori results could be explained by underlying line emission caused by a spherical occulting cloud with a temperature T$_{\rm ex} \sim$~7000~K and a radius of the order of the UX~Ori corotation radius, $R_{\rm cloud}/R_{*} \sim$~1.6 \citep[see][for details of the assumptions]{rodgers2002}.
However, only XY~Per presents Balmer line ratios which could be adjusted using this scenario, namely gas at a T$_{\rm ex} \sim$~6600 K at approximately the corotation radius $R_{\rm cloud}/R_{*} \sim$~1.6.
This is not applicable for the rest of the stars, where it is most likely that the Balmer lines are always saturated.

\begin{table*}
\caption
[Estimated $<R_{\rm max,\,line} / R_{\rm max,\,ref.line}>$ ratios among the lines of the \ion{Na}{i}~D doublet and the \ion{Fe}{ii} 42 triplet for each star]
{Estimated $<R_{\rm max,\,line} / R_{\rm max,\,ref.line}>$ ratios among the lines of the \ion{Na}{i}~D doublet and the \ion{Fe}{ii} 42 triplet for each star.
The theoretical ratios (gf$_{\rm line}$~/~gf$_{\rm ref.line}$) and the line taken as reference are given in the last Col.
The \ion{Fe}{ii}~5169~\AA\ line is included for comparison purposes though its ratios are likely affected by the blend with \ion{Mg}{i}~5167~\AA, \ion{Fe}{i}~5167~\AA\ and \ion{Mg}{i}~5173~\AA.}
\label{haebe_multiplets}
\small
\centerline{
\begin{tabular}{lllllll}
\hline
\hline
Line                 & BF Ori          & SV Cep          & WW Vul          &
                       XY Per          & Theor. & Reference            \\
\hline
\ion{Na}{i} D1       & 0.83 $\pm$ 0.11 & 0.73 $\pm$ 0.15 & 0.81 $\pm$ 0.12 &
                       0.63 $\pm$ 0.07 & 0.50   & \ion{Na}{i} D2       \\
\ion{Fe}{ii} 4924\AA & 0.94 $\pm$ 0.16 & 0.72 $\pm$ 0.08 & 0.79 $\pm$ 0.07 &
                       0.86 $\pm$ 0.17 & 0.69   & \ion{Fe}{ii} 5018\AA \\
\ion{Fe}{ii} 5169\AA & 1.09 $\pm$ 0.03 & 1.07 $\pm$ 0.15 & 1.06 $\pm$ 0.13 &
                       1.07 $\pm$ 0.19 & 1.25   & \ion{Fe}{ii} 5018\AA \\
\hline
\end{tabular}}
\end{table*}

\subsection{Origin of the variable circumstellar gas clumps detected in hydrogen and metallic lines}
\label{haebe_origin_cs_gas}

Excluding the fact that redshifted events seem to be detected at larger velocities than blueshifted events (a fact deserving further observations and also a theoretical explanation), the gas is observed in the same absorption lines with similar $R$ and 
velocity dispersions, i.e., there are no fundamental differences in the behaviour of the infalling and outflowing gas in all the observed stars, including UX~Ori. This result suggests that their physical conditions are 
rather similar and that they probably originate at similar distances from the 
star.
Roughly solar abundances were found by \citet{natta2000} for an event of redshifted gas in UX~Ori, which was further supported by \citetalias{mora2002}.
This led to the conclusion that the clumps of CS gas in UX~Ori are non metal-rich.
Since high velocity gas is observed simultaneously in the Balmer and metallic 
lines in BF~Ori, SV~Sep, XY~Per and 
in most of the detected events in WW~Vul, a similar conclusion very likely holds also for the CS gas in these stars (but see next Sect. for the metallic events in WW~Vul).
Unlike the detected events in UX~Ori, the gas in those stars is often optically thick, as suggested by the saturation of the absorption features.
This might indicate that high density gas is more frequently observed in  BF~Ori, SV~Sep, WW~Vul and XY~Per than in UX~Ori.     
 
\citetalias{mora2002} compares the dynamics of the gaseous clumps in UX~Ori with the predictions of magnetospheric accretion models \citep[see][for a very good basic description of this theory]{hartmann1998} and different wind models \citep[e.g.][]{goodson1997,shu2000,konigl2000}.
The present data  do not add any new substantial aspect to that discussion, only that, with very few exceptions, outflowing gas displays smaller velocities than infalling gas.
Thus, to avoid repetition we refer to that paper, stressing the need for further theoretical efforts to explain the complex circumstellar environment of HAe stars, at least to the level of understanding achieved for the less massive T~Tauri stars.

\subsection{The intriguing case of WW Vul}
\label{haebe_wwvul}

\begin{figure}
\centerline{
\includegraphics[clip=true,angle=-90,width=0.75\hsize]
                {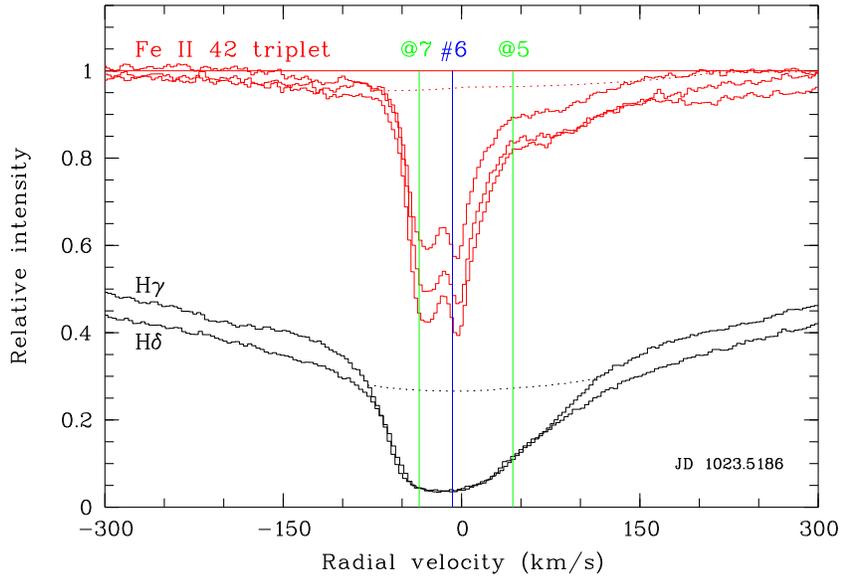}}
\caption
[Original spectrum (i.e. prior to the synthetic stellar spectrum subtraction)
as observed in the JD~1023.52 spectrum of WW~Vul]
{Original spectrum (i.e. prior to the synthetic stellar spectrum subtraction) 
as observed in the JD~1023.52 spectrum of WW~Vul.
Wavelengths have been converted into radial velocities. 
Plotted lines are H$\gamma$, H$\delta$ and the three lines of the 
\ion{Fe}{ii} 42 triplet.
The photospheric synthetic spectrum used to form $R$ is shown in dotted lines for H$\gamma$ and \ion{Fe}{II}~4924\AA\ (being very similar for the other lines).
The synthetic spectra reveal the presence of emission in the wings of H$\gamma$ and a little in the blue wing of \ion{Fe}{II}~4924\AA.
The average radial velocity $<$$v$$>$ of the events identified after the photospheric spectrum removal and $R$ multigaussian fit is indicated using the same colour and label conventions of Fig.~\ref{haebe_master_wwvul}.
There is 1 blueshifted TAC (\#6, blue colour) observed both in metallic and hydrogen lines, note that $\Delta v_{\ion{H}{i}} \gg \Delta v_{\ion{Fe}{ii}}$.
There are also 2 metallic TACs (@5, @7, green) observed only in metallic lines.
Event @7 can be clearly identified by visual inspection as a relatively narrow 
peak in the \ion{Fe}{ii} lines.
In events \#6 and @7 $<$$v$$>$ does not exactly correspond with the local minima in the \ion{Fe}{ii} spectrum because the whole $R$ profile is used in the multigaussian fits.
(This figure is available in color in electronic form).}
\label{haebe_wwvul_metallic}
\end{figure}

WW~Vul is a very interesting case, somewhat different from the other stars.
In all the spectra we have obtained, in addition to events detected in both metallic and hydrogen lines, as it is always the case in the other objects, we see also metallic absorption features (both blueshifted and redshifted) that do not seem to have a counterpart in the hydrogen lines.
They appear as broad high velocity wings in the $R$ profiles, e.g. events \#5 and \#8 in the JD 1023.52 and 1112.37 spectra, as well as relatively narrow distinct peaks, e.g. \#7 in JD~1023.52 (Fig.~\ref{haebe_wwvul_spectra}).
The observed outflowing events are narrower and have larger $R_{\rm max}$ values than the infalling ones. 
To ensure that the metallic events are not artificially introduced by the multigaussian fit we have proceeded in two different ways:
1. We have carefully inspected the original spectra, i.e., before subtraction of the photospheric components.
Fig.~\ref{haebe_wwvul_metallic} shows several lines as observed in the original (without photospheric subtraction) JD~1023.52 WW~Vul spectrum.
In particular, the relatively narrow, blueshifted, metallic event \#7 is clearly seen on a simple visual inspection as an absorption peak with no distinctly separate hydrogen counterpart.
We do note that there is hydrogen absorption covering the velocity range of event \#7 and that the hydrogen lines are quite broad and likely to be saturated at this time.
As a result, Gaussian fitting may not properly represent the intrinsic shape of the hydrogen components; however, without advance knowlegde of their intrinsic shape, Gaussian fitting is used here as a default procedure.
2. In general, our approach has been to use a number as small as possible of gaussians to reproduce the $R$ profiles in a self-consistent way, avoiding to overfit the data.

Nevertheless, since we were intrigued by the WW Vul behaviour, we have performed a number of additional numerical tests, to check as well as possible that the apparent lack of a hydrogen counterpart to some event seen in the metallic lines was not an artifact.
First of all, we have added to the \ion{H}{i} identified components more gaussians which take into account the parameters of the absorptions only identifed in the metallic lines.
This led to unphysical results, i.e., adding more gaussians to the \ion{H}{i} lines does not produce kinematic components similar to those observed in the metallic lines, but create spurious \ion{H}{i} components.
For example, in JD~1023.52 we tried to fit 3 components to the \ion{H}{i} $R_{\rm max}$ profiles, i.e. the same number of gaussians obtained from the \ion{Fe}{ii} fit.
The data obtained from the Balmer lines deconvolution are not self-consistent:
i) strange, unrealistic Balmer intensity sequences are obtained, e.g. in event \#7 $R_{\rm max, H\zeta} = R_{\rm max, H\beta} > R_{\rm max, H\gamma} > R_{\rm max, H\delta}$.
ii) the intensity ratios between different events differ widely from line to line (e.g. H$\beta$ is about two times more intense in \#5 than in \#6, H$\gamma$ is only 0.8 times as intense in \#5 as in \#6.
We have then also tried to fix the radial velocity of the gaussian \ion{H}{i} components to the values obtained from the \ion{Fe}{ii} lines in order to improve the fit, but the results were even worst: the method diverged for H$\beta$ (i.e. no fit could be found) and the sequence of intensities for the Balmer series in event \#6 was unphysical ($R_{\rm max, H\zeta} = R_{\rm max, H\delta} = 2.5  R_{\rm max, H\gamma}$, no H$\beta$ component).

One can also argue that the absence of \ion{H}{i} counterparts to the metallic events is a consequence of blending, which is unresolved by the multigaussian deconvolution procedure.
In order to test such possibility, we have generated a composite synthetic $R$ profile consisting of three gaussians with the same $R_{\rm max}$ and $v$ as the \ion{Fe}{ii}~5018 \AA ~JD 1023.52 spectrum and $\Delta v$ = 150 km/s, which is the average velocity dispersion of the only \ion{H}{i} component detected in that date.
We have also added random noise to achieve a S/N ratio worst than that of the H$\beta$ and H$\gamma$ lines of that spectrum; finally, we have applied our multigaussian deconvolution procedure to the synthetic profile.
The result is that the three individual gaussians have been successfully retrieved.
Note that the gaussian parameters we have introduced are test values.
This result suggests that we would have been able of identifying \ion{H}{i} counterparts of the metallic kinematics components if they would exist. Therefore, we consider that the existence of metallic absorptions without obvious  \ion{H}{i} counterparts is a rather firm result.

There are no obvious differences, in terms of kinematics and absorption strength, between the  WW~Vul metallic events and the rest of the events detected in this star and in BF~Ori, SV~Cep, XY~Per and UX Ori. Some remarkable differences appear, however, when the ion column densities causing the events are compared.
Lower limits on the ion column densities causing the absorptions can be estimated according to the following formula \citep{spitzer1978}, which becomes exact when the gas is optically thin:
\begin{equation}
\label{haebe_eqn_column_densities}
N_a \simeq 
\frac{4 \pi \epsilon_0}{e^2} \frac{m_e c^2 W_\lambda}{\pi f \lambda^2}
\end{equation}

Where $N_a$ is the column density of ions in the ground state of the line, $e$ is the elementary charge, $m_e$ is the electron mass, $\epsilon_0$ is the permeability of free space, $c$ is the speed of light, $f$ is the oscillator strength of the line, $\lambda$ is the wavelength of the line and $W_\lambda$ is the equivalent width of the absorption component.
The column densities obtained are similar, within an order of magnitude, for all the metallic events.
An upper limit on the column density of the \ion{H}{i} atoms in the Balmer energy level (n=2) can be estimated if we assume that TACs with an intensity lower than 3 times the noise level of the spectra cannot be detected.
Table~\ref{haebe_column_densities} gives the estimated values for the \ion{Fe}{ii} lines and some Balmer lines in the case of event WW Vul \#7.
Event \#8 in XY~Per has extremely weak saturated \ion{H}{i} features and strong metallic components (\ion{H}{i} line strengths  are lower than or just comparable to the \ion{Fe}{ii} lines); in  this sense, this event is the more similar to the metallic ones among all detected events with \ion{H}{i} and metallic components.
Table~\ref{haebe_column_densities} gives lower limits to the column densities, estimated from the Balmer and \ion{Fe}{ii} lines of event \#8 in XY~Per. 
The values of Table~\ref{haebe_column_densities} nicely shows the difference between the estimated column densities of ions excited to the energy ground level of the \ion{Fe}{ii} 42 triplet and Balmer lines in both type of events, and it could point out to a fundamental difference on the nature and origin of the gas from which the absorptions rise.
However, a definitive statement on this issue requires a deep analysis of the metallic events detected in WW Vul in order to estimate chemical abundances, or abundance ratios among elements.
Such analysis, which is beyond the scope of this paper, needs a NLTE treatment of the spectra; such treatment  would produce model dependent results, since a previous knowledge of physical quantities, as for example volume densities of the gaseous clumps and electron temperature, are required.
Such quantities cannot be estimated from our data in a confident way.

\begin{table}
\caption
[Column density estimates of ions excited to the energy ground level of Balmer and \ion{Fe}{ii} 42 triplet lines for events WW~Vul \#7 (metallic) and XY~Per \#8]
{Column density estimates (according to Eq.~\ref{haebe_eqn_column_densities}) of ions excited to the energy ground level of Balmer and \ion{Fe}{ii} 42 triplet lines for events WW~Vul \#7 (metallic) and XY~Per \#8.}
\label{haebe_column_densities}
\centerline{
\begin{tabular}{lll}
\hline
\hline
                     & WW Vul \#7             & XY Per \#8             \\
\hline
Line                 & $N_a\,({\rm cm}^{-2})$ & $N_a\,({\rm cm}^{-2})$ \\
\hline
\ion{Fe}{ii} 4924\AA & $\ge1.1 \cdot 10^{14}$ & $\ge4.0 \cdot 10^{14}$ \\
\ion{Fe}{ii} 5018\AA & $\ge8.1 \cdot 10^{13}$ & $\ge2.1 \cdot 10^{14}$ \\
\ion{Fe}{ii} 5169\AA & $\ge7.8 \cdot 10^{13}$ & $\ge4.4 \cdot 10^{14}$ \\
H$\beta$             &   $<2.9 \cdot 10^{11}$ & $\ge1.3 \cdot 10^{13}$ \\
H$\gamma$            &   $<1.2 \cdot 10^{12}$ & $\ge4.2 \cdot 10^{13}$ \\
H$\delta$            &   $<3.3 \cdot 10^{12}$ & $\ge9.6 \cdot 10^{13}$ \\
H$\zeta$             &                        & $\ge2.9 \cdot 10^{14}$ \\
\hline
\end{tabular}
}
\end{table}

\section{Concluding remarks}
\label{haebe_conclusions}

We have analyzed optical high resolution spectra of the HAe stars BF Ori, SV Cep, WW Vul and XY Per.
These spectra monitor the stars on time scales of months and days and, as in the case of the previously studied UX Ori \citepalias{mora2002}, they provide observational constraints, which should be considered for any realistic scenario of the gaseous circumstellar disks around intermediate-mass PMS stars.
Our results and conclusions can be summarized as follows:

\begin{enumerate}

\item
The gaseous circumstellar environment of these stars is very complex and active.
The spectra always show circumstellar line absorptions with remarkable variations in their strength and dynamical properties.

\item 
Variable absorption features are, in most  cases, detected simultaneously in hydrogen and in many metallic lines with similar velocities. 
In each case, there are several  kinematic components in each line, both blue-shifted and red-shifted with respect to the systemic velocity, denoting the simultaneous presence of infalling and outflowing gas.
We attribute the variable features detected in both Balmer and metallic lines to gaseous clumps of solar-like composition, evolving dynamically in the circumstellar disks of these objects.
In this respect, the disks around the stars studied in this paper are similar to the UX Ori disk.
Following the conclusions of \citetalias{mora2002} we suggest that these clumps and their dynamical evolution should be investigated in the context of detailed magnetospheric accretion models, similar to those of T Tauri stars.    

\item
The star WW Vul is peculiar and behaves differently from the other stars studied in this paper and also from UX~Ori.
It is the only star that shows, in addition to events seen both in metallic and hydrogen lines, similar to those observed in the other stars, also transient absorption components in metallic lines that do not apparently have any obvious counterpart in the hydrogen lines.
This result, taken at its face-value, would indicate the presence of a metal-rich gas component in the environment of WW~Vul, possibly related to the evaporation of solid bodies. However, any such conclusion is premature.
We think that a series of optical spectra with  better time resolution (hours) and longer monitoring (up to around seven days), spectra in the far UV range - to analyze Lyman and metallic resonance lines - and detailed NLTE models of different CS gas environments are essential for further progress and to provide clues on the origin of these apparently metal-rich  events, in terms of their appearance/disappearance statistics, dynamics, metallicity  and nature. 

\end{enumerate}



\begin{table*}
\caption[Identified Transient Absorption Components (TACs) in BF~Ori]
{Identified Transient Absorption Components (TACs) in BF~Ori.
Column 1 gives the corresponding Balmer or metallic line, Col. 2  gives the Julian Date ($-$2\,450\,000), Col. 3 represents the event assigned to the particular absorptions (see text Sect.~\ref{haebe_circumstellar_contribution}), Cols. 4 to 6 give the parameters of each transient absorption as estimated from the multigaussian fit of the normalized residual absorption $R$: $v$, radial velocity, \deltav, FWHM, and $R_{\rm max}$, the strength of the absorption (peak of the gaussian).
``0'' in Col. 3 corresponds to the narrow IS absorptions, while ``--'' means that the absorption is not associated with a particular event.}
\label{haebe_master_table_bfori}
\centerline{
\tiny
\begin{tabular}[t]{lllrrl}
\hline
\hline
Line & JD & Event & $v$ (\kms) & \deltav\ (\kms) & $R_{\rm max}$ \\
\hline
H$\beta$             & 1112.6324 & 1 &  -8 & 184 & 0.95 \\
H$\gamma$            & 1112.6324 & 1 &   7 & 183 & 0.95 \\
H$\delta$            & 1112.6324 & 1 &   4 & 157 & 0.93 \\
H$\zeta$             & 1112.6324 & 1 &  11 & 120 & 0.79 \\
\ion{Na}{i} D2       & 1112.6324 & 1 &  -1 &  66 & 0.83 \\
\ion{Na}{i} D1       & 1112.6324 & 1 &  -2 &  60 & 0.75 \\
\ion{Fe}{ii} 4924\AA & 1112.6324 & 1 &   3 &  83 & 0.64 \\
\ion{Fe}{ii} 5018\AA & 1112.6324 & 1 &   5 &  92 & 0.71 \\
\ion{Fe}{ii} 5169\AA & 1112.6324 & 1 &   2 &  93 & 0.76 \\
\ion{Ti}{ii} 4444\AA & 1112.6324 & 1 &  -2 &  61 & 0.33 \\
\ion{Ti}{ii} 4572\AA & 1112.6324 & 1 &  -3 &  58 & 0.35 \\
\ion{Sc}{ii} 4247\AA & 1112.6324 & 1 &  -1 &  66 & 0.30 \\
\ion{Ca}{i} 4227\AA  & 1112.6324 & 1 &   3 &  62 & 0.23 \\
H$\epsilon$          & 1112.6324 & --&  30 & 132 & 0.75 \\
\ion{Ca}{ii} K       & 1112.6324 & --& -52 &  70 & 0.54 \\
\ion{Ca}{ii} K       & 1112.6324 & --&  92 & 172 & 0.61 \\
\ion{Ca}{ii} H       & 1112.6324 & --& -40 &  53 & 0.71 \\
H$\beta$             & 1113.6515 & 1 &  27 & 194 & 0.86 \\
H$\gamma$            & 1113.6515 & 1 &  33 & 189 & 0.86 \\
H$\delta$            & 1113.6515 & 1 &  33 & 195 & 0.79 \\
H$\zeta$             & 1113.6515 & 1 &  39 & 139 & 0.65 \\
\ion{Na}{i} D2       & 1113.6515 & 1 &   6 &  57 & 0.80 \\
\ion{Na}{i} D1       & 1113.6515 & 1 &   5 &  51 & 0.69 \\
\ion{Fe}{ii} 4924\AA & 1113.6515 & 1 &  15 &  98 & 0.54 \\
\ion{Fe}{ii} 5018\AA & 1113.6515 & 1 &  20 & 111 & 0.62 \\
\ion{Fe}{ii} 5169\AA & 1113.6515 & 1 &  19 & 116 & 0.66 \\
\ion{Ti}{ii} 4444\AA & 1113.6515 & 1 &  17 &  76 & 0.24 \\
\ion{Ti}{ii} 4572\AA & 1113.6515 & 1 &  13 &  70 & 0.24 \\
\ion{Sc}{ii} 4247\AA & 1113.6515 & 1 &  16 &  83 & 0.23 \\
\ion{Ca}{i} 4227\AA  & 1113.6515 & 1 &  13 &  51 & 0.15 \\
H$\epsilon$          & 1113.6515 & --&  73 & 172 & 0.63 \\
\ion{Ca}{ii} K       & 1113.6515 & --& -74 & 134 & 0.28 \\
\ion{Ca}{ii} K       & 1113.6515 & --&  96 & 167 & 0.59 \\
H$\beta$             & 1209.5542 & 2 &  40 & 218 & 0.83 \\
H$\gamma$            & 1209.5542 & 2 &  40 & 229 & 0.88 \\
H$\delta$            & 1209.5542 & 2 &  40 & 217 & 0.82 \\
H$\epsilon$          & 1209.5542 & 2 &  58 & 176 & 0.60 \\
H$\zeta$             & 1209.5542 & 2 &  58 & 184 & 0.60 \\
\ion{Na}{i} D2       & 1209.5542 & 2 &  55 & 152 & 0.21 \\
\ion{Na}{i} D1       & 1209.5542 & 2 &  38 & 135 & 0.18 \\
\ion{Fe}{ii} 4924\AA & 1209.5542 & 2 &  46 & 188 & 0.36 \\
\ion{Fe}{ii} 5018\AA & 1209.5542 & 2 &  54 & 196 & 0.44 \\
\hline
\end{tabular}
\hspace{0.2cm}
\begin{tabular}[t]{lllrrl}
\hline
\hline
Line & JD & Event & $v$ (\kms) & \deltav\ (\kms) & $R_{\rm max}$ \\
\hline
\ion{Fe}{ii} 5169\AA & 1209.5542 & 2 &  50 & 199 & 0.48 \\
\ion{Ti}{ii} 4444\AA & 1209.5542 & 2 &  38 & 185 & 0.14 \\
\ion{Ti}{ii} 4572\AA & 1209.5542 & 2 &  52 & 189 & 0.13 \\
\ion{Sc}{ii} 4247\AA & 1209.5542 & 2 &  21 &  75 & 0.22 \\
\ion{Na}{i} D2       & 1209.5542 & 0 &   1 &  31 & 0.58 \\
\ion{Na}{i} D1       & 1209.5542 & 0 &   1 &  22 & 0.51 \\
\ion{Fe}{ii} 4924\AA & 1209.5542 & 0 &   4 &  20 & 0.09 \\
\ion{Fe}{ii} 5018\AA & 1209.5542 & 0 &   4 &  18 & 0.11 \\
\ion{Fe}{ii} 5169\AA & 1209.5542 & 0 &   3 &  24 & 0.12 \\
\ion{Ca}{ii} K       & 1209.5542 & --& -95 & 136 & 0.33 \\
\ion{Ca}{ii} K       & 1209.5542 & --&  96 & 172 & 0.71 \\
H$\beta$             & 1210.4571 & 2 &  38 & 155 & 0.95 \\
H$\gamma$            & 1210.4571 & 2 &  25 & 185 & 0.97 \\
H$\delta$            & 1210.4571 & 2 &  29 & 155 & 0.95 \\
H$\epsilon$          & 1210.4571 & 2 &  32 & 130 & 0.79 \\
H$\zeta$             & 1210.4571 & 2 &  23 & 145 & 0.79 \\
\ion{Na}{i} D2       & 1210.4571 & 2 &  13 &  83 & 0.47 \\
\ion{Na}{i} D1       & 1210.4571 & 2 &  16 &  78 & 0.32 \\
\ion{Fe}{ii} 4924\AA & 1210.4571 & 2 &  23 & 110 & 0.52 \\
\ion{Fe}{ii} 5018\AA & 1210.4571 & 2 &  23 & 115 & 0.59 \\
\ion{Fe}{ii} 5169\AA & 1210.4571 & 2 &  22 & 118 & 0.65 \\
\ion{Ti}{ii} 4444\AA & 1210.4571 & 2 &  20 &  87 & 0.22 \\
\ion{Ti}{ii} 4572\AA & 1210.4571 & 2 &  12 & 117 & 0.21 \\
\ion{Sc}{ii} 4247\AA & 1210.4571 & 2 &  14 &  98 & 0.21 \\
H$\beta$             & 1210.4571 & 3 & -52 &  62 & 0.32 \\
H$\gamma$            & 1210.4571 & 3 & -67 &  94 & 0.19 \\
H$\delta$            & 1210.4571 & 3 & -61 &  90 & 0.27 \\
H$\epsilon$          & 1210.4571 & 3 & -65 &  43 & 0.15 \\
H$\zeta$             & 1210.4571 & 3 & -89 &  63 & 0.15 \\
\ion{Ca}{ii} K       & 1210.4571 & 3 & -77 &  92 & 0.47 \\
\ion{Ca}{ii} H       & 1210.4571 & 3 & -67 &  93 & 0.47 \\
\ion{Fe}{ii} 4924\AA & 1210.4571 & 3 & -57 &  77 & 0.08 \\
\ion{Fe}{ii} 5018\AA & 1210.4571 & 3 & -53 &  55 & 0.07 \\
\ion{Fe}{ii} 5169\AA & 1210.4571 & 3 & -54 &  48 & 0.08 \\
\ion{Ti}{ii} 4444\AA & 1210.4571 & 3 & -42 &  46 & 0.09 \\
\ion{Ti}{ii} 4572\AA & 1210.4571 & 3 & -34 &  25 & 0.04 \\
\ion{Na}{i} D2       & 1210.4571 & 0 &  -1 &  15 & 0.41 \\
\ion{Na}{i} D1       & 1210.4571 & 0 &   0 &  14 & 0.47 \\
\ion{Fe}{ii} 4924\AA & 1210.4571 & 0 &   3 &  10 & 0.05 \\
\ion{Fe}{ii} 5018\AA & 1210.4571 & 0 &   2 &  13 & 0.06 \\
\ion{Fe}{ii} 5169\AA & 1210.4571 & 0 &   2 &  13 & 0.06 \\
\ion{Ca}{ii} K       & 1210.4571 & --&  78 & 146 & 0.69 \\
\hline
\end{tabular}}
\end{table*}


\begin{table*}
\caption[Identified Transient Absorption Components (TACs) in SV~Cep]
{Identified Transient Absorption Components (TACs) in SV~Cep. 
Details as for Table~\ref{haebe_master_table_bfori}.}
\label{haebe_master_table_svcep}
\centerline{
\tiny
\begin{tabular}[t]{lllrrl}
\hline
\hline
Line & JD & Event & $v$ (\kms) & \deltav\ (\kms) & $R_{\rm max}$ \\
\hline
H$\beta$             &  950.6668 & 1 &  35 &  52 & 0.58 \\
H$\gamma$            &  950.6668 & 1 &  40 &  45 & 0.35 \\
H$\delta$            &  950.6668 & 1 &  40 &  46 & 0.33 \\
H$\epsilon$          &  950.6668 & 1 &  31 &  50 & 0.38 \\
H$\zeta$             &  950.6668 & 1 &  40 &  80 & 0.28 \\
\ion{Ca}{ii} K       &  950.6668 & 1 &  25 &  27 & 0.42 \\
\ion{Ca}{ii} H       &  950.6668 & 1 &  23 &  16 & 0.38 \\
\ion{Na}{i} D2       &  950.6668 & 1 &  26 &  14 & 0.25 \\
\ion{Fe}{ii} 4924\AA &  950.6668 & 1 &  23 &  25 & 0.15 \\
\ion{Fe}{ii} 5018\AA &  950.6668 & 1 &  25 &  27 & 0.22 \\
\ion{Fe}{ii} 5169\AA &  950.6668 & 1 &  24 &  28 & 0.25 \\
H$\beta$             &  950.6668 & 2 & -17 &  73 & 0.93 \\
H$\gamma$            &  950.6668 & 2 & -11 &  84 & 0.83 \\
H$\delta$            &  950.6668 & 2 & -13 &  79 & 0.67 \\
H$\epsilon$          &  950.6668 & 2 & -17 &  49 & 0.42 \\
H$\zeta$             &  950.6668 & 2 & -22 &  74 & 0.35 \\
\ion{Ca}{ii} K       &  950.6668 & 2 &  -8 &  28 & 0.73 \\
\ion{Ca}{ii} H       &  950.6668 & 2 &  -6 &  24 & 0.66 \\
\ion{Fe}{ii} 4924\AA &  950.6668 & 2 & -11 &  17 & 0.10 \\
\ion{Fe}{ii} 5018\AA &  950.6668 & 2 &  -9 &  20 & 0.15 \\
\ion{Fe}{ii} 5169\AA &  950.6668 & 2 & -10 &  16 & 0.20 \\
\ion{Na}{i} D2       &  950.6668 & 0 &   1 &  16 & 1.04 \\
\ion{Na}{i} D1       &  950.6668 & 0 &   1 &  13 & 1.05 \\
H$\beta$             & 1025.6260 & 3 &  99 & 276 & 0.34 \\
H$\gamma$            & 1025.6260 & 3 &  78 & 217 & 0.48 \\
H$\delta$            & 1025.6260 & 3 &  76 & 219 & 0.43 \\
H$\zeta$             & 1025.6260 & 3 &  77 & 210 & 0.31 \\
\ion{Ca}{ii} K       & 1025.6260 & 3 &  40 & 193 & 0.51 \\
\ion{Na}{i} D2       & 1025.6260 & 3 &  71 & 162 & 0.13 \\
\ion{Na}{i} D1       & 1025.6260 & 3 &  68 & 199 & 0.10 \\
\ion{Fe}{ii} 4924\AA & 1025.6260 & 3 &  51 & 197 & 0.09 \\
\ion{Fe}{ii} 5018\AA & 1025.6260 & 3 &  70 & 245 & 0.12 \\
\ion{Fe}{ii} 5169\AA & 1025.6260 & 3 &  60 & 207 & 0.12 \\
H$\beta$             & 1025.6260 & 4 &  -6 & 147 & 0.67 \\
H$\gamma$            & 1025.6260 & 4 & -21 & 139 & 0.54 \\
H$\delta$            & 1025.6260 & 4 & -16 & 133 & 0.48 \\
H$\epsilon$          & 1025.6260 & 4 &  15 & 231 & 0.67 \\
H$\zeta$             & 1025.6260 & 4 & -10 & 134 & 0.40 \\
\ion{Ca}{ii} K       & 1025.6260 & 4 &   0 &  29 & 0.41 \\
\ion{Ca}{ii} H       & 1025.6260 & 4 &  -2 &  19 & 0.66 \\
\ion{Fe}{ii} 4924\AA & 1025.6260 & 4 &  12 &  33 & 0.09 \\
\ion{Fe}{ii} 5018\AA & 1025.6260 & 4 &  12 &  39 & 0.14 \\
\ion{Fe}{ii} 5169\AA & 1025.6260 & 4 &  10 &  39 & 0.15 \\
\ion{Na}{i} D2       & 1025.6260 & 0 &   0 &  17 & 0.95 \\
\ion{Na}{i} D1       & 1025.6260 & 0 &  -0 &  14 & 0.96 \\
H$\beta$             & 1026.6684 & 3 &  80 & 202 & 0.75 \\
H$\gamma$            & 1026.6684 & 3 &  72 & 190 & 0.76 \\
H$\delta$            & 1026.6684 & 3 &  69 & 164 & 0.70 \\
H$\epsilon$          & 1026.6684 & 3 &  87 & 116 & 0.56 \\
H$\zeta$             & 1026.6684 & 3 &  66 & 152 & 0.55 \\
\ion{Ca}{ii} K       & 1026.6684 & 3 &  75 & 115 & 0.53 \\
\ion{Na}{i} D2       & 1026.6684 & 3 &  82 & 130 & 0.16 \\
\ion{Na}{i} D1       & 1026.6684 & 3 &  75 & 127 & 0.12 \\
\ion{Fe}{ii} 4924\AA & 1026.6684 & 3 &  84 &  87 & 0.11 \\
\ion{Fe}{ii} 5018\AA & 1026.6684 & 3 &  87 &  95 & 0.16 \\
\ion{Fe}{ii} 5169\AA & 1026.6684 & 3 &  86 &  83 & 0.17 \\
H$\beta$             & 1026.6684 & 4 &  -8 &  82 & 0.59 \\
H$\gamma$            & 1026.6684 & 4 & -11 &  68 & 0.42 \\
H$\delta$            & 1026.6684 & 4 & -13 &  60 & 0.35 \\
H$\epsilon$          & 1026.6684 & 4 & -18 & 112 & 0.59 \\
H$\zeta$             & 1026.6684 & 4 & -10 &  53 & 0.21 \\
\hline
\end{tabular}
\hspace{0.2cm}
\begin{tabular}[t]{lllrrl}
\hline
\hline
Line & JD & Event & $v$ (\kms) & \deltav\ (\kms) & $R_{\rm max}$ \\
\hline
\ion{Ca}{ii} K       & 1026.6684 & 4 &  -6 &  46 & 0.49 \\
\ion{Ca}{ii} H       & 1026.6684 & 4 & -22 &  20 & 0.32 \\
\ion{Fe}{ii} 4924\AA & 1026.6684 & 4 &   1 &  40 & 0.13 \\
\ion{Fe}{ii} 5018\AA & 1026.6684 & 4 &  -0 &  43 & 0.19 \\
\ion{Fe}{ii} 5169\AA & 1026.6684 & 4 &   0 &  42 & 0.23 \\
\ion{Ca}{ii} K       & 1026.6684 & 0 &  -3 &  12 & 0.29 \\
\ion{Ca}{ii} H       & 1026.6684 & 0 &  -2 &  17 & 0.85 \\
\ion{Na}{i} D2       & 1026.6684 & 0 &   0 &  18 & 0.97 \\
\ion{Na}{i} D1       & 1026.6684 & 0 &  -1 &  14 & 0.99 \\
H$\beta$             & 1113.4730 & 5 & 138 &  38 & 0.04 \\
H$\gamma$            & 1113.4730 & 5 & 113 &  64 & 0.12 \\
H$\delta$            & 1113.4730 & 5 & 114 &  81 & 0.12 \\
H$\epsilon$          & 1113.4730 & 5 &  97 &  41 & 0.13 \\
H$\zeta$             & 1113.4730 & 5 &  85 &  20 & 0.13 \\
\ion{Ca}{ii} K       & 1113.4730 & 5 &  89 &  85 & 0.11 \\
\ion{Fe}{ii} 4924\AA & 1113.4730 & 5 &  87 &  63 & 0.04 \\
\ion{Fe}{ii} 5018\AA & 1113.4730 & 5 &  90 & 292 & 0.05 \\
\ion{Fe}{ii} 5169\AA & 1113.4730 & 5 &  67 & 173 & 0.04 \\
H$\beta$             & 1113.4730 & 6 & -28 & 125 & 0.99 \\
H$\gamma$            & 1113.4730 & 6 & -21 & 117 & 0.94 \\
H$\delta$            & 1113.4730 & 6 & -22 & 106 & 0.86 \\
H$\epsilon$          & 1113.4730 & 6 & -13 &  85 & 0.65 \\
H$\zeta$             & 1113.4730 & 6 & -18 &  95 & 0.52 \\
\ion{Ca}{ii} K       & 1113.4730 & 6 & -19 &  77 & 0.56 \\
\ion{Ca}{ii} H       & 1113.4730 & 6 & -11 &  65 & 0.53 \\
\ion{Na}{i} D2       & 1113.4730 & 6 & -26 &  12 & 0.14 \\
\ion{Na}{i} D1       & 1113.4730 & 6 & -27 &  30 & 0.07 \\
\ion{Fe}{ii} 4924\AA & 1113.4730 & 6 & -15 &  78 & 0.15 \\
\ion{Fe}{ii} 5018\AA & 1113.4730 & 6 & -14 &  70 & 0.17 \\
\ion{Fe}{ii} 5169\AA & 1113.4730 & 6 & -23 &  77 & 0.18 \\
\ion{Ca}{ii} K       & 1113.4730 & 0 &  -3 &  15 & 0.44 \\
\ion{Ca}{ii} H       & 1113.4730 & 0 &  -1 &  12 & 0.50 \\
\ion{Na}{i} D2       & 1113.4730 & 0 &  -1 &  16 & 1.04 \\
\ion{Na}{i} D1       & 1113.4730 & 0 &  -1 &  14 & 1.03 \\
\ion{Fe}{ii} 4924\AA & 1113.4730 & 0 &   2 &  13 & 0.09 \\
\ion{Fe}{ii} 5018\AA & 1113.4730 & 0 &  -1 &  14 & 0.11 \\
\ion{Fe}{ii} 5169\AA & 1113.4730 & 0 &  -2 &  23 & 0.16 \\
H$\beta$             & 1209.3372 & 7 & 198 & 349 & 0.33 \\
H$\gamma$            & 1209.3372 & 7 & 164 & 315 & 0.32 \\
H$\delta$            & 1209.3372 & 7 & 134 & 334 & 0.30 \\
H$\epsilon$          & 1209.3372 & 7 & 141 & 298 & 0.29 \\
H$\zeta$             & 1209.3372 & 7 & 143 & 313 & 0.24 \\
\ion{Ca}{ii} K       & 1209.3372 & 7 & 172 & 269 & 0.19 \\
\ion{Fe}{ii} 4924\AA & 1209.3372 & 7 & 136 & 324 & 0.05 \\
\ion{Fe}{ii} 5018\AA & 1209.3372 & 7 & 156 & 253 & 0.07 \\
\ion{Fe}{ii} 5169\AA & 1209.3372 & 7 & 167 & 278 & 0.06 \\
H$\beta$             & 1209.3372 & 8 &  -4 &  99 & 0.97 \\
H$\gamma$            & 1209.3372 & 8 &   3 & 111 & 0.77 \\
H$\delta$            & 1209.3372 & 8 &   6 & 110 & 0.60 \\
H$\epsilon$          & 1209.3372 & 8 &   6 & 100 & 0.51 \\
H$\zeta$             & 1209.3372 & 8 &   9 &  99 & 0.46 \\
\ion{Ca}{ii} K       & 1209.3372 & 8 &  12 & 115 & 0.40 \\
\ion{Na}{i} D2       & 1209.3372 & 8 &  40 &  96 & 0.07 \\
\ion{Na}{i} D1       & 1209.3372 & 8 &  33 &  51 & 0.06 \\
\ion{Fe}{ii} 4924\AA & 1209.3372 & 8 &  19 &  76 & 0.09 \\
\ion{Fe}{ii} 5018\AA & 1209.3372 & 8 &  16 &  78 & 0.13 \\
\ion{Fe}{ii} 5169\AA & 1209.3372 & 8 &  17 &  74 & 0.15 \\
\ion{Ca}{ii} K       & 1209.3372 & 0 &  -1 &  18 & 0.43 \\
\ion{Ca}{ii} H       & 1209.3372 & 0 &   1 &  22 & 0.77 \\
\ion{Na}{i} D2       & 1209.3372 & 0 &   1 &  20 & 0.87 \\
\ion{Na}{i} D1       & 1209.3372 & 0 &   1 &  15 & 0.93 \\
\hline
\end{tabular}}
\end{table*}


\begin{table*}
\caption[Identified Transient Absorption Components (TACs) in WW~Vul]
{Identified Transient Absorption Components (TACs) in WW~Vul.
Details as for Table~\ref{haebe_master_table_bfori}.}
\label{haebe_master_table_wwvul}
\centerline{
\tiny
\begin{tabular}[t]{lllrrl}
\hline
\hline
Line & JD & Event & $v$ (\kms) & \deltav\ (\kms)  & $R_{\rm max}$ \\
\hline
H$\beta$             &  950.6176 & 1 &  131 &  95 & 0.10 \\
H$\gamma$            &  950.6176 & 1 &   71 & 162 & 0.41 \\
H$\delta$            &  950.6176 & 1 &   58 & 174 & 0.44 \\
H$\epsilon$          &  950.6176 & 1 &   59 & 164 & 0.40 \\
H$\zeta$             &  950.6176 & 1 &   69 & 168 & 0.38 \\
\ion{Ca}{ii} K       &  950.6176 & 1 &   74 & 170 & 0.34 \\
\ion{Fe}{ii} 4924\AA &  950.6176 & 1 &   40 & 186 & 0.16 \\
\ion{Fe}{ii} 5018\AA &  950.6176 & 1 &   45 & 203 & 0.20 \\
\ion{Fe}{ii} 5169\AA &  950.6176 & 1 &   44 & 207 & 0.19 \\
\ion{Fe}{ii} 4924\AA &  950.6176 & 2 &  -13 &  36 & 0.17 \\
\ion{Fe}{ii} 5018\AA &  950.6176 & 2 &  -12 &  41 & 0.22 \\
\ion{Fe}{ii} 5169\AA &  950.6176 & 2 &  -12 &  41 & 0.26 \\
H$\beta$             &  950.6176 & 3 &  -25 & 110 & 1.27 \\
H$\gamma$            &  950.6176 & 3 &  -34 & 113 & 0.93 \\
H$\delta$            &  950.6176 & 3 &  -34 & 106 & 0.73 \\
H$\epsilon$          &  950.6176 & 3 &  -30 &  87 & 0.61 \\
H$\zeta$             &  950.6176 & 3 &  -31 & 104 & 0.62 \\
\ion{Ca}{ii} K       &  950.6176 & 3 &  -28 &  89 & 0.61 \\
\ion{Ca}{ii} H       &  950.6176 & 3 &  -11 &  58 & 0.98 \\
\ion{Na}{i} D2       &  950.6176 & 3 &  -34 &  91 & 0.16 \\
\ion{Na}{i} D1       &  950.6176 & 3 &  -22 &  59 & 0.12 \\
H$\beta$             &  950.6176 & 4 & -104 &  58 & 0.63 \\
H$\gamma$            &  950.6176 & 4 &  -97 &  55 & 0.43 \\
H$\delta$            &  950.6176 & 4 &  -90 &  55 & 0.39 \\
H$\zeta$             &  950.6176 & 4 &  -82 &  49 & 0.25 \\
\ion{Ca}{ii} K       &  950.6176 & 4 &  -86 &  42 & 0.31 \\
\ion{Ca}{ii} H       &  950.6176 & 4 &  -74 &  55 & 0.63 \\
\ion{Fe}{ii} 4924\AA &  950.6176 & 4 &  -67 &  45 & 0.09 \\
\ion{Fe}{ii} 5018\AA &  950.6176 & 4 &  -68 &  46 & 0.12 \\
\ion{Fe}{ii} 5169\AA &  950.6176 & 4 &  -66 &  41 & 0.13 \\
\ion{Ca}{ii} K       &  950.6176 & 0 &   -4 &  23 & 0.21 \\
\ion{Ca}{ii} H       &  950.6176 & 0 &   -1 &   7 & 0.09 \\
\ion{Na}{i} D2       &  950.6176 & 0 &   -2 &  22 & 0.86 \\
\ion{Na}{i} D1       &  950.6176 & 0 &   -1 &  16 & 0.80 \\
H$\beta$             &  951.6232 & 1 &   93 &  75 & 0.33 \\
H$\gamma$            &  951.6232 & 1 &   72 & 109 & 0.53 \\
H$\delta$            &  951.6232 & 1 &   42 & 136 & 0.68 \\
H$\epsilon$          &  951.6232 & 1 &   70 &  91 & 0.34 \\
\ion{Ca}{ii} K       &  951.6232 & 1 &   30 & 142 & 0.57 \\
\ion{Na}{i} D2       &  951.6232 & 1 &   59 &  62 & 0.14 \\
\ion{Na}{i} D1       &  951.6232 & 1 &   67 &  43 & 0.12 \\
\ion{Fe}{ii} 4924\AA &  951.6232 & 1 &   45 & 112 & 0.18 \\
\ion{Fe}{ii} 5018\AA &  951.6232 & 1 &   44 & 111 & 0.26 \\
\ion{Fe}{ii} 5169\AA &  951.6232 & 1 &   46 & 113 & 0.27 \\
\ion{Ca}{ii} K       &  951.6232 & 2 &  -21 &  53 & 0.36 \\
\ion{Ca}{ii} H       &  951.6232 & 2 &  -12 &  46 & 0.74 \\
\ion{Na}{i} D2       &  951.6232 & 2 &  -14 &  59 & 0.50 \\
\ion{Na}{i} D1       &  951.6232 & 2 &  -10 &  58 & 0.36 \\
\ion{Fe}{ii} 4924\AA &  951.6232 & 2 &  -20 &  42 & 0.12 \\
\ion{Fe}{ii} 5018\AA &  951.6232 & 2 &  -20 &  52 & 0.21 \\
\ion{Fe}{ii} 5169\AA &  951.6232 & 2 &  -23 &  66 & 0.33 \\
H$\beta$             &  951.6232 & 3 &  -30 & 134 & 1.10 \\
H$\gamma$            &  951.6232 & 3 &  -39 & 136 & 0.96 \\
H$\delta$            &  951.6232 & 3 &  -53 & 107 & 0.70 \\
H$\epsilon$          &  951.6232 & 3 &  -21 & 129 & 0.79 \\
H$\zeta$             &  951.6232 & 3 &  -49 &  52 & 0.21 \\
\ion{Ca}{ii} K       &  951.6232 & 3 &  -64 &  41 & 0.31 \\
\ion{Ca}{ii} H       &  951.6232 & 3 &  -57 &  50 & 0.57 \\
\ion{Fe}{ii} 4924\AA &  951.6232 & 3 &  -40 & 135 & 0.13 \\
\ion{Fe}{ii} 5018\AA &  951.6232 & 3 &  -63 & 127 & 0.12 \\
\ion{Fe}{ii} 5169\AA &  951.6232 & 3 &  -91 &  84 & 0.10 \\
H$\beta$             &  951.6232 & 4 & -174 &  63 & 0.14 \\
H$\gamma$            &  951.6232 & 4 & -170 &  81 & 0.19 \\
H$\delta$            &  951.6232 & 4 & -156 &  78 & 0.18 \\
H$\zeta$             &  951.6232 & 4 & -137 & 119 & 0.10 \\
\ion{Ca}{ii} K       &  951.6232 & 4 & -122 & 113 & 0.17 \\
\ion{Ca}{ii} H       &  951.6232 & 4 & -139 &  70 & 0.10 \\
\ion{Ca}{ii} K       &  951.6232 & 0 &   -3 &  19 & 0.14 \\
\ion{Ca}{ii} H       &  951.6232 & 0 &   -4 &   9 & 0.07 \\
\ion{Na}{i} D2       &  951.6232 & 0 &   -0 &  15 & 0.54 \\
\ion{Na}{i} D1       &  951.6232 & 0 &    0 &  12 & 0.62 \\
\ion{Fe}{ii} 4924\AA &  951.6232 & 0 &   11 &  11 & 0.04 \\
\ion{Fe}{ii} 5018\AA &  951.6232 & 0 &   11 &   8 & 0.06 \\
\ion{Fe}{ii} 5169\AA &  951.6232 & 0 &    8 &   8 & 0.03 \\
H$\zeta$             &  951.6232 & --&   10 & 163 & 0.75 \\
\ion{Ca}{ii} K       & 1023.5186 & 5 &   41 & 106 & 0.46 \\
\ion{Ca}{ii} H       & 1023.5186 & 5 &   28 &  28 & 0.28 \\
\ion{Fe}{ii} 4924\AA & 1023.5186 & 5 &   37 & 115 & 0.08 \\
\ion{Fe}{ii} 5018\AA & 1023.5186 & 5 &   50 & 132 & 0.11 \\
\ion{Fe}{ii} 5169\AA & 1023.5186 & 5 &   40 & 127 & 0.11 \\
\ion{Ti}{ii} 4444\AA & 1023.5186 & 5 &   63 & 113 & 0.02 \\
H$\beta$             & 1023.5186 & 6 &  -11 & 157 & 1.31 \\
H$\gamma$            & 1023.5186 & 6 &   -9 & 155 & 1.09 \\
H$\delta$            & 1023.5186 & 6 &   -9 & 131 & 0.99 \\
\hline
\end{tabular}
\hspace{0.2cm}
\begin{tabular}[t]{lllrrl}
\hline
\hline
Line & JD & Event & $v$ (\kms) & \deltav\ (\kms)  & $R_{\rm max}$ \\
\hline
H$\epsilon$          & 1023.5186 & 6 &   -3 &  99 & 0.98 \\
H$\zeta$             & 1023.5186 & 6 &   -9 & 101 & 0.85 \\
\ion{Ca}{ii} K       & 1023.5186 & 6 &  -11 &  62 & 0.66 \\
\ion{Ca}{ii} H       & 1023.5186 & 6 &   -6 &  46 & 1.15 \\
\ion{Fe}{ii} 4924\AA & 1023.5186 & 6 &   -6 &  34 & 0.32 \\
\ion{Fe}{ii} 5018\AA & 1023.5186 & 6 &   -6 &  40 & 0.41 \\
\ion{Fe}{ii} 5169\AA & 1023.5186 & 6 &   -5 &  39 & 0.45 \\
\ion{Ti}{ii} 4444\AA & 1023.5186 & 6 &   -6 &  36 & 0.05 \\
\ion{Ti}{ii} 4572\AA & 1023.5186 & 6 &   -7 &  37 & 0.05 \\
\ion{Ca}{ii} K       & 1023.5186 & 7 &  -46 &  28 & 0.41 \\
\ion{Ca}{ii} H       & 1023.5186 & 7 &  -41 &  31 & 0.85 \\
\ion{Na}{i} D2       & 1023.5186 & 7 &  -30 &  25 & 0.50 \\
\ion{Na}{i} D1       & 1023.5186 & 7 &  -23 &  34 & 0.35 \\
\ion{Fe}{ii} 4924\AA & 1023.5186 & 7 &  -35 &  24 & 0.30 \\
\ion{Fe}{ii} 5018\AA & 1023.5186 & 7 &  -35 &  23 & 0.33 \\
\ion{Fe}{ii} 5169\AA & 1023.5186 & 7 &  -35 &  24 & 0.40 \\
\ion{Ti}{ii} 4444\AA & 1023.5186 & 7 &  -34 &  21 & 0.04 \\
\ion{Ti}{ii} 4572\AA & 1023.5186 & 7 &  -34 &  22 & 0.04 \\
\ion{Na}{i} D2       & 1023.5186 & 0 &   -1 &  22 & 0.93 \\
\ion{Na}{i} D1       & 1023.5186 & 0 &    0 &  14 & 0.86 \\
\ion{Fe}{ii} 4924\AA & 1112.3689 & 8 &   86 & 162 & 0.09 \\
\ion{Fe}{ii} 5018\AA & 1112.3689 & 8 &   96 & 195 & 0.13 \\
\ion{Fe}{ii} 5169\AA & 1112.3689 & 8 &  113 & 248 & 0.11 \\
\ion{Ti}{ii} 4444\AA & 1112.3689 & 8 &   91 & 163 & 0.03 \\
\ion{Ti}{ii} 4572\AA & 1112.3689 & 8 &   78 & 224 & 0.03 \\
H$\beta$             & 1112.3689 & 9 &    7 & 156 & 0.98 \\
H$\gamma$            & 1112.3689 & 9 &   17 & 159 & 0.97 \\
H$\delta$            & 1112.3689 & 9 &   17 & 152 & 0.99 \\
H$\epsilon$          & 1112.3689 & 9 &   11 & 174 & 1.04 \\
H$\zeta$             & 1112.3689 & 9 &   24 & 145 & 0.93 \\
\ion{Ca}{ii} K       & 1112.3689 & 9 &   23 & 148 & 0.78 \\
\ion{Na}{i} D2       & 1112.3689 & 9 &   21 &  83 & 0.59 \\
\ion{Na}{i} D1       & 1112.3689 & 9 &   22 &  72 & 0.45 \\
\ion{Fe}{ii} 4924\AA & 1112.3689 & 9 &   18 &  77 & 0.32 \\
\ion{Fe}{ii} 5018\AA & 1112.3689 & 9 &   17 &  82 & 0.39 \\
\ion{Fe}{ii} 5169\AA & 1112.3689 & 9 &   18 &  85 & 0.45 \\
\ion{Fe}{i} 4046\AA  & 1112.3689 & 9 &   32 &  76 & 0.03 \\
\ion{Ti}{ii} 4444\AA & 1112.3689 & 9 &   14 &  57 & 0.08 \\
\ion{Ti}{ii} 4572\AA & 1112.3689 & 9 &   16 &  62 & 0.09 \\
\ion{Sc}{ii} 4247\AA & 1112.3689 & 9 &   35 &  79 & 0.08 \\
\ion{Ca}{i} 4227\AA  & 1112.3689 & 9 &   17 &  66 & 0.03 \\
\ion{Ca}{ii} K       & 1112.3689 & 0 &   -7 &  25 & 0.16 \\
\ion{Ca}{ii} H       & 1112.3689 & 0 &   -5 &  56 & 0.89 \\
\ion{Na}{i} D2       & 1112.3689 & 0 &   -2 &  14 & 0.51 \\
\ion{Na}{i} D1       & 1112.3689 & 0 &   -1 &  13 & 0.62 \\
\ion{Fe}{ii} 4924\AA & 1112.3689 & 0 &    0 &  13 & 0.11 \\
\ion{Fe}{ii} 5018\AA & 1112.3689 & 0 &    1 &  12 & 0.12 \\
\ion{Fe}{ii} 5169\AA & 1112.3689 & 0 &   -0 &  13 & 0.15 \\
\ion{Sc}{ii} 4247\AA & 1112.3689 & 0 &    2 &  31 & 0.07 \\
H$\gamma$            & 1113.3958 & 8 &  129 & 248 & 0.42 \\
H$\delta$            & 1113.3958 & 8 &  106 & 233 & 0.53 \\
H$\epsilon$          & 1113.3958 & 8 &   85 & 257 & 0.62 \\
H$\zeta$             & 1113.3958 & 8 &  106 & 220 & 0.47 \\
\ion{Ca}{ii} K       & 1113.3958 & 8 &  108 & 239 & 0.46 \\
\ion{Na}{i} D2       & 1113.3958 & 8 &   82 & 187 & 0.14 \\
\ion{Na}{i} D1       & 1113.3958 & 8 &  129 & 185 & 0.11 \\
\ion{Fe}{ii} 4924\AA & 1113.3958 & 8 &   80 & 236 & 0.18 \\
\ion{Fe}{ii} 5018\AA & 1113.3958 & 8 &   90 & 263 & 0.24 \\
\ion{Fe}{ii} 5169\AA & 1113.3958 & 8 &  116 & 291 & 0.24 \\
\ion{Fe}{i} 4046\AA  & 1113.3958 & 8 &  127 & 116 & 0.04 \\
\ion{Ti}{ii} 4444\AA & 1113.3958 & 8 &   74 & 231 & 0.06 \\
\ion{Ti}{ii} 4572\AA & 1113.3958 & 8 &  107 & 253 & 0.06 \\
\ion{Sc}{ii} 4247\AA & 1113.3958 & 8 &  123 & 140 & 0.06 \\
\ion{Ca}{i} 4227\AA  & 1113.3958 & 8 &  134 & 109 & 0.03 \\
H$\beta$             & 1113.3958 & 9 &   -4 & 124 & 0.96 \\
H$\gamma$            & 1113.3958 & 9 &   -4 & 121 & 0.73 \\
H$\delta$            & 1113.3958 & 9 &  -11 & 109 & 0.66 \\
H$\epsilon$          & 1113.3958 & 9 &    6 &  68 & 0.44 \\
H$\zeta$             & 1113.3958 & 9 &   -9 & 102 & 0.65 \\
\ion{Ca}{ii} K       & 1113.3958 & 9 &   -4 &  63 & 0.49 \\
\ion{Ca}{ii} H       & 1113.3958 & 9 &   -2 &  58 & 0.79 \\
\ion{Na}{i} D2       & 1113.3958 & 9 &    4 &  48 & 0.74 \\
\ion{Na}{i} D1       & 1113.3958 & 9 &    5 &  43 & 0.76 \\
\ion{Fe}{ii} 4924\AA & 1113.3958 & 9 &    4 &  43 & 0.34 \\
\ion{Fe}{ii} 5018\AA & 1113.3958 & 9 &    4 &  47 & 0.39 \\
\ion{Fe}{ii} 5169\AA & 1113.3958 & 9 &    4 &  49 & 0.46 \\
\ion{Fe}{i} 4046\AA  & 1113.3958 & 9 &    3 &  40 & 0.06 \\
\ion{Ti}{ii} 4444\AA & 1113.3958 & 9 &    6 &  36 & 0.09 \\
\ion{Ti}{ii} 4572\AA & 1113.3958 & 9 &    1 &  48 & 0.09 \\
\ion{Sc}{ii} 4247\AA & 1113.3958 & 9 &    6 &  55 & 0.12 \\
\ion{Ca}{i} 4227\AA  & 1113.3958 & 9 &    6 &  46 & 0.10 \\
H$\beta$             & 1113.3958 & --&  251 & 149 & 0.20 \\
\hline
\\
\end{tabular}}
\end{table*}


\begin{table*}
\caption[Identified Transient Absorption Components (TACs) in XY~Per]
{Identified Transient Absorption Components (TACs) in XY~Per.
Details as for Table~\ref{haebe_master_table_bfori}.}
\label{haebe_master_table_xyper}
\centerline{
\tiny
\begin{tabular}[t]{lllrrl}
\hline
\hline
Line & JD & Event & $v$ (\kms) & \deltav\ (\kms) & $R_{\rm max}$ \\
\hline
H$\beta$             & 1024.6728 & 2 &  -9 & 171 & 0.92 \\
H$\gamma$            & 1024.6728 & 2 &  -4 & 171 & 0.92 \\
H$\delta$            & 1024.6728 & 2 &  -3 & 163 & 0.88 \\
H$\epsilon$          & 1024.6728 & 2 &   2 & 147 & 0.72 \\
H$\zeta$             & 1024.6728 & 2 &  -5 & 146 & 0.70 \\
\ion{Ca}{ii} K       & 1024.6728 & 2 &  -7 & 142 & 0.77 \\
\ion{Ca}{ii} H       & 1024.6728 & 2 & -27 & 104 & 0.63 \\
\ion{Na}{i} D2       & 1024.6728 & 2 & -26 & 100 & 0.16 \\
\ion{Na}{i} D1       & 1024.6728 & 2 & -19 &  92 & 0.10 \\
\ion{Fe}{ii} 4924\AA & 1024.6728 & 2 & -15 & 118 & 0.25 \\
\ion{Fe}{ii} 5018\AA & 1024.6728 & 2 & -14 & 115 & 0.32 \\
\ion{Fe}{ii} 5169\AA & 1024.6728 & 2 & -16 & 114 & 0.35 \\
\ion{Ti}{ii} 4444\AA & 1024.6728 & 2 &   2 & 156 & 0.06 \\
\ion{Ti}{ii} 4572\AA & 1024.6728 & 2 & -11 & 144 & 0.06 \\
\ion{Sc}{ii} 4247\AA & 1024.6728 & 2 &  -3 & 133 & 0.04 \\
\ion{Ca}{ii} K       & 1024.6728 & 0 &  -0 &  18 & 0.23 \\
\ion{Ca}{ii} H       & 1024.6728 & 0 &  -1 &  25 & 0.44 \\
\ion{Na}{i} D2       & 1024.6728 & 0 &  -0 &  19 & 0.93 \\
\ion{Na}{i} D1       & 1024.6728 & 0 &   0 &  17 & 0.98 \\
\ion{Fe}{ii} 4924\AA & 1024.6728 & 0 &  -5 &  27 & 0.06 \\
\ion{Fe}{ii} 5018\AA & 1024.6728 & 0 &  -6 &  23 & 0.06 \\
\ion{Fe}{ii} 5169\AA & 1024.6728 & 0 &  -4 &  27 & 0.07 \\
\ion{Sc}{ii} 4247\AA & 1024.6728 & 0 &  -7 &   9 & 0.01 \\
H$\beta$             & 1025.6948 & 1 &  70 &  72 & 0.37 \\
H$\gamma$            & 1025.6948 & 1 &  74 &  58 & 0.16 \\
H$\delta$            & 1025.6948 & 1 &  66 &  93 & 0.21 \\
H$\epsilon$          & 1025.6948 & 1 &  60 &  59 & 0.23 \\
\ion{Fe}{ii} 4924\AA & 1025.6948 & 1 &  59 &  66 & 0.05 \\
\ion{Fe}{ii} 5018\AA & 1025.6948 & 1 &  73 &  66 & 0.05 \\
\ion{Fe}{ii} 5169\AA & 1025.6948 & 1 &  65 &  57 & 0.04 \\
\ion{Ti}{ii} 4444\AA & 1025.6948 & 1 &  34 & 113 & 0.06 \\
\ion{Ti}{ii} 4572\AA & 1025.6948 & 1 &  44 &  79 & 0.04 \\
\ion{Sc}{ii} 4247\AA & 1025.6948 & 1 &  37 &  88 & 0.04 \\
H$\beta$             & 1025.6948 & 2 & -10 &  90 & 0.90 \\
H$\gamma$            & 1025.6948 & 2 & -22 & 147 & 0.96 \\
H$\delta$            & 1025.6948 & 2 & -27 & 133 & 0.90 \\
H$\epsilon$          & 1025.6948 & 2 & -15 &  96 & 0.63 \\
H$\zeta$             & 1025.6948 & 2 & -11 & 141 & 0.70 \\
\ion{Ca}{ii} K       & 1025.6948 & 2 & -20 & 130 & 0.78 \\
\ion{Ca}{ii} H       & 1025.6948 & 2 & -20 &  94 & 0.73 \\
\ion{Na}{i} D2       & 1025.6948 & 2 & -28 & 104 & 0.24 \\
\ion{Na}{i} D1       & 1025.6948 & 2 & -29 &  89 & 0.16 \\
\ion{Fe}{ii} 4924\AA & 1025.6948 & 2 & -35 & 115 & 0.27 \\
\ion{Fe}{ii} 5018\AA & 1025.6948 & 2 & -29 & 117 & 0.34 \\
\ion{Fe}{ii} 5169\AA & 1025.6948 & 2 & -34 & 113 & 0.38 \\
\ion{Ti}{ii} 4444\AA & 1025.6948 & 2 & -44 &  84 & 0.05 \\
\ion{Ti}{ii} 4572\AA & 1025.6948 & 2 & -42 &  88 & 0.06 \\
\ion{Sc}{ii} 4247\AA & 1025.6948 & 2 & -42 &  69 & 0.03 \\
H$\beta$             & 1025.6948 & 3 & -88 &  82 & 0.65 \\
H$\gamma$            & 1025.6948 & 3 & -89 &  51 & 0.17 \\
H$\delta$            & 1025.6948 & 3 & -85 &  46 & 0.14 \\
H$\zeta$             & 1025.6948 & 3 & -68 &  58 & 0.14 \\
\ion{Ca}{ii} K       & 1025.6948 & 3 & -94 &  40 & 0.17 \\
\ion{Ca}{ii} H       & 1025.6948 & 3 & -85 &  51 & 0.28 \\
\ion{Fe}{ii} 5018\AA & 1025.6948 & 3 & -77 &  28 & 0.03 \\
\ion{Fe}{ii} 5169\AA & 1025.6948 & 3 & -79 &  29 & 0.04 \\
\ion{Ca}{ii} K       & 1025.6948 & 0 &   1 &  16 & 0.27 \\
\ion{Ca}{ii} H       & 1025.6948 & 0 &   1 &  14 & 0.32 \\
\ion{Na}{i} D2       & 1025.6948 & 0 &  -1 &  19 & 0.87 \\
\ion{Na}{i} D1       & 1025.6948 & 0 &  -1 &  18 & 0.96 \\
\ion{Fe}{ii} 4924\AA & 1025.6948 & 0 &   1 &  23 & 0.15 \\
\ion{Fe}{ii} 5018\AA & 1025.6948 & 0 &  -0 &  20 & 0.15 \\
\ion{Fe}{ii} 5169\AA & 1025.6948 & 0 &  -1 &  23 & 0.18 \\
\ion{Ti}{ii} 4444\AA & 1025.6948 & 0 &   1 &  20 & 0.04 \\
\ion{Ti}{ii} 4572\AA & 1025.6948 & 0 &   1 &  28 & 0.03 \\
\ion{Sc}{ii} 4247\AA & 1025.6948 & 0 &   2 &  22 & 0.02 \\
H$\beta$             & 1026.7065 & 1 &  69 &  75 & 0.24 \\
H$\gamma$            & 1026.7065 & 1 &  62 &  81 & 0.23 \\
H$\delta$            & 1026.7065 & 1 &  55 &  96 & 0.24 \\
H$\epsilon$          & 1026.7065 & 1 &  65 &  59 & 0.10 \\
H$\zeta$             & 1026.7065 & 1 &  51 &  72 & 0.20 \\
\ion{Ca}{ii} K       & 1026.7065 & 1 &  49 &  27 & 0.10 \\
\ion{Fe}{ii} 4924\AA & 1026.7065 & 1 &  30 &  76 & 0.10 \\
\ion{Fe}{ii} 5169\AA & 1026.7065 & 1 &  24 &  82 & 0.12 \\
\ion{Ti}{ii} 4444\AA & 1026.7065 & 1 &  32 & 113 & 0.04 \\
\ion{Ti}{ii} 4572\AA & 1026.7065 & 1 &  62 &  68 & 0.02 \\
\ion{Sc}{ii} 4247\AA & 1026.7065 & 1 &  58 &  51 & 0.02 \\
H$\beta$             & 1026.7065 & 2 & -58 & 124 & 0.79 \\
H$\gamma$            & 1026.7065 & 2 & -48 & 117 & 0.78 \\
H$\delta$            & 1026.7065 & 2 & -52 & 105 & 0.76 \\
H$\epsilon$          & 1026.7065 & 2 & -61 &  51 & 0.40 \\
H$\zeta$             & 1026.7065 & 2 & -51 &  84 & 0.57 \\
\hline
\end{tabular}
\hspace{0.2cm}
\begin{tabular}[t]{lllrrl}
\hline
\hline
Line & JD & Event & $v$ (\kms) & \deltav\ (\kms) & $R_{\rm max}$ \\
\hline
\ion{Ca}{ii} K       & 1026.7065 & 2 & -66 &  91 & 0.55 \\
\ion{Ca}{ii} H       & 1026.7065 & 2 & -51 & 101 & 0.55 \\
\ion{Na}{i} D2       & 1026.7065 & 2 & -36 & 102 & 0.26 \\
\ion{Na}{i} D1       & 1026.7065 & 2 & -43 &  64 & 0.18 \\
\ion{Fe}{ii} 4924\AA & 1026.7065 & 2 & -52 &  83 & 0.37 \\
\ion{Fe}{ii} 5018\AA & 1026.7065 & 2 & -56 &  66 & 0.43 \\
\ion{Fe}{ii} 5169\AA & 1026.7065 & 2 & -54 &  82 & 0.45 \\
\ion{Ti}{ii} 4444\AA & 1026.7065 & 2 & -52 &  66 & 0.09 \\
\ion{Ti}{ii} 4572\AA & 1026.7065 & 2 & -53 &  68 & 0.10 \\
\ion{Sc}{ii} 4247\AA & 1026.7065 & 2 & -54 &  68 & 0.06 \\
H$\beta$             & 1026.7065 & 4 &  -3 &  54 & 0.42 \\
H$\gamma$            & 1026.7065 & 4 &   0 &  41 & 0.24 \\
H$\delta$            & 1026.7065 & 4 &   2 &  38 & 0.19 \\
H$\epsilon$          & 1026.7065 & 4 &  -4 &  38 & 0.21 \\
H$\zeta$             & 1026.7065 & 4 &   4 &  42 & 0.20 \\
\ion{Ca}{ii} K       & 1026.7065 & 4 &  -8 &  48 & 0.63 \\
\ion{Ca}{ii} H       & 1026.7065 & 4 &  -4 &  33 & 0.49 \\
\ion{Fe}{ii} 4924\AA & 1026.7065 & 4 &  -3 &  26 & 0.11 \\
\ion{Fe}{ii} 5018\AA & 1026.7065 & 4 &  -1 &  46 & 0.28 \\
\ion{Fe}{ii} 5169\AA & 1026.7065 & 4 &  -5 &  29 & 0.15 \\
\ion{Ti}{ii} 4444\AA & 1026.7065 & 4 &  -1 &  33 & 0.04 \\
\ion{Ti}{ii} 4572\AA & 1026.7065 & 4 &   5 &  50 & 0.05 \\
\ion{Sc}{ii} 4247\AA & 1026.7065 & 4 &  12 &  52 & 0.03 \\
\ion{Na}{i} D2       & 1026.7065 & 0 &  -0 &  17 & 0.89 \\
\ion{Na}{i} D1       & 1026.7065 & 0 &  -0 &  17 & 1.02 \\
H$\beta$             & 1112.4978 & 5 & 131 & 149 & 0.27 \\
H$\gamma$            & 1112.4978 & 5 &  83 & 206 & 0.41 \\
H$\delta$            & 1112.4978 & 5 & 118 & 173 & 0.30 \\
H$\zeta$             & 1112.4978 & 5 & 127 & 182 & 0.20 \\
\ion{Ca}{ii} K       & 1112.4978 & 5 & 119 & 177 & 0.29 \\
\ion{Fe}{ii} 4924\AA & 1112.4978 & 5 & 147 & 139 & 0.07 \\
\ion{Fe}{ii} 5018\AA & 1112.4978 & 5 & 123 & 190 & 0.13 \\
\ion{Fe}{ii} 5169\AA & 1112.4978 & 5 & 147 & 218 & 0.11 \\
\ion{Ti}{ii} 4444\AA & 1112.4978 & 5 & 122 & 209 & 0.02 \\
H$\beta$             & 1112.4978 & 6 &  30 & 129 & 0.77 \\
H$\gamma$            & 1112.4978 & 6 &  18 & 136 & 0.58 \\
H$\delta$            & 1112.4978 & 6 &  14 & 142 & 0.73 \\
H$\epsilon$          & 1112.4978 & 6 &  16 & 189 & 0.84 \\
H$\zeta$             & 1112.4978 & 6 &  17 & 139 & 0.59 \\
\ion{Ca}{ii} K       & 1112.4978 & 6 &  35 & 104 & 0.47 \\
\ion{Na}{i} D2       & 1112.4978 & 6 &  40 & 119 & 0.15 \\
\ion{Na}{i} D1       & 1112.4978 & 6 &  29 & 132 & 0.10 \\
\ion{Fe}{ii} 4924\AA & 1112.4978 & 6 &  26 & 124 & 0.25 \\
\ion{Fe}{ii} 5018\AA & 1112.4978 & 6 &  22 & 114 & 0.26 \\
\ion{Fe}{ii} 5169\AA & 1112.4978 & 6 &  18 & 121 & 0.28 \\
\ion{Fe}{i} 4046\AA  & 1112.4978 & 6 &  39 & 100 & 0.03 \\
\ion{Ti}{ii} 4444\AA & 1112.4978 & 6 &  32 & 117 & 0.06 \\
\ion{Ti}{ii} 4572\AA & 1112.4978 & 6 &  38 & 139 & 0.08 \\
\ion{Sc}{ii} 4247\AA & 1112.4978 & 6 &  36 & 137 & 0.07 \\
\ion{Ca}{i} 4227\AA  & 1112.4978 & 6 &  23 &  66 & 0.05 \\
H$\beta$             & 1112.4978 & 7 & -26 &  57 & 0.58 \\
H$\gamma$            & 1112.4978 & 7 & -24 &  44 & 0.34 \\
H$\delta$            & 1112.4978 & 7 & -21 &  37 & 0.23 \\
H$\zeta$             & 1112.4978 & 7 & -16 &  40 & 0.15 \\
\ion{Ca}{ii} K       & 1112.4978 & 7 & -34 &  33 & 0.53 \\
\ion{Ca}{ii} H       & 1112.4978 & 7 & -25 &  43 & 0.90 \\
\ion{Na}{i} D2       & 1112.4978 & 7 & -23 &  19 & 0.31 \\
\ion{Na}{i} D1       & 1112.4978 & 7 & -22 &  18 & 0.18 \\
\ion{Fe}{ii} 4924\AA & 1112.4978 & 7 & -22 &  24 & 0.18 \\
\ion{Fe}{ii} 5018\AA & 1112.4978 & 7 & -22 &  24 & 0.24 \\
\ion{Fe}{ii} 5169\AA & 1112.4978 & 7 & -22 &  22 & 0.27 \\
\ion{Ti}{ii} 4444\AA & 1112.4978 & 7 & -21 &  20 & 0.04 \\
\ion{Ti}{ii} 4572\AA & 1112.4978 & 7 & -24 &  23 & 0.04 \\
\ion{Ca}{ii} K       & 1112.4978 & 0 &  -2 &  30 & 0.48 \\
\ion{Ca}{ii} H       & 1112.4978 & 0 &   2 &  20 & 0.61 \\
\ion{Na}{i} D2       & 1112.4978 & 0 &   1 &  19 & 0.92 \\
\ion{Na}{i} D1       & 1112.4978 & 0 &   1 &  16 & 0.97 \\
H$\beta$             & 1113.5483 & 5 &  92 &  53 & 0.08 \\
H$\gamma$            & 1113.5483 & 5 &  76 & 107 & 0.26 \\
H$\delta$            & 1113.5483 & 5 &  75 & 123 & 0.28 \\
H$\zeta$             & 1113.5483 & 5 &  82 & 111 & 0.21 \\
\ion{Ca}{ii} K       & 1113.5483 & 5 &  76 &  51 & 0.11 \\
\ion{Fe}{ii} 4924\AA & 1113.5483 & 5 &  85 &  85 & 0.12 \\
\ion{Fe}{ii} 5018\AA & 1113.5483 & 5 &  79 & 136 & 0.14 \\
\ion{Fe}{ii} 5169\AA & 1113.5483 & 5 &  87 &  94 & 0.14 \\
\ion{Ti}{ii} 4444\AA & 1113.5483 & 5 &  82 & 115 & 0.03 \\
\ion{Ti}{ii} 4572\AA & 1113.5483 & 5 &  94 &  83 & 0.03 \\
H$\beta$             & 1113.5483 & 7 & -16 & 149 & 0.99 \\
H$\gamma$            & 1113.5483 & 7 & -23 & 133 & 0.93 \\
H$\delta$            & 1113.5483 & 7 & -25 & 126 & 0.89 \\
H$\epsilon$          & 1113.5483 & 7 &   1 & 133 & 0.51 \\
H$\zeta$             & 1113.5483 & 7 & -19 & 117 & 0.73 \\
\hline
\end{tabular}}
\end{table*}
\begin{table*}
\addtocounter{table}{-1}
\caption{(Cont.)}
\label{haebe_master_table_xyper_cont}
\centerline{
\tiny
\begin{tabular}[t]{lllrrl}
\hline
\hline
Line & JD & Event & $v$ (\kms) & \deltav\ (\kms) & $R_{\rm max}$ \\
\hline
\ion{Ca}{ii} K       & 1113.5483 & 7 & -18 & 107 & 0.71 \\
\ion{Ca}{ii} H       & 1113.5483 & 7 & -22 &  83 & 0.81 \\
\ion{Na}{i} D2       & 1113.5483 & 7 &  -4 &  64 & 0.37 \\
\ion{Na}{i} D1       & 1113.5483 & 7 &  -3 &  56 & 0.22 \\
\ion{Fe}{ii} 4924\AA & 1113.5483 & 7 & -10 &  92 & 0.47 \\
\ion{Fe}{ii} 5018\AA & 1113.5483 & 7 & -12 & 101 & 0.42 \\
\ion{Fe}{ii} 5169\AA & 1113.5483 & 7 & -11 &  90 & 0.60 \\
\ion{Ti}{ii} 4444\AA & 1113.5483 & 7 &  -1 & 106 & 0.12 \\
\ion{Ti}{ii} 4572\AA & 1113.5483 & 7 &  -2 & 106 & 0.13 \\
\ion{Ca}{ii} K       & 1113.5483 & 0 &   1 &  17 & 0.27 \\
\ion{Ca}{ii} H       & 1113.5483 & 0 &   1 &  14 & 0.33 \\
\ion{Na}{i} D2       & 1113.5483 & 0 &   1 &  20 & 0.69 \\
\ion{Na}{i} D1       & 1113.5483 & 0 &   1 &  18 & 0.85 \\
\ion{Fe}{ii} 5018\AA & 1113.5483 & 0 & -14 &  27 & 0.17 \\
H$\beta$             & 1207.3786 & 8 & 141 & 120 & 0.16 \\
H$\gamma$            & 1207.3786 & 8 & 121 & 115 & 0.18 \\
H$\delta$            & 1207.3786 & 8 & 118 & 117 & 0.19 \\
H$\zeta$             & 1207.3786 & 8 & 108 & 129 & 0.17 \\
\ion{Ca}{ii} K       & 1207.3786 & 8 & 151 &  58 & 0.06 \\
\ion{Na}{i} D2       & 1207.3786 & 8 &  43 & 117 & 0.16 \\
\ion{Na}{i} D1       & 1207.3786 & 8 &  59 & 159 & 0.10 \\
\ion{Fe}{ii} 4924\AA & 1207.3786 & 8 &  87 & 149 & 0.17 \\
\ion{Fe}{ii} 5018\AA & 1207.3786 & 8 &  99 & 123 & 0.16 \\
\ion{Fe}{ii} 5169\AA & 1207.3786 & 8 &  54 & 196 & 0.27 \\
\ion{Fe}{i} 4046\AA  & 1207.3786 & 8 &  61 & 188 & 0.04 \\
\ion{Ti}{ii} 4444\AA & 1207.3786 & 8 &  46 & 205 & 0.10 \\
\ion{Ti}{ii} 4572\AA & 1207.3786 & 8 &  83 & 144 & 0.08 \\
\ion{Sc}{ii} 4247\AA & 1207.3786 & 8 &  61 & 144 & 0.10 \\
\ion{Ca}{i} 4227\AA  & 1207.3786 & 8 &  40 & 168 & 0.05 \\
H$\beta$             & 1207.3786 & 9 & -13 & 128 & 0.90 \\
H$\gamma$            & 1207.3786 & 9 &  -9 & 132 & 0.84 \\
H$\delta$            & 1207.3786 & 9 &  -7 & 133 & 0.77 \\
H$\epsilon$          & 1207.3786 & 9 &   2 &  72 & 0.53 \\
H$\zeta$             & 1207.3786 & 9 &  -6 & 127 & 0.56 \\
\ion{Ca}{ii} K       & 1207.3786 & 9 & -18 &  92 & 0.71 \\
\ion{Ca}{ii} H       & 1207.3786 & 9 & -16 &  73 & 0.78 \\
\ion{Na}{i} D2       & 1207.3786 & 9 & -23 &  40 & 0.06 \\
\ion{Fe}{ii} 4924\AA & 1207.3786 & 9 & -13 & 125 & 0.20 \\
\ion{Fe}{ii} 5018\AA & 1207.3786 & 9 &  -4 & 111 & 0.28 \\
\ion{Fe}{ii} 5169\AA & 1207.3786 & 9 & -46 &  47 & 0.10 \\
\ion{Ti}{ii} 4444\AA & 1207.3786 & 9 & -21 &  44 & 0.01 \\
\ion{Ti}{ii} 4572\AA & 1207.3786 & 9 & -25 & 113 & 0.07 \\
\ion{Ca}{ii} K       & 1207.3786 & 0 &   0 &  24 & 0.30 \\
\ion{Ca}{ii} H       & 1207.3786 & 0 &   1 &  22 & 0.44 \\
\hline
\end{tabular}
\hspace{0.2cm}
\begin{tabular}[t]{lllrrl}
\hline
\hline
Line & JD & Event & $v$ (\kms) & \deltav\ (\kms) & $R_{\rm max}$ \\
\hline
\ion{Na}{i} D2       & 1207.3786 & 0 &  -0 &  20 & 0.93 \\
\ion{Na}{i} D1       & 1207.3786 & 0 &   0 &  18 & 1.00 \\
\ion{Fe}{ii} 4924\AA & 1207.3786 & 0 &  -2 &  26 & 0.17 \\
\ion{Fe}{ii} 5018\AA & 1207.3786 & 0 &  -2 &  26 & 0.19 \\
\ion{Fe}{ii} 5169\AA & 1207.3786 & 0 &  -4 &  37 & 0.33 \\
\ion{Sc}{ii} 4247\AA & 1207.3786 & 0 &   4 &  18 & 0.04 \\
H$\gamma$            & 1209.3844 & 8 & 109 &  54 & 0.09 \\
H$\delta$            & 1209.3844 & 8 & 102 &  76 & 0.12 \\
H$\epsilon$          & 1209.3844 & 8 &  69 & 123 & 0.11 \\
H$\zeta$             & 1209.3844 & 8 &  84 &  96 & 0.11 \\
\ion{Na}{i} D2       & 1209.3844 & 8 &  42 & 104 & 0.15 \\
\ion{Na}{i} D1       & 1209.3844 & 8 &  55 &  81 & 0.12 \\
\ion{Fe}{ii} 4924\AA & 1209.3844 & 8 &  36 & 145 & 0.20 \\
\ion{Fe}{ii} 5018\AA & 1209.3844 & 8 &  57 & 161 & 0.19 \\
\ion{Fe}{ii} 5169\AA & 1209.3844 & 8 &  53 & 152 & 0.17 \\
\ion{Fe}{i} 4046\AA  & 1209.3844 & 8 &  44 &  88 & 0.04 \\
\ion{Ti}{ii} 4444\AA & 1209.3844 & 8 &  54 & 110 & 0.11 \\
\ion{Ti}{ii} 4572\AA & 1209.3844 & 8 &  55 & 100 & 0.09 \\
\ion{Sc}{ii} 4247\AA & 1209.3844 & 8 &  49 & 117 & 0.11 \\
\ion{Ca}{i} 4227\AA  & 1209.3844 & 8 &  37 & 128 & 0.07 \\
H$\beta$             & 1209.3844 & 9 & -24 & 124 & 0.92 \\
H$\gamma$            & 1209.3844 & 9 & -18 & 129 & 0.83 \\
H$\delta$            & 1209.3844 & 9 & -14 & 118 & 0.73 \\
H$\epsilon$          & 1209.3844 & 9 &  -7 &  50 & 0.32 \\
H$\zeta$             & 1209.3844 & 9 & -14 & 104 & 0.49 \\
\ion{Ca}{ii} K       & 1209.3844 & 9 & -33 &  89 & 0.71 \\
\ion{Ca}{ii} H       & 1209.3844 & 9 & -32 &  74 & 0.77 \\
\ion{Na}{i} D2       & 1209.3844 & 9 & -32 &  45 & 0.18 \\
\ion{Na}{i} D1       & 1209.3844 & 9 & -32 &  25 & 0.09 \\
\ion{Fe}{ii} 4924\AA & 1209.3844 & 9 & -18 &  70 & 0.25 \\
\ion{Fe}{ii} 5018\AA & 1209.3844 & 9 & -15 &  73 & 0.34 \\
\ion{Fe}{ii} 5169\AA & 1209.3844 & 9 & -19 &  78 & 0.40 \\
\ion{Ti}{ii} 4444\AA & 1209.3844 & 9 & -24 &  67 & 0.08 \\
\ion{Ti}{ii} 4572\AA & 1209.3844 & 9 & -19 &  64 & 0.09 \\
\ion{Sc}{ii} 4247\AA & 1209.3844 & 9 & -20 &  39 & 0.05 \\
\ion{Ca}{ii} K       & 1209.3844 & 0 &   1 &  28 & 0.44 \\
\ion{Ca}{ii} H       & 1209.3844 & 0 &   2 &  27 & 0.62 \\
\ion{Na}{i} D2       & 1209.3844 & 0 &   1 &  19 & 0.94 \\
\ion{Na}{i} D1       & 1209.3844 & 0 &   1 &  19 & 1.03 \\
\ion{Fe}{ii} 4924\AA & 1209.3844 & 0 &   0 &  17 & 0.14 \\
\ion{Fe}{ii} 5018\AA & 1209.3844 & 0 &   0 &  17 & 0.16 \\
\ion{Fe}{ii} 5169\AA & 1209.3844 & 0 &   0 &  22 & 0.18 \\
\ion{Ti}{ii} 4444\AA & 1209.3844 & 0 &   3 &  12 & 0.02 \\
\ion{Ti}{ii} 4572\AA & 1209.3844 & 0 &  -2 &   7 & 0.01 \\
\hline
\end{tabular}}
\end{table*}

\chapter{Conclusions}
\label{conclusions}

In this Chapter, the most relevant conclusions derived from this thesis are summarized.
Most of them have already been given in Sections~\ref{uxori_conclusions} and~\ref{haebe_conclusions}, included in \citetalias{mora2002} (Chapter~\ref{uxori}) and \citetalias{mora2004} (Chapter~\ref{haebe}), respectively.
Some considerations about the spectral database used (Chapters~\ref{observations} and~\ref{reduction}) and the measurements of \vsini\ (Chapter~\ref{vsini}) are also included.

\begin{enumerate}

\item
The excellent quality \'echelle spectra used in this thesis belong to a large database which includes a large number of objects extensively monitored.
This unique set of data has allowed a very detailed study, without precedents, of the TACs in UXOR stars.
All the spectra are reduced and available, under request, to all astronomers who ask for them.

\item
A systematic study of the projected rotational velocities, \vsini, for all the 49 stars in the sample observed with the UES \'echelle spectrograph has been performed.
These measurements, apart from having a high scientific value on their own, have been of great help in the determination of the physical parameters for the stars in this thesis.

\item
Optical high resolution spectra of the HAe UXOR stars BF~Ori, SV~Cep, UX~Ori, WW~Vul and XY~Per have been analysed.
These spectra monitor the stars on time scales of months, days and hours.
Significant activity in the CS disks is always present, which manifests itself in the continuous appearance and disappearance of absorption components detected in hydrogen and in many metallic lines.
These components display a wide range of variability in radial velocity, velocity dispersion and intensity.
This activity is not related to substantial variations of the stellar photosphere.

\item
Variable absorption features are, in most  cases, detected simultaneously in hydrogen and in many metallic lines with similar velocities. 
In each case, there are several  kinematic components in each line, both blue-shifted and red-shifted with respect to the systemic velocity, denoting the simultaneous presence of infalling and outflowing gas.
We attribute the variable features detected in both Balmer and metallic lines to gaseous clumps of solar-like composition, evolving dynamically in the circumstellar disks of these objects.

\item
Blobs undergo accelerations/decelerations of the order of tenths of \ms\ and last for a few days.
Detectable changes in the gas dynamics occur on a time scale of hours, but the intrinsic velocity dispersion of the blobs appears to remain rather constant.
It has been found a positive correlation between the velocity of the clumps and their velocity dispersion.
No noticeable differences are seen in the properties of the infalling and outflowing gas, although infalls generally display larger velocity and velocity dispersion.
Strong indications of a negative correlation between the dispersion velocity and the absorption intensity, during the dynamical evolution of the gaseous blobs, have been found.

\item
The relative intensity of the absorptions for several line multiplets has been studied.
The different saturation levels presented by the multiplets reveal the diversity of physical conditions in the CS disks of the objects.
Most of the gas clumps are detected both in hydrogen and metallic lines.
The relative absorption strength of the transient absorptions have been analysed in the context of NLTE models.
The results suggest that the gas abundances are similar to the solar metallicity, ruling out the evaporation of solid bodies as the general physical origin of the transient absorptions in UXOR stars.
It is suggested that these clumps and their dynamical evolution should be investigated in the context of detailed magnetospheric accretion models, similar to those of T Tauri stars.    
The models should be non-axisymmetric, in order to explain the observed diversity of temporal evolution behaviours.

\item
The star WW Vul is peculiar and behaves differently from the other stars studied in this thesis.
It is the only star that shows, in addition to events seen both in metallic and hydrogen lines, similar to those observed in the other stars, also transient absorption components in metallic lines that do not apparently have any obvious counterpart in the hydrogen lines.
This result, taken at its face-value, would indicate the presence of a metal-rich gas component in the environment of WW~Vul, possibly related to the evaporation of solid bodies.
However, any such conclusion is premature.
We think that a series of optical spectra with better time resolution (hours) and longer monitoring (up to around seven days), spectra in the far UV range - to analyse Lyman and metallic resonance lines - and detailed NLTE models of different CS gas environments are essential for further progress and for providing clues on the origin of these apparently metal-rich events, in terms of their appearance/disappearance statistics, dynamics, metallicity and nature.

\item
Finally, the whole set of data and relations obtained in this thesis provide observational constraints, which should be considered for any realistic scenario of the gaseous circumstellar disks around intermediate-mass PMS stars.

\end{enumerate}

\chapter{Conclusiones}
\label{conclusiones}

En este cap\'{\i}tulo se enumeran las conclusiones m\'as relevantes que se derivan de esta tesis.
La mayor parte de ellas ya han sido expuestas en las secciones~\ref{uxori_conclusions} y~\ref{haebe_conclusions}, correspondientes a los art\'{\i}culos \citetalias{mora2002} (cap\'{\i}tulo~\ref{uxori}) y \citetalias{mora2004} (cap\'{\i}tulo~\ref{haebe}), respectivamente.
Se incluyen adem\'as unas consideraciones acerca del conjunto de datos empleado (cap\'{\i}tulos~\ref{observations} y~\ref{reduction}) y de las medidas de \vsini\ (cap\'{\i}tulo~\ref{vsini}).

\begin{enumerate}

\item
Los espectros \'echelle empleados en esta tesis forman parte de una base de datos de excelente calidad, amplia cobertura temporal y gran n\'umero de objetos.
Esto ha permitido efectuar un estudio de los TACs en estrellas UXOR con un grado de detalle sin precedentes.
Todos los espectros est\'an reducidos y son accesibles, bajo petici\'on previa, a los investigadores que lo soliciten.

\item
Se ha llevado a cabo un estudio sistem\'atico de las velocidades de rotaci\'on proyectadas seg\'un la l\'{\i}nea de visi\'on, \vsini, para el conjunto de las 49 estrellas de la muestra observadas con el espectr\'ografo \'echelle UES.
Estas medidas, adem\'as de tener un gran valor cient\'{\i}fico por s\'{\i} mismas, han facilitado la determinaci\'on de los par\'ametros f\'{\i}sicos para las estrellas de esta tesis.

\item
Se han analizado los espectros visibles de alta resoluci\'on de las siguientes estrellas HAe con comportamiento UXOR: BF~Ori, SV~Cep, UX~Ori, WW~Vul y XY~Per.
Se ha efectuado un seguimiento de estos objetos en escalas temporales de meses, d\'{\i}as y horas.
Los espectros muestran una intensa actividad en los discos CS de estas estrellas, caracterizada por la aparici\'on y desaparici\'on de componentes de absorci\'on en l\'{\i}neas de hidr\'ogeno y met\'alicas.
Estas componentes presentan una gran variabilidad en velocidad radial, anchura e intensidad.
Esta actividad no est\'a relacionada con cambios apreciables en la fotosfera estelar.

\item
Generalmente, se observan en cada espectro varias componentes cinem\'aticas en cada l\'{\i}nea, desplazadas tanto al rojo como al azul.
Esto supone la existencia simult\'anea de gas eyectado y en ca\'{\i}da hacia la estrella.
En general, las componentes de absorci\'on detectadas simult\'aneamente en distintas l\'{\i}neas, se pueden agrupar en funci\'on de la velocidad radial de las mismas.
Se ha atribuido el origen y va\-ria\-bi\-li\-dad de estas absorciones a la evoluci\'on din\'amica (aceleraci\'on o desaceleraci\'on) de condensaciones de gas, en el seno de los discos CS de las estrellas estudiadas.

\item
Estas nubes de gas experimentan aceleraciones o desaceleraciones de valor en torno a d\'ecimas de \ms.
La escala temporal en la cual aparecen y desaparecen estas condensaciones es de unos pocos d\'{\i}as, aunque es posible detectar cambios de velocidad a las pocas horas.
La anchura de las absorciones, relacionada con la dispersi\'on de velocidades de la nube de gas, permanece aproximadamente constante con el tiempo.
Se ha encontrado, sin embargo, una correlaci\'on positiva entre la velocidad de las condensaciones y su dispersi\'on de velocidades.
No se han encontrado grandes diferencias entre las propiedades del gas desplazado al rojo o al azul, aunque las mayores velocidades y dispersiones de velocidades se suelen presentar en el gas desplazado al rojo.
Se han encontrado fuertes indicios de que, durante la evoluci\'on de una nube de gas, el aumento de la dispersi\'on de velocidades conlleva una disminuci\'on de la intensidad de las absorciones y viceversa.

\item
Se ha estudiado la intensidad relativa de las absorciones para l\'{\i}neas pertenecientes a varios multipletes.
Las distintas condiciones de saturaci\'on de los mismos, muestran la diversidad de condiciones f\'{\i}sicas presentes en los discos CS de los objetos estudiados.
La gran mayor\'{\i}a de las condensaciones de gas se detectan tanto en l\'{\i}neas met\'alicas como de hidr\'ogeno.
Un an\'alisis detallado mediante modelos \mbox{NLTE} de este fen\'omeno, sugiere que las abundancias qu\'{\i}micas del gas son similares a la solar.
Esto descarta la evaporaci\'on de cuerpos s\'olidos como el me\-ca\-nis\-mo general de producci\'on de las absorciones CS en estrellas UXOR.
Todo apunta a que el an\'alisis de los resultados de esta tesis se debe realizar en el contexto de modelos detallados de acreci\'on magnetosf\'erica.
Se requiere para ello el desarrollo de modelos te\'oricos con un grado de detalle similar al alcanzado para CTTSs, pero que tengan en cuenta la naturaleza no estacionaria de estos fen\'omenos.

\item
La estrella WW~Vul es distinta del resto de objetos estudiados.
En WW~Vul, al igual que en las dem\'as estrellas, se pueden encontrar eventos detectados tanto en l\'{\i}neas met\'alicas como de hidr\'ogeno.
Pero adem\'as, en WW~Vul se han encontrado componentes met\'alicas sin contrapartida clara en las l\'{\i}neas de hidr\'ogeno.
Este resultado podr\'{\i}a estar relacionado con la presencia de nubes de gas ricas en metales, quiz\'as originadas por la evaporaci\'on de cuerpos s\'olidos.
A pesar de su inter\'es, el hecho de encontrar estas componentes an\'omalas no es una prueba suficiente de la presencia de FEBs en WW~Vul.
Se necesita la combinaci\'on de un fuerte esfuerzo observacional y te\'orico para poder arrojar luz sobre este aspecto.
Por una parte, se deber\'{\i}a efectuar una ambiciosa campa\~na de observaciones espectrosc\'opicas en el visible, con una cobertura temporal amplia (unos siete d\'{\i}as) y densa (espectros cada varias horas), complementada con espectros en el rango UV lejano, que permitiesen analizar las l\'{\i}neas de resonancia met\'alicas y de Lyman.
Por otra parte, se tendr\'{\i}an que desarrollar modelos NLTE representativos de diferentes entornos gaseosos CS y diversas condiciones f\'{\i}sicas.
De esta manera se podr\'{\i}a efectuar un progreso real que proporcionara claves acerca del origen de estos eventos aparentemente ricos en metales, tanto de su estad\'{\i}stica de aparici\'on y desaparici\'on, como de su din\'amica, metalicidad y procedencia.

\item
Por \'ultimo, el conjunto de todos los datos y relaciones obtenidos en esta tesis supone un conjunto de restricciones observacionales, que deben ser tenidas en cuenta en los futuros modelos de discos de gas circunestelares en estrellas PMS de masa intermedia.

\end{enumerate}

\cleardoublepage

\addcontentsline{toc}{chapter}{Bibliography}
\bibliography{bibtex}
\cleardoublepage

\end{document}